\documentclass[11pt,a4paper,epsfig]{article}
\usepackage{color}
\usepackage{amsfonts}
\usepackage{epsfig}

\title{{\bf Lectures on the Mass of Topological Solitons}\\\vspace{2cm} {\it Heat kernel/Zeta function control of one-loop
divergences}}

\author{A. Alonso Izquierdo$^{(a)}$,
W. Garcia Fuertes$^{(b)}$, M.A. Gonzalez Leon$^{(a)}$ \\ M. de la
Torre Mayado$^{(c)}$, J. Mateos Guilarte$^{(d)}$, J.M.
Mu$\tilde{{\rm n}}$oz Casta$\tilde{{\rm n}}$eda$^{(e)}$
\\ {\normalsize {\it $^{(a)}$ Departamento de Matematica
Aplicada}, {\it Universidad de Salamanca, SPAIN}}\\{\normalsize {\it
$^{(b)}$ Departamento de Fisica},{\it Universidad de Oviedo,
SPAIN}}\\ {\normalsize {\it $^{(c)}$ Departamento de Fisica
Fundamental},{\it Universidad de Salamanca,
SPAIN}}\\{\normalsize{\it $^{(d)}$ Departamento de Fisica
Fundamental and IUFFyM}, {\it Universidad de Salamanca,
SPAIN}}\\{\normalsize {\it $^{(e)}$ Departamento de Fisica Teorica},
{\it Universidad de Zaragoza, Spain}}}
\date{03/09/07}

\begin{document}
\maketitle

\begin{abstract}

In this series of lectures a method is developed to compute one-loop
shifts to classical masses of kinks, multi-component kinks, and
self-dual vortices. Canonical quantization is used to show that the
mass shift induced by one-loop quantum fluctuations is the trace of
the square root of the differential operator governing these
fluctuations. Standard mathematical techniques are used to deal with
some powers of pseudo-differential operators. Ultraviolet
divergences are tamed by using generalized zeta function
regularization methods and, then performing zero-point energy and
mass renormalizations. Information about the meromorphic structure
of the generalized zeta function of the second-order fluctuation
operator $K$ around the classical solution is obtained from the
$K$-heat equation kernel via the Mellin transform. In particular,
the high-temperature expansion of the partition function provides
the residua at the poles of the generalized zeta function in terms
of the Seeley coefficients of the asymptotic approximation. In this
way a formula is derived  that allows computation of one-loop mass
shifts for kinks, multi-component kinks, and self-dual
Abrikosov-Nielsen-Olesen vortices. Numerical results for the Seeley
coefficients as well as the mass shifts, obtained by means of a
Mathematica environment implemented on a standard PC, are offered. A
qualitative analysis of the outcome shows a common trend in the mass
shift of the three types of topological defects analyzed. A
comparison with exact results is presented whenever possible, i.e.,
for the kink and the $TK1$ kink, respectively, of the
$\lambda\phi^4$ and $BNRT$ models. One-loop renormalization of the
planar Abelian Higgs model requires use of the Feynman-'t Hooft
renormalizable gauge, in the vacuum sector, or the background gauge,
in vortex sectors. Faddeev-Popov ghosts that restore unitarity are
dealt with in the Hamiltonian framework in a novel fashion.

\end{abstract}

\newpage
\tableofcontents

\clearpage

\section{Introduction}

\subsection{A brief history of soliton quantization}

We start this long Introduction by offering a short and biased
history of classical solitons and their quantization. The emphasis
will be oriented towards the topics to be analyzed in this set of
Lectures, skipping many important aspects of such a broad and
fertile subject.

\begin{itemize}

\item {\it Solitons and solitary waves}

Traditionally, wave phenomena in nature have been distinguished by
their dispersive character, i.e., the property by which propagating
waves eventually fade away in finite time. Fascination with the
soliton phenomenon started with the \lq\lq experimental" observation
of the Scottish engineer Scott-Russell circa 1870 in a Edinburgh
channel:\lq\lq {\it A solitary wave travels without changing its
shape, size, or, speed}", \cite{Scott}.

Linear wave equations only admit traveling or solitary wave
solutions if the dispersion law linking the frequency of the wave
motion with the wave vector of a \lq\lq monochromatic" component is
linear, because in such a case all the waves in a wave packet travel
with the same speed without interferences between them. PDEs of this
type are very rare but very well known: they are essentially
variations of the free-string and massless Dirac equations. Thus,
the impact of the discovery by Korteweg-de Vries around 1905 of
their non-linear PDE describing wave motion in shallow waters in
channels has been enormous. Besides providing a mechanism by which
the non-linearity balances the dispersive character of the KdV
equation, circa 1965 Kruskal, Miura, Lax and others showed that that
this magic equation can be completely solved despite its complexity.
Between the solutions of the KdV equation there are solitary waves
that keep their shape, height and speed during the propagation, thus
providing a theoretical explanation for Scott-Russell traveling
lumps of water. Due to complete integrability, KdV solitary wave
solutions not only keep their shapes in free propagation but also
survive collisions without damage, just as fundamental particles
survive scattering at not too high energies. For this reason, the
non-dispersive solutions of the KdV equation were christened as
solitons.

In fact, the solution of the KdV equation was the main impulse that
led to the creation and development of new ideas and techniques of
extraordinary importance in Mathematical Physics over the last fifty
years, such as the inverse scattering method (Kruskal/Miura), Lax
pairs and non-linear compatibility conditions (P. Lax), classical
spectral transforms (Sakharov), etcetera. Also, old and almost
forgotten methods such as the Backlund transformation (W. Lamb) or
highly sophisticated ideas of algebraic geometry (Novikov/Dubrovine)
found a new playground for application. Most remarkably, similar
unexpected properties were discovered in other non-linear PDE, such
as the non-linear Schrodinger equation and the sine-Gordon equation.
Amazingly, both equations govern the dynamics of real physical
systems (thus displaying the soliton phenomenon): the non-linear
Schrodinger equation governs some phenomena in non-linear optics;
the sine-Gordon equation (discovered in geometry) explains the
Josephson effect in semi-conductor physics, etcetera .

\item {\it Topological defects in condensed matter physics and cosmology}

In several of these systems and in other higher-dimensional
relatives, there are topological reasons for the strong stability of
solitary waves. By this statement we mean that non-linear PDE
equations of this type are sometimes the variational Euler-Lagrange
equations of some Lagrangian functional and, in fewer cases that are
essentially one-dimensional, also admit a Hamiltonian formulation
amenable to a sum of infinite angle-action variables. As a general
feature, the configuration space is the sum of several (frequently
infinite) topologically disconnected sub-spaces. Because temporal
evolution is a homotopy transformation, field configurations in
different topological sectors cannot evolve into each other. For
this reason, lumps arising as absolute minima of the energy in each
topological sector are called topological defects. In a more complex
physical system, in superconductors of Type II (some alloys below
the critical temperature) magnetic flux tubes were discovered by
Abrikosov circa 1957 and were understood by him to be topological
defects arising in the Ginzburg-Landau phenomenological theory of
superconductivity.

Similar topological defects forming tubes along a central line were
also discovered in liquid crystals and quantum fluids by other Nobel
laureates such as de Gennes and Legget, respectively in 1973 and
1978, who also found domain walls, point defects and textures in
these exotic materials. The topological and group theoretical roots
of these extended structures arising in nematic and cholesteric
liquid crystals or in phases A and B of helium 3 have been studied
in depth by Mermin, Michel and others.

More recently, Kibble and others, circa 1989, studied how Cosmology
would be affected by the existence of domain walls in the Universe
itself. Following this line of research, around 1990 Vilenkin and
Shellard proposed that possible effects of cosmic strings in stellar
and galactic formation and structure should be addressed.

\item {\it Classical/quantum lumps in field theory and elementary particle
physics}

The main theme where these highly stable lumps of energy will
attract our interest is quantum field theory. Many field theoretical
models at the heart of our present understanding of elementary
particles and their interactions have topological defects between
the solutions of their classical counterparts. Because hadrons,
particles interacting via strong subnuclear forces, are of two types
-heavy (baryons), and light (mesons)- it was tempting to think of
them respectively as quantum solitons and light quanta. This point
of view was pioneered by Skyrme and Finkelstein as early as the
sixties. The first author even proposed a variation that encompasses
solitons on the (at that time fashionable) Gell-Mann/Levy sigma
model of strong interactions . In the Skyrme model, the solitons,
usually referred to as Skyrmions, would describe the classical limit
of baryons whereas mesons were associated with light quanta.

Needless to say that a puzzling  question arose: what is the nature
of the quantum field states that are the descendants of classical
lumps? What do solitons look like in quantum field theory? The first
attempts to explore this territory concentrated on studying the
quantum $\lambda (\phi)^4_2$ and sine-Gordon kinks. In 1974
Dashen-Hasslacher-Neveu succeeded in computing the one-loop
correction to the classical mass of these solitary waves by
developing the $\hbar$-expansion of these (1+1)-dimensional field
theories. Moreover, in the second case, where periodic in time
soliton-antisoliton solutions (breathers) exist, DHN generalized the
Bohr-Sommerfeld quantization procedure to field theory, obtaining
the semi-classical spectrum of these new types of bound states. Two
years later, Comtet, Cahill, and Glauber provided a closed formula
for the expectation value of the normal ordered Hamiltonian in
quantum soliton states of these one-dimensional systems. The CCG
formula accounts for the bound states of the second-order
fluctuation operator around the classical kinks and exactly
reproduces the DHN results for static solitons.

Other techniques for the quantization of non-linear waves were soon
developed. To mention but a few: 1) Goldstone and Jackiw related the
semi-classical expansion to approximations working in molecular and
many-body physics. 2) Christ and Lee used a collective coordinates
method. 3) Cahill unveiled a variational/coherent state approach. 4)
Faddeev and Korepin profited from the fact that the sine-Gordon
equation is a completely integrable Hamiltonian system with an
infinite number of degrees of freedom to invent a completely new
field: Solving quantum infinite systems by means of the quantum
spectral transform. 5) Coleman, besides writing a priceless review
on the subject, showed that the quantum soliton of the sine-Gordon
theory was no more than the fundamental fermion of the massive
Thirring model. Two revolutions were sparked: a) Solitons, despite
arising in bosonic theories are fermions (like baryons). b)
Dualities between different models at different regimes of the
parameters exist. 6) Mandelstam discovered the (non-local) creation
operator of the sine-Gordon soliton.

In the midst of all this excitement, further fuel was added to the
fire by three new findings:

1) In 1973 Nielsen and Olesen rediscovered Abrikosov magnetic tubes
in a different system. The Abelian Higgs model supports topological
defects that are mathematically identical to Abrikosov vortices in a
relativistic context. Immediate interest in NO vortices was kindled
because they were thought of as field theoretical models of dual
strings, popular in those days in hadron physics.

2) Looking for a non-Abelian cousin of ANO vortices, also in 1994 't
Hooft at CERN and Polyakov in Russia independently found extended
objects in the Georgi-Glashow model. 't Hooft-Polyakov magnetic
monopoles are not tubes of magnetic flux but, instead, proper
solitons (or point defects); their energy density is localized
mainly in a finite 3D ball with exponentially decaying tails, except
for an Abelian long-range (${1\over r}$) potential, thus resembling
magnetic monopoles from afar.

Abelian ANO vortices have been shown to induce half-integer angular
momentum quantum numbers on an electrically charged particle and a
change in statistics (Wilczek), whereas 'tHP magnetic monopoles also
carry spin ${1\over 2}$ (Jackiw-Rebbi, 't Hooft-Hasenfratz) from a
spin/isospin mechanism.

3) In Moscow, in 1975 a Russian quartet -Belavin, Polyakov,
Schwartz, and Tyupkin- also discovered proper solitons (up to scale
invariance) in pure Yang-Mills gauge theory, without any interaction
with any kind of matter, in (1+4)-dimensions. Because there is no
physically sensible space-time of 5 dimensions, BPST solitons are
considered in 4-dimensional Euclidean space. In this context, the
fourth coordinate is understood as \lq\lq imaginary" time,
suggesting a change of name to BPST instantons and a different
physical r$\hat{\rm o}$le: being classical minima of the Euclidean
Yang-Mills action, instantons dominate the semi-classical expansion
of the Euclidean YM integral functional. These topological solutions
thus provide the leading approximation to the tunnel effect
amplitude between classical vacua and build the YM vacuum as a Bloch
wave.

Therefore, topological defects dress different physical disguises in
different dimensions of the space-time in which they live. This,
which determines when a given topological defect is a domain wall
(surface defect), string (line defect), particle (point defect), or
texture (instanton), lies at the core of the p-brane scan of
Townsend.

\item {\it Multicomponent kinks}

Advances in the study of multi-component kinks/solitary waves/domain
walls have been achieved over the past thirty years. Derrick's
theorem forbids the existence of soliton-like solutions in scalar
field theories with (1+d)-dimensional space-times if $d>1$. There
are no obstructions, however, to the existence of kinks in theories
with $N$ interacting scalar fields, provided that the space-time is
the ${\mathbb R}^{1,1}$ two-dimensional Minkowski space.

 In 1976 Montonen, and independently Sarker,Trullinger, and Bishop
 proposed a model with two real scalar fields and field interactions such
 that the old $\lambda\phi^4$ kink belongs to the space of static
 solutions of finite energy of this field theoretical model. There
 was however an important novelty: another kink was found, such that the two
 components of the field profile were not zero. To distinguish between the
 two kinds of solitary waves, the old kinks were denoted as $TK1$ - one-component topological - kinks
 whereas the new kinks were referred to as $TK2$ - two-component topological - kinks. Rapidly, Rajaraman and
 Weinberg, using the so called trial orbit method, identified a
 special member of a third class of $NTK2$. The whole manifold of
 non-topological two-component - $NTK2$ - kinks was identified slightly later
 by numerical integration, but the deep reason for their existence was
 unveiled by Magyari and Thomas, who showed that the system of two
 ODEs
 to be solved in the search for kinks is a two-dimensional integrable
 mechanical system: the Garnier system discovered in 1915. The Garnier system is not only integrable but
 Hamilton-Jacobi separable, and Ito took profit from this fact to
 analytically calculate all the kink solutions of the so-called MSTB
 model. For more than one scalar field, simple topological arguments do not
 ensure lump
 stability, but Ito and Tasaki classified
 stable and non-stable kinks by using the sophisticated Morse index theorem.
 Overlooking the difficulty of controlling the spectrum of the second
-order fluctuation operator (Hessian), in this case a $2\times 2$
 matrix Schrodinger operator, one of us (JMG) developed the complete
 Morse theory of the configuration space of the MSTB
 model \'{a} la Bott . Models in the same class as the MSTB model were addressed by
 the AAI, MAGL, JMG trio at the turn of the last century. The kink
 varieties were identified and their stability was unveiled in a series
 of papers.

(1+1)-dimensional models of a complex scalar field with potential
energy equal to the square of the norm of the gradient of a
holomorphic function are very interesting because of the possibility
of supersymmetric extensions; in fact, these models are obtained by
dimensional reduction of ${\cal N}=1$ supersymmetric models of one
chiral superfield. Vafa et al, in 1989, thoroughly studied all the
stable kinks arising in these models, whereas Townsend analyzed the
balance between kink masses. Again, the AAI, MAGL, JMG trio explored
the same system by taking a real analytic point of view. Another
interesting model, in this case coming from the dimensional
reduction of a ${\cal N}=1$ supersymmetric theory of two chiral
superfields, was addressed by Bazeia, Nascimento, Ribeiro, and
Toledo -henceforth the BNRT model- in 1995. The kink equations are
not completely integrable but, Shifman and Voloshin found a complete
family of $TK2$ kink solutions whereas Bazeia and collaborators
studied the stability properties. Although the analogous mechanical
system is not completely integrable, some of us discovered that for
some values of the mass of the second boson integrability holds and
all the kinks can be found.

\item {\it Recent advances in soliton quantization}

In 1994, the spectacular solution of ${\cal N}=2$ supersymmetric
Yang-Mills theory by Seiberg-Witten provided, as an aside, an exact
formula for the quantum mass of BPS monopoles. The same result in
the low energy domain was derived ten years later, using more
down-to-earth methods, by the Stony Brook/Wien group formed by
Rebhan, van Nieuwenhuizen, and Wimmer. This work followed previous
investigations by the same team, together with Goldhaber, about
computations of mass shifts induced by one-loop fluctuations on
supersymmetric kinks. The extreme elusiveness of this issue did not
prevent these authors from identifying the old DHN formula as being
based on a regularization method that sets a cutoff in the number of
fluctuation modes to be counted, rather than the conventional energy
cutoff. Another group from Minnesota University addressed the same
problem by using high-derivative regularization, with SUSY being
preserved by boundary conditions to find similar results.
Phase-shift analysis by an MIT group -Jaffe, Graham, and
collaborators- also led to some advances, in this case in a purely
bosonic setting. Finally, in 2003 van Nieuwenhuizen, Rebhan, and
Wimmer, and independently Vassilevich, succeeded in computing the
one-loop mass shift to the mass of the supersymmetric Abelian
vortex.

\end{itemize}

\subsection{A brief history of heat kernel/zeta function regularization methods}

\begin{itemize}

\item {\it Zeta function regularization and the heat kernel expansion}

The method of zeta function regularization was invented by Dowker
and Critchley and, independently by Hawking, circa 1976.
Implementation of standard regularization/renormalization procedures
in Quantum Field Theory on curved space-time backgrounds led to the
introduction of this regularization method as the best suited
technique to combine second quantization phenomena with general
relativity. Vacuum expectation values of spatial integrals of the
energy-momentum tensor are essentially given by the trace of the
square root of some differential operator of Laplace type. Simili
modo, the partition function of Euclidean quantum field theories is
a functional integral that, up to one-loop order in the
$\hbar$-expansion, is the inverse of the square root of the
determinant of another differential operator of Laplace type times
the exponential of the Euclidean action over $\hbar$.

Traces and determinants of powers of elliptic operators can only be
defined by means of a process of analytic continuation that mimics
the definition of the Riemann zeta function as a meromorphic
function, giving formal meaning to strictly divergent series in some
region (${\rm Re}\, {\rm s}<1$) of the complex plane. By replacing
natural numbers by eigenvalues (hopefully forming a discrete
spectrum), generalized zeta functions associated with differential
operators are defined. There is a general theory of elliptic
pseudo-differential operators that characterizes the conditions
under which the generalized zeta functions are meromorphic
functions, and values away from poles of the zeta function, and
derivatives of zeta, are taken as \lq\lq regularized" definitions of
traces and logarithms of determinants of (complex powers of
pseudo-)differential operators.

In interesting physical, cases the pertinent differential operators
are those ruling small quantum fluctuations in gravitational,
Yang-Mills, or solitonic classical backgrounds. Generically, the
spectral information in these situations is grossly insufficient for
identifying the generalized zeta function in terms of known spectral
functions. Fortunately, B. and C. de Witt had already proposed, in
the mid sixties, use of the high-temperature expansion of the kernel
of the generalized heat equation provided by the differential
operator of Laplace type to unveil the meromorphic structure of the
generalized zeta function. To achieve this goal, one takes advantage
of the link between generalized heat and zeta functions via Mellin
transforms, such that the residua at the poles of the generalized
zeta function are proportional to the Seeley coefficients of the
heat kernel expansion.

\item {\it Heat kernel proof of index theorems}

It is remarkable that almost simultaneously, starting around 1970,
parallel ideas were applied by Atiyah, Patodi, and Bott to construct
the heat kernel proof of the Atiyah-Singer index theorem. The
context was mathematically much more precise, considering the index
of the Dirac operator acting on sections of spin bundles tensored
with vector bundles on compact spin manifolds with or without
boundary. The theorem identifies the index of an elliptic operator
with some characteristic classes of the base manifold: typically the
A-genus times the Chern character.

In contrast to physical situations, generalized zeta functions are
well defined because the spectrum of elliptic operators on spaces of
sections in bundles with a compact manifold without boundary as the
base space is discrete. On open spaces, characteristic of physical
problems, one must impose a rapidly decaying behavior (exponential)
at infinity in such a way that the elliptic operator will act on
$L^2$ spaces of functions. Alternatively, one could consider the
same problem for manifolds with boundary with spectral boundary
conditions \'{a} la Atiyah-Patodi-Singer and allow the boundary to
go to infinity to recover the usual situation in physical problems.

\item {\it Generalized zeta functions and heat equation kernels in
physics}

On the physical side, heat kernels and zeta functions have proved to
be of great use in the analysis of gravitational and gauge anomalies
arising in the one-loop approximation to the effective action. The
computation of quantum effects around gravitational, Yang-Mills, or
other classical backgrounds using heat kernel/zeta function
regularization has been a important theme in theoretical physics
over the last thirty years. Although many researchers have
contributed to these developments, we particularly mention the
Leipzig/Barcelona group of Bordag, Elizalde, Kirsten, Vassilevich
and collaborators.

In particular Bordag and Vassilevich, together with two members of
the Stony Brook/Viena group, van Niewenhuizen and Goldhaber, used
these techniques to compute the one-loop mass shift to
supersymmetric kinks. Starting almost at the same time, we applied
heat kernel/zeta function methods to re-work one-loop mass shifts
for $\lambda\phi^4$ and sine-Gordon kinks in a purely bosonic
setting. Having established the method, we succeeded in computing
mass shifts for other kinks in models with a single real scalar kink
and non-Posch-Teller Schrodinger operators governing quadratic
fluctuations. Moreover, the generalization to models with several
scalar fields having multi-component kinks was reported by us in a
series of papers. The generalization for dealing with matrix
differential operators was the key step that allowed us to compute
the one-loop mass shift for Abelian self-dual vortices.

\end{itemize}

\subsection{Chart of aims}

Research on the quantum descendants of classical topological defects
can be classified within two broad areas, although with important
(1+1)-dimensional exceptions.
\begin{enumerate}

\item  In ordinary field theories, the most effective approach is to develop semi-classical analyses
or $\hbar$-expansions around the classical soliton solutions,
generalizing the old WKB approximation method of quantum mechanics
to quantum fields. This strategy has so far been fully successful in
computing one-loop corrections to classical observables only for
sine-Gordon and $\lambda\phi^4$ kinks and sine-Gordon
multi-solitons.

\item After the seminal paper of Olive and Witten identifying solitons as BPS states in theories with extended
supersymmetry,  taking advantage of this fact much more detailed
information on quantum corrections to supersymmetric solitons has
been acquired. In parallel, the conventional semi-classical
expansion has been used to estimate the mass shift for SUSY kinks,
vortices and magnetic monopoles, although great care is needed in
combining supersymmetry with suitable boundary conditions.

\item In integrable (1+1)-dimensional field theories such as the
sine-Gordon system, full information on quantum solitons is
available due to the existence of an infinite number of conserved
charges. Also in this case, the identification of solitons as
coherent states is enlightening because normal ordering is
sufficient to achieve full renormalization, and the expectation
values of operators in coherent states behave as their classical
counter-parts.

\end{enumerate}

Our goal in this set of lectures is to develop semi-classical (weak
coupling approximation) analyses for multi-component kinks, arising
in multi-scalar field theory, and Abrikosov-Nielsen-Olesen vortices,
arising in the Abelian Higgs model. The one-loop mass shift is
essentially the trace of the square root of the second-order
fluctuation operator (Hessian), modulo some (infinite)
renormalizations. Because the spectrum of the Hessian is generally
unknown in these cases, we are forced to use asymptotic techniques
to deal with the generalized zeta function of these second-order
matrix differential operators. We shall describe our method as
applied to multicomponent kinks in Sections \S. 5 and 6, whereas a
conceptually identical but much more technically complex procedure
is developed in Sections \S. 7 and 8 to compute the one loop mass
shift of ANO vortices. To explain all the subtleties of our approach
in as simple a context as possible, in Sections \S. 3 and 4 we fully
address the problem of computing the (very well known) one-loop mass
shift of the $\lambda\phi^4$ kink. As a bonus, comparison with
solidly established results obtained by other procedures will
provide a precision test for our method.

In Section \S. 2 we offer a summary of heat equation kernels,
asymptotic (high-temperature) expansions, and generalized zeta
functions for a very broad class of differential operators of the
type that we are going to handle. The connection of these concepts
and techniques with the formulas arising in our physical
calculations is explained in Appendices II, III, and IV.

\subsection{One-loop quantum corrections to soliton masses and the Casimir effect}

The problem of computing quantum corrections to the mass of
topological defects is closely related to the Casimir effect. The
field profile distorts the spectrum of quantum fluctuations around
the ground state in a similar manner to the plates of a capacitor in
a vacuum. The Casimir effect measures the quantum energy of the
vacuum when two plates are present with respect to the same quantity
without plates. The quantum correction to the mass of a topological
defect measures the quantum energy of the topological defect in its
ground state with respect to the quantum vacuum energy. These
problems lie at the heart of the conceptual foundations of quantum
mechanics: there is nothing more quantum mechanical than the
non-zero energy of nothing!. Throughout these lecture notes we shall
refer to such things as kink Casimir or vortex Casimir energies by
analogy with the quantum energy of the Casimir set-up. To justify
such an abuse of language, we include Appendix I to describe the
Casimir effect.

\subsection{Note on the bibliography}

We shall present the bibliographical References in a global and
non-detailed way, except in the cases where specific and new results
are discussed. It is understood that standard books and monographs
contain precise bibliographical information. Also, recent References
are chosen insofar that they have been used in the elaboration of
these Lectures.

\begin{itemize}

\item {\it Classical papers, lectures and treatises on solitons}

Important classical papers on the foundations of the matter are:
\cite{Dashen}, \cite{Kor}, and \cite{Glauber}. A complete collection
of timely mid-seventies works about soliton quantization and
semi-classical methods can be found in Reference \cite{Man}. Seminal
lectures on classical lumps and their quantum descendants are those
of Sidney Coleman in Jaca and Erice, see \cite{Coleman}. In
Reference \cite{BCM} a review is offered emphasizing the homotopical
nature of topological solitons. The earlier monographic books are
those of Rajaraman \cite{Rajaraman} and Drazin \cite{Drazin}. The
Rajaraman treatise aims to address both the physically and
mathematically relevant aspects of extended states in quantum field
theory. The Drazin goal is rather mathematical; i.e., the
application of techniques of integrable systems as the inverse
scattering method to find soliton and multi-soliton solutions. An
important book on vortices and monopoles of the highest mathematical
rigor is monograph \cite{jatb}. More recent treatises such as the
books by Vilenkin/Shellard or Manton/Sutcliffe thoroughly address
the issues, with emphasis on Cosmology in the former, and
Mathematical Physics in the latter. We also mention the earlier
papers on Abrikosov-Nielsen-Olesen vortices \cite{Abrikosov} and
\cite{Nielsen} because (quantization of) these extended objects is
the main concern of these Lectures.

\item {\it Bibliography on generalized zeta functions and heat
kernel methods}

The zeta function regularization method started with papers
\cite{Dowker} and \cite{Hawking} in response to the need to compute
quantum effects on curved backgrounds. Even before it was used as a
regulator of a physical observable, the generalized zeta function
arose as Mellin transforms of heat kernels, giving particle
propagation in curved spaces. Comparison with particle propagators
in Euclidean time led B. de Witt, \cite{DeWitt}, to study the
asymptotic high-temperature (short-time) expansion of the heat
kernel. See also \cite{Stone} for a modern review on this important
mathematical tool. Over the last thirty years this broader field,
the physical applications  of heat kernel expansions and generalized
zeta functions, has become  one of the most important subjects of
Mathematical Physics. Good References are the monographs
\cite{Elizalde}, \cite{Kirsten}, and \cite{Vassilevich}. On the
mathematical front, we choose \cite{Gilkey} and \cite{Roe} as our
favorite References. The link between the coefficients of the
high-temperature heat kernel expansion and the Korteweg-de Vries
conserved charges is explained, for example, in \cite{Perelomov}.

\item {\it The 1976-1989 period}

This period started with two papers addressing the quantization of
supersymmetric kinks \cite{Schonfeld}, \cite{Adda} whereas the
seminal paper of Olive and Witten \cite{Olive} recognized the link
between BPS solitons and extended supersymmetry. Also, the issue of
kink quantization was addressed in the paper \cite{O'Brien},
although in this case it was applied to the exotic $\lambda\phi^6$
kink discovered in \cite{Lohe}. Important advances in our analytical
knowledge of vortex and multi-vortex scalar and vector field
profiles were achieved in papers \cite{de Vega}, \cite{Jacobs}, and
\cite{Weinberg}. In a almost simultaneous development, the MSTB
model was introduced in papers \cite{Montonen} and \cite{Sarker}.
This model is a system of two one-dimensional scalar fields having a
rich variety of two-component kinks that was first investigated in
\cite{Rajaraman1} using the trial orbit method. The integrability of
the kink equations was unveiled in \cite{Magyari}, although this
fact was not fully exploited until Ito showed that the mechanical
system is Hamilton-Jacobi separable \cite{Ito}. The kink stability
issue was elucidated in \cite{Tasaki} by applying the Morse index
theorem, and one of us developed the full Morse theory of this
problem in \cite{Guilarte} and \cite{Guilarte1}.

\item {\it Recent papers on multi-component kinks}

Over the last decade  many works have been devoted to investigating
kink or solitary wave solutions in systems, supersymmetric or not,
with two or more scalar fields. It is important to mention this
research because computation of one-loop mass shifts for
multi-component kinks was the intermediate landmark that allowed us
to fulfil the same task for self-dual vortices. Besides the MSTB
model, which is not discussed in these Lectures, another interesting
field theoretical model with two real scalar fields was first
described in \cite{Bazeia} and \cite{Bazeia1}. One-component and
two-component stable kinks with the same energy were discovered very
soon after. Shifman and Voloshin found in \cite{Voloshin} that these
kinks belonged to a continuous family, all of them degenerate in
energy, and hence stable kinks. The AAI, MAGL, JMG trio discovered
that, for special values of the second boson mass, the static
equations are fully integrable and the whole kink variety was
studied in \cite{Gonzalez}. The trick is to realize that the
mechanical problem is Hamilton-Jacobi separable in either Cartesian
or parabolic coordinates when the pseudo-Goldstone boson mass is
either $2m$ or ${m\over 2}$, as was shown in \cite{Alonso}. Other
systems with two scalar fields have been considered, for instance in
\cite{Bazeia2}, where kink solutions are discussed in either planar
or cylindrical Minkowskian space-time. References analyzing kink
solutions in models with three scalar fields are \cite{Gonzalez1}
and \cite{Bazeia3}. In the case of systems of a complex scalar
field, holomorphic superpotentials are naturally connected with
extended supersymmetry and automatically provide ${\cal N}=2$ BPS
kinks, see \cite{Alonso1}, \cite{Fendley}, and \cite{Cecotti}, the
latter reference offering a thorough analysis of this topic. A
recent review dealing with these developments and other interesting
soliton phenomena is \cite{Bazeia5}.

\item {\it Recent papers on soliton quantization}

In the second half of the nineties, much attention was drawn to the
study of Casimir energies in different geometries, see e.g.
\cite{Bordag}, \cite{Bordag1}, and \cite{Kirsten1}, a problem close
to computing kink ground-state energies. The issue of quantum
corrections to SUSY kinks was revisited from different viewpoints in
\cite{Rebhan}, \cite{Shifman1}, \cite{Rebhan1}, \cite{Graham}, and
\cite{Graham1}. A deeper understanding of the several different
regularization methods used became available after the work of the
Stony Brook/Wien, Minnesota, and M.I.T. groups, see also
\cite{Wimmer}. Another regularization method was applied to the SUSY
kink by a Stony Brook/Leipzig collaboration based on heat
kernel/zeta function methods in \cite{Rebhan4}. Almost at the same
time, several of us applied heat kernel/zeta function technology to
calculate one-loop mass shifts to the mass of many one-component
purely bosonic kinks in the sine-Gordon, $\lambda\phi^4$, a
variation of the sinh-Gordon, and $\lambda\phi^6$ models, see
\cite{Guilarte2}. This work was followed by similar calculations
applied to one-component and two-component kinks in the MSTB and
BNRT models in \cite{Guilarte3}, and \cite{Guilarte4}. After a paper
by Bordag on the fermionic vacuum energy in a vortex background,
\cite{Bordag3}, the one-loop mass shift to the SUSY vortex was
calculated in \cite{Vassilevich1} and \cite{Rebhan4}. In both
papers, \cite{Guilarte5} and \cite{Guilarte6}, we were able to
compute the same quantity for purely bosonic self-dual vortices. The
one-loop renormalization program in the Abelian Higgs model can be
found in Reference \cite{Casta} and is suitable for the goals
addressed in this work. In \cite{Guilarte7} a more or less unified
formula is offered, giving the one-loop mass shift of kinks and
self-dual vortices as a truncated series involving the Seeley
coefficients starting from the second one. Although the
Seiberg-Witten solution of ${\cal N}=2$ SUSY Yang-Mills allows us to
know the mass of quantum BPS states in any energy regime, the recent
interesting papers \cite{Rebhan3} and \cite{Rebhan5} provide a more
detailed knowledge of one-loop mass shifts of ${\cal N}=2$ SUSY
monopoles.

\end{itemize}

\subsection{Note on units and dimensions}

Throughout this work, we shall use a system of units where the speed
of light is the unit of velocity: $c=1$. The Planck constant,
however, will be kept explicit because we shall perform
semi-classical computations. Thus, the dimension of $\hbar$ is
$[\hbar]=ML$, mass $\times$ length. These are also the dimensions of
the Boltzman constant $[k_B]=ML$, whereas particle masses and
temperature have dimensions of inverse length: $[m]=[T]=L^{-1}$.

\subsection{Brazil lectures}

This work is a written outgrowth of a series of three two hour
Lectures given by one of us, J. M. G., at the Physics Department of
Paraiba University in Joao Pessoa (Brazil) during the third week of
July 2005. The material presented at each of those Lectures is
contained respectively in Sections \S. 3-4, \S. 5-6, and \S. 7-8 and
readers wishing to become acquainted with the physical aspects of
semi-classical soliton mass shifts can skip reading the rest. We
have sketched some brief historical notes in the Introduction to
place the matter in perspective, at the request of Roberto Menezes,
without pretensions of completeness or high precision. We also
include a Section, \S. 2, where heat equations, heat kernel
expansions, and generalized zeta functions are discussed at a higher
level of Mathematical rigor. Appendices II, III, and IV are included
to establish contact between the spectral functions described in
Section \S. 2 and the physicist's version of the same functions used
in the core of the Lectures.

\section{Generalized zeta functions and heat equation kernels}

Let us focus on elliptic operators of the general form:
\[
K=K_0+\vec{Q}(\vec{x})\cdot \vec{\nabla} + V(\vec{x}) \qquad ,
\qquad K_0=(-\bigtriangleup+c^2)\cdot {\mathbb I} \qquad , \qquad
\lim_{|\vec{x}|\rightarrow \infty}V(\vec{x})=0
\]
acting on the Hilbert space of functions ${\cal H}=\oplus_{A=1}^N
L^2_A({\mathbb T}^d)$. Here: (a) ${\mathbb T}^d$ is a toroidal
variety, the direct product of d ${\mathbb S}^1$ circles of radius
$R=\frac{mL}{2\pi}$. (b) ${\mathbb I}$ is the $N\times N$ unit
matrix. (c) $V(\vec{x}): {\mathbb T}^d\rightarrow {\mathbb
Mat}_{{\mathbb R}}(N)$ is a map from ${\mathbb T}^d$ to the set of
$N\times N$ matrices with real coefficients. (d)
$\vec{Q}(\vec{x})\cdot \vec{\nabla}: {\mathbb T}^d\rightarrow
T({\mathbb T}^d)\otimes{\mathbb Mat}_{{\mathbb R}}(N)$ is a map from ${\mathbb T}^d$ to
the tensor product of the tangent space to
${\mathbb T}^d$ times ${\mathbb Mat}_{{\mathbb R}}(N)$. (e)
$\vec{\nabla}$ and $\bigtriangleup=\vec{\nabla}\cdot\vec{\nabla}$
are respectively the gradient and Laplacian operators in ${\mathbb
T}^d$. (f) $c^2$ is a constant. Assuming that the spectrum of $K$
is definite positive,
\[
K \, f_n(\vec{x})= \lambda_n \, f_n(\vec{x}) \qquad , \qquad
\lambda_n \in {\mathbb R}> 0 \qquad ,
\]
the generalized zeta function associated to $K$ is defined as:
\[
\zeta_K(s)={\rm Tr}\, K^{-s}=\sum_{{\rm Spec}\, K} \,\,
{1\over\lambda_n^s} \qquad , \qquad s\in{\mathbb C} \qquad ,
\]
where $s$ is a complex parameter. Via the Mellin transform
\[
\zeta_K(s)={1\over\Gamma(s)}\cdot \int_0^\infty \, d\beta \, \beta^{s-1} \,
{\rm Tr} \, e^{-\beta K}={1\over\Gamma(s)}\cdot \sum_{{\rm Spec}\,
K} \, \int_0^\infty \, d\beta \, \beta^{s-1} \, e^{-\beta \lambda_n}
\]
the generalized zeta function is related to the partition (heat)
function $h(K)={\rm Tr}\, e^{-\beta K}$ of the generalized heat
equation:
\[
\sum_{B=1}^N\left({\partial\over\partial\beta}\cdot
\delta^{AB}+K^{AB}\right)F^B(\vec{x},\beta)=0^A \qquad \qquad ,
\qquad \qquad \beta=\frac{\hbar m}{k_B T} \qquad .
\]
The partition function is the integral of the $K$-heat equation
kernel on the diagonal sub-space of ${\mathbb T}^d\times{\mathbb
T}^d$:
\[
{\rm Tr}\, e^{-\beta K}={\rm tr} \,\int \, d{\rm vol}_{{\mathbb
T}^d} \, K_K(\vec{x},\vec{x};\beta)=\sum_{A=1}^N\sum_{B=1}^N
\,\delta^{AB} \, \int \, d{\rm vol}_{{\mathbb T}^d} \,
K_{K}^{BA}(\vec{x},\vec{x};\beta) \qquad ,
\]
whereas the kernel itself is the solution of the $K$-heat equation
\begin{equation}
\sum_{C=1}^N \, \left({\partial\over\partial\beta}\cdot
\delta^{AC}+K^{AC}\right)K_{K}^{CB}(\vec{x},\vec{y};\beta)=0^{AB}
\qquad , \qquad K_{K}^{AB}(\vec{x},\vec{y};0)=\delta^{AB}\cdot
\delta^{(d)}(\vec{y}-\vec{x})\label{eq:Khe}
\end{equation}
with unit source at infinite temperature.

\subsection{Heat kernel and generalized zeta function for Klein-Gordon operators}

The spectrum of $K_0$ - an $N\times N$ diagonal matrix of
$d$-dimensional Klein-Gordon operators-
\begin{eqnarray*}
&& \sum_{B=1}^N \, K_0^{AB} \cdot {\rm
exp}\{i\frac{\vec{n}^{(B)}\cdot\vec{x}}{R}\}\cdot
u^B=\lambda_n^{(A)}\cdot {\rm
exp}\{i\frac{\vec{n}^{(A)}\cdot\vec{x}}{R}\}\cdot u^A \qquad ,
\qquad \lambda_n^{(A)}=\frac{\vec{n}^{(A)}\cdot\vec{n}^{(A)}}{R^2}+c^2 \\
&& \vec{n}^{(A)}=\sum_{k=1}^d \, n_k^{(A)}\cdot\vec{e}_k \qquad ,
\qquad \vec{e}_k\cdot\vec{e}_j=\delta_{kj} \qquad , \qquad
n_k^{(A)}\in{\mathbb Z} \qquad , \qquad u^A\cdot u^B=\delta^{AB}
\end{eqnarray*}
provides the spectral resolution of the $K_0$-heat kernel
\begin{eqnarray*}
K_{K_0}^{AB}(\vec{x},\vec{y};\beta)&=&\delta^{AB}\cdot\sum_{{\rm
Spec}K_0^{AA}}\,\, {\rm
exp}\left\{-\beta(\frac{\vec{n}^{(A)}\cdot\vec{n}^{(A)}}{R^2}+c^2)\right\}\cdot{\rm
exp}\left\{i\frac{\vec{n}^{(A)}\cdot(\vec{x}-\vec{y})}{R}\right\}\\&=&
\delta^{AB}\cdot e^{-c^2\beta}\cdot{\rm
exp}\left\{-\frac{|\vec{x}-\vec{y}|^2}{4\beta}\right\}\cdot\sum_{{\rm
Spec}\, K_0^{AA}}\,{\rm exp}\left\{-{\beta\over
R^2}|\vec{n}^{(A)}+i{R\over
2\beta}(\vec{y}-\vec{x})|^2\right\}\qquad ,
\end{eqnarray*}
and the Poisson summation formula
\[
\sum_{\vec{n}^{(A)}\in{\mathbb Z}^d} \, {\rm
exp}\left\{-t|\vec{n}^{(A)}+\vec{v}|^2\right\}=\left({\pi\over
t}\right)^\frac{d}{2} \cdot \sum_{\vec{l}^{(A)}\in{\mathbb Z}^d}\,
{\rm exp}\left\{-{\pi^2 \vec{l}^{(A)}\cdot\vec{l}^{(A)}\over
t}-2\pi i\vec{l}^{(A)}\cdot\vec{v}\right\} \, \, ; \, \,
t={\beta\over R^2} \quad , \quad \vec{v}=i{R\over 2\beta}
(\vec{y}-\vec{x})
\]
leads to the formula:
\[
K_{K_0}^{AB}(\vec{x},\vec{y};\beta)=\delta^{AB}\cdot
e^{-c^2\beta}\, \cdot \, \left({\pi R^2\over\beta}\right)^{{d\over
2}}\, \cdot \, {\rm
exp}\left\{-\frac{|\vec{y}-\vec{x}|^2}{4\beta}\right\}\,
 \cdot \, \sum_{\vec{l}^{(A)}\in{\mathbb Z}^d} \, {\rm exp}\left\{-\frac{\pi R\vec{l}^{(A)} \cdot [\pi
R\vec{l}^{(A)}-(\vec{y}-\vec{x})]}{\beta}\right\} \qquad .
\]
On the other hand, the generalized zeta function is:
\[
\zeta_{K_0}(s)={\rm Tr}\, \left[-(\bigtriangleup+c^2)\cdot{\mathbb
I}\right]=\sum_{A=1}^N\, \sum_{\vec{n}^{(A)}\in{\mathbb Z}^d}\,
\frac{1}{[\frac{\vec{n}^{(A)}\cdot\vec{n}^{(A)}}{R^2}+c^2]^s}=\sum_{A=1}^N
\, E(s,c^2|\sum_{k=1}^d {1\over R^2}\vec{e}_k\, u^A\,) \qquad .
\]
Via the Mellin transform, the Epstein zeta function can be written
in the form:
\[
E(s,c^2|\sum_{k=1}^d \, {1\over R^2}\vec{e}_k \, u^A \,)=
{1\over\Gamma(s)}\cdot \sum_{\vec{n}^{(A)}\in{\mathbb Z}^d}\,
\int_0^\infty \, d\beta \, \beta^{s-1}\cdot{\rm
exp}\{-\beta(\frac{\vec{n}^{(A)}\cdot\vec{n}^{(A)}}{R^2}+c^2)\}
\qquad .
\]
Again the Poisson summation formula
\[
\sum_{\vec{n}^{(A)}\in{\mathbb Z}^d}\,\, {\rm exp}\{-{\beta\over
R^2}\cdot\, (\vec{n}^{(A)}\cdot\vec{n}^{(A)})\}=\left({\pi
R^2\over\beta}\right)^{d\over 2}\cdot
\sum_{\vec{l}^{(A)}\in{\mathbb Z}^d} \, \, {\rm exp}\{-\frac{\pi^2
R^2}{\beta}\cdot\, (\vec{l}^{(A)}\cdot\vec{l}^{(A)})\}
\]
allows us to write the Epstein zeta function in terms of the
integral representation of Kelvin functions:
\[
K_{-\nu}(z)={1\over 2}\cdot\left({z\over 2}\right)^{-\nu} \cdot \,
\int_0^\infty \, dt \, t^{\nu-1}\, e^{-t-{z^2\over 4t}}
\]
\begin{eqnarray*}
&&E(s,c^2|\sum_{k=1}^d \, {1\over R^2}\vec{e}_k \, u^A
\,)=\\&=&{\pi^{d\over 2}R^d\over\Gamma(s)}\cdot   \int_0^\infty \,
d\beta \, \beta^{s-\frac{d+2}{2}} \, e^{-\beta c^2}+ {\pi^{d\over
2}R^d\over\Gamma(s)}\,  \cdot \sum_{\vec{l}^{(A)}\in{\mathbb
Z}^d-\{\vec{0}\}} \, \int_0^\infty \, d\beta \,
\beta^{s-\frac{d+2}{2}} \, e^{-\beta c^2}\, {\rm
exp}\{-{\pi^2 R^2\over\beta}\cdot (\vec{l}^{(A)}\cdot\vec{l}^{(A)})\}\\
&=& \pi^{d\over 2}R^dc^{d-2s}\cdot\frac{\Gamma(s-{d\over
2})}{\Gamma(s)}+ \frac{2\pi^s c^{d-{s\over 2}}R^{s+{d\over
2}}}{\Gamma(s)}\cdot \sum_{\vec{l}^{(A)}\in{\mathbb
Z}^d/\vec{0}}\, (\vec{l}^{(A)}\cdot\vec{l}^{(A)})^{{1\over
2}(s-{d\over 2})} \cdot K_{{d\over 2}-s}\left(2\pi
cR(\vec{l}^{(A)}\cdot\vec{l}^{(A)})\right)\qquad .
\end{eqnarray*}
At the infinite volume $R\rightarrow\infty$ limit, only the first
term survives:
\[
\lim_{R\rightarrow\infty} \, E(s,c^2|\sum_{k=1}^d \, {1\over
R^2}\vec{e}_k \, u^A \,)=\pi^{d\over
2}R^dc^{d-2s}\cdot\frac{\Gamma(s-{d\over 2})}{\Gamma(s)} \qquad .
\]

\subsection{High-temperature (asymptotic) expansion of the K-heat equation kernel}

In fact, not only when $R\rightarrow\infty$ do Kelvin integrals go
to zero but, also, $K_{-\nu}(z)$  becomes negligible when
$\beta\rightarrow 0$, i.e., at the high-temperature limit:
\[
K_{K_0}^{AB}(\vec{x},\vec{y};\beta)=\delta^{AB}\cdot
e^{-c^2\beta}\, \cdot \, \left({\pi R^2\over\beta}\right)^{{d\over
2}}\, \cdot \, {\rm
exp}\left\{-\frac{|\vec{y}-\vec{x}|^2}{4\beta}\right\}\cdot\{1+{\cal
O}(e^{-{C\over\beta}})\}
\]
shows the asymptotic behavior of the \lq\lq free" heat kernel.

To find the $K$-heat kernel, we plug in the ansatz
\[
K_{K}^{AB}(\vec{x},\vec{y};\beta)=\sum_{C=1}^N \,
C^{AC}_K(\vec{x},\vec{y};\beta)\cdot
K_{K_0}^{CB}(\vec{x},\vec{y};\beta)
\]
in equation (\ref{eq:Khe}), leading to:
\begin{eqnarray}
&&\sum_{C=1}^N\, \left\{\frac{\partial}{\partial\beta}\cdot
\delta^{AC}+\frac{x_k-y_k}{\beta}\cdot\left(\delta^{AC}\partial_k-{1\over
2}Q_k^{AC}(\vec{x})\right)-\right.\nonumber
\\&-&\left.\delta^{AC}\cdot\bigtriangleup+Q_k^{AC}(\vec{x}).\partial_k+V^{AC}(\vec{x})\right\}\cdot
C^{CB}_K(\vec{x},\vec{y};\beta)=0^{AB} \qquad ; \qquad
C^{AB}_K(\vec{x},\vec{y};0)=\delta^{AB}\,\,\, \quad \,
\label{eq:Khe1}\,
\end{eqnarray}
In the high-temperature limit one can write the heat kernel as the
asymptotic series:
\begin{eqnarray}
K^{AB}_K(\vec{x},\vec{y};\beta)&=&  e^{-c^2\beta}\, \cdot \,
\left({\pi R^2\over\beta}\right)^{{d\over 2}}\, \cdot \, {\rm
exp}\{-\frac{|\vec{x}-\vec{y}|^2}{4\beta}\}\cdot\sum_{C=1}^N\delta^{AC}\sum_{n=0}^\infty
\, c_n^{CB}(\vec{x},\vec{y};K)\cdot \beta^n \label{eq:asy}
\\ C^{AB}_K(\vec{x},\vec{y};\beta)&=&\sum_{n=0}^\infty
\, c_n^{AB}(\vec{x},\vec{y};K)\cdot \beta^n \nonumber \qquad ,
\end{eqnarray}
if $C^{AB}_K(\vec{x},\vec{y};\beta)$ is written as the power
expansion above. Plugging (\ref{eq:asy}) into (\ref{eq:Khe1}), one
obtains the recurrence relation between the Seeley densities
$c_n^{AB}(\vec{x},\vec{y};K)$:
\begin{eqnarray}
&&\sum_{C=1}^N \, \left[n\cdot \delta^{AC}+(x_k-y_k)\cdot
(\delta^{AC}\partial_k-{1\over 2}Q_k^{AC}(\vec{x}))\right]\,
c_n^{CB}(\vec{x},\vec{y};K)\nonumber\\&=&\sum_{C=1}^N \, \left[
\delta^{AC}\cdot\bigtriangleup -Q_k^{AC}(\vec{x})\partial_k
-V^{AC}(\vec{x})\right]\, c_{n-1}^{CB}(\vec{x},\vec{y};K)
\end{eqnarray}
starting from: $c_0^{AB}(\vec{x},\vec{y};K)=\delta^{AB}$.

Let us introduce the following notation:
\begin{eqnarray*}
{}^{(\alpha_1,\alpha_2, \cdots
,\alpha_d)}C_n^{AB}(\vec{x})&=&\lim_{\vec{y}\rightarrow \vec{x}}
\frac{\partial^{\alpha_1+\alpha_2+\cdots +\alpha_d}\cdot
c_n^{AB}(\vec{x},\vec{y};K)}{\partial x_1^{\alpha_1}\partial
x_2^{\alpha_2}\cdots \partial x_d^{\alpha_d}} \hspace{1cm} ,
\hspace{1cm} c_n^{AB}(\vec{x},\vec{x};K)={}^{(0,0,\cdots
,0)}C_n^{AB}(\vec{x})\\ 0&\leq & \sum_{k=1}^d \, \alpha_k \, \leq
d \qquad , \qquad \alpha_1, \alpha_2, \cdots , \alpha_d=0,1,2,
\ldots , d \qquad \qquad .
\end{eqnarray*}
Thus, in the $\vec{y}\rightarrow\vec{x}$ limit the recurrence
relations between densities and partial derivatives of densities
can be written in the compact form:
\newpage
\begin{eqnarray*} &&(n+1+\sum_{k=1}^d \, \alpha_k)
{}^{(\alpha_1,\alpha_2,\cdots ,
\alpha_d)}C_{n+1}^{AB}(\vec{x})=\\&=& {}^{(\alpha_1+2,\alpha_2,
\cdots ,\alpha_d)}C_{n}^{AB}(\vec{x})+ {}^{(\alpha_1,\alpha_2+2,
\cdots , \alpha_d)}C_{n}^{AB}(\vec{x})+ \cdots
+{}^{(\alpha_1,\alpha_2, \cdots ,
\alpha_d+2)}C_{n}^{AB}(\vec{x})-\\&-&\sum_{D=1}^N
\sum_{r_1=0}^{\alpha_1}\sum_{r_2=0}^{\alpha_2}\cdots
\sum_{r_d=0}^{\alpha_d} {\alpha_1 \choose r_1} {\alpha_2 \choose
r_2}\cdots {\alpha_d \choose r_d} \left[ \frac{\partial^{r_1+r_2+
\cdots +r_d}\cdot Q^{AD}_1(\vec{x})}{\partial x_1^{r_1}\partial
x_2^{r_2}\cdots\partial x_d^{r_d}}\cdot
{}^{(\alpha_1-r_1+1,\alpha_2-r_2, \cdots
,\alpha_d-r_d)}C_{n}^{DB}(\vec{x})\right.+\\&&+\frac{\partial^{r_1+r_2+
\cdots +r_d}\cdot Q^{AD}_2(\vec{x})}{\partial x_1^{r_1}\partial
x_2^{r_2}\cdots\partial x_d^{r_d}}\cdot
{}^{(\alpha_1-r_1,\alpha_2-r_2+1, \cdots
,\alpha_d-r_d)}C_{n}^{DB}(\vec{x})+\cdots\\&&\cdots + \left.
\frac{\partial^{r_1+r_2+ \cdots +r_d}\cdot
Q^{AD}_d(\vec{x})}{\partial x_1^{r_1}\partial
x_2^{r_2}\cdots\partial x_d^{r_d}}\cdot
{}^{(\alpha_1-r_1,\alpha_2-r_2, \cdots
,\alpha_d-r_d+1)}C_{n}^{DB}(\vec{x})
\right]+  \\
&&+\frac{1}{2}\sum_{D=1}^N
\sum_{r_1=0}^{\alpha_1-1}\sum_{r_2=0}^{\alpha_2} \cdots
\sum_{r_d=0}^{\alpha_d} \alpha_1{\alpha_1-1 \choose r_1} {\alpha_2
\choose r_2}\cdots {\alpha_d \choose r_d}\cdot \\&&\cdot
\frac{\partial^{r_1+r_2+\cdots +r_d}\cdot
Q^{AD}_1(\vec{x})}{\partial x_1^{r_1}\partial
x_2^{r_2}\cdots\partial x_d^{r_d}}\cdot
{}^{(\alpha_1-1-r_1,\alpha_2-r_2, \cdots ,\alpha_d-r_d)}C_{n+1}^{DB}(\vec{x})+ \\
&&+\frac{1}{2}\sum_{D=1}^N
\sum_{r_1=0}^{\alpha_1}\sum_{r_2=0}^{\alpha_2-1} \cdots
\sum_{r_d=0}^{\alpha_d} \alpha_2{\alpha_1 \choose r_1} {\alpha_2-1
\choose r_2}\cdots {\alpha_d \choose r_d}\cdot \\&&\cdot
\frac{\partial^{r_1+r_2+\cdots +r_d}\cdot
Q^{AD}_2(\vec{x})}{\partial x_1^{r_1}\partial
x_2^{r_2}\cdots\partial x_d^{r_d}}\cdot
{}^{(\alpha_1-r_1,\alpha_2-1-r_2, \cdots
,\alpha_d-r_d)}C_{n+1}^{DB}(\vec{x})+ \cdots \\ &&
\cdots+\frac{1}{2}\sum_{D=1}^N
\sum_{r_1=0}^{\alpha_1}\sum_{r_2=0}^{\alpha_2} \cdots
\sum_{r_d=0}^{\alpha_d-1} \alpha_d{\alpha_1 \choose r_1} {\alpha_2
\choose r_2}\cdots {\alpha_d-1 \choose r_d}\cdot \\&&\cdot
\frac{\partial^{r_1+r_2+\cdots +r_d}\cdot
Q^{AD}_d(\vec{x})}{\partial x_1^{r_1}\partial
x_2^{r_2}\cdots\partial x_d^{r_d}}\cdot
{}^{(\alpha_1-r_1,\alpha_2-r_2, \cdots
,\alpha_d-1-r_d)}C_{n+1}^{DB}(\vec{x})+
\\&&-\sum_{D=1}^N
\sum_{r_1=0}^{\alpha_1}\sum_{r_2=0}^{\alpha_2}\cdots\sum_{r_d=0}^{\alpha_d}\,
{\alpha_1 \choose r_1}{\alpha_2\choose r_2}\cdots {\alpha_d\choose
r_d} \cdot\frac{\partial^{r_1+r_2+\cdots r_d}
V^{AD}(\vec{x})}{\partial x_1^{r_1}\partial x_2^{r_2}\cdots\partial
x_d^{r_d}}\cdot {}^{(\alpha_1-r_1,\alpha_2-r_2,\cdots ,
\alpha_d-r_d)}C_n^{DB}(\vec{x})
\end{eqnarray*}
to be solved starting from
\[
c_0^{AB}(\vec{x},\vec{x};K)=\delta^{AB} \Rightarrow\left\{
\begin{array}{c} {}^{(\alpha_1,\alpha_2, \cdots , \alpha_d)}C_0^{AB}(\vec{x})=0 , \, {\rm
if}\,
\alpha_k \neq 0, \forall  k=1,2, \cdots , d \\
{}^{(0,0, \cdots , 0)}C_0^{AA}(\vec{x})=1 , \, A=1,2, \cdots ,N
\end{array}\right. \qquad .
\]
\subsection{Asymptotics of the partition function and generalized
\\
zeta function meromorphy}

Defining
\[
{\rm tr}\, c_n(K)=\sum_{A=1}^N \, \int \, d{\rm vol}_{{\mathbb
T}^d}\, \, {}^{(0,0, \cdots , 0)}C_n^{AA}(\vec{x}) \qquad ,
\]
the asymptotic expansion of the partition function reads:
\[
{\rm Tr} \, e^{-\beta K}=e^{-c^2\beta}\cdot \left(\frac{\pi
R^2}{\beta}\right)^{{d\over 2}}\cdot \sum_{n=0}^\infty \, {\rm tr}
\, c_n(K)
\]
Via the Mellin transform, one writes the generalized zeta function
as the sum of meromorphic and entire functions of s:
\begin{eqnarray}
\zeta_K(s)&=& (\pi R^2)^{{d\over 2}}\cdot {1\over\Gamma(s)}\cdot
\sum_{n=0}^\infty \, {\rm tr}\, c_n(K) \, \int_0^1 \, d\beta
\,\beta^{s+n-1-{d\over 2}}e^{-c^2\beta} + {1\over\Gamma(s)}\cdot
\int_1^\infty \, d\beta \,\beta^{s-1}\, {\rm Tr} \, e^{-\beta
K}\nonumber\\&=& (\pi R^2)^{d\over 2}\cdot \sum_{n=0}^\infty \,
{1\over c^{2s+2n-d}} \cdot{\rm tr}\,
c_n(K)\cdot\frac{\gamma[s+n-{d\over 2},c^2]}{\Gamma(s)}+
{1\over\Gamma(s)}\cdot \int_1^\infty \, d\beta \, \beta^{s-1}\, {\rm
Tr} \, e^{-\beta K}\label{eq:mer} \qquad .
\end{eqnarray}
One can show that $B(s,K)={1\over\Gamma(s)}\cdot \int_1^\infty \,
d\beta \, \beta^{s-1}\, {\rm Tr} \, e^{-\beta K}$ is a entire
function of $s$ (holomorphic in the whole complex $s$-plane
${\mathbb C}$).
\[
b(s,K)=(\pi R^2)^{d\over 2}\cdot \sum_{n=0}^\infty \, {1\over
c^{2s+2n-d}} \cdot{\rm tr}\, c_n(K)\cdot\frac{\gamma[s+n-{1\over
2},c^2]}{\Gamma(s)} \qquad ,
\]
however, is meromorphic, with poles at the poles of the incomplete
Euler ${\rm Gamma}$ functions: $\gamma[s+n-{d\over 2},c^2]$.

\section{{\bf The $\lambda(\phi^4)$-model on a line}}

In the $\lambda(\phi^4)_2$-model the action
\[
S=\int \, dy^2 \, \left\{{1\over 2}\frac{\partial\psi}{\partial
y^\mu}\frac{\partial\psi}{\partial y_\mu}-{\lambda\over
4}(\psi^2(y_0,y)-{m^2\over\lambda})^2\right\}
\]
governs the dynamics of the scalar field $\psi(y_0,y): {\mathbb
R}^{1,1}\rightarrow {\mathbb R}$. We choose the metric
$g_{\mu\nu}={\rm diag}(1,-1)$ in (1+1)-dimensional ${\mathbb
R}^{1,1}$ Minkowskian space-time. In our systems of units the
dimension of the field and the coupling constant are respectively:
$[\psi]=M^{{1\over 2}}L^{{1\over 2}}$, $[\lambda]=M^{-1}L^{-3}$. In
terms of non-dimensional space-time coordinates and fields
\[
y^\mu\rightarrow y^\mu={\sqrt{2}\over m}\cdot x^\mu \hspace{1.5cm} ;
\hspace{1.5cm} \psi(y^\mu)\rightarrow
\psi(y^\mu)={m\over\sqrt{\lambda}}\cdot \phi(x^\mu) \qquad ,
\]
the action functional and the field equations of the
$\lambda(\phi)^4_2$ model read:
\[
S={m^2\over\lambda}\int \, dx^2 \, \left\{{1\over
2}\frac{\partial\phi}{\partial x^\mu}\frac{\partial\phi}{\partial
x_\mu}-{1\over 2}(\phi^2(x_0,x)-1)^2\right\}
\]
\[
\frac{\partial^2\phi}{\partial
x_0^2}(x_0,x)-\frac{\partial^2\phi}{\partial
x^2}(x_0,x)=2\phi(x_0,x)(1-\phi^2(x_0,x)) \qquad .
\]

The shift of the scalar field from the homogeneous stable
solution, $\phi(x^\mu)=1+H(x^\mu)$, leads to the action
\[
S=\frac{m^2}{\lambda}\int \, d^2x \, \left\{\left[{1\over
2}\partial_\mu H\partial^\mu H -2H^2(x^\mu)
\right]-\left[2H^3(x^\mu)+{1\over 2}H^4(x^\mu)\right]\right\} \qquad
,
\]
which shows the spontaneous symmetry breakdown of the internal
parity ${\mathbb Z}_2$ symmetry. The Feynman rules are thus
obtained in terms of the Higgs propagator as well as three-valent
and four-valent Higgs self-coupling vertices:
\begin{table}[h]
\begin{center}
\caption{Propagator}
\begin{tabular}{lccc} \\ \hline
\textit{Particle} & \textit{Field} & \textit{Propagator} & \textit{Diagram} \\
\hline \\ Higgs & $H(x^\mu)$ & $\displaystyle\frac{i \lambda \hbar
}{m^2(k_0^2-k^2-4+i\varepsilon)}$ &
\parbox{2cm}{\epsfig{file=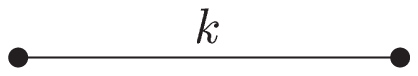,width=2cm}}
\\[0.5cm] \hline
\end{tabular}
\end{center}
\end{table}
\begin{table}[hbt]
\begin{center}
\caption{Third- and fourth-order vertices }
\begin{tabular}{clcl} \\ \hline
\textit{Vertex} & \textit{Weight} & \textit{Vertex} & \textit{Weight} \\
\hline \\
\parbox{2.3cm}{\epsfig{file=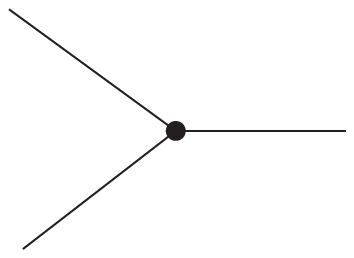,width=2cm}} &
$\displaystyle -12i\frac{m^2}{\hbar \lambda} $  &
\parbox{2.3cm}{\epsfig{file=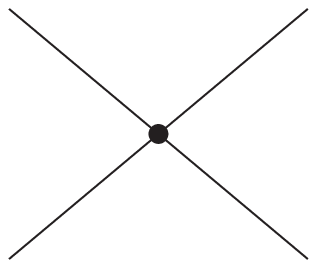,width=2cm}} &
$\displaystyle -12i\frac{m^2}{\hbar \lambda}$  \\[0.7cm] \hline
\end{tabular}
\end{center}
\end{table}

\subsection{Plane waves and vacuum energy}

The general solution of the linearized field equations
\[
\frac{\partial^2\delta H}{\partial
x_0^2}(x_0,x)-\frac{\partial^2\delta H}{\partial
x^2}(x_0,x)+4\delta H(x_0,x)=0
\]
governing the small fluctuations of the Higgs field is:
\[
\delta
H(x_0,x)=\frac{\sqrt{\lambda}}{m}\cdot\sqrt{\frac{\sqrt{2}\hbar}{mL}}\sum_k{1\over\sqrt{2\omega(k)}}
\left\{a(k)e^{-ik_0x_0+ikx}+a^*(k)e^{ik_0x_0-ikx}\right\} \qquad ,
\]
where $k_0=\omega(k)=\sqrt{k^2+4}$, and the dispersion relation
$k_0^2-k^2-4=0$ holds:
\[
K_0e^{ikx}=\omega^2(k)e^{ikx} \hspace{1.5cm} , \hspace{1.5cm}
K_0=-{d^2\over dx^2}+4 \qquad .
\]
We choose a normalization interval of non-dimensional \lq\lq length"
$\frac{mL}{\sqrt{2}}$,
$I=[-\frac{mL}{2\sqrt{2}},\frac{mL}{2\sqrt{2}}]$, and we impose
periodic boundary conditions on the plane waves such that
$k{mL\over\sqrt{2}}=2\pi n$, $n\in{\mathbb Z}$, and the spectral
density of $K_0$ is: $ \rho_{K_0}(k)=\frac{dn}{dk}={1\over
2\pi}{mL\over\sqrt{2}}$. This is tantamount to considering the
$d=1$, $N=1$ case of Section \S.2, although the radius
$R=\frac{mL}{2\sqrt{2}\pi}$ of the spatial circle is slightly
modified to fit in with the conventions most frequently used in the
literature on kinks.

From the classical free Hamiltonian
\begin{eqnarray*}
H^{(2)}&=&{m^3\over\sqrt{2}\lambda}\int \, dx \, \left\{{1\over
2}\frac{\partial\delta H}{\partial x_0}\cdot\frac{\partial\delta H
}{\partial x_0}+{1\over 2}\frac{\partial\delta H}{\partial
x}\cdot\frac{\partial\delta H}{\partial x}+\delta
H(x_0,x)\cdot\delta H(x_0,x) \right\}\\&=&\sum_k \, \hbar {m\over
2\sqrt{2}} \omega(k)\left(\frac{}{}a^*(k)a(k)+a(k)a^*(k)\right)
\qquad ,
\end{eqnarray*}
one obtains the quantum free Hamiltonian:
\[
\hat{H}^{(2)}_0=\sum_k \, \hbar {m\over\sqrt{2}}
\omega(k)\left(\hat{a}^\dagger(k)\hat{a}(k)+{1\over 2}\right)
\]
via canonical quantization: $[{\hat a}(k),{\hat
a}^\dagger(q)]=\delta_{kq}$. The vacuum energy is:
\[
\Delta E_0={\hbar m\over 2\sqrt{2}}\sum_k\omega(k)={\hbar m\over
2\sqrt{2}}{\rm Tr} K_0^{{1\over 2}} \hspace{1cm} .
\]

\subsection{The one-loop mass renormalization counter-term}

The Higgs tadpole and the Higgs self-energy are
ultraviolet-divergent in the one-loop order of the
$\hbar$-expansion:
\[
-6i\cdot I(4)=-6i\cdot \int \, \frac{d^2k}{(2\pi)^2} \, \cdot
\frac{i}{(k_0^2-k^2-4+i\varepsilon)} =\hspace{2.2cm}
\parbox{3cm}{\epsfig{file=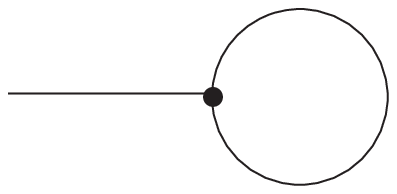,width=3cm}}\hspace{1cm}
\]
\[
-6i\cdot I(4)=-6i\cdot \int \,\frac{dk}{4\pi}\, \cdot
\frac{1}{\sqrt{k^2+4}}=-6i\cdot \frac{\sqrt{2}}{mL}\cdot {1\over
2}\sum_n\frac{1}{\sqrt{{n^2\over R^2}+4}} =\hspace{1cm}
\parbox{3cm}{\epsfig{file=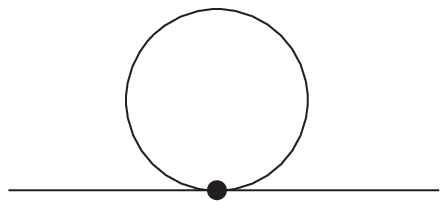,width=3cm}}\hspace{1cm}
\]
A combinatorial factor of ${1\over 2}$ has been taken into account
in both graphs. The Lagrangian density of counter-terms ${\cal
L}_{C.T.}=3\hbar \left(\phi^2(x)-1\right)\cdot I(4)$, giving the
vertices in Table 3,
\begin{table}[hbt]
\begin{center}
\caption{One-loop counter-terms }  \hspace{2cm}
\begin{tabular}{ccc} \\ \hline
\textit{Diagram} & \hspace{1.5cm} & \textit{Weight} \\
\hline
\parbox{4cm}{\epsfig{file=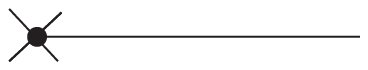,width=4cm}} & &
$\displaystyle 6iI(4) $
\\
\parbox{4cm}{\epsfig{file=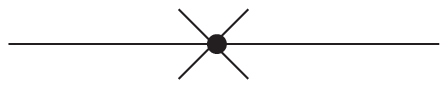,width=4cm}} & &
$\displaystyle 6i I(4)$  \\[0.5cm] \hline \\[0.25cm]
\end{tabular}
\end{center}
\end{table}
must be added to exactly cancel the divergences above.

\subsection{$\lambda(\phi^4)_2$ kinks}

The configuration space of the classical $\lambda(\phi^4)_2$-model
\[
{\cal C}=\{\phi(x)\in {\rm Maps}({\mathbb R},{\mathbb R})/
E[\phi]<+\infty\}
\]
is non-connected: ${\cal C}={\cal C}_{+\,+}\sqcup{\cal
C}_{-\,-}\sqcup{\cal C}_{+\,-}\sqcup{\cal C}_{-\,+}$. The energy for
time-independent configurations
\[
E={m^3\over\sqrt{2}\lambda}\int \, dx \, \left\{{1\over
2}\frac{d\phi}{dx}\cdot\frac{d\phi}{dx}+{1\over
2}(1-\phi^2)^2\right\}
\]
is finite if and only if
\[
 \displaystyle\lim_{x\rightarrow\pm\infty}\frac{d\phi}{dx}=0
\hspace{0.5cm} , \hspace{0.5cm}
\displaystyle\lim_{x\rightarrow\pm\infty}\phi(x)=\left\{\begin{array}{c}
\phi_+=+1
\\ \phi_-=-1\end{array}\right. \qquad ,
\]
and the four non-connected components of ${\cal C}$ are classified
by the behavior of the scalar field at $x=\pm\infty$. The
Bogomolny splitting of the static energy - the energy for time
independent field configurations -
\begin{eqnarray*}
E&=&{m^3\over\sqrt{2}\lambda}\int \, dx \, {1\over
2}\left(\frac{d\phi}{dx}\mp(1-\phi^2)\right)^2 \pm
{m^3\over\sqrt{2}\lambda}\cdot \left(\phi-{\phi^3\over
3}\right)\left|_{\phi(-\infty)}^{\phi(\infty)}\right.
\end{eqnarray*}
shows that the absolute minima of $E$ in each sector of ${\cal C}$
satisfy the first-order equations:
\[
{d\phi\over dx}=\pm(1-\phi^2) \hspace{2cm} .
\]
Besides the homogeneous solutions $\phi_\pm=\pm 1$ of energy
$E(\phi_\pm)=0$, there are kinks or traveling wave solutions
\[
\phi_K(x)=\pm{\rm tanh}(x-a) \hspace{0.5cm} , \hspace{0.5cm}
\phi_K(x_0,x)=\pm{\rm tanh}(\frac{x-a-vt}{\sqrt{1-v^2}}) \qquad ,
\qquad \varepsilon_K(x)=\frac{1}{{\rm cosh}^2(x-a)}
\]
of energy $E(\phi_K)=\frac{4m^3}{3\sqrt{2}\lambda}$.
\begin{center}
\noindent\begin{tabular}{p{4.5cm}}
\epsfig{file=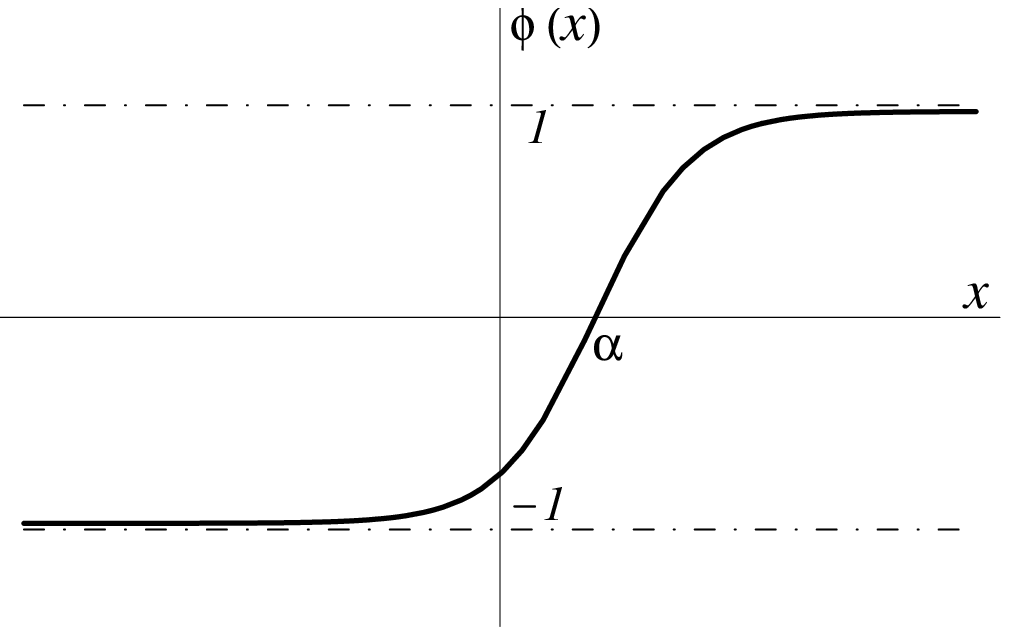,height=3cm}
\end{tabular}\hspace{2cm}
\noindent\begin{tabular}{p{6cm}}
\epsfig{file=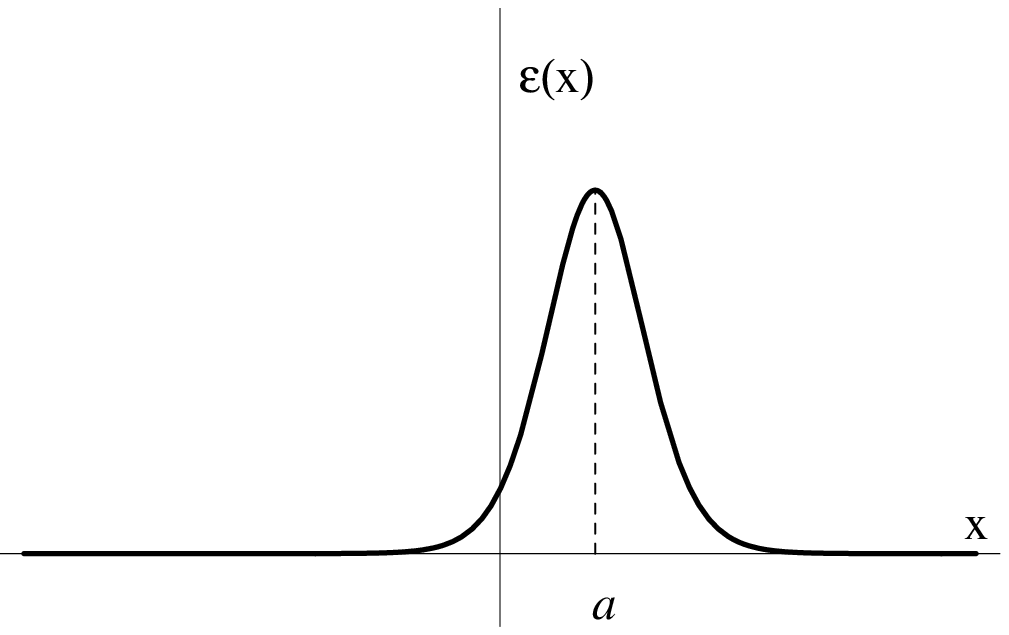,height=4cm}
\end{tabular}
\end{center}

\subsection{Kink Casimir energy}

Small kink deformations $\phi(x)=\phi_K(x)+\delta\phi(x)$ are
still solutions of the first-order equations if $\delta\phi(x)\in
{\rm Ker}\, D$
\[
D\delta\phi(x)=\left(-{d\over
dx}+2\phi_K(x)\right)\delta\phi(x)=\left(-{d\over dx}+2{\rm
tanh}x\right)\delta\phi(x)=0 \qquad .
\]
Note that
\[
K^-=D^\dagger D=-{d^2\over dx^2}+4-{2\over{\rm cosh}^2x}
\hspace{2cm} , \hspace{2cm} K=D D^\dagger=-{d^2\over
dx^2}+4-{6\over{\rm cosh}^2x} \qquad .
\]

Moreover, the shift of the Higgs field from the stable kink
solution, $\phi(x^\mu)=\phi_K(x)+H(x^\mu)$, leads to the action:
\begin{eqnarray*}
S=&-&\frac{4m^3}{3\sqrt{2}\lambda}\lim_{T\rightarrow\infty}\int_{-\frac{T}{2}}^\frac{T}{2}\,
dx_0  \\&+&\frac{m^2}{\lambda}\int \, d^2x \,
\left\{\left[{1\over 2}\partial_\mu H\partial^\mu H -(2-{3\over{\rm
cosh}^2x})H^2(x^\mu) \right]-\left[2{\rm tanh}x H^3(x^\mu)+{1\over
2}H^4(x^\mu)\right]\right\} \qquad .
\end{eqnarray*}
Thus, the Feynman rules in the kink sector must be modified. Both
the Higgs propagator and the trivalent Higgs self-energy vertex are
strongly influenced by the kink.

The general solution of the linearized field equations
\[
\frac{\partial^2\delta H}{\partial
x_0^2}(x_0,x)-\frac{\partial^2\delta H}{\partial
x^2}(x_0,x)+\left(4-{6\over{\rm cosh}^2x}\right)\delta H
(x_0,x)=\left[\frac{\partial^2}{\partial x_0^2}+K\right]\delta
H(x_0,x)=0
\]
for small fluctuations $\phi(x_0,x)=\phi_K(x)+\delta H(x_0,x)$ on
the kink background
\begin{eqnarray*}
\delta H
^\prime(x_0,x)&=&\frac{\sqrt{\lambda}}{m}\cdot\sqrt{\frac{\sqrt{2}\hbar}{mL}}({1\over\sqrt{2\sqrt{3}}}
A_3e^{-i\sqrt{3}x_0}+{1\over\sqrt{2\sqrt{3}}}A^*_3e^{i\sqrt{3}x_0})f_3(x)+
\\&+&\frac{\sqrt{\lambda}}{m}\cdot\sqrt{\frac{\sqrt{2}\hbar}{mL}}\sum_k{1\over\sqrt{2\varepsilon(k)}}
\left(A(k)e^{-i\varepsilon
x_0}f_\varepsilon(x)+A^*(k)e^{i\varepsilon
x_0}f_\varepsilon^*(x)\right)
\end{eqnarray*}
is written in terms of the positive eigenfunctions of the
$K$-operator:
\[
Kf_\varepsilon(x)=\left[-{d^2\over dx^2}+4-{6\over{\rm
cosh}^2x}\right]f_\varepsilon(x)=\varepsilon^2f_\varepsilon(x)\qquad
, \qquad k\in {\mathbb R}
\]

\begin{center}
\begin{tabular}{l|c} \hline
Eigenvalues &  Eigenfunctions \\[0.4cm]  \hline $\varepsilon^2=0$ &
$f_0(x)={1\over {\rm cosh}^2x}$ \\[0.4cm] $\varepsilon_3^2=3$ &
$f_3(x)=\frac{{\rm sinh}x}{{\rm cosh}^2x}$\\[0.4cm]
 $\varepsilon^2=k^2+4$ & $f_\varepsilon(x)=e^{ikx}(3{\rm
tanh}^2x-1-3ik{\rm tanh}x-k^2)$\\[0.4cm] \hline
\end{tabular}
\end{center}
The choice of periodic boundary conditions in the interval
$I=[-{mL\over 2\sqrt{2}}, {mL\over 2\sqrt{2}}]$
\[
k\frac{mL}{\sqrt{2}}+\delta(k)=2\pi n
\]
gives the following phase shifts and spectral density:
\[
\rho_K(k)={1\over 2\pi}(\frac{mL}{\sqrt{2}}+\frac{d\delta(k)}{dk})
\qquad , \qquad \delta(k)=-2{\rm arctan}\frac{3k}{2-k^2} \qquad .
\]
Thus, the sums are over the solutions of the transcendental
equations
\begin{equation}
k-\frac{n}{R}=\frac{1}{\pi R}\cdot {\rm arctan}\frac{3k}{2-k^2}
\qquad , \qquad n\in {\mathbb Z} \qquad . \label{eq:kinkspec}
\end{equation}
The classical free Hamiltonian for kink fluctuations becomes
\begin{eqnarray*}
H^{(2)}&=&\frac{2m^3}{\sqrt{2}\lambda}\int \, dx \,
\left[\frac{\partial\delta H}{\partial x_0}\frac{\partial\delta
H}{\partial x_0}+ \, \delta H(x_0,x)K\delta
H(x_0,x)\right]=\\&=&\frac{\hbar
m}{2\sqrt{2}}\left\{\sqrt{3}(A^*_3A_3+A_3A_3^*)+\sum_k\varepsilon(k)(A^*(k)A(k)+A(k)A^*(k))\right\}
\qquad ,
\end{eqnarray*}
and, after canonical quantization,
\[
[{\hat A}_3,{\hat A}_3^\dagger]=1 \hspace{2cm} , \hspace{2cm}
[{\hat A}(k),{\hat A}^\dagger(q)]=\delta_{kq}
\]
one obtains the quantum free Hamiltonian
\[
\hat{H}^{(2)}=\hbar {m\over\sqrt{2}}
\left(\sqrt{3}(\hat{A}_3^\dagger\hat{A}_3+{1\over
2})+\sum_k\varepsilon(k)(\hat{A}^\dagger(k)\hat{A}(k)+{1\over
2})\right)
\]
and the kink Casimir energy
\[
\Delta E(\phi_K)=\frac{\hbar
m}{2\sqrt{2}}(\sqrt{3}+\sum_k\varepsilon(k))=\frac{\hbar
m}{2\sqrt{2}}{\rm Tr}K^{{1\over 2}}
\]
when all the positive modes are non-occupied.

In sum, the kink semi-classical energy -one-loop order- receives
three contributions:
\begin{enumerate}

\item The classical energy,
$E(\phi_K)=\frac{4m^3}{3\sqrt{2}\lambda}$.

\item The kink Casimir energy -zero point energy renormalization-
\[
\Delta M_K^C=\Delta E(\phi_K)- \Delta E_0=\frac{\hbar
m}{2\sqrt{2}}\left({\rm Tr}K^{{1\over 2}}-{\rm Tr}K_0^{{1\over
2}}\right) \qquad .
\]

\item The contribution of ${\cal L}_{C.T.}$ to the one-loop kink
mass, which is:
\[
\Delta M_K^R=-3\frac{\hbar m}{\sqrt{2}}\cdot I(4)\cdot \int \, dx
\,\left(\phi_K^2(x)-\phi_\pm^2\right)=6\frac{\hbar m}{\sqrt{2}}\cdot
I(4) \qquad .
\]
\end{enumerate}
Therefore, the one-loop kink mass shift and the semi-classical
kink energy  are the divergent quantities:
\[
\Delta M_K=\Delta M_K^C+\Delta M_K^R \hspace{2cm} ,\hspace{2cm}
E_S(\phi_K)=E(\phi_K)+\Delta M_K \qquad .
\]

\section{The kink heat kernel and generalized zeta function}

\subsection{Zeta function regularization}

We regularize the ultraviolet divergent kink and vacuum energies
$\Delta E(\phi_K)$, $\Delta E_0$ in terms of their generalized
zeta functions:
\[
\Delta M_K^C(s)={\hbar\over 2} (2{\mu^2\over m^2})^s\mu
\left(\zeta^*_K(s)-\zeta_{K_0}(s)\right) \qquad .
\]
Here
\[
\zeta^*_K(s)=\frac{1}{\varepsilon_3^{2s}}+\sum_{k}\frac{1}{\varepsilon(k)^{2s}}
\hspace{2cm} , \hspace{2cm}
\zeta_{K_0}(s)=\sum_{k}\frac{1}{\omega(k)^{2s}} \qquad ,
\]
and $\mu$ is a parameter of dimensions $L^{-1}$, necessary to keep
the dimension of $\Delta M_K^C(s)$ independent from the complex
variable $s$. The star means that the zero mode does not enter the
kink generalized zeta function. Therefore,
\[
\Delta M_K^C=\lim_{s\rightarrow -{1\over 2}}\Delta M_K^C(s)={\hbar
m\over 2\sqrt{2}}\left(\zeta^*_K(-{1\over 2})-\zeta_{K_0}(-{1\over
2})\right) \qquad ,
\]
and the divergences reappear at $s=-{1\over 2}$, which is a pole of
$\Delta M_K^C(s)$. $\Delta M_K^C(s)$, however, is a meromorphic
function of $s$.

$\Delta M_K^R$ can also be regularized in terms of zeta functions.
Note that the divergent integral $I(4)$ can be expressed as the
limit:
\begin{equation}
I(4)=-\lim_{s\rightarrow -{1\over 2}}\frac{1}{\mu L}\cdot
\frac{\Gamma(s+1)}{\Gamma(s)}\cdot\left(\frac{2\mu^2}{m^2}\right)^{s+1}\cdot
\zeta_{K_0}(s+1) \label{eq:reg1}
\end{equation}
when the system is defined in the interval
$[-\frac{mL}{2\sqrt{2}},\frac{mL}{2\sqrt{2}}]$. Thus,
\[
\Delta
M_K^R(s)=-\frac{6\hbar}{L}\cdot\left(\frac{2\mu^2}{m^2}\right)^{s+{1\over
2}}\cdot\frac{\Gamma(s+1)}{\Gamma(s)}\cdot\zeta_{K_0}(s+1) \qquad ,
\]
and
\[
\Delta M_K^R=\lim_{s\rightarrow -{1\over 2}} \Delta
M_K^R(s)=\frac{3\hbar}{L}\cdot\zeta_{K_0}({1\over 2})\qquad .
\]

Another, more direct, regularization for $I(4)$ is possible:
\begin{equation}
I(4)=\lim_{s\rightarrow {1\over 2}}\frac{1}{2\mu
L}\cdot\left(\frac{2\mu^2}{m^2}\right)^{s}\cdot \zeta_{K_0}(s)
\qquad . \label{eq:reg2}
\end{equation}
The problem is that
\[
\Delta
M_K^R(s)=\frac{3\hbar}{L}\cdot\left(\frac{2\mu^2}{m^2}\right)^{s-{1\over
2}}\cdot\zeta_{K_0}(s)
\]
and $\Delta M_K^R=\lim_{s\rightarrow{1\over 2}} \Delta M_K^R(s)$
arise at a different point in the complex $s$-plane from the point
where $\bigtriangleup M_K^C$ is obtained.

\subsection{The Dashen-Hasslacher-Neveu (DHN) exact formula}

The partition and generalized zeta functions for the vacuum operator
$K_0$ are in the $R\rightarrow\infty$ limit \footnote{In Appendix I
it is shown how this limit can be safely taken, leaving no remnants,
when PBC are chosen.}:
\[
{\rm Tr}e^{-\beta
K_0}=\frac{mL}{2\sqrt{2}\pi}\int_{-\infty}^{+\infty} \, dk
\,e^{-\beta(k^2+4)}=\frac{mL}{\sqrt{8\pi\beta}}\cdot e^{-4\beta}
\]
\[ \zeta_{K_0}(s)={mL\over\sqrt{8\pi}}\cdot {1\over\Gamma
(s)}\cdot \int_0^\infty \, d\beta \, \beta^{s-{3\over 2}}\,
e^{-4\beta}={mL\over\sqrt{8\pi}}\cdot {1\over 2^{2s-1}}\cdot
\frac{\Gamma(s-{1\over 2})}{\Gamma(s)}\qquad .
\]
The poles of $\zeta_{K_0}(x)$ are thus the poles of the Euler
Gamma function $\Gamma (s-{1\over 2})$: $s-{1\over 2}=0,-1,-2,
\cdots , -n, \cdots$. The vacuum energy reads:
\[
\Delta E(\phi_\pm)=\lim_{s \rightarrow
-\frac{1}{2}}\frac{\hbar}{2} \left( \frac{2 \mu^2}{m^2} \right)^s
\mu \cdot \zeta_{K_0}(s)=\lim_{s \rightarrow
-\frac{1}{2}}\frac{\hbar}{2} \left( \frac{2 \mu^2}{m^2} \right)^s
\mu\cdot {mL\over\sqrt{8\pi}}\cdot {1\over 2^{2s-1}}\cdot
\frac{\Gamma(s-{1\over 2})}{\Gamma(s)}\qquad .
\]

The partition and generalized zeta functions for the kink operator
$K$ can also be given analytically:
\begin{eqnarray*}
{\rm Tr}^* e^{-\beta
K}&=&e^{-3\beta}+\frac{mL}{2\sqrt{2}\pi}\int_{-\infty}^{+\infty}
\, dk \,e^{-\beta(k^2+4)}+{1\over 2\pi}\int_{-\infty}^{+\infty} \,
dk \, \frac{d\delta(k)}{dk} \,
e^{-\beta(k^2+4)}\\&=&\frac{mL}{\sqrt{8\pi\beta}}\cdot
e^{-4\beta}+e^{-3\beta}(1-{\rm Erfc}\sqrt{\beta})-{\rm
Erfc}2\sqrt{\beta}
\end{eqnarray*}
\begin{eqnarray*}
\zeta^*_{K}(s)&=&\zeta_{K_0}(s)+\frac{1}{\Gamma(s)}\left[\frac{1}{3^s}+\frac{1}{2\pi}\int_{-\infty}^{+\infty}\,
dk \, \frac{d\delta}{dk}(k)\cdot\frac{1}{(k^2+4)^s}\right]\\
&=&\zeta_{K_0}(s)+\frac{\Gamma(s+{1\over
2})}{\sqrt{\pi}\Gamma(s)}\left[{2\over 3^{s+{1\over 2}}}\cdot {}_2
F_1[{1\over 2},s+{1\over 2},{3\over 2},-{1\over 3}]-{1\over
4^s}{1\over s}\right]\qquad ,
\end{eqnarray*}
respectively in terms of complementary error functions ${\rm
Erfc}x$ and hypergeometric Gauss functions ${}_2F_1[a,b,c;d]$:
\[
{}_2 F_1[{1\over 2},s+{1\over 2},{3\over 2},-{1\over
3}]=\frac{\Gamma({3\over 2})}{\Gamma({1\over 2})\Gamma(s+{1\over
2})}\cdot \sum_{l=0}^\infty \,
\frac{(-1)^l}{3^ll¡}\cdot\frac{\Gamma(l+{1\over
2})\Gamma(s+l+{1\over 2})}{\Gamma(l+{3\over 2})} \qquad .
\]
Thus, besides the poles of $\zeta_{K_0}(s)$, $\zeta_{K}(s)$ has
poles at: $s+l+{1\over 2}=0,-1,-2, \cdots , -n, \cdots$.

The renormalized kink Casimir energy
\begin{eqnarray*}
\Delta M_K^C&=& \frac{\hbar m}{2\sqrt{2\pi}} \lim_{\varepsilon
\rightarrow 0}\left( \frac{2 \mu^2}{m^2} \right)^{\varepsilon}
\frac{\Gamma(\varepsilon)}{\Gamma(-\frac{1}{2}+\varepsilon)}
\left[ \frac{2}{3^\varepsilon} \, {}_2
F_1[{\textstyle\frac{1}{2}},\varepsilon,
{\textstyle\frac{3}{2}},-{\textstyle\frac{1}{3}}]-
\frac{1}{(-\frac{1}{2}+\varepsilon)\,4^{-\frac{1}{2}+\varepsilon}}
\right]\\ &=& \frac{\hbar m}{2\sqrt{2}\pi} \lim_{\varepsilon
\rightarrow 0} \left[ -\frac{3}{\varepsilon}+2 +\ln \frac{3}{4}-3
\ln \frac{2 \mu^2}{m^2}
-{}_2F_1'[{\textstyle\frac{1}{2}},0,{\textstyle\frac{3}{2}},-{\textstyle\frac{1}{3}}]
+o(\varepsilon) \right]\\&=&-\frac{\hbar
m}{2\sqrt{2}\pi}\lim_{\varepsilon\rightarrow
0}\left[{3\over\varepsilon}+3\ln\frac{2\mu^2}{m^2}-\frac{\pi}{\sqrt{3}}\right]
\end{eqnarray*}
still has a pole; zero-point vacuum energy renormalization is not
sufficient. The special values
\[
{}_2F_1[{1\over 2},0,{3\over 2};-{1\over 3}]=1 \qquad , \qquad
{}_2F_1^\prime[{1\over 2},0,{3\over 2};-{1\over
3}]=2-\frac{\pi}{\sqrt{3}}-\ln\frac{4}{3}
\]
have been taken into account in the derivation above.

Does this result agree with the corresponding
Dashen-Hasslacher-Neveu (DHN) formula obtained via the Stony
Brook/Wien mode number regularization method:
\[
\bigtriangleup M_K^C=\frac{\hbar
m}{2\sqrt{2}}\left[\sqrt{3}+{1\over
2\pi}\lim_{\Lambda\rightarrow\infty}\int_{-\Lambda}^{+\Lambda}\,
dk \,
\frac{d\delta}{dk}(k)\cdot\sqrt{k^2+4}-{2(2+1)\over\pi}\right]
\]
\[
\Delta M_K^R=\frac{3\hbar
m}{2\sqrt{2}\pi}\lim_{\Lambda\rightarrow\infty}\int_{-\Lambda}^\Lambda
\, \frac{dk}{\sqrt{k^2+4}} \qquad ?
\]

The zeta function regularization procedure (\ref{eq:reg1}) for
$\Delta M_K^R$ provides the result:
\begin{eqnarray*}
\Delta M_K^R&=&-\frac{6\hbar}{L}\lim_{s\rightarrow -{1\over
2}}\left(\frac{2\mu^2}{m^2}\right)^{s+{1\over
2}}\cdot\frac{mL}{\sqrt{8\pi}}\cdot\frac{1}{4^{s+{1\over 2}}}\cdot
\frac{\Gamma(s+{1\over 2})}{\Gamma(s)}\\&=&-\frac{3 \hbar
m}{\sqrt{2\pi}} \lim_{\varepsilon \rightarrow 0} \left( \frac{2
\mu^2}{m^2} \right)^\varepsilon \frac{4^{-\varepsilon}
\Gamma(\varepsilon)}{\Gamma(-\frac{1}{2}+\varepsilon)}\\&=&\frac{3\hbar
m}{2 \sqrt{2}\pi} \lim_{\varepsilon \rightarrow 0} \left[
\frac{1}{\varepsilon}+ \ln \frac{2 \mu^2}{m^2} -\ln
4+(\psi(1)-\psi(-{1\over 2})) +o(\varepsilon)
\right]\\&=&\frac{\hbar
m}{2\sqrt{2}\pi}\lim_{\varepsilon\rightarrow
0}\left[{3\over\varepsilon}+3\ln\frac{2\mu^2}{m^2}-2(2+1)\right]
\end{eqnarray*}
because the difference of the digamma functions is:
$\psi(1)-\psi(-{1\over 2})=\ln 4-2$. Thus, both methods give the well known answer for the one-loop kink mass
shift:
\[
\Delta M_K=\Delta M_K^C+\Delta M_K^R=\frac{\hbar
m}{2\sqrt{6}}-\frac{3\hbar m}{\pi\sqrt{2}} \qquad .
\]
In the DHN derivation, however, when the mode number regularization
cutoff is used, the second (negative) summand in the formula comes
from the kink Casimir energy $\Delta M_K^C$. The first summand is
due to the non-exact cancelation between the ultraviolet divergences
arising in $\Delta M_K^C$ and the induced energy $\Delta M_K^R$ by
mass renormalization added to the contribution of the bound state.
In our zeta function regularization procedure, the origin of the two
terms is more clear: the first summand comes from the finite piece
in $\Delta M_K^C$ at the physical point $s=-{1\over 2}$ but the
other piece is found in the regularization of $\Delta M_K^R$; i.e.,
$\Delta M_K^R$ does not exactly cancel the divergence of $\Delta
M_K^C$ but the renormalization process leaves finite reminders in
the kink Casimir energy and the mass renormalization counter-terms
induced, providing the correct answer.

The alternative regularization of $I(4)$ (\ref{eq:reg2}) applied
to $\Delta M_K^R$ leads to the result:
\begin{eqnarray*}
\Delta M_K^R&=&\frac{3\hbar}{L}\lim_{s\rightarrow {1\over
2}}\left(\frac{2\mu^2}{m^2}\right)^{s-{1\over
2}}\cdot\frac{mL}{\sqrt{8\pi}}\cdot\frac{1}{4^{s-{1\over 2}}}\cdot
\frac{\Gamma(s-{1\over 2})}{\Gamma(s)}\\&=&\frac{3 \hbar
m}{\sqrt{8\pi}} \lim_{\varepsilon \rightarrow 0} \left( \frac{2
\mu^2}{m^2} \right)^\varepsilon \frac{4^{-\varepsilon}
\Gamma(\varepsilon)}{\Gamma(\frac{1}{2}+\varepsilon)}\\&=&\frac{3\hbar
m}{2 \sqrt{2}\pi} \lim_{\varepsilon \rightarrow 0} \left[
\frac{1}{\varepsilon}+ \ln \frac{2 \mu^2}{m^2} -\ln
4+(\psi(1)-\psi({1\over 2})) +o(\varepsilon)
\right]\\&=&\frac{\hbar
m}{2\sqrt{2}\pi}\lim_{\varepsilon\rightarrow
0}\left[{3\over\varepsilon}+3\ln\frac{2\mu^2}{m^2}\right]
\end{eqnarray*}
because the difference of the digamma functions is:
$\psi(1)-\psi({1\over 2})=\ln 4$. Thus, using the latter
regularization for $I(4)$ we would obtain
\[
\Delta M_K=\Delta M_K^C+\Delta M_K^R=\frac{\hbar m}{2\sqrt{6}}
\qquad.
\]
This (bad) result was achieved in the literature on the matter by
regularizing the ultraviolet divergences by means of a cutoff in the
energy, rather than in the number of modes. Also, one could give
this answer without taking into account the zero mode in the CCG
formula, see section \S. 6.2 .

\subsection{The high-temperature expansion of the partition function}

Even without complete knowledge of the spectral data of the kink
fluctuation operator $K$, it would be possible to obtain an
approximate formula for the one-loop mass shift from the
high-temperature expansion of the partition function.

The heat equation kernel for the $K_0$-heat equation
\[
\left(\frac{\partial}{\partial\beta}-\frac{\partial^2}{\partial
x^2}+4 \right)K_{K_0}(x,y;\beta)=0 \quad , \quad
K_{K_0}(x,y;0)=\delta(x-y)
\]
of the vacuum fluctuation operator $K_0=-\frac{d^2}{dx^2}+4$, for
$\beta$ small, is:
\[
K_{K_0}(x,y;\beta)=\frac{e^{-4\beta}}{\sqrt{4\pi\beta}}\cdot
e^{-\frac{(x-y)^2}{4\beta}}\qquad .
\]
The kink fluctuation operator $K$ is the Schrodinger operator
\[
K=-\frac{d^2}{dx^2}+4+V(x)\hspace{1.5cm} , \hspace{1.5cm}
V(x)=6\phi_K^2(x)-6=-{6\over {\rm cosh}^2x} \qquad ,
\]
whereas the corresponding $K$-heat equation kernel
\[
\left(\frac{\partial}{\partial\beta}-\frac{\partial^2}{\partial
x^2}+4+V(x) \right)K_K(x,y;\beta)=0 \quad , \quad
K_K(x,y;0)=\delta(x-y)
\]
can be written in the form
\[
K_K(x,y;\beta)=K_{K_0}(x,y;\beta)\cdot C_K(x,y;\beta)
\]
if $C_K(x,y;\beta)$ satisfies the transfer equation
\[
\left(\frac{\partial}{\partial\beta}+\frac{x-y}{\beta}\frac{\partial}{\partial
x}-\frac{\partial^2}{\partial x^2}+V(x) \right)C_K(x,y;\beta)=0
\quad , \,\, C_K(x,y;0)=1 \qquad ,
\]
and it is set to be unity at infinite temperature.

Solving the transfer equation as a power series in $\beta$
\[
C_K(x,y;\beta)=\sum_{n=0}^\infty \, c_n(x,y;K)\, \beta^n
\hspace{1.5cm} , \hspace{1.5cm} c_0(x,y;K)=1 \qquad ,
\]
the PDE equation becomes tantamount to the recurrence relations:
\[
(n+1)c_{n+1}(x,y;K)+(x-y)\frac{\partial c_{n+1}}{\partial
x}(x,y;K)+V(x)c_n(x,y;K)=\frac{\partial^2 c_n}{\partial x^2}(x,y;K)
\qquad .
\]
The high-temperature heat equation kernel for $K$ is thus given as:
\[
K_K(x,y;\beta)=\frac{e^{-4\beta}}{\sqrt{4\pi\beta}}\cdot
e^{-\frac{(x-y)^2}{4\beta}}\cdot \sum_{n=0}^\infty \, c_n(x,y;K)\,
\beta^n\qquad .
\]
We are actually interested in the trace of the heat kernel to find
the partition function for small $\beta$. The recurrence relations
become
\[
c_{n+1}(x,x;K)=\frac{1}{n+1}\left[^{(2)}C_n(x)-V(x)c_n(x,x;K)\right]
\]
when $\lim_{y\rightarrow x}$. To deal with this delicate limit, we
have introduced the following notation:
$^{(k)}C_n(x)=\lim_{y\rightarrow x}\frac{\partial^k
c_n(x,y;K)}{\partial x^k}$. Recall that
$^{(k)}C_0(x)=\lim_{y\rightarrow x}\frac{\partial^kc_0}{\partial
x^k}=\delta^{k0}$. We also need (obtained after differentiating the
first recurrence formula $k$-times) recurrence relations among
derivatives:
\[
^{(k)}C_n(x)={1\over
n+k}\left[^{(k+2)}C_{n-1}(x)-\sum_{j=0}^k\left(\begin{array}{c} k
\\ j
\end{array}\right)\frac{d^jV(x)}{d x^j}\cdot ^{(k-j)}C_{n-1}(x)\right]\qquad .
\]
The high-temperature asymptotic expansion of the partition function
reads:
\[
{\rm Tr}e^{-\beta K}=\frac{e^{-4\beta}}{\sqrt{4\pi\beta}}\cdot
\sum_{n=0}^\infty \, c_n(K)\, \beta^n \hspace{2cm} , \hspace{2cm}
c_n(K)=\lim_{L\rightarrow\infty}\int_{-\frac{mL}{2\sqrt{2}}}^{\frac{mL}{2\sqrt{2}}}
\hspace{-0.3cm} dx \, c_{n}(x,x;K)\qquad .
\]
Using the recurrence relations the $c_n(x,x;K)$ densities can be
found -they are the conserved charges of the KdV equation, see
Appendix II- and, via integration over the whole line, the kink
Seeley coefficients are obtained:
\[
c_0(K)=\lim_{L\rightarrow\infty}\frac{mL}{\sqrt{2}}\hspace{0.5cm},\hspace{0.5cm}
c_{n}(K)=\frac{2^{n+1}(1+2^{2n-1})}{(2 n-1)!!}\, , \hspace{0.5cm}
n\geq 1 \qquad .
\]

\subsection{The Mellin transform of the asymptotic expansion}

To obtain the generalized zeta function from the asymptotic
expansion of the partition function, the Mellin transform is split
into two integrals, inside and outside the convergence radius:
\[
\zeta^*_K(s)= \frac{1}{\Gamma(s)}\left[ {1\over\sqrt{4\pi}}\cdot
\sum_{n=0}^{\infty}c_n(K)\cdot\int_0^1 \, d\beta \,
\beta^{s+n-{3\over 2}}\cdot e^{-4\beta} + \int_1^\infty \, d\beta \,
\beta^{s-1} {\rm Tr}^* e^{-\beta K}-\int_0^1 \, d\beta \,
\beta^{s-1}\right]\qquad .
\]
On general grounds, it is possible to show that
\[
B_K(s)={1\over\Gamma(s)}\int_1^\infty \, d\beta \, \beta^{s-1}
{\rm Tr}^* e^{-\beta K} \qquad ,
\]
where the star means that the zero eigenvalue is not accounted
for, is an entire function of s.

Note, however, that the zero mode is included in the heat kernel
high-temperature expansion. Subtraction of the contribution of the
zero mode to $\zeta^*_K(s)$ coming from the high-temperature range
of the Mellin transform is a tricky affair. In fact,
\[
I={1\over\Gamma(s)}\cdot \int_0^1 \, d\beta \, \beta^{s-1}
\]
is an improper integral if ${\rm Re} s<0$. We define this integral
in the spirit of zeta function regularization as:
\[
\lim_{\varepsilon\rightarrow 0} \, {1\over\Gamma(s)}\cdot \int_0^1
\, d\beta \,
\beta^{s-1}e^{-\varepsilon\beta}=\lim_{\varepsilon\rightarrow 0}
\, {1\over\varepsilon^s}\cdot
{\gamma[s,\varepsilon]\over\Gamma(s)} \qquad .
\]
Near the physical value $\varepsilon=0$, the behavior of the Euler
incomplete Gamma function is such that the regularized integral is:
\[
\gamma[s,\varepsilon]\equiv{1\over s}\cdot \varepsilon^s -{1\over
s+1}\cdot \varepsilon^{s+1} \qquad \Rightarrow \qquad I^R={1\over
s\Gamma(s)}={1\over\Gamma(s+1)}\qquad .
\]
The zeta function regularization procedure directly provides the
value for the improper integral that would be obtained if the
divergent integral ${\rm I}$ had been renormalized by adding another
(related) divergent integral:
\[
I^R={1\over\Gamma(s)}\lim_{c\rightarrow 0}\left\{\int_c^1 \,
d\beta \, \beta^{s-1}+\int_{-\infty}^c \, d\beta \,
\beta^{s-1}\right\} \qquad .
\]

A similar strategy must be adopted for computing the zeta function
of the vacuum (Klein-Gordon) operator to find:
\begin{eqnarray*}
\zeta_{K_0}(s)&=&\frac{mL}{\sqrt{8\pi}}\cdot
\frac{1}{\Gamma(s)}\cdot \left[\int_0^1 \, d\beta \,
\beta^{s-{3\over 2}}e^{-4\beta}+\int_1^\infty \, d\beta \,
\beta^{s-{3\over 2}}e^{-4\beta}
\right]\\&=&\frac{mL}{\sqrt{8\pi}}\cdot \frac{1}{\Gamma(s)}\cdot
{1\over 4^{s-{1\over 2}}}\cdot \left[\gamma[s-{1\over
2},4]+\Gamma[s-{1\over 2},4]\right] \qquad .
\end{eqnarray*}
The incomplete Euler Gamma function $\gamma[s-{1\over 2},4]$ has
poles at $s-{1\over 2}=0,-1,-2,-3, \cdots $ but its complementary
$\Gamma[s-{1\over 2},4]$ is an  entire function of $s$.

By the same token the generalized zeta function of the kink
fluctuation operator reads:
\[
\zeta^*_K(s)= \frac{1}{\Gamma(s)}\left[ {1\over\sqrt{4\pi}}\cdot
\sum_{n=0}^{N_0}c_n(K)\cdot \frac{\gamma[s+n-{1\over
2},4]}{4^{s+n-{1\over 2}}}+{1\over\sqrt{4\pi}}\cdot
\sum_{n=N_0+1}^{\infty}c_n(K)\cdot \frac{\gamma[s+n-{1\over
2},4]}{4^{s+n-{1\over 2}}}-{1\over s}\right]+B_K(s)
\]
$\gamma[s+n-{1\over 2},4]$ has poles at $s+n-{1\over 2}=0,-1,-2,-3,
\cdots $ and a large but finite number $N_0$ is chosen to separate
the contribution of the high-order coefficients.
\[
b^{N_0}_K(s)={1\over\sqrt{4\pi}}\cdot
\sum_{n=N_0+1}^{\infty}c_n(K)\cdot \frac{\gamma[s+n-{1\over
2},4]}{4^{s+n-{1\over 2}}}
\]
is holomorphic, however, for ${\rm Re}s>-N_0-1$.

\subsection{The high-temperature one-loop kink mass shift formula}

Neglecting the (very small) contribution of the entire functions,
the kink Casimir energy becomes
\[
\Delta M_K^C \simeq \frac{\hbar}{2}\cdot \lim_{s\rightarrow -{1\over
2}}\left(\frac{2\mu^2}{m^2}\right)^s\cdot \mu \cdot
\frac{1}{\Gamma(s)}\cdot
\left[{1\over\sqrt{4\pi}}\sum_{n=1}^{N_0}c_n(K)\frac{\gamma[s+n-{1\over
2},4]}{4^{s+n-{1\over 2}}} -{1\over s}\right] \qquad ,
\]
i.e. the zero-point vacuum energy renormalization takes care of the
term coming from $c_0(K)$.

The other correction due to the mass renormalization counter-terms
can also be arranged into meromorphic and entire parts:
\[
\Delta M_K^R=-\frac{\hbar \mu}{2\sqrt{4\pi}}\cdot c_1(K) \cdot
\lim_{s\rightarrow -{1\over
2}}\left(\frac{2\mu^2}{m^2}\right)^{s+{1\over 2}}\cdot
\frac{1}{4^{s+{1\over 2}}\Gamma(s)}\cdot  \left[\gamma[s+{1\over
2},4]+\Gamma[s+{1\over 2},4] \right]
\]
The mass renormalization term exactly cancels the $c_1(K)$
contribution. Our minimal subtraction scheme fits the following
renormalization prescription: for theories with only massive
fluctuations, the quantum corrections should vanish at the limit
where all the masses go to infinity.

We end with the high-temperature one-loop kink mass shift formula:
\[
\Delta M_K=-\frac{\hbar
m}{4\sqrt{2\pi}}\cdot
\left[\frac{1}{\sqrt{4\pi}}\cdot\sum_{n=2}^{N_0}\, c_n(K)\cdot
\frac{\gamma[n-1,4]}{4^{n-1}}+2\right]\qquad .
\]
A good test for the appropriate regularization prescription chosen
for the zero mode subtraction is the value
\[
\bigtriangleup M_K^{(0)}=-\frac{\hbar m}{2\sqrt{2\pi}}=-0.199471
\hbar m \qquad ,
\]
in perfect agreement (up to a factor ${\hbar\over\sqrt{2}}$ due to
different conventions) with the result obtained by Glauber, Comtet,
and Cahill, in 1976.

\subsection{Mathematica calculations}

Computational limitations impose a practical bound on the choice of
$N_0$. Knowledge of, say, ${}^{(0)}C_2$ requires computation of 9
densities:
\[
\begin{array}{ccccc} {}^{(4)}C_0 & {}^{(3)}C_0 & {}^{(2)}C_0 & {}^{(1)}C_0& {}^{(0)}C_0 \\
 & & {}^{(2)}C_1 & {}^{(1)}C_1 & {}^{(0)}C_1 \\ & & & & {}^{(0)}C_2
 \end{array} \qquad \qquad .
\]
In general, the evaluation of ${}^{(0)}C_n(x)$ requires previous
calculation of
\[
1+3+5+7+\cdots + 2n-1+2n+1=(n+1)^2
\]
${}^{(k)}C_0(x)$ densities. In fact, because ${}^{(0)}C_0=1$ and
${}^{(1)}C_0={}^{(2)}C_0=\cdots={}^{(2n)}C_0=0$ there would be a
need to compute only $n^2$ coefficients, but the computer ignores
this circumstance.

Observe that
\[
\Delta M_K \cong  -0.199471 \hbar m+D_{N_0} \hbar m \qquad \qquad
, \qquad \qquad  D_{N_0}=- \sum_{n=2}^{N_0} c_n( K)\,
\frac{\gamma[n-1,4]}{8 \sqrt{2} \pi\, 4^{n-1} }
\]
is far from the exact result without adding $D_{N_0} \hbar m$:
$\Delta M_K= -0.471113 \hbar m $.

In the following Table Mathematica, calculations of the Seeley
coefficients and partial sums $D_{N_0}$ are shown up to $N_0=10$
\begin{center}
\begin{tabular}{|c|c|c|c|}
\\[-0.5cm] \hline
$n$ & $c_n (K)$ & $N_0$ & $D_{N_0}$ \\ \hline 2 & 24.0000 & 2 &
-0.165717 \\ 3 & 35.2000 & 3 &-0.221946 \\ 4 & 39.3143 & 4
&-0.248281  \\ 5 & 34.7429 & 5 &-0.261260  \\ 6 & 25.2306 & 6 &
-0.267436  \\ 7 & 15.5208 & 7 & -0.270186  \\ 8 & 8.27702 & 8 &
-0.271317 \\ 9 & 3.89498 & 9 & -0.271748 \\ 10 & 1.63998 & 10
&-0.271900 \\ \hline
\end{tabular} \qquad \qquad ,
\end{center}
giving the very good result: $D_{10}=-0.271900 \hbar m$. Finally,
we find
\[
\Delta M_K \cong -0.471371 \hbar m \qquad ,
\]
with an error with respect to the DHN result of $0.0002580 \hbar m$.

\begin{eqnarray*}
&& \frac{\hbar m}{2}[B_{K} (-{\textstyle\frac{1}{2}})-B_{
K_0}(-{\textstyle\frac{1}{2}})]+\frac{3 \hbar m}{\sqrt{2} }
B_{K_0}({\textstyle\frac{1}{2}}) \\ &&=\frac{\hbar m}{2
\sqrt{2\pi}} \int_1^\infty d\beta \left( -\frac{e^{-3 \beta}}{2
\beta^{\frac{3}{2}}}+\frac{e^{-3\beta} {\rm Erfc \sqrt{\beta}}}{2
\beta^{\frac{3}{2}}}+\frac{{\rm
Erfc}\,2\sqrt{\beta}}{2\beta^{\frac{3}{2}}}+\frac{3e^{-4\beta}}{\sqrt{\pi}\beta}
\right)\\&& \approx  0.00032792 \hbar m
\end{eqnarray*}
is almost the total error. The deviation from the total error is
$-\frac{\hbar m}{4\sqrt{2\pi}}b_K^{N_0}(-\frac{1}{2})
 \approx 10^{-4} \hbar m\,$.

\section{{\bf The BNRT-model on a line }}

In this Section we analyze a model in (1+1)-dimensional ${\mathbb
R}^{1,1}$ space-time of two scalar fields with dynamics governed by
the action functional:
\[
S=\int dy^2 \left\{{1\over
2}\sum_{a=1}^2\frac{\partial\psi_a}{\partial
y^\mu}\frac{\partial\psi_a}{\partial y_\mu}-\left({\lambda\over
2}(\psi_1^2-{m^2\over\lambda})^2+\frac{\nu^2}{8}\psi_2^4+
\nu\sqrt{\lambda}\psi_2^2(\psi_1^2-{m^2\over\lambda})\right)\right\}\qquad
.
\]
This model was originally discussed by Bazeia, Nascimento, Ribeiro,
and Toledo (BNRT). The main novelty with respect to the model
studied in the previous Section is that there are two real scalar
fields in this system that can be assembled into a \lq\lq isospin"
vector field:
\[
\vec{\psi}(y_0,y)=\sum_{a=1}^2\,\psi_a(y_0,y)\vec{e}_a:{\mathbb
R}^{1,1}\rightarrow \, {\mathbb R}^2 \qquad , \qquad
\vec{e}_a\cdot\vec{e}_b=\delta_{ab} \quad ,\quad a,b=1,2 \quad ,
\]
where $\vec{e}_1,\vec{e}_2$ form a orthonormal basis in the target
space ${\mathbb R}^2$. The dimensions of the fields and the
coupling constants are respectively: $[\psi_a]=M^{{1\over
2}}L^{{1\over 2}}$, $[\lambda]=[\nu^2]=M^{-1}L^{-3}$, and
$[m]=L^{-1}$. In terms of non-dimensional space-time coordinates,
fields, and parameters
\[
y^\mu={1\over m}\cdot x^\mu \hspace{1.5cm} ; \hspace{1.5cm}
\psi_a(y^\mu)=2{m\over\sqrt{\lambda}}\cdot \phi_a(x^\mu)\qquad ;
\qquad \sigma^2=\frac{\nu^2}{\lambda} \qquad ,
\]
the action functional and the field equations of the BNRT model
read:
\[
S={4m^2\over\lambda}\int \, dx^2 \, \left\{{1\over
2}\sum_{a=1}^2\frac{\partial\phi_a}{\partial
x^\mu}\frac{\partial\phi_a}{\partial x_\mu}-{1\over
8}(4\phi_1^2+2\sigma\phi_2^2-1)^2-2\sigma^2\phi_1^2\phi_2^2\right\}
\]
\[
\frac{\partial^2\phi_1}{\partial
x_0^2}(x_0,x)-\frac{\partial_1^2\phi_1}{\partial
x^2}(x_0,x)=2\phi_1(x_0,x)(1-2\sigma(\sigma+1)\phi_2^2(x_0,x)-4\phi_1^2(x_0,x))
\]
\[
\frac{\partial^2\phi_2}{\partial
x_0^2}(x_0,x)-\frac{\partial_2^2\phi_2}{\partial
x^2}(x_0,x)=\sigma\phi_2(x_0,x)(1-2\sigma\phi_2^2(x_0,x)-4(\sigma+1)\phi_1^2(x_0,x))\qquad .
\]
There are four homogeneous stable solutions:
\[
1.  \quad \vec{\phi}^{\,\pm(1)}(x_0,x)=\pm {1\over
2}\cdot\vec{e}_1 \hspace{3cm} 2. \quad
\vec{\phi}^{\,\pm(2)}(x_0,x)=\pm\frac{1}{\sqrt{2\sigma}}\cdot\vec{e}_2
\qquad .
\]
To quantize, we choose the $\vec{\phi}^{\,+(1)}$ vacuum (because it
is the asymptotic value of the generic kinks of the model) and shift
the fields from it
\[
\phi_1(x^\mu)={1\over 2}+H(x^\mu)\hspace{1cm} , \hspace{1cm}
\phi_2(x^\mu)=G(x^\mu)
\]
to write the action in the form:
\begin{eqnarray*}
S&=&\frac{4m^2}{\lambda}\int \, d^2x \, \left\{{1\over
2}\partial_\mu H\partial^\mu H -2H^2(x^\mu)\right\}+\left\{{1\over
2}\partial_\mu G\partial^\mu G -{\sigma^2\over 2}G^2(x^\mu)
\right\}-\\&-& \frac{4m^2}{\lambda}\int \, d^2x
\,\left\{4H^3(x^\mu)+2\sigma(\sigma+1)H(x^\mu)G^2(x^\mu)\right\}-\\&-&
\frac{4m^2}{\lambda}\int \, d^2x \,\left\{
2H^4(x^\mu)+2\sigma(\sigma+1)H^2(x^\mu)G^2(x^\mu)+{\sigma^2\over
2}G^4(x^\mu)\right\} \qquad ,
\end{eqnarray*}
showing the spontaneous symmetry breaking of the ${\mathbb
Z}_2\times{\mathbb Z}_2$ vierergroup generated by the internal
reflections $\phi_1\rightarrow -\phi_1$ and $\phi_2\rightarrow
-\phi_2$. The Feynman rules are thus obtained in terms of Higgs
and (pseudo) Goldstone propagators and three-valent and
four-valent vertices:
\begin{table}[h]
\begin{center}
\caption{Propagators}
\begin{tabular}{lccc} \\ \hline
\textit{Particle} & \textit{Field} & \textit{Propagator} & \textit{Diagram} \\
\hline \\ Higgs & $H(x)$ & $\displaystyle\frac{i \lambda \hbar
}{4m^2(k_0^2-k^2-4+i\varepsilon)}$ &
\parbox{2cm}{\epsfig{file=qsdv1.ps,width=2cm}}
\\[0.5cm]
Goldstone & $G(x)$ & $\displaystyle\frac{i\lambda
\hbar}{4m^2(k_0^2-k^2-\sigma^2+i\varepsilon)}$ &
\parbox{2cm}{\epsfig{file=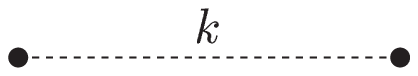,width=2cm}} \\[0.5cm] \hline
\end{tabular}
\end{center}
\end{table}

\begin{table}[hbt]
\begin{center}
\caption{Third- and fourth-order vertices }
\begin{tabular}{clcl} \\ \hline
\textit{Vertex} & \textit{Weight} & \textit{Vertex} & \textit{Weight} \\
\hline \\
\parbox{2.3cm}{\epsfig{file=qsdv5.ps,width=2cm}} &
$\displaystyle -96i\frac{m^2}{\hbar \lambda} $  &
\parbox{2.3cm}{\epsfig{file=qsdv6.ps,width=2cm}} &
$\displaystyle -192i\frac{m^2}{\hbar \lambda}$ \\[0.5cm]
\parbox{2.3cm}{\epsfig{file=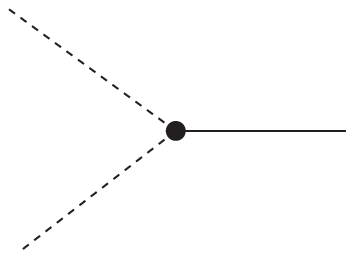,width=1.7cm}} &
$\displaystyle -16\sigma(\sigma+1)i\frac{m^2}{\lambda\hbar }$  &
\parbox{2.3cm}{\epsfig{file=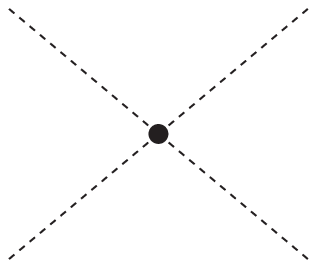,width=1.7cm}} &
$\displaystyle -48\sigma^2i\frac{m^2}{\lambda\hbar }$\\[0.5cm]
 & & \parbox{2.3cm}{\epsfig{file=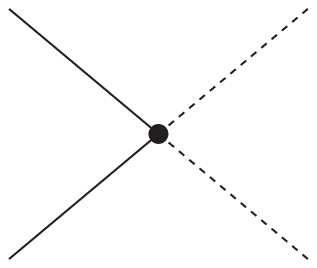,width=1.7cm}} &
$\displaystyle -32\sigma(\sigma+1)i\frac{m^2}{\lambda \hbar }$  \\[0.7cm] \hline
\end{tabular}
\end{center}
\end{table}

\subsection{Plane waves and vacuum energy}

The general solution of the linearized field equations
\[
\frac{\partial^2\delta H}{\partial
x_0^2}(x_0,x)-\frac{\partial^2\delta H}{\partial x^2}(x_0,x)+4\delta
H(x_0,x)=0
\]
\[
\frac{\partial^2\delta G}{\partial
x_0^2}(x_0,x)-\frac{\partial^2\delta G}{\partial
x^2}(x_0,x)+\sigma^2\delta G(x_0,x)=0
\]
governing the small fluctuations of the Higgs and (pseudo)Goldstone
fields is:
\begin{eqnarray*}
\delta H(x_0,x)&=&\frac{\sqrt{\lambda}}{2m}\cdot\sqrt{{\hbar\over
mL}}\sum_k{1\over\sqrt{2\omega(k)}}
\left\{a_1(k)e^{-ik_0x_0+ikx}+a_1^*(k)e^{ik_0x_0-ikx}\right\}
\\\delta
G(x_0,x)&=&\frac{\sqrt{\lambda}}{2m}\cdot\sqrt{\frac{\hbar}{mL}}\sum_q{1\over\sqrt{2\gamma(q)}}
\left\{a_2(q)e^{iq_0x_0-iqx}+a_2^*(q)e^{-iq_0x_0+iqx}\right\}
\end{eqnarray*}
where $k_0=\omega(k)=\sqrt{k^2+4}$,
$q_0=\gamma(q)=\sqrt{q^2+\sigma^2}$, and the dispersion relations
$k_0^2-k^2-4=0$, $q_0^2-q^2-\sigma^2=0$ hold:
\[
K_0\left(\begin{array}{c} e^{ikx} \\ o
\end{array}\right)=\omega^2(k)\left(\begin{array}{c} e^{ikx} \\ o
\end{array}\right)\quad , \quad K_0\left(\begin{array}{c} 0 \\e^{iqx}
\end{array}\right)=\gamma^2(q)\left(\begin{array}{c} 0 \\ e^{iqx}
\end{array}\right) \quad , \quad K_0=\left(\begin{array}{cc} -{d^2\over dx^2}+4 & 0 \\ 0 & -{d^2\over
dx^2}+\sigma^2\end{array} \right) \quad .
\]
We choose a normalization interval of non-dimensional \lq\lq length"
$mL$, $I=[-\frac{mL}{2},\frac{mL}{2}]$, and impose PBC on the plane
waves so that: $k\cdot mL=2\pi n$ , $q\cdot mL=2\pi r$ with
$n,r\in{\mathbb Z}$. Thus, the spectral density of $K_0$
is:
\[
\rho_{K_0}(k)={\rm tr}\left(\begin{array}{cc} \frac{dn}{dk} & 0
\\ 0 & \frac{dr}{dk} \end{array}\right)={1\over \pi}mL \qquad .
\]
This is tantamount to considering the $d=1$, $N=2$ case of Section
\S.2.

From the classical free Hamiltonian
\begin{eqnarray*}
H^{(2)}&=&{m^3\over\lambda}\int \, dx \, \left\{{1\over
2}\left(\frac{\partial\delta H}{\partial
x_0}\cdot\frac{\partial\delta H}{\partial x_0}+{1\over
2}\frac{\partial\delta H}{\partial x}\cdot\frac{\partial\delta
H}{\partial x}\right)+{1\over 2}\left(\frac{\partial\delta
G}{\partial x_0}\cdot\frac{\partial\delta G}{\partial x_0}+{1\over
2}\frac{\partial\delta G}{\partial x}\cdot\frac{\partial\delta
G}{\partial x}\right)\right.\\&+&\left. 2\delta H\cdot\delta
H+{\sigma^2\over 2}\delta G\cdot\delta G\right\}\\&=&\sum_k \, \hbar
{m\over 2}\left[
\omega(k)(a_1^*(k)a_1(k)+a_1(k)a_1^*(k))+\gamma(k)(a_2^*(k)a_2(k)+a_2(k)a_2^*(k))\right]\qquad
.
\end{eqnarray*}
One goes to the quantum free Hamiltonian:
\[
\hat{H}^{(2)}_0=\sum_k \, \hbar m\left[
\omega(k)\left(\hat{a}_1^\dagger(k)\hat{a}_1(k)+{1\over
2}\right)+\gamma(k)\left(\hat{a}_2^\dagger(k)\hat{a}_2(k)+{1\over
2}\right)\right]
\]
via canonical quantization:
$[\hat{a}_b(k),\hat{a}^\dagger_c(q)]=\delta_{bc}\delta_{kq}$. The
vacuum energy is:
\[
\Delta E_0={\hbar m\over 2}\sum_k\omega(k)+{\hbar m\over
2}\sum_k\gamma(k)={\hbar m\over 2}{\rm Tr} K_0^{{1\over 2}}
\]

\subsection{One-loop mass renormalization counter-terms}

There are three ultraviolet divergent graphs in one-loop order of the
$\hbar$-expansion contributing to:
\begin{itemize}
\item The Higgs boson tadpole:
\begin{eqnarray*}
&&-12i\cdot I(4)-2\sigma(\sigma+1)i\cdot
I(\sigma^2)=\hspace{1cm}\parbox{3cm}{\epsfig{file=qsdv15.ps,width=3cm}}+
\parbox{3.5cm}{\epsfig{file=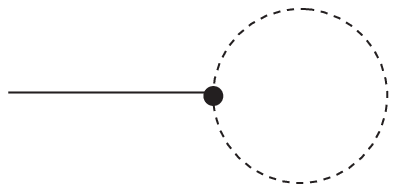,width=3.5cm}}\\&&-12i\cdot
\int \, \frac{d^2k}{(2\pi)^2} \, \cdot
\frac{i}{(k_0^2-k^2-4+i\varepsilon)}-2\sigma(\sigma+1)i\cdot \int \,
\frac{d^2k}{(2\pi)^2} \, \cdot
\frac{i}{(k_0^2-k^2-\sigma^2+i\varepsilon)}
\end{eqnarray*}

\item The Higgs boson self-energy

\begin{eqnarray*}
&&-24i\cdot I(4)-4\sigma(\sigma+1)\cdot
I(\sigma^2)=\hspace{1cm}\parbox{3cm}{\epsfig{file=qsdv19.ps,width=3cm}}+
\parbox{4cm}{\epsfig{file=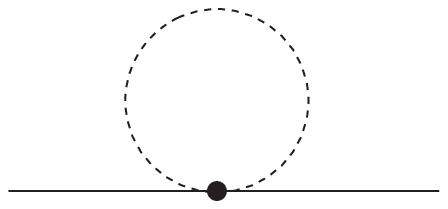,width=3cm}}\\&=&-24i\cdot
\int \,\frac{dk}{4\pi}\, \cdot
\frac{1}{\sqrt{k^2+4}}-4\sigma(\sigma+1)i\cdot \int
\,\frac{dk}{4\pi}\, \cdot \frac{1}{\sqrt{k^2+\sigma^2}}
\end{eqnarray*}
\item The Goldstone boson self-energy:

\begin{eqnarray*} &&-4\sigma(\sigma+1)i\cdot I(4)-6\sigma^2i\cdot
I(\sigma^2)=\hspace{1cm}\parbox{3cm}{\epsfig{file=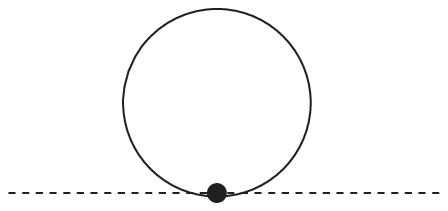,width=3cm}}
+
\parbox{3cm}{\epsfig{file=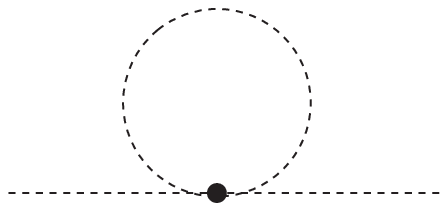,width=3cm}}\\&=&-4\sigma(\sigma+1)i\cdot \int \,\frac{dk}{4\pi}\,
\cdot \frac{1}{\sqrt{k^2+4}}-6\sigma^2i\cdot \int
\,\frac{dk}{4\pi}\, \cdot \frac{1}{\sqrt{k^2+\sigma^2}}\qquad .
\end{eqnarray*}
\end{itemize}
Due care is needed to take into account a combinatorial factor of
${1\over 2}$ in all these graphs. The Lagrangian density of
counter-terms:
\begin{eqnarray*} {\cal L}_{C.T.}&=&{\hbar\over 2}\left[6\cdot
I(4)+\sigma(\sigma+1)\cdot I(\sigma^2)\right]
\left(4\phi_1^2(x^\mu)-1\right)+\\&+&{\hbar\over 2}
\left[2(\sigma+1)\cdot I(4)+3\sigma \cdot
I(\sigma^2)\right]2\sigma\phi_2^2(x^\mu)\qquad ,
\end{eqnarray*}
giving the vertices in the next Table, must be added to exactly
cancel the divergences above.

\begin{table}[hbt]
\begin{center}
\caption{One-loop counter-terms }
\begin{tabular}{ccc} \\ \hline
\textit{Diagram} & \hspace{0.3cm} & \textit{Weight} \\
\hline
\parbox{4cm}{\epsfig{file=qsdv30.eps,width=4cm}} & &
$\displaystyle 2i(6I(4)+\sigma(\sigma+1)I(\sigma^2))$
\\
\parbox{4cm}{\epsfig{file=qsdv31.eps,width=4cm}} & &
$\displaystyle 4i(6I(4)+\sigma(\sigma+1)I(\sigma^2))$ \\
 \parbox{4cm}{\epsfig{file=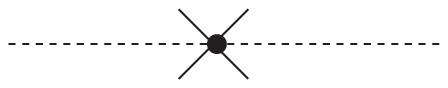,width=4cm}} & &
$\displaystyle 2i\sigma (2(\sigma+1)I(4)+3\sigma I(\sigma^2))$ \\[0.5cm] \hline \\[0.25cm]
\end{tabular}
\end{center}
\end{table}

\subsection{Moduli space of BNRT kinks}

The configuration space of the classical BNRT model
\[
{\cal C}=\{\phi_a(x)\in {\rm Maps}({\mathbb R},{\mathbb R}^2)/
E[\phi_a]<+\infty\}
\]
is non-connected:
\[
{\cal C}={\cal C}^{I\, J}_{+\,+}\sqcup {\cal C}^{I\, J}_{-\,-}\sqcup
{\cal C}^{I\, J}_{+\,-}\sqcup {\cal C}^{I\, J}_{-\,+} \qquad ,
\qquad \,\,  \,\,  I,J=1,2 \qquad .
\]
The energy for time-independent configurations
\[
E={4m^3\over\lambda}\int \, dx \, \left\{{1\over
2}\sum_{a=1}^2\frac{d\phi_a}{dx}\cdot\frac{d\phi_a}{dx}+{1\over
8}(1-2\sigma\phi_2^2-4\phi_1^2)^2+2\sigma^2\phi_1^2\phi_2^2
\right\}
\]
is finite if and only if
\[
\lim_{x\rightarrow\pm\infty}\frac{d\vec{\phi}}{dx}=0 \qquad ,
\qquad
\lim_{x\rightarrow\pm\infty}\vec{\phi}(x)=\left\{\begin{array}{c}
\vec{\phi}^{+(I)}
\\ \vec{\phi}^{-(J)} \end{array}\right.
\]
and the sixteen non-connected components of ${\cal C}$ are
classified by the behavior of the scalar field at
$x\rightarrow\pm\infty$.

\noindent\begin{figure}[htbp] \centerline{\epsfig{file=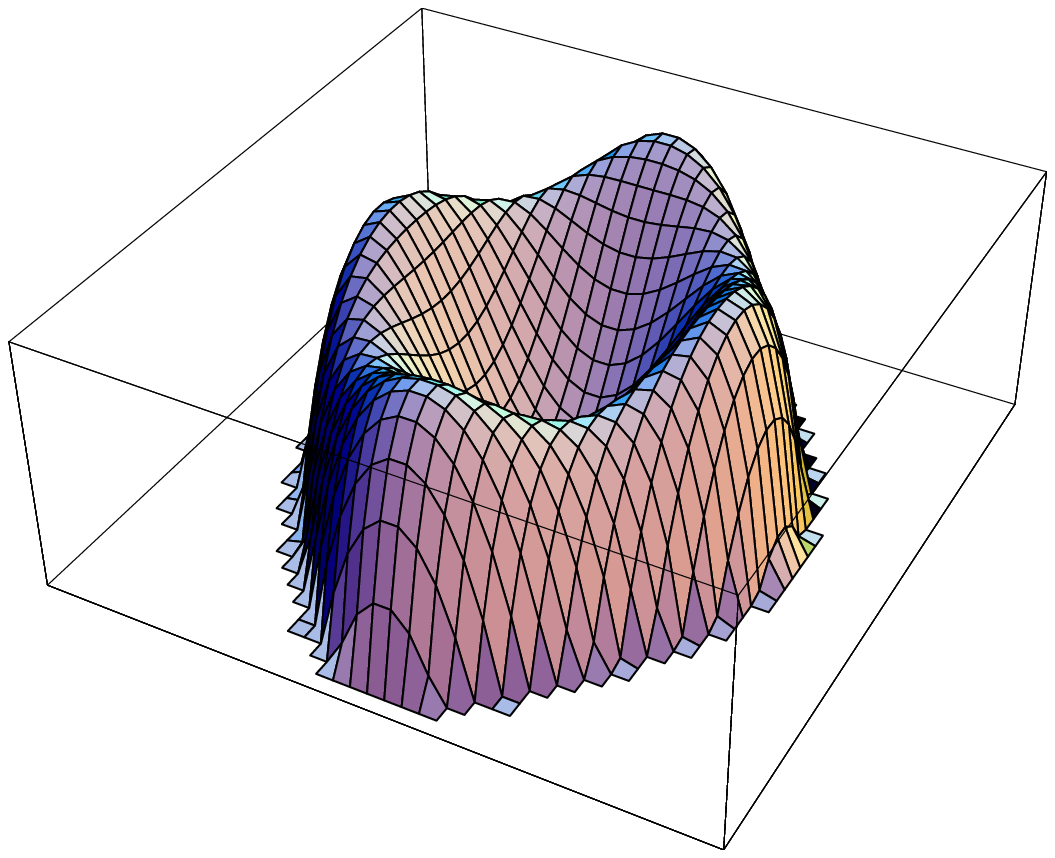,
height=4cm} \hspace{1.2cm} \epsfig{file=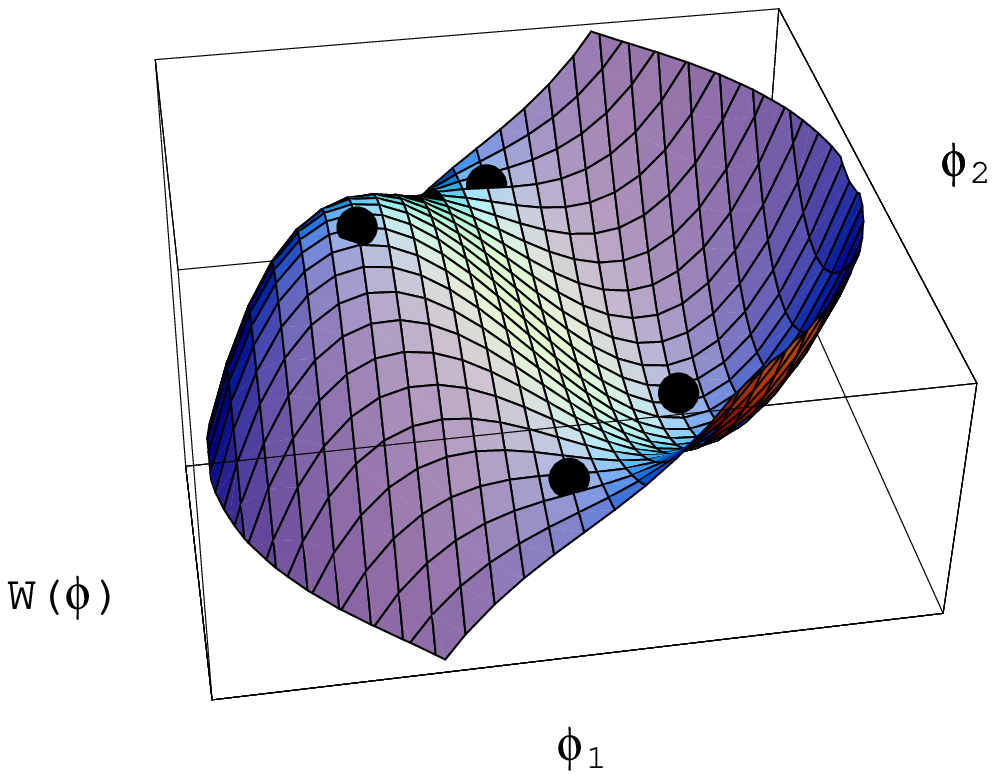,height=4cm}}
\caption{\small \it The $-U(\phi_1,\phi_2)$ potential (left) and
the superpotential $W(\phi_1,\phi_2)$ (right)}
\end{figure}
The existence of the superpotential
\[
W(\phi_a)=2\left({1\over 3}\phi_1^3-{1\over 4}\phi_1+{\sigma\over
2}\phi_1\phi_2^2 \right) \qquad , \qquad U(\phi_1,\phi_2)={1\over
2}\sum_{a=1}^2\, \frac{\partial
W}{\partial\phi_a}\cdot\frac{\partial W}{\partial\phi_a}
\]
allows for the Bogomolny splitting of the static energy:
\begin{equation}
E={4m^3\over\lambda}\int \, dx \, {1\over
2}\left[\sum_{a=1}^2\left(\frac{d\phi_a}{dx}\mp\frac{\partial
W}{\partial\phi_a}\right)\left(\frac{d\phi_a}{dx}\mp\frac{\partial
W}{\partial\phi_a}\right)\right] \pm {4m^3\over\lambda}\cdot
\left\{W(\phi_a(+\infty))-W(\phi_a(-\infty))\right\}\qquad ,
\label{eq:grf}
\end{equation}
showing that the absolute minima of $E$ in each sector of ${\cal
C}$ satisfy the first-order equations:
\begin{equation}
{d\phi_a\over dx}=\pm\frac{\partial W}{\partial\phi_a}\qquad ,
\qquad \left\{\begin{array}{c} \qquad
\displaystyle\frac{d\phi_1}{dx}=(-1)^\alpha
(2\phi_1^2+\sigma\phi_2^2-{1\over 2}) \\ \\
\hspace{-1cm}\displaystyle\frac{d\phi_2}{dx}=(-1)^\beta
2\sigma\phi_1\phi_2
\end{array}\right. \quad , \quad \alpha,\beta=0,1 \label{eq:grffo} \qquad .
\end{equation}

\subsubsection{Kink flow lines}

 The solutions of (\ref{eq:grffo}) are the flow lines of the gradient of $W$, and those starting and ending
 at critical points of $W$ are the kink orbits.

\begin{figure}[htbp]
\centerline{\epsfig{file=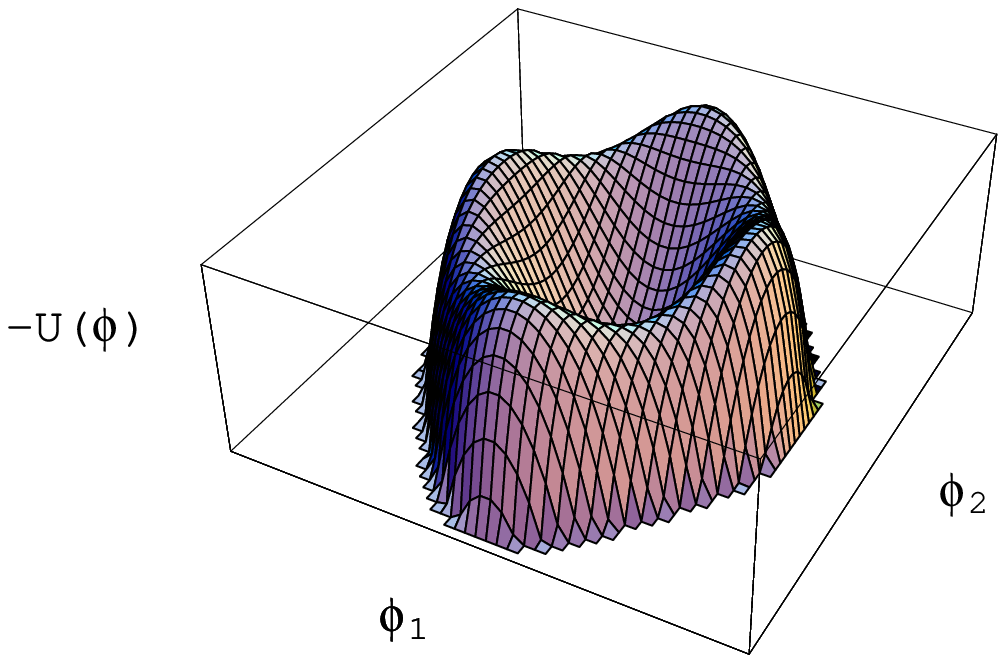, height=3cm}
\hspace{1cm}\epsfig{file=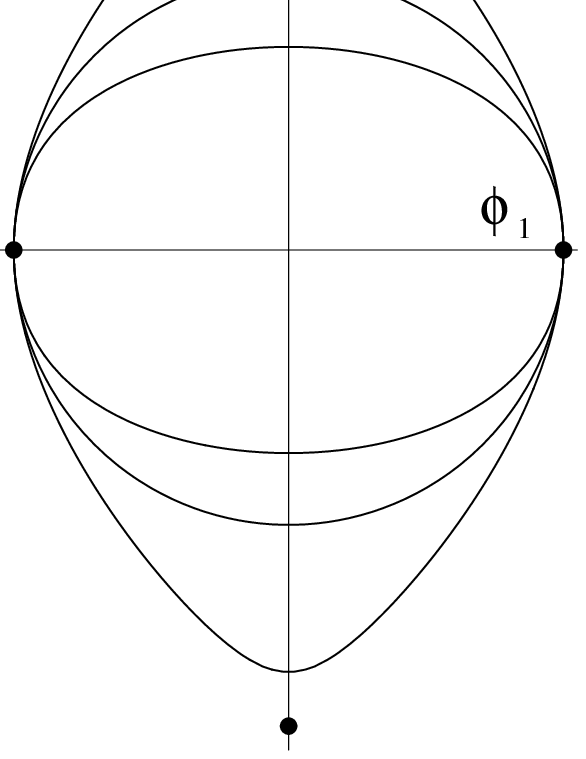, height=3cm} \hspace{1cm}
\epsfig{file=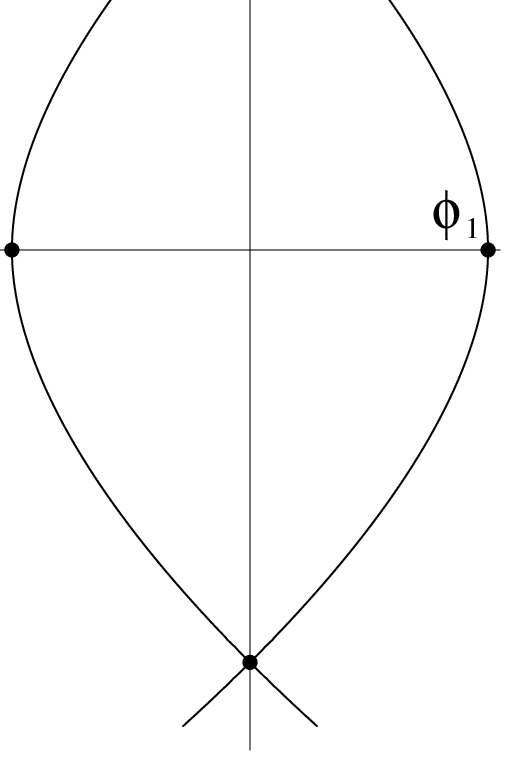,height=3cm}
\hspace{1cm}\epsfig{file=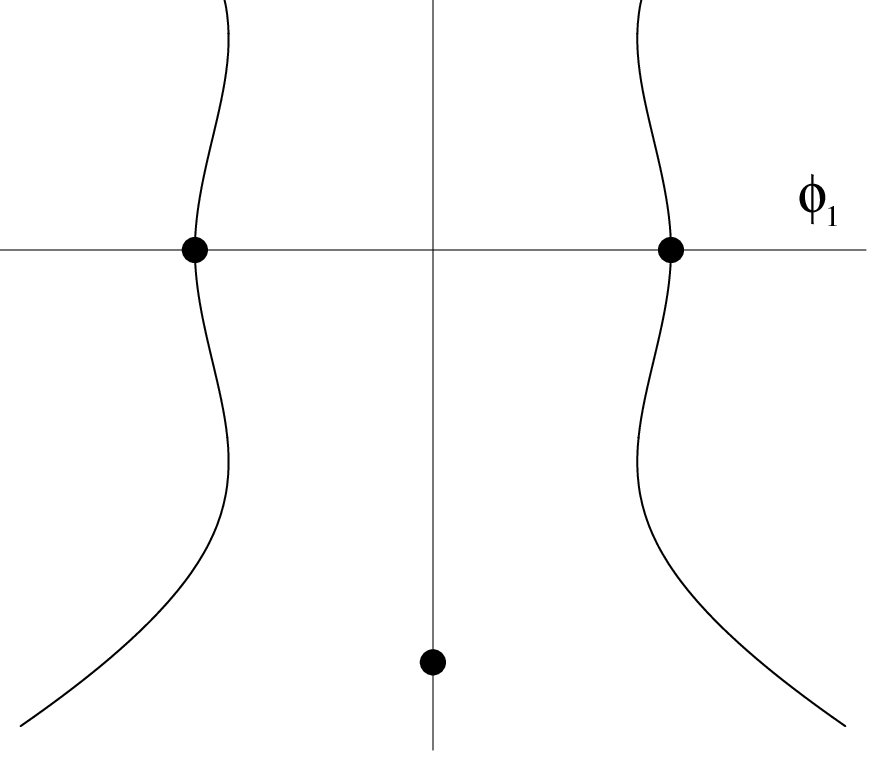, height=3cm}}
\caption{\small \it The $-U(\phi)=-{1\over 2}{\partial W\over
\partial \phi_1}{\partial W\over
\partial \phi_1} -{1\over 2}{\partial W\over
\partial \phi_2}{\partial W\over
\partial \phi_2}$ potential (left). Flow-lines: in the ranges $c
\in (-\infty,c^S)$ (middle left), $c=c^S$ (middle right), and $c
\in (c^S ,\infty)$ (right)}.
\end{figure}
The kink flow lines can be obtained analytically by integrating
\[
(-1)^\alpha\frac{d\phi_1}{2\phi_1^2+\sigma\phi_2^2-{1\over
2}}+(-1)^\beta\frac{d\phi_2}{2\sigma\phi_1\phi_2}=0 \qquad ,
\]
to find, after use of the integrating factor
$|\phi_2|^{-{2\over\sigma}}$,
\[
\phi_1^2 + \frac{\sigma}{2(1-\sigma)} \phi_2^2 = \frac{1}{4}+
\frac{c}{2\sigma} |\phi_2|^{2\over\sigma} \hspace{1cm} , \,
\sigma\neq 1 \, , \hspace{1cm} c\in (-\infty, {1\over
4}{\sigma\over 1-\sigma}(2\sigma)^{{1+\sigma\over\sigma}})
\]
\[
\phi_1^2-\phi_2^2\left({c\over 2}+\ln|\phi_2| \right)={1\over 4}
\hspace{1cm}, \,\sigma=1 \, , \hspace{1cm} c\in (-\infty,-1+\ln
2)\qquad .
\]
Thus, there exits a family of two-component topological kinks -TK2
kinks- parametrized by the integration constant $c$, all of them
with the same energy:
\[
E(\phi_1^{\pm(1)},\phi_2^{\pm(1)})=0 \qquad , \qquad
E(\phi_1^K,\phi_2^K)=\frac{4m^3}{\lambda}|W(\phi_1^{\pm(1)},\phi_2^{\pm(1)})-W(\phi_1^{\mp(1)},\phi_2^{\mp(1)})|=\frac{4m^3}{3\lambda}\qquad .
\]
Note that there is a critical value $c^S={1\over 4}{\sigma\over
1-\sigma}(2\sigma)^{{1+\sigma\over\sigma}}$ if $\sigma\neq 1$, or,
$c^S=-1+\ln 2$, if $\sigma=1$ beyond which the flow lines go to
infinity and do not correspond to kink orbits because
$W(|\vec{\phi}|(\infty))=\infty$ and the energy becomes infinite.

In general, the kink profile, the dependence on $x$ of the fields
for a given kink orbit, cannot be expressed analytically. There are,
however, two special kink orbits for which this is possible:
\begin{eqnarray*} && c=-\infty \quad , \quad
\left\{\begin{array}{c}\phi_1^{{\rm TK1}}(x)=(-1)^{\alpha}{1\over
2}{\rm tanh}(x-a)\\ \phi_2^{{\rm TK1}}(x)=0\end{array}\right.\quad ;
\,\, \alpha=0,1 \, , \, a\in {\mathbb R}\\&&\\&&\\&& c=0 \quad ,
\quad \left\{\begin{array}{c}\phi_1^{{\rm
TK2}}(x)=(-1)^{\alpha_1}{1\over 2}{\rm tanh}[2(1-\sigma)(x-a)]
\\ \phi_2^{{\rm TK2}}(x)=(-1)^{\alpha_2}\sqrt{{1-\sigma\over
\sigma}}{\rm sech}[2(1-\sigma)(x-a)]\end{array}\right. \qquad ,
\qquad \alpha_1,\alpha_2=0,1 \qquad .
\end{eqnarray*}
If $c=-\infty$, the orbit is a segment on the abscissa axis and one
finds the very well known $\lambda(\phi)^4_2$ kink buried in the
kink variety of the BNRT model. Since only one of the two components
of the scalar field is different from zero, this kind of kink is
termed TK1 kinks -one-component topological kinks- in the context of
the BNRT model.

When $c=0$ things become more interesting. The kink orbit is a
half-ellipse (if $\sigma\neq 1$) and the two components of the
scalar field are not zero for this kind of topological -TK2- kink.
Note that in both cases the kink profile is a function of a real
parameter $a\in{\mathbb R}$, obeying the freedom of setting the kink
center. This is a general feature shared with other kinks for which
analytical expressions giving the kink profiles are not available.
Therefore, the moduli space of kinks in the BNRT model is the
two-dimensional manifold ${\cal M}_K^2=(-\infty,c^S]\times
(-\infty,\infty)$ with coordinates $(c,a)$. That is, each kink in
the BNRT model is determined by its orbit ($c\in (-\infty,c^S]$) and
its center ($a\in
 (-\infty,\infty)$).

 We close this subsection by enumerating the main features of non-generic
 kink orbits.
 \begin{enumerate}

\item  $c=-\infty$: The second component of the scalar field is zero
and the kink orbit is invariant under the $\phi_2\rightarrow
-\phi_2$ transformation. These orbits belongs to the ${\cal
C}_\pm^{11}$ or ${\cal C}^{11}_\mp$ sectors.

\item  $c=0$: The kink orbits are half-ellipses also living in the
${\cal C}_\pm^{11}$ or ${\cal C}^{11}_\mp$ sectors.

\item  $c=c^S$: These orbits are the boundary between bounded and
unbounded flow lines. They belong to ${\cal C}_\pm^{12}$, ${\cal
C}^{12}_\mp$, ${\cal C}_\pm^{21}$ or ${\cal C}^{21}_\mp$ sectors.

 \end{enumerate}

\subsubsection{Kink profiles from integrable systems}

The search for finite energy static solutions in (1+1)-dimensional
theories of two scalar fields is tantamount to investigating finite
action trajectories in a mechanical system of two degrees of
freedom. The potential energy of the mechanical system is equal to
minus the potential energy density of the field  theory. The
mechanical analogue in this sense to the BNRT model is a Liouville
integrable system if $\sigma={1\over 2}$ and $\sigma={1\over 2}$.
Liouville systems are Hamilton-Jacobi separable and all the
trajectories can be found. Here we shall describe only the
$\sigma={1\over 2}$ case, where the HJ equation is separable using
elliptic coordinates. Standard application of the HJ procedure
provides the following formulas for all the kink
profiles{\footnote{There is an open problem related with the search
of kink profiles in this model: for $\sigma=3$, $\sigma={1\over 3}$,
$\sigma=4$, and $\sigma={1\over 4}$ the kink profiles can be
analytically expressed in terms of elliptic functions, although
these solutions are not given in the literature.}}:
\[
\phi_1^{{\rm TK}} [x;a,b]=(-1)^{{\alpha_1}}\left(\frac{1}{2}
\frac{\sinh\left(
 (x-a)\right)}{\cosh\left( (x-a)\right)+b^2}\right)\hspace{0.5cm} , \hspace{0.5cm} \phi_2^{{\rm TK}}
[x;a,b]=(-1)^{{\alpha_2}}\left(\frac{b}{\sqrt{b^2+\cosh\left((x-a)\right)}}\right)\qquad .
\]
The variety of kinks depends on two integration constants: $a,b\in
(-\infty,\infty)$. The second one,
$b=\pm\sqrt{\frac{1}{\sqrt{1-4c}}}$, is determined by the constant
$c\in (-\infty,{1\over 4})$, giving the kink orbit. In Figure 3 it
is observed that upon increasing $b$ a splitting into two kinks
arises.

\begin{figure}[htbp]
\centerline{ \epsfig{file=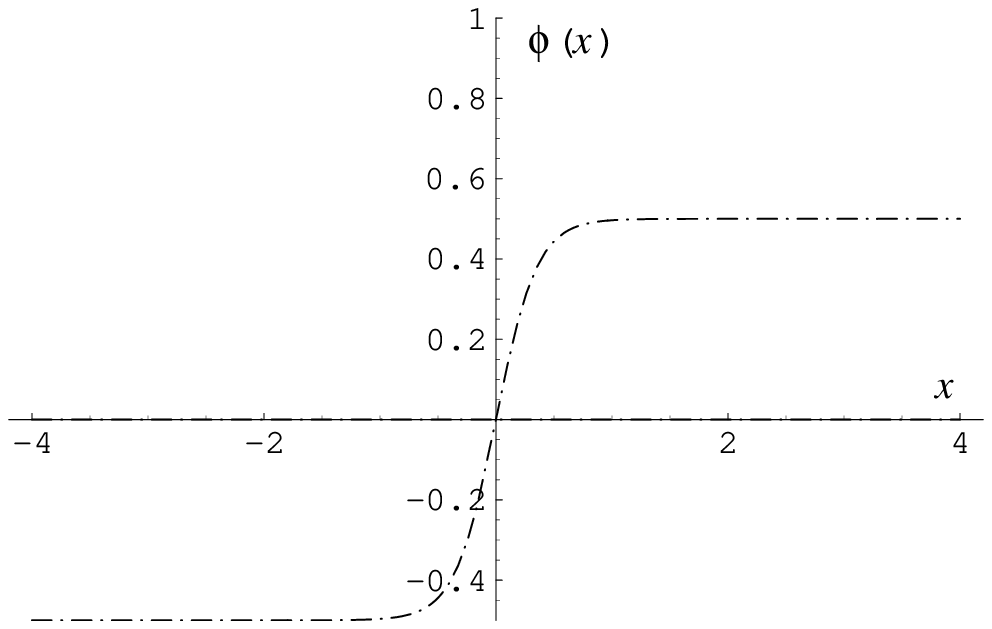, height=2.5cm}
\epsfig{file=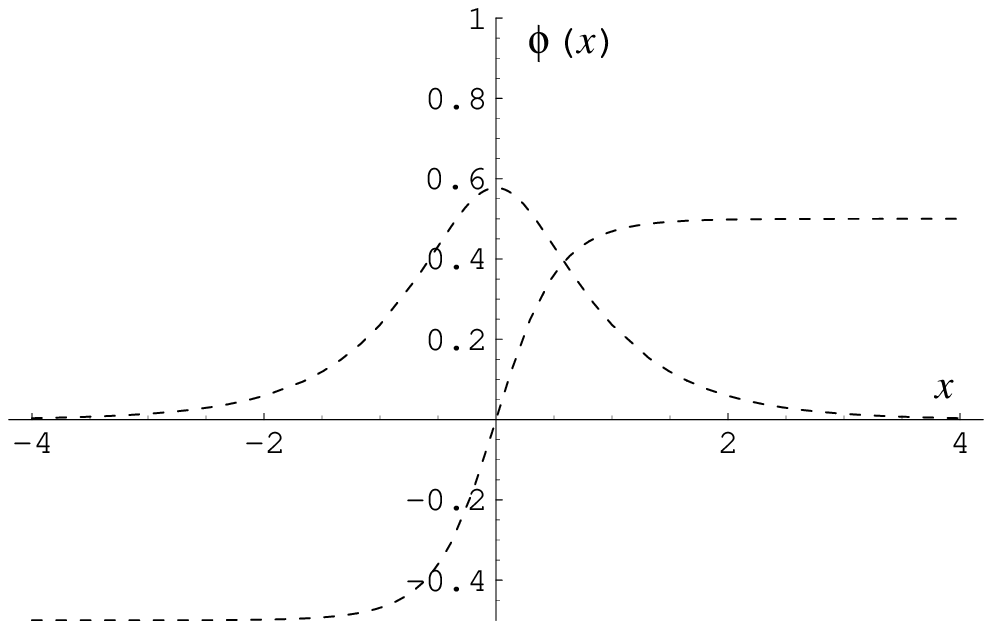, height=2.5cm} \epsfig{file=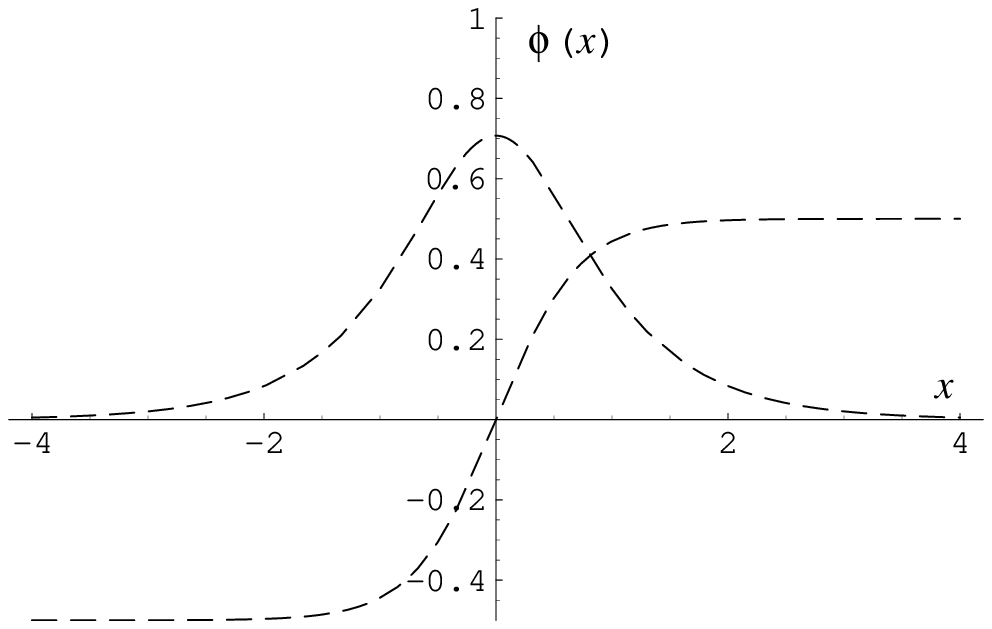,
height=2.5cm} \epsfig{file=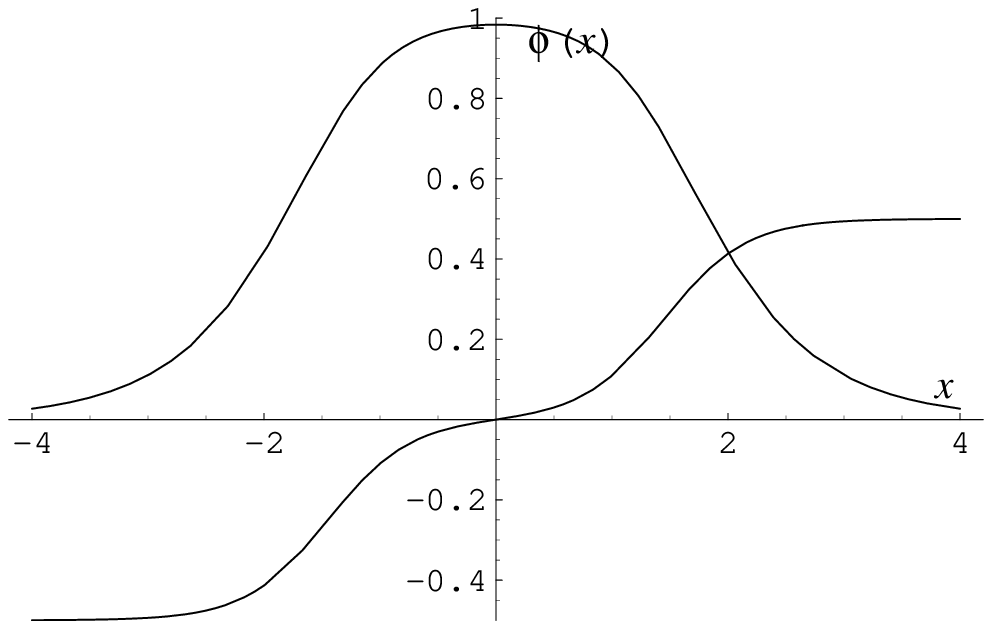, height=2.5cm}}
\caption{\small Solitary waves  corresponding to: {\it (a)} $b=0$,
{\it (b)} $b=\sqrt{0.5}$, {\it (c)} $b=1$ and {\it (d)}
$b=\sqrt{30}$.}
\end{figure}

This phenomenon is better understood by studying how the energy
density varies with $b$. The energy density
\[
{\cal E}^{\rm TK}[x;0,b]=\sum_{a=1}^2\frac{\partial\phi_a^{\rm
TK}}{\partial x}.\frac{\partial\phi_a^{\rm TK}}{\partial
x}=\frac{4+7b^2{\rm cosh}[x]+4b^4{\rm cosh}[2x]+b^2{\rm
cosh}[3x]}{2(b^2+{\rm cosh}[x])^4}
\]
is shown in Figure 4.

\begin{figure}[htbp]
\centerline{ \epsfig{file=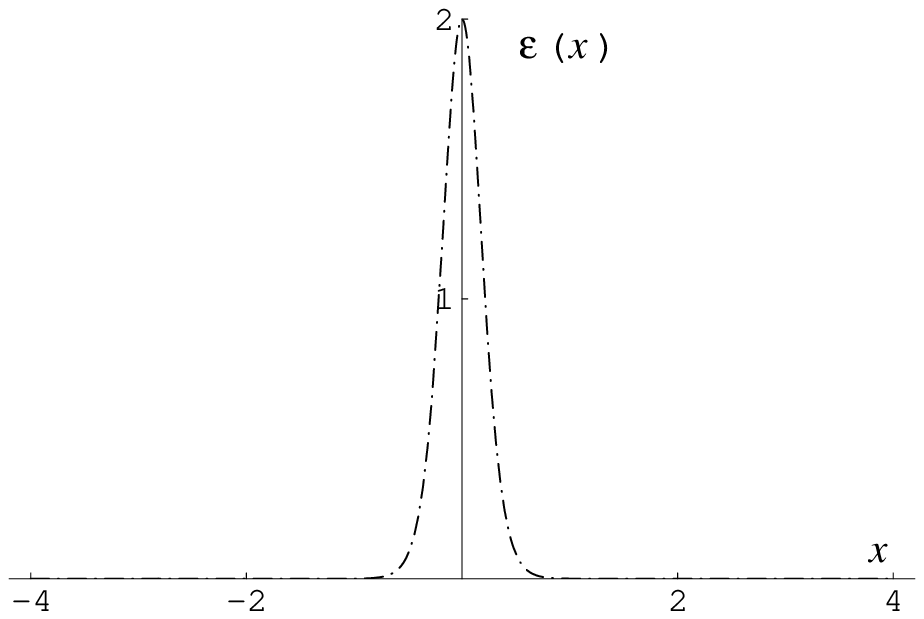, height=2.5cm}
\epsfig{file=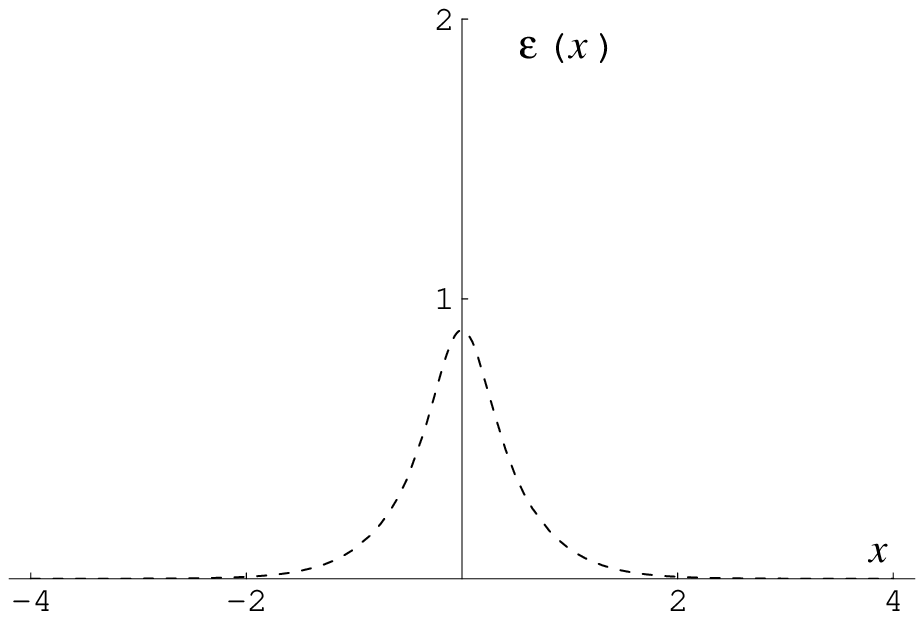, height=2.5cm}
\epsfig{file=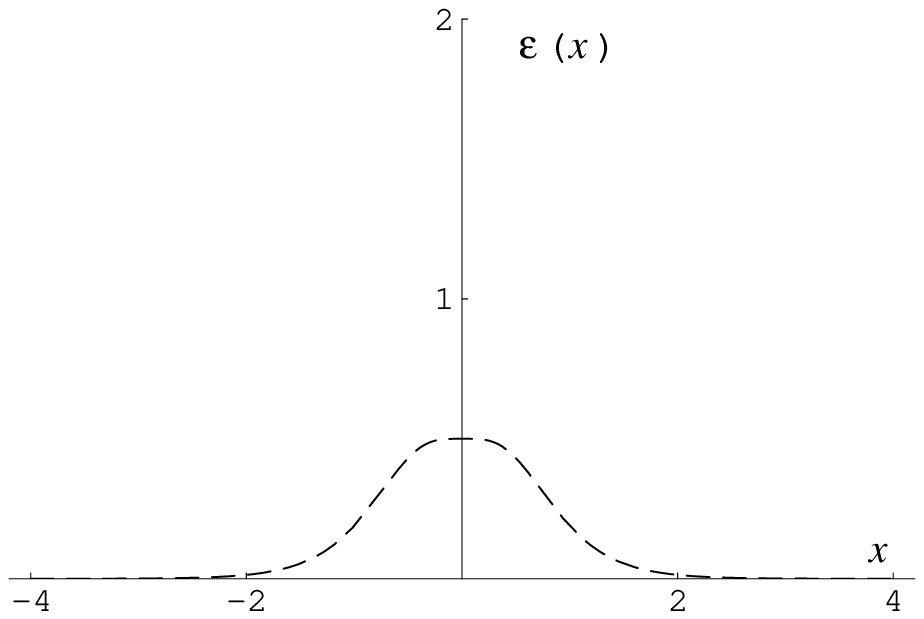, height=2.5cm}
\epsfig{file=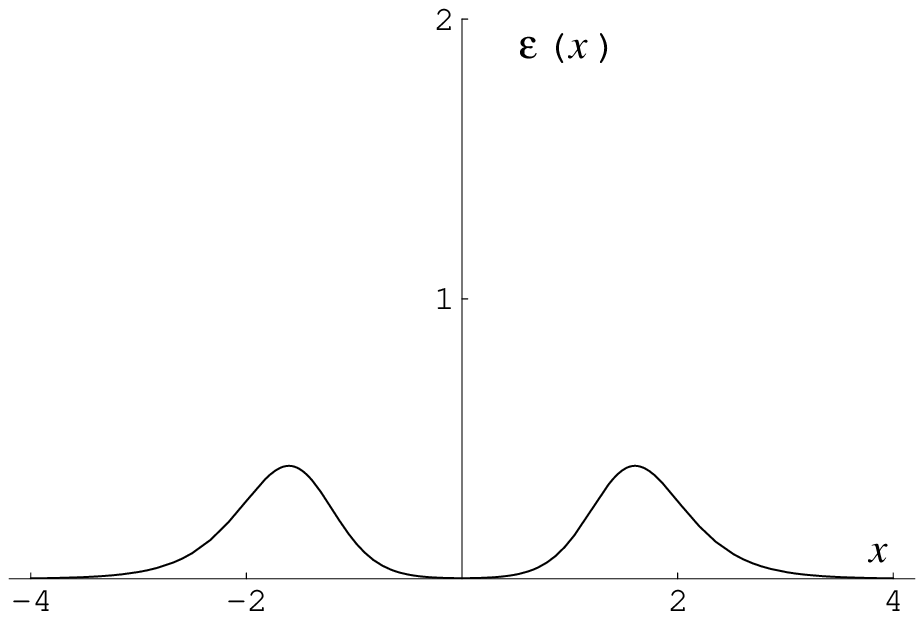, height=2.5cm} } \caption{\small
Energy density ${\cal E}^K[x;0,b]$ for {\it (a)} $b=0$, {\it (b)}
$b=\sqrt{0.5}$, {\it (c)} $b=1$ and {\it (d)} $b=\sqrt{30}$.}
\end{figure}
The critical points of ${\cal E}^{\rm TK}[x;0,b]$
\[
\frac{\partial{\cal E}^{\rm TK}}{\partial x}[x;0,b]=\frac{2 \sinh
x}{(b^2+\cosh x)^5}\ P_3(\cosh x) \hspace{0.5cm}, \hspace{0.5cm}
\frac{\partial{\cal E}^{\rm TK}}{\partial x}[x;0,b]=0
\]
are the origin $x=0 \simeq \sinh x=0, \, \forall b^2$ and the roots
of the third-order polynomial $P_3(\cosh x)$:
\[
P_3(\cosh x)= -b^2\cosh^3x-b^4 \cosh^2 x+(-3b^2+4b^6) \cosh x+5b^4-4
\qquad ,
\]
which under the change of variables $P_3(\cosh x)=-b^2\tilde{P}(u)$,
$u^2(x)=-1+\cosh x$  becomes:
\[
\tilde{P}(u)=(u^2)^3+(b^2+3) (u^2)^2-(4b^4-2b^2-6) (u^2)-4\left(
b^4+b^2-1-\frac{1}{b^2}\right) \qquad .
\]
In sum, use of the Cardano-Vieta formulas to solve cubic algebraic
equations prompts us to the following conclusions:
\begin{enumerate}
\item If $b^2\in [0,1]$ there are no real roots $P_3$. The only critical point of the energy density
, a maximum, is the origin and only one lump of energy is carried by
these kinks:
\[
\tilde{P}(u(x))\neq 0, \forall x , u(x)\in {\mathbb R} \qquad ,
\qquad {\cal E}^{\rm TK}[0;0,b]=\frac{2}{(b^2+1)^2} \qquad .
\]

\item If $b^2>1$, $\tilde{P}(u)=0$ has two real solutions: $u_\pm=\pm\sqrt{r(b^2)}\in{\mathbb
R}$. The points
\[
x=\pm m(b^2) \qquad , \qquad m(b^2)=\frac{1}{2\sqrt{2}} {\rm
arccosh} (1+r(b^2))
\]
are the maxima of ${\cal E}^{\rm TK}$ on the real line (the origin
is now a minimum) and  these kink profiles carry two lumps of energy
. Understanding these two lumps as fundamental particles, $a$ is the
center of mass whereas $b$ is the relative coordinate of this system
of two particles that become glued when $b^2<1$.

\end{enumerate}

\subsubsection{Kink profiles: numerical methods}

For generic values of $\sigma$ one must rely on numerical
integration methods to find the profiles of the kink solutions. We
do this by solving the first-order equations by standard numerical
methods with \lq\lq initial" conditions:
\[
\phi_1(0)=0 \qquad , \qquad \frac{\sigma}{2(1-\sigma)}\phi_2^2(0)
-\frac{c}{2\sigma}|\phi_2(0)|^{\frac{2}{\sigma}}=\frac{1}{4} \qquad
.
\]
The reasons for this choice are twofold: (1) For any kink solution,
$\phi_1(x)$ always has a zero. Translational invariance allows us to
set the zero at $x=0$; (2) To ensure that we will find a numerical
kink solution, we fix $\phi_2(0)$ on a kink orbit for a given value
of $\sigma$ and arbitrary choices of $c$.

The numerical method provides us with a succession of points of the
kink solution generated by an interpolation polynomial. The plots of
the numerical results show that the behavior derived analytically
for the kink profiles when $\sigma={1\over 2}$ is generic. For any
value of $\sigma$ the kink profiles are composed of two kinks. The
parameter $c$ giving the orbit is related to the kink separation. In
some range of $c$ the two kinks melt into a single kink. The precise
value of $c$ at which this happens depends on the value of $\sigma$.
$\sigma={1\over 2}$ is singled out, because in this case $c=0$ is
the value where two kinks fuse into a single kink, or, viceversa, a
single kink splits into two kinks.

\subsection{TK2 kink Casimir energy}

Small kink deformations $\phi_a(x)=\phi_a^{\rm
TK}(x;c)+\delta\phi_a(x)$ are still solutions of the first-order
equations if $\delta\phi_a(x)\in {\rm Ker}\, D(c)$:
\[
D(c)\delta\vec{\phi}(x)=\left(\begin{array}{cc} -{d\over
dx}-4\bar{\phi}_1(x;c) & -2\sigma\bar{\phi}_2(x;c) \\
-2\sigma\bar{\phi}_2(x;c) & -{d\over dx}-2\sigma\bar{\phi}_1(x;c)
\end{array}\right)\left(\begin{array}{c} \delta\phi_1(x) \\ \delta\phi_2(x)\end{array}\right)=
\left(\begin{array}{c} 0 \\ 0 \end{array}\right) \qquad .
\]
We shall use the notation $\vec{\phi}^{\rm
TK}(x;c)=\bar{\phi}_1^{\rm TK}(x;c)\vec{e}_1+\bar{\phi}_2^{\rm
TK}(x;c)\vec{e}_2$ because no analytic expressions for the kink
profiles are known (except for some special values of $\sigma$). The
$c$ parameter tells us what kink orbit is chosen. Note that:
\[
K^-(c)=\left(\begin{array}{cc}
-\frac{d^2}{dx^2}+8\bar{\phi}_1^2(x;c)+4\sigma(\sigma-1)\bar{\phi}_2^2(x;c)+2
& 8\sigma\bar{\phi}_1(x;c)\bar{\phi_2}(x;c) \\
8\sigma\bar{\phi}_1(x;c)\bar{\phi_2}(x;c) &
-\frac{d^2}{dx^2}+4\sigma(\sigma-1)\bar{\phi}_1^2(x;c)+2\sigma^2\bar{\phi}_2^2(x;c)+\sigma\end{array}\right)
\]
\[
K(c)=\left(\begin{array}{cc}
-\frac{d^2}{dx^2}+24\bar{\phi}_1^2(x;c)+4\sigma(\sigma+1)\bar{\phi}_2^2(x;c)-2
& 8\sigma(\sigma+1)\bar{\phi}_1(x;c)\bar{\phi_2}(x;c) \\
8\sigma(\sigma+1)\bar{\phi}_1(x;c)\bar{\phi_2}(x;c) &
-\frac{d^2}{dx^2}+4\sigma(\sigma+1)\bar{\phi}_1^2(x;c)+6\sigma^2\bar{\phi}_2^2(x;c)-\sigma\end{array}\right)
\quad ,
\]
if $\alpha=\beta$. For $\alpha\neq\beta$, $K^-(c)=D^\dagger(c)
D(c)$ and $K(c)=D(c)D^\dagger(c)$ are exchanged.

Moreover, the shift of the Higgs and Goldstone fields from the
stable kink solution, $\vec{\phi}(x^\mu)=\vec{\phi}^{\rm
TK}(x;c)+H(x^\mu)\vec{e}_1+G(x^\mu)\vec{e}_2$, causes the action in
the kink sector to be the complicated expression:
\begin{eqnarray*}
S&=&-\frac{4m^3}{3\lambda}\lim_{T\rightarrow\infty}\int_{-\frac{T}{2}}^\frac{T}{2}\,
dx_0 \,+ \frac{m^2}{\lambda}\int  d^2x \left[{1\over
2}\partial_\mu H\partial^\mu H
-\left(12\bar{\phi}^2_1(x;c)+2\sigma(\sigma+1)\bar{\phi}_2^2(x;c)-1\right)H^2(x^\mu)\right]\\&+&
\frac{m^2}{\lambda}\int  d^2x \left[{1\over 2}\partial_\mu
G\partial^\mu G
-\left(2\sigma(\sigma+1)\bar{\phi}^2_1(x;c)+3\sigma^2\bar{\phi}_2^2(x;c)-{\sigma\over
2}\right)G^2(x^\mu)\right]\\&-& {m^2\over\lambda}\int  d^2x
\left[8\sigma(\sigma+1)\bar{\phi}_1(x;c)\bar{\phi}_2(x;c)H(x^\mu)G(x^\mu)\right.+\\&+&
\left. 2H^4(x^\mu)
+2\sigma(\sigma+1)H^2(x^\mu)G^2(x^\mu)+{\sigma^2\over
2}G^4(x^\mu)\right]\\&-& {m^2\over\lambda}\int  d^2x
\left[8\bar{\phi}_1(x;c)H^3(x^\mu)+4\sigma(\sigma+1)\bar{\phi}_1(x;c)H(x^\mu)G^2(x^\mu)
+\right.\\&&\hspace{2.2cm}\left.
+4\sigma(\sigma+1)\bar{\phi}_2(x;c)H^2(x^\mu)G(x^\mu)+2\sigma^2\bar{\phi}_2(x;c)G^3(x^\mu)\right]
\qquad .
\end{eqnarray*}
Both the Higgs and Goldstone propagators, as well as all the
trivalent vertices, are distorted by the kink. In the background of
a TK2 kink, Higgs and Goldstone particles can transform into each
other by means of the bivalent vertex induced.

Thus, the classical energy for small fluctuations
$\vec{\phi}(x_0,x)=\vec{\phi}^{\rm TK}(x;c)+\delta
H(x_0,x)\vec{e}_1+\delta G(x_0,x)\vec{e}_2$ reads:
\begin{eqnarray*}
H^{(2)}&=&\frac{2m^3}{\lambda}\int \, dx \,
\left[\frac{\partial\delta H}{\partial
x_0}\cdot\frac{\partial\delta H}{\partial x_0}+\delta
H(x_0,x)K_{11}(c)\delta H(x_0,x)+ \delta H(x_0,x)K_{12}(c)\delta
G(x_0,x)\right. +\\&+& \left. \delta G(x_0,x)K_{21}(c)\delta
H(x_0,x)+\delta G(x_0,x)K_{22}(c)\delta G(x_0,x)\frac{}{}\right] \qquad ,
\end{eqnarray*}
where $K_{ab}(c)$ are the matrix elements of the second order
fluctuation operator, which we rewrite, together with his
supersymmetric partner, in the form:
\[
K(c)=D(c)D^\dagger(c)=\left(\begin{array}{cc} -{d^2\over
dx^2}+4+V_{11}(x;c) & V_{12}(x;c)
\\ V_{21}(x;c) & -{d^2\over
dx^2}+\sigma^2+V_{22}(x;c)\end{array}\right) \qquad ,
\]
\[
\qquad K^-=D^\dagger(c)D(c)=\left(\begin{array}{cc} -{d^2\over
dx^2}+4+V_{11}^-(x;c) & V_{12}^-(x;c)
\\ V_{21}^-(x;c) & -{d^2\over
dx^2}+\sigma^2+V_{22}^-(x;c)
\end{array}\right)\qquad
,
\]
\begin{eqnarray*}
V_{11}(x;c)&=&24\bar{\phi}_1^2(x;c)+4\sigma(\sigma+1)\bar{\phi}_2^2(x;c)-6
\\
V_{12}(x;c)&=&8\sigma(\sigma+1)\bar{\phi}_1(x;c)\bar{\phi}_2(x;c)=V_{21}(x;c)\\
V_{22}(x;c)&=&4\sigma(\sigma+1)\bar{\phi}_1^2(x;c)+6\sigma^2\bar{\phi}_2^2(x;c)-\sigma(\sigma+1)\\
V_{11}^-(x;c)&=&8\bar{\phi}_1^2(x;c)+4\sigma(\sigma-1)\bar{\phi}_2^2(x;c)-2
\\
V_{12}^-(x;c)&=&8\sigma\bar{\phi}(x;c)\bar{\phi}_2(x;c)=V_{21}^-(x;c)\\
V_{22}^-(x;c)&=&4\sigma(\sigma-1)\bar{\phi}_1^2(x;c)+2\sigma^2\bar{\phi}_2^2(x;c)-\sigma(\sigma+1)
\end{eqnarray*}
For instance, the family of Hessian operators for the topological
kinks found in the $\sigma={1\over 2}$ case can be written
explicitly:
\[ K(b)=
\left( \begin{array}{cc} -\frac{d^2}{dx^2}+\frac{3b}{\cosh
x+b}+\frac{6\sinh^2x}{(\cosh x+b)^2}-2 & \frac{6\sqrt{b}\sinh
x}{(\cosh x+b)^{\frac{3}{2}}}
\\\frac{ 6\sqrt{b}\sinh x}{(\cosh x+b)^{\frac{3}{2}}} &
-\frac{d^2}{dx^2}+\frac{3}{2}\frac{b}{\cosh x+b}+\frac{3}{4}
\frac{\sinh^2x}{(\cosh x+b)^2}-\frac{1}{2}\end{array} \right)
\qquad .
\]

\begin{figure}[htbp] \centerline{
\epsfig{file=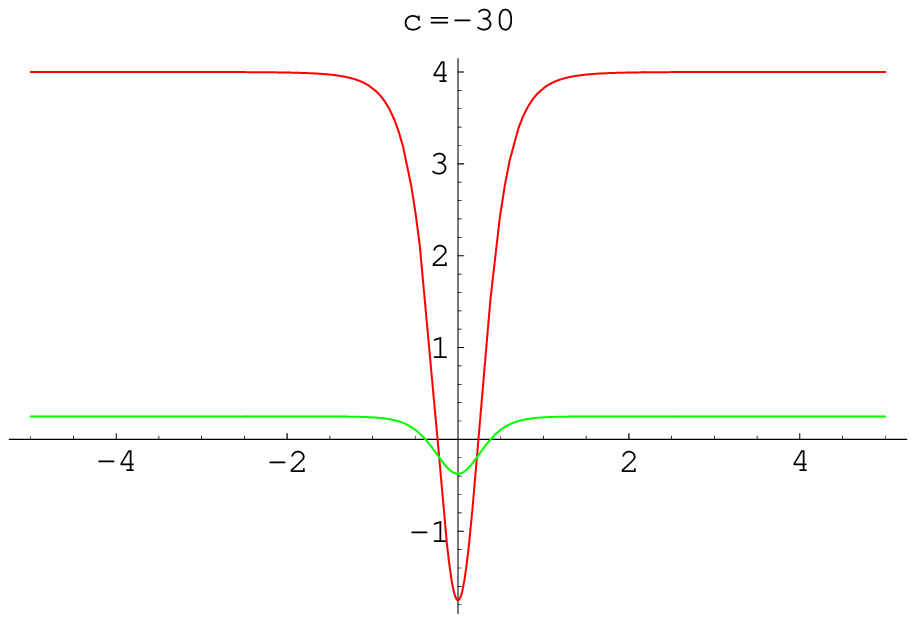,height=2cm}\quad
\epsfig{file=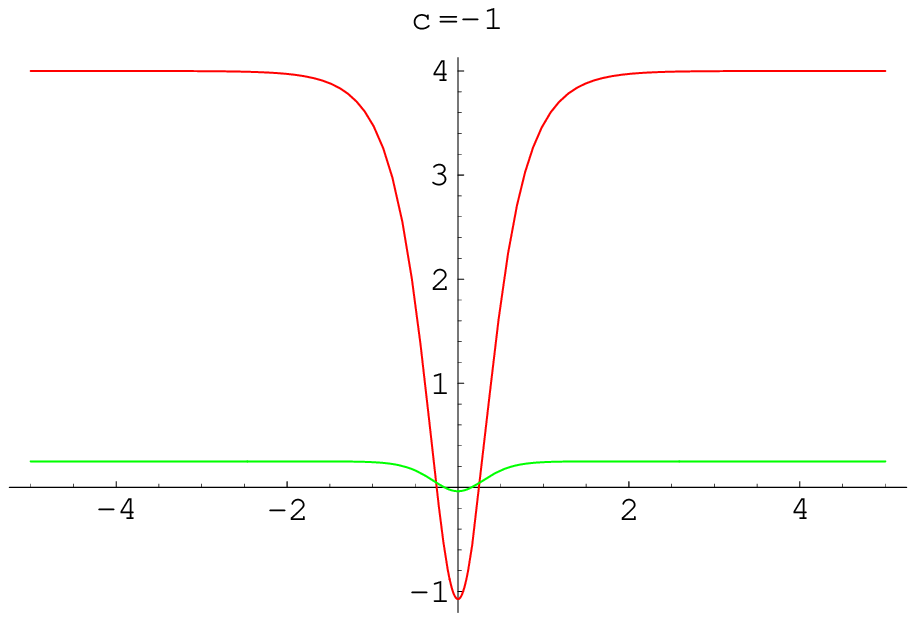,height=2cm}}
\centerline{\epsfig{file=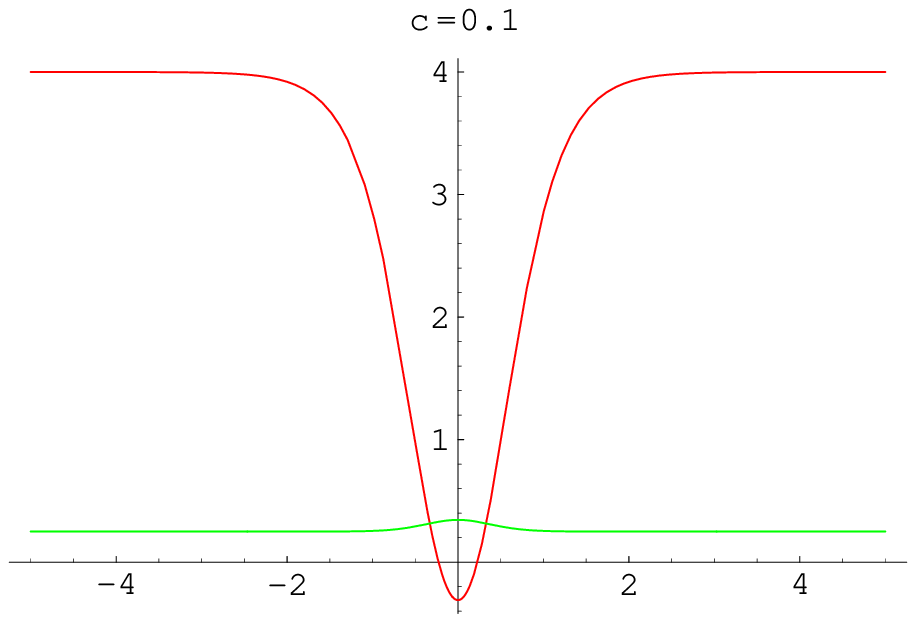,height=2cm}\quad
\epsfig{file=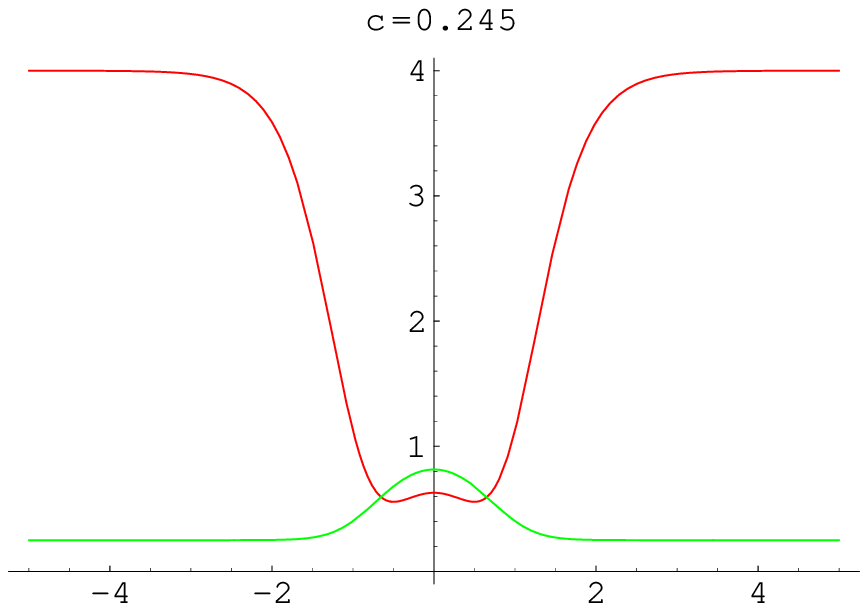,height=2cm}\quad
\epsfig{file=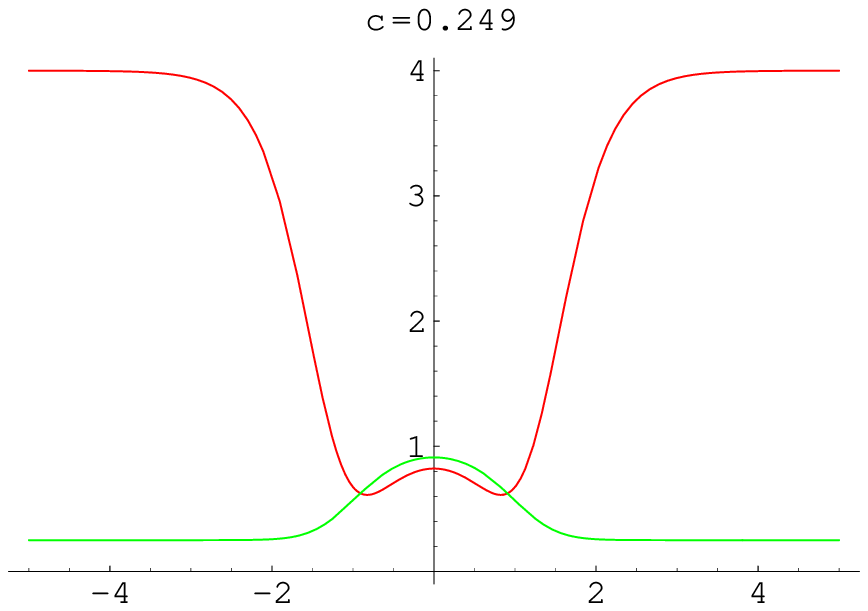,height=2cm}} \caption{\small \it Diagonal
components $V_{11}(x)$ (red) and $V_{22}(x)$ (green) of the
potential for c=-30, c=-1, c=0.1, c=0.245 and c=0.249.}
\end{figure}
In any case, the spectral resolution of
\[
K(c)f_\varepsilon(x)=\varepsilon^2 f_\varepsilon(x) \hspace{1cm} ,
\hspace{1cm} \lim_{x\rightarrow\pm\infty}K(c)=K_0
\]
for any value of $\sigma$ has the following features:
\begin{enumerate}
\item The kernel of $K(c)$ is of dimension two and the orthogonal basis is provided
by the two eigenfunctions (zero modes):
\[
\varepsilon=0 \qquad ; \qquad f_0^{(1)}(x)=\left(\begin{array}{c}
\frac{\partial\phi_1^{\rm TK}}{\partial x}(x) \\[0.2cm]
\frac{\partial\phi_2^{\rm TK}}{\partial x}(x)
\end{array}\right) \qquad \qquad  , \qquad \qquad  f_0^{(2)}(x)=\left(\begin{array}{c}
\frac{\partial\phi_1^{\rm TK}}{\partial c}(x) \\[0.2cm]
\frac{\partial\phi_2^{\rm TK}}{\partial c}(x)
\end{array}\right) \qquad ,
\]
obeying the fact that motion in the kink moduli space costs no
energy: all the kink solutions are in neutral equilibrium.

\item There are bound states in a number, $N(c)$, that depends on
$c$.
\[
\varepsilon^2=\varepsilon_l>0, \,\,  l=1,2, \cdots , N(c) \quad ;
\quad f_{\varepsilon_l}(x)=\left(\begin{array}{c}
f_1^{\varepsilon_l}(x) \\f_2^{\varepsilon_l}(x)\end{array}\right)
\quad , \quad \int \, dx \,
f_{\varepsilon_l}^{T}(x)f_{\varepsilon_m}(x)=\delta^{lm} \quad .
\]

\item There are two branches of scattering states:
\[
(1)\quad \varepsilon^2=k^2+4 \quad ; \quad
f_\varepsilon^{(1)}(x;k))=e^{-ikx}\left(\begin{array}{c}
u_1^{(1)}(x;k) \\
u_2^{(1)}(x;k)
\end{array}\right) \quad , \quad
\lim_{x\rightarrow\pm\infty}\left(\begin{array}{c}u_1^{(1)}(x;k)=u_1^\pm(k)\\u_2^{(1)}(x;k)=0\end{array}
\right)\quad .
\]
In terms of non-explicitly known phase shifts -determined by
$u_1^\pm(k)$-, Periodic Boundary Conditions in the interval
$I=[-\frac{mL}{2},\frac{mL}{2}]$ give the spectral densities in
${\rm Spec}K(c)$:
\[
k\cdot mL+\delta_1(k)=2\pi n \qquad , \qquad \rho_1^{\rm
TK}(k)={1\over 2\pi}(mL+\frac{d\delta_1(k)}{dk}) \qquad .
\]

\[
(2) \quad \lambda^2=q^2+\sigma^2 \quad ; \quad
f_\lambda^{(2)}(x;q))=e^{-iqx}\left(\begin{array}{c}
u_1^{(2)}(x;q) \\
u_2^{(2)}(x;q)
\end{array}\right)\quad , \quad
\lim_{x\rightarrow\pm\infty}\left(\begin{array}{c}
u_1^{(2)}(x;q)=0 \\
u_2^{(2)}(x;q)=u_2^\pm(q)\end{array}\right)\quad .
\]
The phase shifts are now read from $u_2^\pm(q)$ and the PBC give the
spectral densities in the other branch of ${\rm Spec}K(c)$:
\[
q\cdot mL+\delta_2(q)=2\pi n \qquad , \qquad \rho_2^{\rm
TK}(q)={1\over 2\pi}(mL+\frac{d\delta_2(q)}{dq}) \qquad .
\]
\end{enumerate}
The eigenfunctions in the continuous spectrum of $K(c)$ also satisfy
orthogonality conditions: {\footnote{There is a very subtle
possibility. If $\varepsilon_{N(c)}=\varepsilon(k=0)\,\, or \,\,
\lambda(q=0)$, i.e. if the last eigenvalue in the discrete spectrum
coincides with the threshold of any branch of the continuous
spectrum, half-zero modes enter the game. The Levinson theorem in
one dimension forces us to include a weight of ${1\over 2}$ in the
contribution to the energy of those states.}}
\[
\int \, dx \,
f_\varepsilon^{\dagger(I)}(x;k)f_{\varepsilon^\prime}^{(J)}(x;q)=
\delta^{IJ}\delta_{kq} \qquad , \qquad I,J=1,2 \qquad .
\]

The general solution to the linearized field equations
\begin{eqnarray*}
&&\frac{\partial^2\delta H}{\partial x_0^2}-\frac{\partial^2\delta
H}{\partial
x^2}+(24\bar{\phi}_1^2(x;c)+4\sigma(\sigma+1)\bar{\phi}_2^2(x;c)-2)\delta
H(x^\mu)+8\sigma(\sigma+1)\bar{\phi}_1(x;c)\bar{\phi}_2(x;c)\delta
G(x^\mu)=0\\&&\frac{\partial^2\delta G}{\partial
x_0^2}-\frac{\partial^2\delta G}{\partial
x^2}+8\sigma(\sigma+1)\bar{\phi}_1(x;c)\bar{\phi}_2(x;c)\delta
H(x^\mu)+(4(\sigma(\sigma+1)\bar{\phi}_1^2(x;c)+6\sigma^2\bar{\phi}_2^2(x;c)-\sigma)\delta
G(x^\mu)=0
\end{eqnarray*}
is a linear combination of the eigen-functions of $K(c)$
\footnote{The eigenfunctions belonging to the kernel of $K(c)$ are
excluded because they do no contribute to the energy. The prime in
the fields refers to this exclusion.}:
\begin{eqnarray*}
\left(\begin{array}{c}\delta H^\prime (x_0,x)\\\delta
G^\prime(x_0,x)\end{array}\right)&=&\frac{\sqrt{\lambda}}{2m}\cdot
\sqrt{\frac{\hbar}{mL}}\cdot\left\{\sum_{l=1}^{N(c)}\frac{1}{\sqrt{2\sqrt{\varepsilon_l}}}
\left(A^l e^{-i\varepsilon_lx_0}+A^{*l}
e^{i\varepsilon_lx_0}\right)f_{\varepsilon_l}(x)+\right. \\
&+&\sum_k\frac{1}{\sqrt{2\varepsilon(k)}}
\left\{A^{(1)}(k)e^{-i\varepsilon(k)x_0}f_\varepsilon^{(1)}(x)+
A^{*(1)}(k)e^{i\varepsilon(k)x_0}f_\varepsilon^{(1)*}(x)\right\}+\\&+&\left.
\sum_q\frac{1}{\sqrt{2\lambda(q)}}
\left\{A^{(2)}(q)e^{-i\lambda(q)x_0}f^{(2)}_\lambda(x)+
A^{*(2)}(q)e^{i\lambda(q)x_0}f_\lambda^{*(2)}(x)\right\}\right\}\qquad
.
\end{eqnarray*}
The classical free Hamiltonian for kink fluctuations becomes:
\begin{eqnarray*}
H^{(2)}&=&\frac{\hbar
m}{2}\left\{\sum_{l=1}^{N(c)}\sqrt{\varepsilon_l}(A^{*l}A^l+A^lA^{*l})+
\sum_k\varepsilon(k)(A^{(1)*}(k)A^{(1)}(k)+A^{(1)}(k)A^{(1)*}(k))+\right.
\\&+&\left.\sum_q\lambda(q)(A^{(2)*}(q)A^{(2)}(q)+A^{(2)}(q)A^{(2)*}(q))\right\}
\end{eqnarray*}
and, after canonical quantization,
\[
[A^{l},A^{m\dagger}]=\delta_{lm} \qquad , \qquad
[A^{(I)}(k),A^{(J)\dagger}(q)]=\delta^{IJ}\delta_{kq}
\]
one obtains the quantum free Hamiltonian
\[
\hat{H}^{(2)}=\hbar
m\left\{\sum_{l=1}^{N(c)}\sqrt{\varepsilon_l}\left(\hat{A}^{l\dagger}\hat{A}^l+{1\over
2}\right)+
\sum_k\varepsilon(k)\left(\hat{A}^{(1)\dagger}(k)\hat{A}^{(1)}(k)+{1\over
2}\right)+\sum_k\lambda(k)\left(\hat{A}^{(2)\dagger}(k)\hat{A}^{(2)}(k)+{1\over
2}\right)\right\}
\]
and the kink Casimir energy
\[
\Delta E(\vec{\phi}^{\rm TK}(c))={\hbar m\over 2}
\left(\sum_{l=1}^{N(c)}\sqrt{\varepsilon_l}+\sum_k\varepsilon(k)+\sum_q\lambda(q)\right)={\hbar
m\over 2}{\rm Tr}K(c)^{{1\over 2}}
\]
when all the positive modes are non-occupied.

In sum, the TK2 kink semi-classical energy -one-loop order- receives
three contributions:
\begin{enumerate}

\item The classical energy $E(\vec{\phi}^{\rm TK}(c))=\frac{4m^3}{3\lambda}$.

\item The TK2 kink Casimir energy -zero point energy
renormalization-
\[
\Delta M_K^C(c)=\Delta E(\vec{\phi}^{\rm TK}(c))-\Delta
E(\vec{\phi}^{\pm(1)})={\hbar m\over 2}\left({\rm Tr}K(c)^{{1\over
2}}-{\rm Tr}K_0^{{1\over 2}}\right)\qquad .
\]
\item The contribution of ${\cal L}_{C.T.}$ to one-loop TK2 kink
masses is:
\begin{eqnarray*}
\Delta M_K^R(c)=&-&\frac{\hbar m}{2}\int \, dx\left\{\left[6\cdot
I(4)+\sigma(\sigma+1)\cdot I(\sigma^2)\right][(\phi_1^{\rm
TK}(x,c))^2-(\phi_1^{(1)})^2]\right\}-\\&-&\frac{\hbar m}{2}\int \,
dx\left\{\left[4\sigma(\sigma+1)\cdot I(4)+6\sigma^2\cdot
I(\sigma^2)\right][(\phi_2^{\rm
TK}(x,c))^2-(\phi_2^{(1)})^2]\right\}\qquad .
\end{eqnarray*}
\end{enumerate}
Therefore, the one-loop TK2 kink mass and the semi-classical kink
energy are the divergent quantities:
\[
\Delta M_K(c)=\Delta M_K^C(c)+\Delta M_K^R(c) \qquad , \qquad
E_S(\vec{\phi}^{\rm TK}(c)=E(\vec{\phi}^{\rm TK}(c))+\Delta M_K(c)\qquad .
\]

\section{The TK2 kink heat kernel and generalized zeta function}

\subsection{Zeta function regularization}

We regularize the ultraviolet divergent TK2 kink and vacuum energies
in terms of their generalized zeta functions:
\[
\Delta M_K^C(c,s)={\hbar\over 2} ({\mu^2\over m^2})^s\mu
\left(\zeta_K(c)(s)-\zeta_{K_0}(s)\right) \qquad .
\]
Here, $s$ is the non-dimensional complex parameter already
introduced in the previous Lecture; the auxiliary parameter $\mu$ of
$L^{-1}$ dimensions has also been used in the previous Lecture to
keep the dimensions in order in the regularization procedure, and
the generalized zeta functions are the series:
\[
\zeta_{K(c)}(s)=\sum_{l=1}^{N(c)}\,
{1\over\varepsilon_l^{2s}}+\sum_k\,{1\over\varepsilon(k)^{2s}}+\sum_q\,{1\over\lambda(q)^{2s}}
\qquad , \qquad
\zeta_{K_0}(s)=\sum_k\,{1\over\omega(k)^{2s}}+\sum_q\,{1\over\gamma(q)^{2s}}\qquad
.
\]
Therefore,
\[
\Delta M_K^C(c)=\lim_{s\rightarrow -{1\over 2}} \Delta M_K^C(c,s)
={\hbar m\over 2}\left(\zeta_K(c)(-{1\over 2})-\zeta_{K_0}(-{1\over
2})\right) \qquad ,
\]
and the divergences reappear at $s=-{1\over 2}$, which is a pole of
the meromorphic function $\Delta M_K^C(c,s)$ of the complex
parameter s.

$\Delta M_K^R(c)$ can also be regularized in terms of generalized
zeta functions. Both divergent integrals $I(4)$ and $I(\sigma^2)$
when the system is considered in the interval $I=-[{mL\over
2},{mL\over 2}]$ become the divergent series
\[
I(4)={1\over 2}{1\over mL}\sum_n\,\frac{1}{\sqrt{{n^2\over R^2}+4}}
\qquad , \qquad I(\sigma^2)={1\over 2}{1\over
mL}\sum_n\,\frac{1}{\sqrt{{n^2\over R^2}+\sigma^2}} \qquad .
\]
Thus, they are obtained as the limits:
\begin{eqnarray*}
I(4)&=&-\lim_{s\rightarrow -{1\over 2}}\frac{1}{\mu L}\cdot
\frac{\Gamma(s+1)}{\Gamma(s)}\cdot\left(\frac{\mu^2}{m^2}\right)^{s+1}\cdot
\zeta_{K_0^{11}}(s+1) \quad , \quad K_0^{11}=-\frac{d^2}{dx^2}+4
\\
I(\sigma^2)&=&-\lim_{s\rightarrow -{1\over 2}}\frac{1}{\mu L}\cdot
\frac{\Gamma(s+1)}{\Gamma(s)}\cdot\left(\frac{\mu^2}{m^2}\right)^{s+1}\cdot
\zeta_{K_0^{22}}(s+1) \quad , \quad
K_0^{22}=-\frac{d^2}{dx^2}+\sigma^2\qquad .
\end{eqnarray*}
The regularized contribution of the mass renormalization
counter-terms to the kink mass is:
\begin{eqnarray*}
\Delta M_K^R(c,s)&=&\frac{\hbar }{2 L}\lim_{s\rightarrow -{1\over
2}}\,
\frac{\Gamma(s+1)}{\Gamma(s)}\cdot\left(\frac{\mu^2}{m^2}\right)^{s+\frac{1}{2}}\cdot
\\&\cdot&\left\{[6\zeta_{K_0^{11}}(s+1)
+\sigma(\sigma+1)\zeta_{K_0^{22}}(s+1)]\int \, dx \,
\left[(\phi_1^{\rm TK}(x,c))^2-(\phi_1^{(1)})^2\right]\right.
+\\&+&\left. [4\sigma(\sigma+1)\zeta_{K_0^{11}}(s+1)
+6\sigma^2\zeta_{K_0^{22}}(s+1)]\int \, dx \, \left[(\phi_2^{\rm
TK}(x,c))^2-(\phi_2^{(1)})^2\right]\right\} \qquad .
\end{eqnarray*}
Therefore,
\begin{eqnarray*}
\Delta M_K^C(c)&=&\lim_{s\rightarrow -{1\over 2}}\Delta
M_K^R(c,s)\\&=&-\frac{\hbar}{4 L}\left\{[6\zeta_{K_0^{11}}({1\over
2}) +\sigma(\sigma+1)\zeta_{K_0^{22}}({1\over 2})]\int \, dx \,
\left[(\phi_1^{\rm TK}(x,c))^2-(\phi_1^{(1)})^2\right]\right.
+\\&+&\left. [4\sigma(\sigma+1)\zeta_{K_0^{11}}({1\over 2})
+6\sigma^2\zeta_{K_0^{22}}({1\over 2})]\int \, dx \,
\left[(\phi_2^{\rm TK}(x,c))^2-(\phi_2^{(1)})^2\right]\right\}
\qquad .
\end{eqnarray*}

\subsection{The Cahill-Comtet-Glauber (CCG) exact formula}

The goal in this sub-Section is to compute the one-loop mass shift
for the one-component TK1 topological kink:
\[
\phi_1^{{\rm TK}1}(x;c=-\infty)={1\over 2}{\rm tanh}x \hspace{1.5cm}
, \hspace{1.5cm} \phi_2^{{\rm TK}1}(x;c=-\infty)=0 \qquad .
\]
The Kink fluctuation operator -$K(-\infty)=K$ for short- is diagonal
for any value of $\sigma$:
\[
K=\left(\begin{array}{cc} K^{11} & 0
\\ 0 & K^{22}\end{array}\right)=\left(\begin{array}{cc}
-\frac{d^2}{dx^2}+4-\frac{6}{{\rm cosh}^2x} & 0 \\ 0 &
-\frac{d^2}{dx^2}+\sigma^2-\frac{\sigma(\sigma+1)}{{\rm
cosh}^2x}\end{array}\right) \qquad .
\]
Both diagonal entries $K^{11}$ and $K^{22}$ are Posch-Teller
Schrodinger operators with very well known spectra.

\noindent {\bf I. The spectrum of $K^{11}$:}

There are three types of eigenfunctions
\begin{enumerate}
\item {\it Bound states:}
\begin{eqnarray*}
&&\varepsilon_0=0 \hspace{1.5cm} , \hspace{1.5cm}
\psi_0(x)={1\over{\rm cosh}^2x}\\&&\varepsilon_1=3 \hspace{1.5cm} ,
\hspace{1.5cm} \psi_3(x)={{\rm sinh}x\over{\rm cosh}^2x}
\end{eqnarray*}

\item {\it Scattering states:}
\[
\varepsilon^2(k)=k^2+4 \hspace{1.5cm} , \hspace{1.5cm}
\psi_k(x)=e^{ikx}P_2({\rm tanh}x;k)\quad , \quad
P_2(z;k)=3z^2-1-3ikz-k^2 \qquad ,
\]
with phase shifts:
\[
\delta_1(k)=-2{\rm arctan}\frac{3k}{2-k^2} \qquad .
\]
\item {\it Half-bound state}:
\[
\varepsilon_{1\over 2}=4 \hspace{1.5cm} , \hspace{1.5cm}
\psi_{(k=0)}(x)=P_2({\rm tanh}x;0) \qquad .
\]

\end{enumerate}

\noindent {\bf II. The spectrum of $K^{22}$:}

\begin{enumerate}
\item {\it Bound states:} $l=0,1,2, \cdots , N , \qquad N=I[\sigma]$
\[
\varepsilon_l=(2\sigma-l)l  \hspace{1.5cm} , \hspace{1.5cm}
\psi_l(x)=\frac{1}{({\rm cosh}x)^{\sigma-l}}\,
{}_2F_1[-l,2\sigma-l,\sigma-l+1;{1\over 2}(1+{\rm tanh}x)]
\]

\item {\it Scattering states:}
\[
\varepsilon=q^2+\sigma^2 \hspace{1.5cm} , \hspace{1.5cm}
\psi_q(x)=e^{iqx}\,
{}_2F_1[-\sigma,\sigma,1-iq;\frac{e^x}{e^x+e^{-x}}] \qquad .
\]
From these wave functions, one reads the following transmission and
reflection scattering coefficients:
\[
T(q)=\frac{\Gamma(\sigma+1-iq)\Gamma(-\sigma-iq)}{\Gamma(1-iq)\Gamma(-iq)}\qquad
, \qquad R(q)=\frac{\Gamma(\sigma+1-iq)\Gamma(-\sigma-iq)\Gamma(iq)}
{\Gamma(1+\sigma)\Gamma(-\sigma)\Gamma(-iq)}\qquad .
\]
Henceforth, the phase shifts
\[
\delta_2(q)=\delta_2^+(q)+\delta_2^-(q) \qquad ; \qquad
\delta_2^\pm(q)={1\over 4}{\rm arctan}\left(\frac{{\rm Im}(T(q)\pm
R(q))}{{\rm Re}(T(q)\pm R(q)}\right)
\]
identify these scattering processes.

If $\sigma=N \in {\mathbb N}$ is a natural number, $R(q)=0$,
$\delta_2^+(q)=\delta_2^-(q)$, and the total phase shift
$\delta_2(q)=\delta_2^+(q)+\delta_2^-(q)$ is:
\[
\delta_2(q)=\frac{1}{2}{\rm arctan}\left(\frac{{\rm
Im}\prod_{n=0}^{N-1}(q^2-(N-n)^2+2iq(N-n))}{{\rm
Re}\prod_{n=0}^{N-1}(q^2-(N-n)^2+2iq(N-n))}\right)
\]
\item {\it Half-bound state:}
If $\sigma=I[\sigma]=N \in {\mathbb N}$
\[
\varepsilon_{1\over 2}=N^2 \qquad \quad , \qquad \quad
\psi_{q=0}(x)={}_2F_1[-\sigma,\sigma ,1; \frac{1}{2}(1+{\rm tanh}x)]
\]
also belongs to the spectrum.
\end{enumerate}

Thus, one expects that the one-loop TK1 mass shift can be calculated
exactly from this complete spectral information. We distinguish,
however, two different situations according to whether or not
$\sigma$ is a (positive) integer.

\subsubsection{$\sigma=N\in{\mathbb N}$: one-loop TK1 mass shift from bound
states}

If $\sigma$ is a natural number, the reflection scattering
coefficient is zero for both $K^{11}$ and $K^{22}$. The
Cahill-Comtet-Glauber (CCG) formula can be applied. This formula
gives the one-loop mass shift of one-dimensional solitons from the
energies of their bound states. Applied to the TK1 kink of the BNRT
model it reads:
\begin{equation}
\Delta M(\vec{\phi}^{\, TK1})=-\frac{\hbar
m}{\pi}\left(\sum_{i=0}^1\, 2({\rm sin}\theta_i-\theta_i{\rm
cos}\theta_i)+\sum_{l=0}^{N-1}\, N({\rm sin}\alpha_l-\alpha_l{\rm
cos}\alpha_l)\right) \label{eq:ccg} \qquad .
\end{equation}
The angles are defined in terms of the eigenvalues of the bound
states of $K^{11}$ and $K^{22}$:
\[
\theta_0={\rm arccos}({0\over 2})={\pi\over 2} \quad , \quad
\theta_1={\rm arccos}({\sqrt{3}\over 2})={\pi\over 6} \quad , \quad
\alpha_l={\rm arccos}(\frac{\sqrt{(2N-l)l}}{N}) \quad .
\]
Thus, for the first five cases we obtain:
\begin{enumerate}

\item $\sigma=1$
\[
\alpha_0={\pi\over 2} \quad , \quad {\rm sin}\theta_0={\rm
sin}\alpha_0=1 \qquad , \qquad {\rm sin}\theta_1={1\over 2}
\]
\[
\Delta M(\vec{\phi}^{\, TK1})=(-{3\over\pi}+{1\over 2\sqrt{3}})\hbar
m-{1\over\pi}\hbar m=-0.984564 \hbar m\qquad .
\]

\item $\sigma=2$
\[
\alpha_1={\pi\over 6} \qquad \qquad , \qquad \Delta M(\vec{\phi}^{\,
TK1})=(-{3\over\pi}+{1\over 2\sqrt{3}})\hbar m-({3\over\pi}-{1\over
2\sqrt{3}})\hbar m=-1.33251 \hbar m \qquad .
\]

\item $\sigma=3$
\[
\alpha_1={\rm arccos}({\sqrt{5}\over 3}) \quad , \quad {\rm
sin}\alpha_1={2\over 3} \qquad , \qquad \alpha_2={\rm
arccos}(2{\sqrt{2}\over 3}) \quad , \quad {\rm sin}\alpha_2={1\over
3}
\]
\[
\Delta M(\vec{\phi}^{\, TK1})=(-{3\over\pi}+{1\over 2\sqrt{3}})\hbar
m-({6\over\pi}-{1\over\pi}(\sqrt{5}{\rm arccos}({\sqrt{5}\over
3})+2\sqrt{2}{\rm arccos}(2{\sqrt{2}\over 3})))\hbar m=-1.75076
\hbar m \qquad .
\]

\item $\sigma=4$
\begin{eqnarray*}
\alpha_1={\rm arccos}({\sqrt{7}\over 4}) \quad , \quad {\rm
sin}\alpha_1={3\over 4} \qquad &,& \qquad \alpha_2={\rm
arccos}(2{\sqrt{3}\over 4}) \quad , \quad {\rm
sin}\alpha_2={2\over 4} \\ \alpha_3={\rm arccos}({\sqrt{15}\over
4}) \quad &,& \quad {\rm sin}\alpha_3={1\over 4}
\end{eqnarray*}
\begin{eqnarray*}
\Delta M(\vec{\phi}^{\, TK1})&=&(-{3\over\pi}+{1\over
2\sqrt{3}})\hbar m-({10\over\pi}-{1\over\pi}(\sqrt{7}{\rm
arccos}({\sqrt{7}\over 4})+2\sqrt{3}{\rm arccos}({2\sqrt{3}\over
4})+ \\&+&\sqrt{15}{\rm arccos}({\sqrt{15}\over 4})))\hbar
m=-2.24628 \hbar m\qquad .
\end{eqnarray*}

\item $\sigma=5$
\begin{eqnarray*}
\alpha_1={\rm arccos}({\sqrt{9}\over 5}) \quad , \quad {\rm
sin}\alpha_1={4\over 5} \qquad &,& \qquad \alpha_2={\rm
arccos}(2{\sqrt{4}\over 5}) \quad , \quad {\rm sin}\alpha_2={3\over
5} \\ \alpha_3={\rm arccos}({\sqrt{21}\over 5}) \quad , \quad {\rm
sin}\alpha_3={2\over 5} \quad &,& \quad \alpha_4={\rm
arccos}({2\sqrt{6}\over 5}) \quad , \quad {\rm sin}\alpha_4={1\over
5}
\end{eqnarray*}
\begin{eqnarray*}
\Delta M(\vec{\phi}^{\, TK1})&=&(-{3\over\pi}+{1\over
2\sqrt{3}})\hbar m-({15\over\pi}-{1\over\pi}(\sqrt{9}{\rm
arccos}({\sqrt{9}\over 5})+2\sqrt{4}{\rm arccos}({2\sqrt{4}\over
5})+
\\&+&\sqrt{21}{\rm arccos}({\sqrt{21}\over 5})+2\sqrt{6}{\rm arccos}({2\sqrt{6}\over 5})))\hbar
m=-2.82180 \hbar m\qquad .
\end{eqnarray*}

\end{enumerate}

\subsubsection{$\sigma\in{\mathbb R}^+$: one-loop TK1 mass shift from zeta functions}

The partition and generalized zeta functions for the vacuum operator
$K_0$ are at the $R\rightarrow\infty$ limit
respectively:{\footnote{In Appendix I it has been shown how this
limit can be safely taken, leaving no remnants, when $PBC$ are
chosen}}
\begin{eqnarray*}
{\rm Tr}e^{-\beta
K_0}&=&\frac{mL}{2\pi}\left[\int_{-\infty}^{+\infty} \, dk
\,e^{-\beta(k^2+4)}+\int_{-\infty}^{+\infty} \, dq
\,e^{-\beta(q^2+\sigma^2)}\right]=\frac{mL}{\sqrt{4\pi\beta}}\cdot\left[
e^{-4\beta}+e^{-\sigma^2\beta}\right]\\
\zeta_{K_0}(s)&=&{mL\over\sqrt{4\pi}}\cdot {1\over\Gamma (s)}\cdot
\int_0^\infty \, d\beta \, \beta^{s-{3\over 2}}\,
\left[e^{-4\beta}+e^{-\sigma^2\beta}\right]={mL\over\sqrt{4\pi}}\cdot
\left[{1\over 2^{2s-1}}+{1\over \sigma^{2s-1}}\right]\cdot
\frac{\Gamma(s-{1\over 2})}{\Gamma(s)} \qquad .
\end{eqnarray*}
The poles of $\zeta_{K_0}(x)$ are thus the poles of the Euler Gamma
function $\Gamma (s-{1\over 2})$: $s-{1\over 2}=0,-1,-2, \cdots ,
-n, \cdots$. The vacuum energy reads:
\[
\Delta E(\vec{\phi}^{\pm(1)})=\lim_{s \rightarrow
-\frac{1}{2}}\frac{\hbar}{2} \left( \frac{\mu^2}{m^2} \right)^s \mu
\cdot \zeta_{K_0}(s)=\lim_{s \rightarrow
-\frac{1}{2}}\frac{\hbar}{2} \left( \frac{ \mu^2}{m^2} \right)^s
\mu\cdot {mL\over\sqrt{4\pi}}\cdot \left[{1\over 2^{2s-1}}+{1\over
\sigma^{2s-1}}\right]\cdot \frac{\Gamma(s-{1\over
2})}{\Gamma(s)}\qquad .
\]
The partition function for the kink operator $K$ accounts for the
half-bound state in a subtle way:
\begin{eqnarray*}
{\rm Tr}^* e^{-\beta K}&=&{\rm Tr}^*e^{-\beta K^{11}}+{\rm
Tr}^*e^{-\beta K^{22}}={\rm Tr}e^{-\beta K_0}+e^{-3\beta}+{1\over
\pi}\int_{0}^{+\infty} \, dk \, \frac{d\delta_1(k)}{dk} \,
e^{-\beta(k^2+4)}\\&+&\sum_{l=0}^{N-1}e^{-l(2\sigma-l)\beta}+{\aleph\over
2}e^{- N(2\sigma-N)\beta}+{1\over \pi}\int_{0}^{+\infty} \, dq \,
\frac{d\delta_2(q)}{dq} \, e^{-\beta(q^2+\sigma^2)} \qquad ,
\end{eqnarray*}
and the weight of the last bound state is $\aleph=1$, if it is
buried in the threshold of the continuous spectrum;
$I[\sigma]=N=\sigma$, or $\aleph=0$, if it is not: $I[\sigma]\neq
\sigma$.

Accordingly, via the Mellin transform, the generalized zeta function
of the kink operator $K$ reads:

\begin{equation}
\zeta_{K}(s)=\zeta_{K_0}(s)+{1\over 3^s}+\int_0^\infty
 \frac{d\delta_1(k)}{dk} {dk\over
\pi(k^2+4)^s}+\sum_{l=0}^{N-1}{1\over l^s(2\sigma-l)^s}+{\aleph\over
2N^s(2\sigma-N)^s}+\int_0^\infty \frac{d\delta_2(q)}{dq}
{dq\over\pi(q^2+\sigma^2)^s} \quad . \label{eq:ztk2}
\end{equation}
Both in the partition function and in the generalized zeta function
half-zero bound states contribute half as much as bound states.

Because in the general case when $\sigma\neq N$ analytical
expressions for the $TK1$ kink generalized zeta function are not
available, we shall rely on the DHN procedure, very well tested in
the paradigmatic $\lambda \phi^4$ and sine-Gordon kinks. In this
framework, the $TK1$ kink Casimir energy, where the contribution of
the half-bound state of the vacuum operator $K_0$ - $\,$
${\sigma\over 2}$ $\,$ - is always present, can be written as:

\begin{eqnarray*}
\Delta M_K^C(\vec{\phi}^{TK1})&=&\Delta
E(\vec{\phi}^{\,{\rm TK}1})-\Delta
E(\vec{\phi}^{\,\pm(1)})\\&=&\lim_{s \rightarrow
-\frac{1}{2}}\frac{\hbar}{2} \left( \frac{ \mu^2}{m^2} \right)^s \mu
\left[ \zeta_K (s)-\zeta_{K_0}(s) -\left({\pi\over
2(2+1)}\right)^{2s}-\left({\pi\over
\sigma(\sigma+1)}\right)^{2s}\right]\\&=&{\hbar m\over
2}\left[\sqrt{3}+{1\over\pi}\int_0^\infty \, dk \,
\frac{d\delta_1(k)}{dk}
\sqrt{k^2+4}-{2(2+1)\over\pi}\right]+\\&+&{\hbar m\over
2}\left[\sum_{l=0}^{N-1}\,\sqrt{l(2\sigma-l)}+{\aleph\over 2}\sqrt{
N(2\sigma-N)}-{\sigma\over 2}+{1\over\pi}\int_0^\infty \, dq \,
\frac{d\delta_2(q)}{dq} {1\over
(q^2+\sigma^2)^s}-{\sigma(\sigma+1)\over\pi}\right]
\end{eqnarray*}
is still divergent. Zero-point vacuum energy renormalization is not
enough. Note that we have subtracted a finite piece to use the
mode-number regularization method.

The contribution to the one-loop $TK1$ kink mass, induced by the
mass renormalization counter-terms through the Lagrangian density
${\cal L}_{C.T.}$, is:
\begin{eqnarray*}
\Delta M_K^R(\vec{\phi}^{TK1})&=&-2\hbar m\left[
6I(4)+\sigma(\sigma+1)I(\sigma^2\right]\int dx \left((\phi_1^{{\rm
TK}1})^2(x)-(\phi_1^{\pm(1)})^2\right)\\&=&\hbar
m\left[6I(4)+\sigma(\sigma+1)I(\sigma^2)\right]=\hbar m
\left[{3\over\pi}\int_0^\infty \, dk \, {1\over(k^2+4)^{{1\over
2}}}+\frac{\sigma(\sigma+1)}{2\pi}\int_0^\infty \, dq \,
{1\over(q^2+\sigma)^{{1\over 2}}}\right] \quad .
\end{eqnarray*}
 The divergent integrals in $\Delta M_K^R(\vec{\phi}_K^{TK1})$ can also be regularized
 by means of zeta function methods:
\[
I(4)=\lim_{s\rightarrow {1\over 2}}\frac{1}{2\mu L}\cdot
\left(\frac{\mu^2}{m^2}\right)^{s}\cdot
\zeta_{K_0^{11}}(s)=\lim_{s\rightarrow {1\over 2}}
\cdot\left(\frac{\mu^2}{m^2}\right)^{s-{1\over 2}}\cdot
\int_0^\infty \, dk \, \frac{1}{(k^2+4)^{s}}
\]
\[
I(\sigma^2)=\lim_{s\rightarrow {1\over 2}}\frac{1}{2\mu L}\cdot
\left(\frac{\mu^2}{m^2}\right)^{s}\cdot
\zeta_{K_0^{22}}(s)=\lim_{s\rightarrow {1\over 2}}
\left(\frac{\mu^2}{m^2}\right)^{s-{1\over 2}}\cdot \int_0^\infty \,
dq \, \frac{1}{(q^2+\sigma^2)^{s}} \qquad .
\]
We have used the second choice of zeta function regularization
proposed in Section \S 4 because the difference in finite
renormalization is included here in $\Delta M_K^C$. Therefore, the
regularized induced energy is:

\begin{eqnarray*} \bigtriangleup M_K^R(\vec{\phi}^{TK1};s)&=&\frac{\hbar
m}{2}\left({\mu^2\over m^2}\right)^{s}{1\over 2\mu
L}\left[6\zeta_{K_0^{11}}(s)+\sigma(\sigma+1)\zeta_{K_0^{22}}(s)\right]\\&=&
\frac{\hbar m}{2}\left({\mu^2\over m^2}\right)^{s-{1\over
2}}\left[3\int_0^\infty dk \frac{1}{(k^2+4)^{s}}
+{\sigma(\sigma+1)\over 2}\int_0^\infty dq
\frac{1}{(q^2+\sigma^2)^{s}}\right]
\end{eqnarray*}
and $\bigtriangleup M_K^R(\vec{\phi}^{TK1};{1\over 2})=\Delta
M_K^R(\vec{\phi}^{TK1})$.

In the next Table exact results are shown for several values of
$\sigma$, obtained through numerical integration of the above
formulas.

{\small\begin{center}
\begin{tabular}{|c|c|} \\[-0.6cm] \hline
$\sigma$ & $\Delta M({\rm TK1})/\hbar m$ \\ \hline $0.4$ &
$-0.799335$ \\ $0.5$ & $ -0.829892$ \\ $0.6$ & $-0.860369$ \\
$0.7$ & $ -0.890955$ \\ $0.8$ & $-0.921788$ \\ $0.9$ & $
-0.952966$ \\ $0.99$ & $ -0.981384$ \\ $1.00$ & $ -0.984565$ \\
$1.01$ & $ -0.98775$ \\ $1.1$ & $ -1.01664$ \\ $1.2$ & $ -1.04925$
\\ $1.3$ & $ -1.08242 $ \\ \hline
\end{tabular} \hspace{0.1cm}
\begin{tabular}{|c|c|} \\[-0.6cm] \hline
$\sigma$ & $\Delta M({\rm TK1})/\hbar m$ \\ \hline $1.4$ & $
-1.11618 $ \\ $1.5$ & $ -1.15057 $ \\ $1.6$ & $ -1.18559 $ \\
$1.7$ & $ -1.22128 $ \\ $1.8$ & $ -1.25765 $ \\ $1.9$ & $ -1.2947
$ \\ $1.99$ & $ -1.32865 $ \\ $2.0$ & $-1.33251$  \\ $2.01$ & $
-1.33627 $ \\ $2.1$ &$ -1.37094 $\\ $2.2$ & $ -1.41013 $ \\ $2.3$
& $-1.45005 $ \\ \hline
\end{tabular} \hspace{0.1cm}
\begin{tabular}{|c|c|} \\[-0.6cm] \hline
$\sigma$ & $\Delta M({\rm TK1})/\hbar m$ \\ \hline $2.4$ &
$-1.4907 $ \\ $2.5$ & $ -1.53212$ \\ $2.6$ & $-1.57427 $ \\ $2.7$
& $ -1.61717 $ \\ $2.8$ & $-1.65316 $ \\ $2.9$ & $ -1.70527 $ \\
$2.99$ & $-1.74592 $ \\ $3.0$ & $ -1.75077 $\\ $3.01$ & $ -1.75503
$
\\ $3.1$ & $-1.79644 $  \\ $3.2$ & $ -1.84319 $  \\ $3.3$ & $
-1.89071 $  \\ \hline
\end{tabular}
\end{center}}

Departure from the reflectionless case (captured by the CCG
formula) is thus measured.

\subsection{The high-temperature expansion of the partition function}

For any other $TK2$ kink, knowledge of the spectrum of the kink
fluctuation operator $K(c)$ is grossly insufficient to compute the
generalized zeta function. Thus, we need to use asymptotic methods
to obtain sufficiently good approximations to kink generalized zeta
functions.

The heat equation kernel for the $K_0$-heat equation
\[
\left(\begin{array}{cc}\frac{\partial}{\partial\beta}-\frac{\partial^2}{\partial
x^2}+4 & 0 \\ 0 &
\frac{\partial}{\partial\beta}-\frac{\partial^2}{\partial
x^2}+\sigma^2
\end{array}\right)K_{K_0}(x,y;\beta)=0 \quad , \quad
K_{K_0}(x,y;0)=\left(\begin{array}{cc} \delta(x-y) & 0\\0&
\delta(x-y)
\end{array} \right) \quad ,
\]
of the vacuum fluctuation operator $K_0$, for small $\beta$ is:
\[
K_0=\left(\begin{array}{cc} -{d^2\over dx^2}+4 & 0
\\ 0 & -{d^2\over
dx^2}+\sigma^2\end{array}\right) \qquad , \qquad
K_{K_0}(x,y;\beta)=\left(\begin{array}{cc}\frac{e^{-4\beta}}{\sqrt{4\pi\beta}}\cdot
e^{-\frac{(x-y)^2}{4\beta}}&0\\0&\frac{e^{-\sigma^2\beta}}{\sqrt{4\pi\beta}}\cdot
e^{-\frac{(x-y)^2}{4\pi\beta}}\end{array}\right) \qquad .
\]

The $TK2$ kink fluctuation operator is the $2\times 2$ matrix
Schrodinger differential operator
\[
K(c)=\left(\begin{array}{cc} -{d^2\over dx^2}+4+V^{11}(x;c) &
V^{12}(x;c)
\\ V^{21}(x;c) & -{d^2\over
dx^2}+\sigma^2+V^{22}(x;c)\end{array}\right)=K_0+V(x;c)\qquad ,
\]
whereas the corresponding heat equation kernel
\[
\left\{\left(\begin{array}{cc}
\frac{\partial}{\partial\beta}-\frac{\partial^2}{\partial x^2}+4 &
0 \\ 0 & \frac{\partial}{\partial\beta}-\frac{\partial^2}{\partial
x^2}+\sigma^2
\end{array}\right)+V(x;c) \right\}K_{K(c)}(x,y;\beta)=0\qquad ,
\]
with
\[
K_{K(c)}(x,y;0)=\left(\begin{array}{cc} \delta(x-y) & 0\\
0& \delta(x-y)\end{array} \right)  ,
\]
can be written in the form
\[
K_{K(c)}(x,y;\beta)=C_K(x,y;\beta)\cdot K_{K_0}(x,y;\beta) \qquad
, \qquad C_K(x,y;0)=\left(\begin{array}{cc} 1 & 0 \\
0 & 1
\end{array}\right)
\]
if the $2\times 2$ matrix $C_K(x,y;\beta)$ satisfies the transfer
equation
\[
\left(\frac{\partial}{\partial\beta}+\frac{x-y}{\beta}\frac{\partial}{\partial
x}-\frac{\partial^2}{\partial
x^2}+\left(4(\delta^{A1}-\delta^{B1})+\sigma^2(\delta^{A2}-\delta^{B2})\right)\right)C^{AB}_K(x,y;\beta)+\sum_{C=1}^2V^{AC}(x;c)
C^{CB}_K(x,y;\beta)=0^{AB} \, \, ,
\]
and if it is set to the unit matrix at infinite temperature.

Solving the transfer equation as a power series in $\beta$
\[
C^{AB}_K(x,y;\beta)=\sum_{n=0}^\infty \, c_n^{AB}(x,y;K)\, \beta^n
\hspace{1.5cm} , \hspace{1.5cm} c_0^{AB}(x,y;K)=\delta^{AB}
\]
the PDE transfer system of equations becomes tantamount to the
system of recurrence relations: {\normalsize\begin{eqnarray*}
(n+1)c_{n+1}^{AB}(x,y;K)&+&(x-y)\frac{\partial
c_{n+1}^{AB}}{\partial
x}(x,y;K)=\\&=&\left(\frac{\partial^2}{\partial
x^2}-\left(4(\delta^{A1}-\delta^{B1})+\sigma^2(\delta^{A2}-\delta^{B2})\right)\right)c_n^{AB}(x,y;K)-\sum_{C=1}^2V^{AC}(x)c_n^{CB}(x,y;K)
\, \, . \end{eqnarray*}}

The matrix elements of the heat equation kernel for $K(c)$ have
the high-temperature asymptotic form
\[
K_{K(c)}^{11}(x,y;\beta)=\frac{e^{-4\beta}}{\sqrt{4\pi\beta}}\cdot
e^{-\frac{(x-y)^2}{4\beta}}\cdot \sum_{n=0}^\infty
c_n^{11}(x,y;K)\beta^n \, ; \,
K_{K(c)}^{12}(x,y;\beta)=\frac{e^{-\sigma^2\beta}}{\sqrt{4\pi\beta}}\cdot
e^{-\frac{(x-y)^2}{4\beta}}\cdot \sum_{n=0}^\infty
c_n^{12}(x,y;K)\beta^n
\]
\[
K_{K(c)}^{22}(x,y;\beta)=\frac{e^{-\sigma^2\beta}}{\sqrt{4\pi\beta}}\cdot
e^{-\frac{(x-y)^2}{4\beta}}\cdot \sum_{n=0}^\infty
c_n^{22}(x,y;K)\beta^n \, ; \,
K_{K(c)}^{12}(x,y;\beta)=\frac{e^{-4\beta}}{\sqrt{4\pi\beta}}\cdot
e^{-\frac{(x-y)^2}{4\beta}}\cdot \sum_{n=0}^\infty
c_n^{21}(x,y;K)\beta^n \quad .
\]

Actually, we are interested in the trace of the heat kernel, both
matrix and functional, to find the partition function for small
$\beta$. The recurrence relations become
\[
^{(0)}C_{n+1}^{AB}(x)=\frac{1}{n+1}\left[^{(2)}C_n^{AB}(x)-\left(4(\delta^{A1}-
\delta^{B1})+\sigma^2(\delta^{A2}-\delta^{B2})\right)^{(0)}C_n^{AB}(x)-\sum_{C=1}^2V^{AC}(x)^{(0)}C_n^{CB}(x)\right]
\]
when $y\rightarrow x$. To deal with this delicate point, we have
introduced the following notation:
\[
^{(k)}C_n^{AB}(x)=\lim_{y\rightarrow x}\frac{\partial^k
c_n^{AB}}{\partial x^k}(x,y;K)\qquad , \qquad
 ^{(k)}C_0^{AB}(x)=\lim_{y\rightarrow
x}\frac{\partial^kc_0^{AB}}{\partial
x^k}(x,y;K)=\delta^{k0}\delta^{AB} \qquad .
\]
We also need (obtained after differentiating the first recurrence
formula $k$-times) recurrence relations among the derivatives:
\begin{eqnarray*}
^{(k)}C_{n+1}^{AB}(x)&=&{1\over
n+k+1}\left[^{(k+2)}C_{n}^{AB}(x)-\left(4(\delta^{A1}-\delta^{B1})+\sigma^2(\delta^{A2}-\delta^{B2})\right)^{(k)}C_n^{AB}(x)-\right.\\
&-&\left.\sum_{j=0}^k\sum_{C=1}^2\left(\begin{array}{c} k
\\ j
\end{array}\right)\frac{d^jV^{AC}(x)}{d x^j}\cdot
^{(k-j)}C_{n}^{CB}(x)\right] \qquad .
\end{eqnarray*}

The high temperature asymptotic expansion of the partition
function reads:
\[
{\rm Tr}e^{-\beta K}={1\over\sqrt{4\pi\beta}}\cdot
\sum_{n=0}^\infty\,
\left[e^{-4\beta}c_n^{11}(K)+e^{-\sigma^2}c_n^{22}(k)\right]\beta^n
\qquad , \qquad
c_n^{AB}(K)=\lim_{L\rightarrow\infty}\int_{-{mL\over 2}}^{{mL\over
2}}\, dx \, c_n^{AB}(x,x;K) \qquad .
\]
Using the recurrence relations, the $c_n^{AB}(x,x;K)$ densities can
be found. They are the conserved charges of a generalized matrix KdV
equation, see Appendix III.

\subsection{The Mellin transform of the asymptotic expansion}

We now write the generalized zeta function of $K_0$ as the Mellin
transform of the $K_0$ partition function split into two integrals:
\begin{eqnarray*}
\zeta_{K_0}(s)&=&\frac{mL}{\sqrt{4\pi}}\cdot
\frac{1}{\Gamma(s)}\cdot \left[\int_0^1 \, d\beta \,
\beta^{s-{3\over 2}}\left[
e^{-4\beta}+e^{-\sigma^2\beta}\right]+\int_1^\infty \, d\beta \,
\beta^{s-{3\over 2}}\left[e^{-4\beta}+e^{-\sigma^2\beta}\right]
\right]\\&=&\frac{mL}{\sqrt{4\pi}}\cdot \frac{1}{\Gamma(s)}\cdot
\left\{{1\over 4^{s-{1\over 2}}}\cdot \left[\gamma[s-{1\over
2},4]+\Gamma[s-{1\over 2},4]\right]+{1\over \sigma^{2s-1}}\cdot
\left[\gamma[s-{1\over 2},\sigma^2]+\Gamma[s-{1\over
2},\sigma^2]\right]\right\}
\end{eqnarray*}
The incomplete $\gamma[s-{1\over 2},4]$ and $\gamma[s-{1\over
2},\sigma^2]$ have poles at $s-{1\over 2}=0,-1,-2,-3, \cdots $ but
their complementary functions $\Gamma[s-{1\over 2},4]$ and
$\Gamma[s-{1\over 2},\sigma^2]$ are entire functions of $s$.

To obtain the generalized zeta function from the asymptotic
expansion of the $K$ partition function the Mellin transform is also
split into two integrals, inside and outside the convergence radius:
\begin{eqnarray*}
&&\zeta^*_K(s)= \frac{1}{\Gamma(s)}\left[ -2\int_0^1 \, d\beta \,
\beta^{s-1}+{1\over\sqrt{4\pi}}\cdot
\sum_{n=0}^{N_0}\left(c_n^{11}(K)\cdot \frac{\gamma[s+n-{1\over
2},4]}{4^{s+n-{1\over 2}}}+c_n^{22}(K)\cdot \frac{\gamma[s+n-{1\over
2},\sigma^2]}{\sigma^{2s+2n-1}}\right)+\right.\\&+&\left.
{1\over\sqrt{4\pi}}\cdot \sum_{n=N_0+1}^{\infty}\left(
c_n^{11}(K)\cdot \frac{\gamma[s+n-{1\over 2},4]}{4^{s+n-{1\over
2}}}+c_n^{22}(K)\cdot \frac{\gamma[s+n-{1\over
2},\sigma^2]}{\sigma^{2s+2n-1}}\right)+{1\over\sqrt{4\pi}}\cdot
\int_1^\infty \, d\beta \, \beta^{s-1} {\rm Tr}^{*} e^{-\beta
K}\right]
\end{eqnarray*}
The two zero modes have been not accounted for and the incomplete
Gamma functions $\gamma[s+n-{1\over 2},4]$ and $\gamma[s+n-{1\over
2},\sigma^2]$ have poles at $s+n-{1\over 2}=0,-1,-2,-3, \cdots $. A
large but finite number $N_0$ is chosen to separate the contribution
of the high-order coefficients
\[
b^{N_0}_K(-{1\over 2})={1\over\sqrt{4\pi}}\cdot
\sum_{n=N_0+1}^{\infty}\left(c^{11}_n(K)\cdot
\frac{\gamma[s+n-{1\over 2},4]}{4^{s+n-{1\over 2}}}+c^{22}_n(K)\cdot
\frac{\gamma[s+n-{1\over 2},\sigma^2]}{\sigma^{2s+2n-1}}\right)
\]
which are holomorphic functions of $s$ for ${\rm Re}s>-N_0-1$. $
B_K(s)={1\over\sqrt{4\pi}}\cdot \int_1^\infty \, d\beta \,
\beta^{s-1} {\rm Tr}^* e^{-\beta K}$, however, is a entire function
of $s$.

\subsection{The high-temperature one-loop TK2 kink mass shift formula}

Neglecting the (very small) contribution of the entire functions,
the $TK2$ kink Casimir energy becomes:
\[
\Delta M_{K(c)}^C\simeq{\hbar\over 2}\lim_{s\rightarrow -{1\over
2}}\left({\mu^2\over
m^2}\right)^s\cdot\mu\cdot\frac{1}{\Gamma(s)}\cdot
\left[{1\over\sqrt{4\pi}}\sum_{n=1}^{N_0}\left(
c_n^{11}(K(c))\frac{\gamma[s+n-{1\over 2},4]}{4^{s+n-{1\over
2}}}+c_n^{22}(K(c))\frac{\gamma[s+n-{1\over
2},\sigma^2]}{\sigma^{2s+2n-1}}\right)-{2\over s}\right]
\]
i.e., the zero-point vacuum energy renormalization takes care of the
term coming from $c^{11}_0(K(c))$ and $c^{22}_0(K(c))$.

The other correction due to the mass renormalization counter-terms
can also be arranged into meromorphic and entire parts:
\begin{eqnarray*}
\Delta M_K^R=-\frac{\hbar \mu}{2\sqrt{4\pi}} \cdot
\lim_{s\rightarrow -{1\over
2}}\left(\frac{\mu^2}{m^2}\right)^{s+{1\over 2}}\cdot
{1\over\Gamma(s)} &\cdot &\left\{\frac{c_1^{11}(K)(c)}{4^{s+{1\over
2}}} \cdot \left[\gamma[s+{1\over 2},4]+\Gamma[s+{1\over 2},4]
\right]+ \right.\\&+&\left. \frac{ c_1^{22}(K(c))}{\sigma^{2s+1}}
\cdot \left[\gamma[s+{1\over 2},\sigma^2]+\Gamma[s+{1\over
2},\sigma^2]\right]\right\}
\end{eqnarray*}
The mass renormalization terms exactly cancel the contributions of
$c^{11}_1(K(c))$ and $c^{22}_1(K(c))$. Our minimal subtraction
scheme fits in with the following renormalization prescription: in
theories with only massive fluctuations quantum corrections vanish
at the limit where all the masses go to infinity.

We end with the high-temperature one-loop $TK2$ kink mass shift
formula:
\[
\Delta M_K(c)=-\frac{\hbar
m}{4\sqrt{\pi}}\cdot
\left[\frac{1}{\sqrt{4\pi}}\cdot\sum_{n=2}^{N_0}\,\left(
c^{11}_n(K(c))\cdot
\frac{\gamma[n-1,4]}{4^{n-1}}+c^{22}_n(K(c))\cdot
\frac{\gamma[n-1,\sigma^2]}{\sigma^{2n-2}}\right)+4\right]
\]

In this case, the subtraction of the two zero modes contributes to
the mass shift in the $c$-independent quantity:
\[
\Delta M_{K(c)}^{(0)}=-{\hbar m\over\sqrt{\pi}}=-0.56419 \hbar m
\qquad ,
\]
i.e., for each kink in the $TK2$ family we must subtract the same
quantity to discard zero mode effects at the one-loop level.

\subsection{Mathematica calculations}
Computational limitations put a practical bound on the choice of
$N_0$. Knowledge of, say, ${}^{(0)}C_2$ requires computation of 36
(or $9N^2$ in field theories of $N$ scalar fields) densities:
\[
\begin{array}{ccccc} {}^{(4)}C_0 & {}^{(3)}C_0 & {}^{(2)}C_0 & {}^{(1)}C_0& {}^{(0)}C_0 \\
 & & {}^{(2)}C_1 & {}^{(1)}C_1 & {}^{(0)}C_1 \\ & & & & {}^{(0)}C_2 \end{array}
\]
In general, evaluation of ${}^{(0)}C_n(x)$ requires previous
calculation of
\[
4(1+3+5+7+\cdots + 2n-1+2n+1)=4(n+1)^2 \qquad , \qquad
N^2(1+3+5+7+\cdots + 2n-1+2n+1)=N^2(n+1)^2 \, \, \, \, .
\]
This count could be slightly abbreviated bearing in mind that the
$4(2n+1)$ coefficients in the upper row are fixed by the initial
conditions of the recurrence relation.

\subsubsection{TK1 topological kinks}

The previous formulas can be applied to the $c=-\infty$ case for
several values of $\sigma$; i.e., to compute, by means of the
asymptotic method, the $TK1$ kink mass shift, to find -using
Mathematica- the results shown in the next Table.

{\large\begin{center}
\begin{tabular}{|c|c|} \\[-0.6cm] \hline
$\sigma$ & $\Delta M_{K(-\infty)}/\hbar m$ \\ \hline $0.5$ & $
-0.962386 $ \\ $0.6$ & $ -0.970537 $ \\ $0.7$ & $ -0.981183 $ \\
$0.8$ & $ -0.994487 $ \\ $0.9$ & $ -1.01053 $ \\ $1.0$ & $ -1.0293
$ \\ \hline
\end{tabular} \hspace{0.1cm}
\begin{tabular}{|c|c|} \\[-0.6cm] \hline
$\sigma$ & $\Delta M_{K(-\infty)}/\hbar m$ \\ \hline $1.1$ & $
-1.05073 $
\\ $1.2$ & $ -1.07468$ \\ $1.3$ & $ -1.10097 $ \\ $1.4$
& $ -1.12939 $ \\ $1.5$ & $ -1.15971 $ \\ $1.6$ & $ -1.19174 $ \\
\hline
\end{tabular} \hspace{0.1cm}
\begin{tabular}{|c|c|} \\[-0.6cm] \hline
$\sigma$ & $\Delta M_{K(-\infty)}/\hbar m$ \\ \hline $1.7$ & $
-1.22526 $
\\ $1.8$ & $ -1.2599 $ \\ $1.9$ & $ -1.29571 $ \\ $2.0$
& $ -1.33324 $  \\ $2.1$ & $ -1.37074 $ \\ $2.2$ & $ -1.41007 $ \\
\hline
\end{tabular}
\end{center}}
By comparing these numbers with those obtained by the exact
procedures of Cahill-Comtet-Glauber and Dashen-Hasslacher-Neveu we
show that the error of the asymptotic method decrease with
increasing $\sigma$, see next Table and Figure:
\begin{center}
\begin{tabular}{|c|c|c|c|c|} \hline
Value of $\sigma$ & $\sigma=0.9$ & $\sigma=1.5$ & $\sigma=2.0$ &
$\sigma=2.2$ \\ \hline  Relative Error &
6.0  \% &  0.79 \% & 0.055 \% & 0.004  \% \\
\hline
\end{tabular}
\end{center}
\begin{figure}[htbp]
\centerline{\epsfig{file=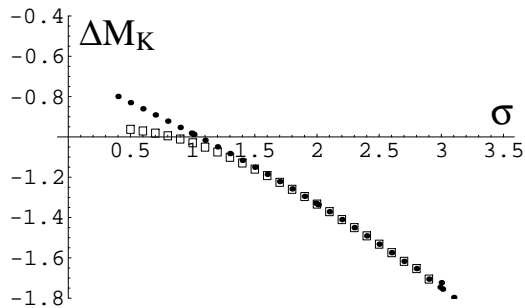,height=4cm}} \caption{\small
{\it One-loop correction to the one-component topological kink (TK1)
mass in units of $\hbar m$. $\bullet$,  DHN formula . $\Box$,
asymptotic series.}}
\end{figure}

\subsubsection{Elliptic two-component TK2 topological kinks}

Along as $\sigma<1$, $c=0$ is a kink orbit. This kink orbit is a
half-ellipse in the $\phi_1$-$\phi_2$ plane and the corresponding
kinks are $TK2$ kinks for which the kink fluctuation operator reads:
\[
K(0)=\left(\begin{array}{cc} -{d^2\over
dx^2}+4-\frac{2(1+2\sigma^2)}{{\rm cosh}^2[2(1-\sigma)x]} &
4\bar{\sigma}\sqrt{\sigma(\sigma+1)} \frac{{\rm
sinh}[2(1-\sigma)x]}{{\rm
cosh}^2[2(1-\sigma)x]}\\4\bar{\sigma}\sqrt{\sigma(\sigma+1)}
\frac{{\rm sinh}[2(1-\sigma)x]}{{\rm cosh}^2[2(1-\sigma)x]} &
-{d^2\over dx^2}+ \sigma^2-\frac{\sigma(5-7\sigma)}{{\rm
cosh}^2[2(1-\sigma)x]}\end{array}\right)\qquad .
\]
The $ TK2(0)$ Seeley coefficients are computed for several values of
$\sigma$ in the $\sigma \in [0.96,1)$ range. Together with the
one-loop ${\rm TK}2(0)$ mass quantum corrections according to the
asymptotic method, they are shown in the next Tables: {\small
\begin{center}
\begin{tabular}{|c|c|c|c|c|} \hline
&
\multicolumn{2}{|c|}{$\sigma=0.96$}&\multicolumn{2}{|c|}{$\sigma=0.97$}
\\ \hline
$n$ & $c_n^{11}(K(0))$ & $c_n^{22}(K(0))$ & $c_n^{11}(K(0))$ &
$c_n^{22}(K(0))$
\\ & \rule{1.6cm}{0cm} & \rule{1.6cm}{0cm} & \rule{1.6cm}{0cm} &
\rule{1.6cm}{0cm} \\[-0.4cm]
\hline
1&12.6806 & 3.83333& 12.5025& 3.87629\\
2&27.2626 & 2.91223& 26.3923& 2.84445 \\
3&43.0467 & 0.929709 &40.8995&0.971713\\
4&51.8330 & 0.431151 &48.3164&0.391887\\
5&49.4842 & -0.00452478 & 45.2317& 0.0193729\\
6&38.8786& 0.0514525 & 34.8351& 0.0380063 \\
7& 25.8993& -0.0156285 &22.7417&-0.0090224\\
8&14.9659&0.00713977& 12.8764& 0.00456185 \\ \hline
\end{tabular}
\end{center}

{\small\begin{center}
\begin{tabular}{|c|c|c|c|c|} \hline
&
\multicolumn{2}{|c|}{$\sigma=0.98$}&\multicolumn{2}{|c|}{$\sigma=0.99$}
\\ \hline
$n$ & $c_n^{11}(K(0))$ & $c_n^{22}(K(0))$ & $c_n^{11}(K(0))$ &
$c_n^{22}(K(0))$
\\ & \rule{1.6cm}{0cm} & \rule{1.6cm}{0cm} & \rule{1.6cm}{0cm} &
\rule{1.6cm}{0cm} \\[-0.4cm]
\hline
1&12.3299 & 3.91837& 12.1624& 3.9596\\
2&25.5597 & 2.78109& 24.7630& 2.7219 \\
3&38.8820 & 1.00819 &36.9849&1.03968\\
4&45.0733 & 0.358188 &42.0798&0.329352\\
5&41.3849 & 0.0389505 & 37.9012& 0.0548776\\
6&31.2486& 0.0273206 & 28.0632& 0.0188943 \\
7& 19.9962& -0.00450157 &17.6055&-0.000842262\\
8&11.0960&0.00265278& 9.57626& 0.0003676\\ \hline
\end{tabular}
\end{center}}}

\begin{center}
\begin{tabular}{|c|c|} \\[-0.6cm] \hline
$\sigma$ & $\Delta M_{K(0)}\hbar m$ \\ \hline
$0.96$ & $ -1.06082 $ \\
$0.97$ & $ -1.05253 $ \\
$0.98$ & $ -1.04422 $ \\
$0.99$ & $ -1.03624 $ \\
\hline
\end{tabular}
\end{center}

\subsubsection{One-loop breaking of classical kink degeneracy}

To end this Section we compute one-loop mass shifts for ${\rm TK}2$
kink families by means of the asymptotic approximation. In the next
Tables we offer results for $\sigma=1.5$, $\sigma=2$, and
$\sigma=2.5$ and several values of $c$ between $-30$ and a value
very close to $c^S$. {\small
\begin{center}
{\begin{tabular}{cc} $\sigma=1.5$ & \\ \hline $c$ & $\Delta M$ \\
\hline $-30$ & $-1.16009$ \\ $-27.5$ & $-1.16017$ \\ $-25$ &
$-1.16128$ \\ $-22.5$ & $-1.16042$ \\ $-20$ & $-1.16061$ \\
$-17.5$ & $-1.16088$ \\ $-15$ & $-1.16128$ \\ $-12.5$ & $-1.16193$
\\ $-10$ & $-1.16313$ \\ $-7.5$ & $-1.16597$ \\ $-5$  & $-1.18205$
\\ $-4.6801886$ & $-1.24345$ \\ $-4.68018860186678332$ &
$-1.25103$ \\ \hline
 & \\
 & \\
\end{tabular}} \hspace{2cm} {\begin{tabular}{cc} $\sigma=2.0$ & \\ \hline $c$ & $\Delta M$ \\
\hline $-30$ & $-1.33281$ \\ $-27.5$ & $-1.33281$ \\ $-25$ &
$-1.33281$ \\ $-22.5$ & $-1.33281$ \\ $-20$ & $-1.33281$ \\
$-17.5$ & $-1.33281$ \\ $-15$ & $-1.33281$ \\ $-12.5$ & $-1.33281$
\\ $-10$ & $-1.33281$ \\ $-7.5$ & $-1.33281$ \\ $-5$  & $-1.33280$
\\ $-4.001$  & $-1.33280$ \\ $-4.00001$  & $-1.33280$ \\ \hline &
\\ & \\
\end{tabular}}
\end{center}

 \begin{tabular}{cc} $\sigma=2.5$ & \\ \hline
 $c$ & $\Delta M$ \\ \hline
$-30$ & $-1.52784$ \\ $-27.5$ & $-1.52782$ \\ $-25$ & $-1.52780$
\\ $-22.5$ & $-1.52778$ \\ $-20$ & $-1.52774$ \\ $-17.5$ &
$-1.52769$
\\ $-15$ & $-1.52760$ \\ $-12.5$ & $-1.52744$ \\ $-10$ &
$-1.52711$ \\ $-7.5$ & $-1.52626$ \\ $-5$  & $-1.52285$ \\ $-4$  &
$-1.52168$ \\ $-3.97$ & $-1.52915$ \\ $-3.96594571$ & $-1.55402$
\\ $-3.96594570565808127$ & $-1.56127$ \\ \hline
\end{tabular}\hspace{0.5cm}
\begin{tabular}{c}
\epsfig{file=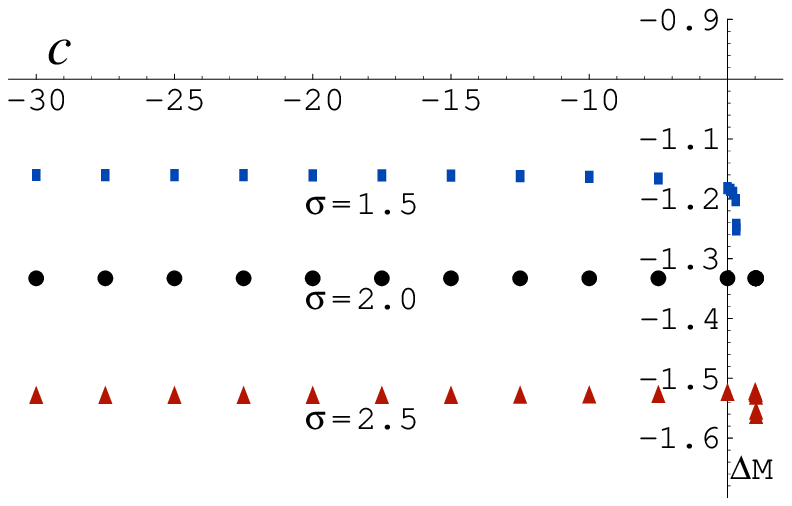,height=5.5cm} \\ {\large \it The One-loop
Quantum}
\\[0.3cm] {\large Mass Correction in the cases} \\[0.3cm] {\large $\sigma=1.5$, $\sigma=2.0$ and
$\sigma=2.5$}
\end{tabular}
}

The behavior is the same for $\sigma=1.5$  and $\sigma=2.5$: the
classical degeneracy of kink energies survives one-loop quantum
fluctuations for values of $c$ lower than the critical values where
$TK2$ kinks start to split into two lumps. The one-loop mass shifts
for $TK2$ kinks formed by two lumps are remarkably higher (in
absolute value) and increase with kink separation. For the value
$\sigma=2$, kink energy degeneracy is not lifted by one-loop
fluctuations. The reason is that the BNRT model for this value of
$\sigma$ is no more than two $\lambda\phi^4$ independent models if
appropriate linear combinations of $\phi_1$ and $\phi_2$ are chosen.


\section{The planar Abelian Higgs model}

Given a complex scalar field and a $U(1)$ gauge potential
\[
\psi(y^\mu):{\mathbb R}^{1,2}\longrightarrow {\mathbb C}
\hspace{1.5cm} ; \hspace{1.5cm} B^\mu(y^\mu)\frac{\partial}{\partial
y^\mu}: T{\mathbb R}^{1,2}\longrightarrow Lie {\mathbb U}(1) \qquad
,
\]
the action for the planar Abelian Higgs  model reads:
\begin{eqnarray*}
S&=&\int \, d^3y \, \left[-{1\over 4}G_{\mu\nu}G^{\mu\nu}+{1\over
2}(\nabla_\mu\psi)^* \nabla^\mu\psi-\frac{\lambda}{8}(\psi^*\psi-v^2)^2\right]\\
\nabla_\mu\psi&=&(\frac{\partial}{\partial y^\mu}+ieB_\mu)\psi
\hspace{2cm} , \hspace{2cm} G_{\mu\nu}=\frac{\partial
B_\nu}{\partial y^\mu}-\frac{\partial B_\mu}{\partial y^\nu}
\qquad ,
\end{eqnarray*}
where the volume and the metric tensor in (2+1)-dimensional
${\mathbb R}^{1,2}$ Minkowski space are given below, together
with the dimensions of the fields and parameters:
\begin{eqnarray*}
&& d^3y=dy^0dy^1dy^2 \hspace{3cm} , \hspace{1.5cm} g_{\mu\nu}={\rm
diag}(1,-1,-1) \\ && [\psi]=[B_\mu]=[v]=M^{{1\over 2}}
\hspace{1.5cm} , \hspace{1.5cm} [e]=[\lambda^{{1\over
2}}]=M^{-{1\over 2}}L^{-1} \qquad .
\end{eqnarray*}
Defining non-dimensional coordinates, fields, and parameters,
\[
y^\mu =\frac{1}{ev} x^\mu \hspace{0.5cm}, \hspace{0.5cm}
\psi=v\phi=v(\phi_1+i\phi_2)\hspace{0.5cm}, \hspace{0.5cm} B_\mu= v
A_\mu \hspace{0.2cm} , \hspace{0.2cm} \kappa^2=\frac{\lambda}{e^2}
\qquad ,
\]
the action and the field equations are:
\begin{eqnarray*} S&=& \frac{v}{e}\int d^3 x \left[ -\frac{1}{4}
F_{\mu \nu} F^{\mu \nu}+\frac{1}{2} (D_\mu \phi)^* D^\mu \phi -
\frac{\kappa^2}{8} (\phi^* \phi(x^\mu)-1)^2
\right]\\D_\mu\phi&=&(\frac{\partial}{\partial x^\mu}+iA_\mu)\phi
\hspace{2cm} , \hspace{2cm} F_{\mu\nu}=\frac{\partial
A_\nu}{\partial x^\mu}-\frac{\partial A_\mu}{\partial x^\nu}
\end{eqnarray*}
\[
\partial_\mu
F^{\mu\nu}=i\left[(D^\nu\phi)^*\phi-\phi^*D^\nu\phi\right]\qquad ,
\qquad \qquad D_\mu D^\mu\phi
=\frac{\kappa^2}{4}\phi(1-\phi^*\phi) \qquad .
\]

\subsection{Feynman rules in the Feynman-'t Hooft R-gauge}

There is $U(1)$-gauge symmetry
\[
\phi(x^\mu)\longrightarrow \phi^\prime(x^\mu)=e^{i\alpha
(x^\mu)}\phi(x^\mu) \qquad , \qquad   A_\mu(x^\mu)\longrightarrow
A^\prime_\mu(x^\mu)=A_\mu(x^\mu)+\frac{\partial\alpha}{\partial
x^\mu}(x^\mu)
\]
and if $\Lambda\in [0,2\pi]$ is a constant angle the vacuum orbit of
the gauge group is:
\[
\phi^V(x^\mu)=e^{i\Lambda} \quad , \quad
(\phi^\prime)(x^\mu)=e^{i(\Lambda+\alpha(x^\mu))}  \qquad ; \qquad
A_\mu^V(x^\mu)=0_\mu  \quad , \quad
(A^\prime_\mu)^V(x^\mu)=\frac{\partial\alpha}{\partial
x^\mu}(x^\mu) \qquad .
\]
We shift the scalar field away from the vacuum in $H(x^\mu)$-Higgs
and $G(x^\mu)$-Goldstone fields:
\[
\phi(x^\mu)= 1+H(x^\mu)+i G(x^\mu) \Rightarrow
\left\{\begin{array}{c} (1+H)^\prime(x^\mu)={\rm
cos}\alpha(x^\mu)(1+H(x^\mu))-{\rm sin}\alpha(x^\mu)G(x^\mu) \\
G^\prime(x^\mu)={\rm sin}\alpha(x^\mu)(1+H(x^\mu))+{\rm
cos}\alpha(x^\mu)G(x^\mu)\end{array}\right.\qquad .
\]
The choice of the Feynman-'t Hooft R-gauge
\[
R(A_\mu , G)=\partial_\mu A^\mu(x^\mu) - G(x^\mu) \qquad , \qquad
S_{{\rm g.f.}}=-{1\over 2}\int \, d^3x \, \left(\partial_\mu
A^\mu(x^\mu) - G(x^\mu)\right)^2 \qquad
\]
needs a Faddeev-Popov determinant to restore unitarity, which
amounts to introducing a complex ghost field. Because
\[
R(A^\prime_\mu,G^\prime)\simeq
R(A_\mu,G)+\left(\Box-1-H(x^\mu)\right)\cdot \delta\alpha(x^\mu)
\qquad ,
\]
we find
\begin{eqnarray*}
{\rm Det}\frac{\delta R}{\delta\alpha}&=&\int \,
[d\chi^*(x^\mu)][d\chi(x^\mu)] \, {\rm exp}\left(iS_{{\rm
ghost}}[\chi^*,\chi]\right)\\&=&\int [d\chi^*(x^\mu)][d\chi(x^\mu)]
{\rm exp}\left\{i\int  d^3x
\chi^*(x^\mu)\left(\Box-1-H(x^\mu)\right)\chi(x^\mu)\right\} \qquad
.
\end{eqnarray*}
All this together allows us to write the action in the form
\begin{eqnarray*}
S+S_{{\rm g.f.}}+S_{{\rm ghost}}&=&{v\over e}\int \, d^3x \,
\left[ -\frac{1}{2} A_\mu
[-g^{\mu\nu}(\Box +1)]A_\nu + \partial_\mu\chi^*\partial^\mu \chi- \chi^*\chi \right.\\
&+&\frac{1}{2}\partial_\mu G\partial^\mu G-\frac{1}{2} G^2+
\frac{1}{2}\partial_\mu H\partial^\mu H-\frac{\kappa^2}{2}  H^2\\
&-&{\kappa^2\over 2}H (H^2+G^2)+A_\mu (\partial^\mu H
G-\partial^\mu G H)+H (A_\mu A^\mu -\chi^* \chi)\\
&-&\left. \frac{\kappa^2}{8} (H^2+G^2)^2 +\frac{1}{2}(G^2+H^2)
A_\mu A^\mu \right] \qquad ,
\end{eqnarray*}
which encodes the Feynman rules shown in Tables 7 and 8. It should
be noted that fermionic (ghost) loops carry a $(-1)$ factor.

\begin{table}[hp]
\begin{center}
\caption{Propagators}
\begin{tabular}{lccc} \\ \hline
\textit{Particle} & \textit{Field} & \textit{Propagator} & \textit{Diagram} \\
\hline \\ Higgs & $H(x)$ & $\displaystyle\frac{i e \hbar
}{v(k^2-\kappa^2+i\varepsilon)}$ &
\parbox{2.0cm}{\epsfig{file=qsdv1.ps,width=2cm}} \\[0.5cm]
Goldstone & $G(x)$ & $\displaystyle\frac{ie
\hbar}{v(k^2-1+i\varepsilon)}$ &
\parbox{2.0cm}{\epsfig{file=qsdv2.ps,width=2cm}} \\[0.5cm]
Ghost & $\chi(x)$ & $\displaystyle\frac{ie
\hbar}{v(k^2-1+i\varepsilon)}$ &
\parbox{2.0cm}{\epsfig{file=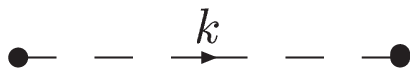,width=1.95cm}} \\[0.5cm]
Vector Boson & $A_\mu(x)$  &  {\Large $\frac{-ie \hbar g^{\mu\nu}
}{v(k^2-1+i\varepsilon)}$} &
\parbox{2.0cm}{\epsfig{file=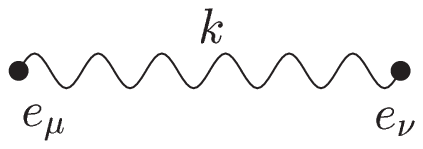,width=1.95cm}}
\\[0.5cm] \hline
\end{tabular}
\end{center}
\vspace{-0.6cm}
\begin{center}
\caption{Third- and fourth-order vertices }
\begin{tabular}{clcl} \\ \hline
\textit{Vertex} & \textit{Weight} & \textit{Vertex} & \textit{Weight} \\
\hline \\
\parbox{2cm}{\epsfig{file=qsdv5.ps,width=1.5cm}} &
$\displaystyle -3i\kappa^2\frac{v}{\hbar e} $  &
\parbox{2cm}{\epsfig{file=qsdv6.ps,width=1.5cm}} &
$\displaystyle -3i\kappa^2\frac{v}{\hbar e}$ \\[0.5cm]
\parbox{2cm}{\epsfig{file=qsdv7.ps,width=1.5cm}} &
$\displaystyle -i\kappa^2\frac{v}{\hbar e}$  &
\parbox{2cm}{\epsfig{file=qsdv8.ps,width=1.5cm}} &
$\displaystyle -3i\kappa^2\frac{v}{\hbar e}$ \\[0.5cm]
 \parbox{2cm}{\epsfig{file=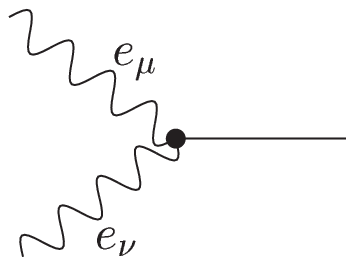,width=1.5cm}} &
$\displaystyle 2i\frac{v}{\hbar e} g^{\mu \nu}$  &
\parbox{2cm}{\epsfig{file=qsdv10.ps,width=1.5cm}} &
$\displaystyle -i\kappa^2\frac{v}{\hbar e}$
\\[0.5cm]
 \parbox{2cm}{\epsfig{file=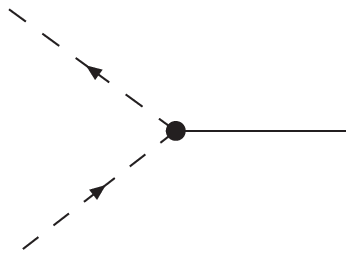,width=1.5cm}} &
$\displaystyle -i\frac{v}{\hbar e}$ &
\parbox{2cm}{\epsfig{file=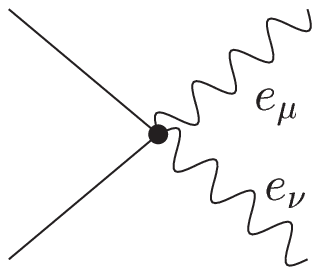,width=1.5cm}} &
$\displaystyle 2 i\frac{v}{\hbar e}g^{\mu \nu}$\\[0.5cm]
\parbox{2cm}{\epsfig{file=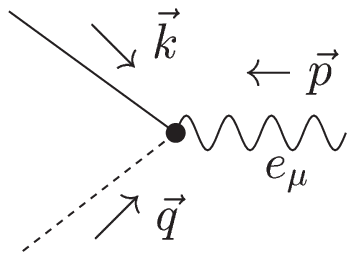,width=1.5cm}} &
$\displaystyle (k^\mu-q^\mu)\frac{v}{\hbar e}$  &
\parbox{2cm}{\epsfig{file=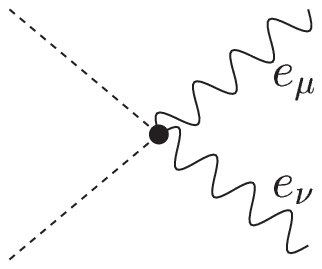,width=1.5cm}} &
$\displaystyle 2 i \frac{v}{\hbar e} g^{\mu \nu}$ \\[0.7cm] \hline
\end{tabular}
\end{center}
\end{table}

\subsection{Plane waves and vacuum energy}

\subsubsection{Vector bosons}

The general solution of the linearized equation for small
fluctuations of the vector field around the vacuum
\[
\left(\frac{\partial^2}{\partial
x_0^2}-\vec{\nabla}\vec{\nabla}+1\right)\delta
A_\mu(x_0,\vec{x})=0  \quad  , \quad A_\mu(x_0,\vec{x})\simeq
0_\mu+\delta A_\mu(x_0,\vec{x})
\]
is the plane wave expansion
\[
\delta A_\mu(x_0,\vec{x})=\left(\frac{\hbar^{{1\over
2}}}{e^{{1\over 2}}v^{{3\over 2}}L}\right) \cdot
\sum_{\vec{k}}\sum_{\alpha}\frac{1}{\sqrt{2\omega(\vec{k})}}\left[
a^*_\alpha(\vec{k})e^\alpha_\mu(k)e^{ikx}+a_\alpha(\vec{k})e^\alpha_
\mu(k)e^{-ikx}\right]
\]
if the dispersion relation $k_0^2-\vec{k}\vec{k}-1=0$ holds. Here,
we denote
\[
kx=k_\mu x^\mu=k_0x_0-\vec{k}\vec{x} \qquad , \qquad
\omega(\vec{k})=+\sqrt{\vec{k}\vec{k}+1} \qquad ,
\]
and consider periodic boundary conditions on a square of area
$m^2L^2$, $m=e v$,  such that:
\[
k_i=\frac{2\pi}{mL}n_i \quad , \quad  n_i\in{\mathbb Z} \quad ,
i=1,2 \qquad .
\]
The polarization vectors $e^\alpha_\mu(k)$, $\alpha=0,1,2$,
satisfy the ortho-normality condition:
\[
e^\alpha(k)\cdot
e^\beta(k)=e^{\alpha\mu}(k)e^\beta_\mu(k)=-(-1)^{\delta^{\alpha
0}}\delta^{\alpha\beta}\qquad .
\]
The linear classical Hamiltonian
\begin{eqnarray*}
H^{(2)}[\delta A_\mu ]&=&\frac{v^2}{2}\int \, d^2x \, \delta
A^\mu(x_0,\vec{x})\left[\frac{\partial^2}{\partial
x_0^2}+\vec{\nabla}\vec{\nabla}-1\right]\delta
A_\mu(x_0,\vec{x})\\&=&\frac{\hbar m}{2}\sum_{\vec{k}}\sum_\alpha
\omega(\vec{k})\left[(-1)^{\delta_{\alpha
0}}(a^*_\alpha(\vec{k})a_\alpha(\vec{k})+
a_\alpha(\vec{k})a^*_\alpha(\vec{k}))\right]
\end{eqnarray*}
leads, via canonical quantization
\[
[\hat{a}_\alpha(\vec{k}),\hat{a}_\alpha^\dagger
(\vec{q})]=(-1)^{\delta_{\alpha
0}}\delta_{\alpha\beta}\delta_{\vec{k}\vec{q}} \qquad \qquad ,
\]
to the quantum Hamiltonian for free massive vector bosons:
\[
H^{(2)}[\delta \hat{A}_\mu]=\sum_{\vec{k}}\sum_\alpha \hbar m
\omega(\vec{k})\left((-1)^{\delta_{\alpha
0}}\hat{a}_\alpha^\dagger(\vec{k})\hat{a}_\alpha(\vec{k})+{1\over
2})\right) \qquad .
\]
The contribution to the vacuum energy of the massive vector bosons
is:
\[
\Delta E^{(1)}_0=\sum_{\vec{k}}\sum_\alpha {\hbar m\over 2}
\omega(\vec{k})=\frac{3\hbar m}{2}{\rm
Tr}[-\vec{\nabla}\vec{\nabla}+1]^{{1\over 2}} \qquad .
\]

\subsubsection{Higgs bosons}

The linearized field equations for small fluctuations of the Higgs
field around the vacuum,
\[
\left(\frac{\partial^2}{\partial
x_0^2}-\vec{\nabla}\vec{\nabla}+\kappa^2\right)\delta
H(x_0,\vec{x})=0 \qquad ,
\]
are solved by the plane wave Higgs expansion
\begin{eqnarray*}
&&\delta
H(x_0,\vec{x})=\frac{1}{vL}\sqrt{\frac{\hbar}{ev}}\sum_{\vec{k}}\frac{1}{\sqrt{2\omega(\vec{k})}}
\left[a^*(\vec{k})e^{ikx}+a(\vec{k})e^{-ikx}\right]\\
&& k_0^2-\vec{k}\vec{k}-\kappa^2=0 \hspace{1.5cm} , \hspace{1.5cm}
\omega(\vec{k})=+\sqrt{\vec{k}\vec{k}+\kappa^2}\qquad .
\end{eqnarray*}
The classical energy of the Higgs plane waves
\begin{eqnarray*}
H^{(2)}[\delta H]&=&\frac{v^2}{2}\int \, d^2x \,
\left[\left(\frac{\partial\delta H}{\partial
x_0}\right)^2+\vec{\nabla}\delta H\vec{\nabla}\delta H+\kappa^2
\delta H\delta H\right]=\frac{\hbar
m}{2}\sum_{\vec{k}}\omega(\vec{k})\left[a^*(\vec{k})a(\vec{k})+a(\vec{k})a^*(\vec{k})\right]
\end{eqnarray*}
becomes -through the canonical quantization
$[\hat{a}(\vec{k}),\hat{a}^\dagger(\vec{q})]=\delta_{\vec{k}\vec{q}}$-
the quantum Hamiltonian for the Higgs bosons:
\[
H^{(2)}[\delta\hat{H}]=\hbar
m\sum_{\vec{k}}\omega(k)\left(\hat{a}^\dagger(\vec{k})\hat{a}(\vec{k})+{1\over
2}\right)\qquad .
\]
The contribution to the vacuum energy of the Higgs bosons is thus:
\[
\Delta E^{(2)}_0=\sum_{\vec{k}}{\hbar m\over 2}
\omega(\vec{k})=\frac{\hbar m}{2}{\rm
Tr}[-\vec{\nabla}\vec{\nabla}+\kappa^2]^{{1\over 2}} \qquad .
\]
\subsubsection{Goldstone bosons}

Simili modo, the linearized field equations for small fluctuations
of the Goldstone field around the vacuum
\[
\left(\frac{\partial^2}{\partial
x_0^2}-\vec{\nabla}\vec{\nabla}+1\right)\delta G(x_0,\vec{x})=0
\]
are solved in terms of Goldstone plane waves:
\begin{eqnarray*}
&&\delta
G(x_0,\vec{x})=\frac{1}{vL}\sqrt{\frac{\hbar}{ev}}\sum_{\vec{k}}\frac{1}{\sqrt{2\omega(\vec{k})}}
\left[a^*(\vec{k})e^{ikx}+a(\vec{k})e^{-ikx}\right]\\
&& k_0^2-\vec{k}\vec{k}-1=0 \hspace{1.5cm} , \hspace{1.5cm}
\omega(\vec{k})=+\sqrt{\vec{k}\vec{k}+1} \qquad .
\end{eqnarray*}
The classical energy of the Goldstone plane waves
\begin{eqnarray*}
H^{(2)}[\delta G]&=&\frac{v^2}{2}\int \, d^2x \,
\left[\left(\frac{\partial\delta G}{\partial
x_0}\right)^2+\vec{\nabla}\delta G\vec{\nabla}\delta G+ \delta
G\delta G\right]=\frac{\hbar
m}{2}\sum_{\vec{k}}\omega(\vec{k})\left[a^*(\vec{k})a(\vec{k})+a(\vec{k})a^*(\vec{k})\right]
\end{eqnarray*}
is promoted through canonical quantization
$[\hat{a}(\vec{k}),\hat{a}(\vec{q})]=\delta_{\vec{k}\vec{q}}$ to
the quantum free Hamiltonian
\[
H^{(2)}[\delta\hat{H}]=\hbar
m\sum_{\vec{k}}\omega(k)\left(\hat{a}^\dagger(\vec{k})\hat{a}(\vec{k})+{1\over
2}\right)\qquad ,
\]
such that the contribution to the vacuum energy of the Goldstone
bosons is:
\[
\Delta E^{(3)}_0=\sum_{\vec{k}}{\hbar m\over 2}
\omega(\vec{k})=\frac{\hbar m}{2}{\rm
Tr}[-\vec{\nabla}\vec{\nabla}+1]^{{1\over 2}} \qquad .
\]

\subsubsection{Ghost particles}

The contribution of ghosts to the vacuum energy is more tricky. The
solution of the linearized field equations for small fluctuations of
the Ghost field around the vacuum
\[
\left(\frac{\partial^2}{\partial
x_0^2}-\vec{\nabla}\vec{\nabla}+1\right)\delta \chi(x_0,\vec{x})=0
\]
is also a plane wave (ghost) expansion:
\begin{eqnarray*}
&&\delta
\chi(x_0,\vec{x})=\frac{1}{vL}\sqrt{\frac{\hbar}{ev}}\sum_{\vec{k}}\frac{1}{\sqrt{2\omega(\vec{k})}}\left[c(\vec{k})e^{-ikx}+d^*(\vec{k})e^{ikx}\right]\\
&& k_0^2-\vec{k}\vec{k}-1=0 \hspace{1.5cm} , \hspace{1.5cm}
\omega(\vec{k})=+\sqrt{\vec{k}\vec{k}+1} \qquad .
\end{eqnarray*}
The coefficients, however, are Grassman variables -classical
cousins of Fermi fields- satisfying the anti-commutation
relations:

\begin{eqnarray*}
&&c^2(\vec{k})=d^2(\vec{k})=0 \, , \, \forall \vec{k}
\hspace{1.5cm} , \hspace{1.5cm}
c(\vec{k})d(\vec{q})+d(\vec{q})c(\vec{k})=0
\\&&
c(\vec{k})c(\vec{q})+c(\vec{q})c(\vec{k})=d(\vec{k})d(\vec{q})+d(\vec{k})d(\vec{k})=0\,
, \, \forall
\vec{k},\vec{q}\\&&c(\vec{k})c^*(\vec{q})+c^*(\vec{q})c(\vec{k})
=d(\vec{k})d^*(\vec{q})+d^*(\vec{k})d(\vec{k})=0 \qquad .
\end{eqnarray*}
The classical energy of ghost plane waves looks familiar up to a
sign
\begin{eqnarray*}
&&H^{(2)}[\delta \chi]=v^2\int \, d^2x \, \left[\frac{\partial\delta
\chi^*}{\partial x_0}\frac{\partial\delta \chi}{\partial
x_0}+\vec{\nabla}\delta \chi^*\vec{\nabla}\delta \chi+ \delta
\chi^*\delta \chi\right]\\&=&\frac{\hbar
m}{2}\sum_{\vec{k}}\omega(\vec{k})\left[c^*(\vec{k})c(\vec{k})
+d^*(\vec{k})d(\vec{k})-c(\vec{k})c^*(\vec{k})-d(\vec{k})d^*(\vec{k})\right]
\qquad ,
\end{eqnarray*}
but canonical quantization proceeds by the anti-commutators
\[
\{\hat{c}^\dagger(\vec{k}),\hat{c}(\vec{q})\}=\{\hat{d}^\dagger(\vec{k}),\hat{d}(\vec{q})\}=\delta_{\vec{k}\vec{q}}\qquad
,
\]
and the free quantum Hamiltonian is:
\[
H^{(2)}[\delta\hat{\chi}]=\hbar
m\sum_{\vec{k}}\omega(k)\left(\hat{c}^\dagger(\vec{k})\hat{c}(\vec{k})+\hat{d}^\dagger(\vec{k})\hat{d}(\vec{k})-1\right)\qquad .
\]

Thus, the contribution to the vacuum energy of Ghosts is negative:
\[
\Delta E^{(4)}_0=-\sum_{\vec{k}}\hbar m \omega(\vec{k})=-\hbar
m{\rm Tr}[-\vec{\nabla}\vec{\nabla}+1]^{{1\over 2}}
\]
Preserving unitarity, the ghosts exactly cancel the contribution
to the vacuum energy of the non physical temporal vector bosons
and Goldstone bosons.

In sum, the vacuum energy in the planar AHM
\[
\Delta E_0=\sum_{r=1}^4 \Delta E^{(r)}_0=\hbar m{\rm
Tr}[-\vec{\nabla}\vec{\nabla}+1]^{{1\over 2}}+{\hbar m\over 2}{\rm
Tr}[-\vec{\nabla}\vec{\nabla}+\kappa^2]^{{1\over 2}}
\]
is due only to Higgs particles and transverse massive vector
bosons, as it should be.

\subsection{One-loop mass renormalization counter-terms}

Denoting as ${\rm I(c^2)}$ the divergent integral
\[
I(c^2)=\int \, \frac{d^3k}{(2\pi)^3} \cdot
\frac{i}{k^2-c^2+i\varepsilon}\qquad ,
\]
the AHM encompasses the following one-loop divergent graphs:
\begin{enumerate}
\item Higgs boson tadpole:
\vspace{0.4cm}

\centerline{
\parbox{2.0cm}{\epsfig{file=qsdv15.ps,width=2.0cm}} $+$
\parbox{2.0cm}{\epsfig{file=qsdv16.ps,width=2.0cm}} $+$
\parbox{2.0cm}{\epsfig{file=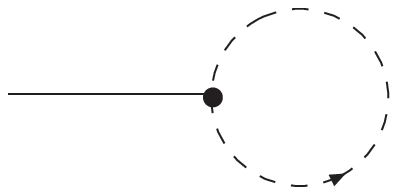,width=2.0cm}} $+$
\parbox{2.0cm}{\epsfig{file=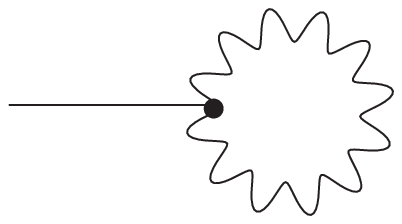,width=2.0cm}} $=$}
\vspace{-0.6cm}
\[
= - 2 i ( \kappa^2 + 1) I(1) + \mbox{finite part}
\]

\item Higgs boson self-energy:
\vspace{0.4cm}

\centerline{
\parbox{2.0cm}{\epsfig{file=qsdv19.ps,width=2.0cm}} $+$
\parbox{2.0cm}{\epsfig{file=qsdv20.ps,width=2.0cm}} $+$
\parbox{2.0cm}{\epsfig{file=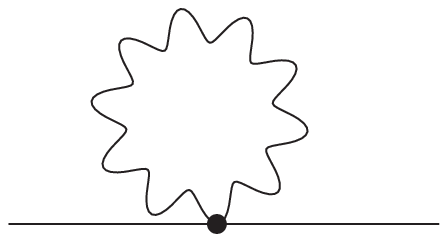,width=2.0cm}} $+$
\parbox{2.0cm}{\epsfig{file=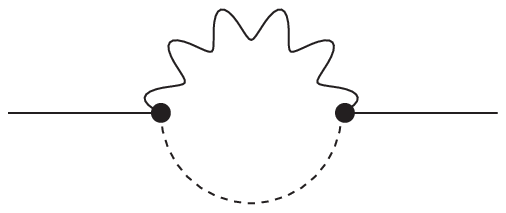,width=2.0cm}} $=$}
\vspace{-0.2cm}
\[
= - 2 i ( \kappa^2+ 1) I(1) + \mbox{finite part}
\]

\item Goldstone boson self-energy:
\vspace{0.4cm}

\centerline{
\parbox{2.0cm}{\epsfig{file=qsdv23.ps,width=2.0cm}} $+$
\parbox{2.0cm}{\epsfig{file=qsdv24.ps,width=2.0cm}} $+$
\parbox{2.0cm}{\epsfig{file=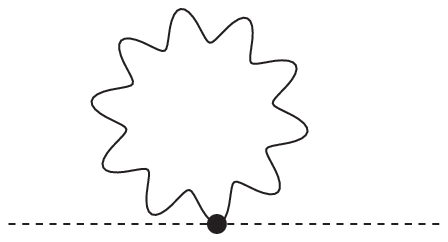,width=2.0cm}} $+$
\parbox{2.0cm}{\epsfig{file=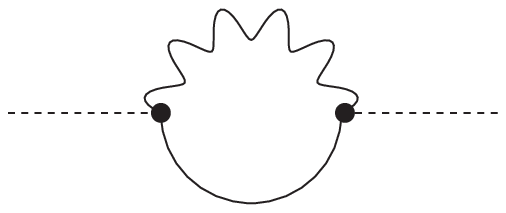,width=2.0cm}} $=$}
\vspace{-0.2cm}
\[
= - 2 i ( \kappa^2+ 1) I(1) + \mbox{finite part}
\]

\item  Vector boson self-energy:
\vspace{0.4cm}

\centerline{
\parbox{2.0cm}{\epsfig{file=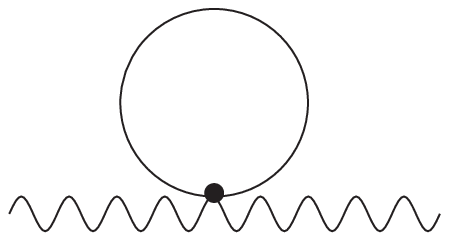,width=2.0cm}} $+$
\parbox{2.0cm}{\epsfig{file=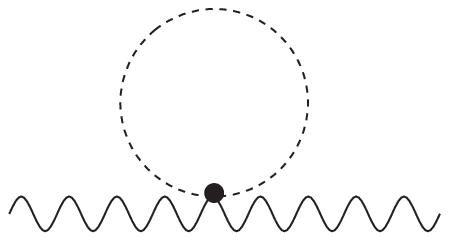,width=2.0cm}} $+$
\parbox{2.0cm}{\epsfig{file=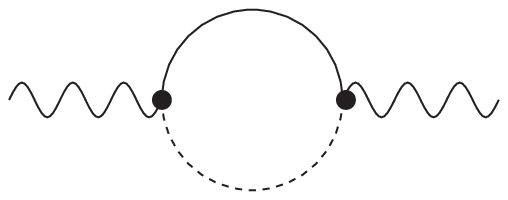,width=2.0cm}} $=$}
\vspace{-0.2cm}
\[
=  2 i I(1)g^{\mu\nu} + \mbox{finite part}
\]
\end{enumerate}
These calculations are performed in a detailed manner in Appendix
\S. 12.4.

All the finite parts are proportional to ${\rm I(\kappa^2)}-{\rm
I(1)}$ so that with a minimal substraction scheme we can get rid off
all these one-loop divergences by adding the counter-terms
\[
{\cal L}_{c.t.}^S = \hbar(\kappa^2+1) I(1) \left[|\phi|^2-1
\right]  \hspace{2cm} , \hspace{2cm} {\cal L}_{c.t.}^A = -\hbar
I(1) A_\mu A^\mu
\]
to the Lagrangian.

In (2+1)-dimensions however, the graphs above are the only
divergent diagrams in the system (the theory is
super-renormalizable). Thus, the diagrams coming from these
counter-terms completely cancel any divergence (not only to
one-loop order) arising in the vacuum sector of the model:

\begin{table}[hbt]
\begin{center}
\begin{tabular}{ccc} \\ \hline
\textit{Diagram} & \hspace{0.3cm} & \textit{Weight} \\
\hline
\parbox{2.3cm}{\epsfig{file=qsdv30.eps,width=2.3cm}} & &
$\displaystyle 2i(\kappa^2+1)I(1) $
\\
\parbox{2.3cm}{\epsfig{file=qsdv31.eps,width=2.3cm}} & &
$\displaystyle 2i(\kappa^2+1) I(1)$ \\
 \parbox{2.3cm}{\epsfig{file=qsdv32.eps,width=2.3cm}} & &
$\displaystyle 2i(\kappa^2+1) I(1)$
\\
 \parbox{2.3cm}{\epsfig{file=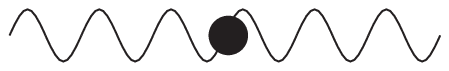,width=2.3cm}} & &
$\displaystyle -2i I(1)g^{\mu\nu}$ \\[0.5cm] \hline \\[0.25cm]
\end{tabular}
\end{center}
\end{table}
We stress, however, that renormalization in the planar Abelian Higgs
model requires more than merely normal ordering, contrarily to
(1+1)-dimensional scalar field theory, where normal order is
sufficient to cope with ultraviolet divergences. In this system
there are divergences due to graphs with two vertices that are not
suppressed by normal ordering.

\subsection{Self-dual Abrikosov-Nielsen-Olesen vortices}

There are also topological solitons in the AHM: solutions of the
static field equations
\[
\partial_i F_{ij} = J_j \hspace{1.3cm} ; \hspace{1.3cm}  D_i D_i
\phi= \frac{\kappa^2}{4}\phi(\phi^*\phi-1)  \qquad ,
\]
where $J_j= \frac{i}{2} \left( \phi^* D_j \phi - (D_j \phi)^* \,
\phi \right)$ is the electric current, of finite energy:
\[
E= \int d^2 x [\frac{1}{4} F_{ij} F_{ij} + \frac{1}{2} (D_i
\phi)^* D_i \phi + \frac{\kappa^2}{8} (\phi^* \phi-1)^2 ]
\nonumber \qquad .
\]
In fact, the configuration space ${\cal C}$
\[
{\cal C}=\left\{\phi(\vec{x})\in Maps({\mathbb R}^2,{\mathbb
C}),A_i(\vec{x})\in Maps({\mathbb R}^2,T{\mathbb
R}^2)/E(\phi,A_i)<+\infty \right\}
\]
is topologically disconnected. The boundary of the spatial plane
is the sphere of infinite radius $S_\infty^1=\lim_{r\rightarrow
+\infty}\left[x_1^2+x_2^2=r^2\right]$. Finite energy
configurations comply with the asymptotic behavior
\[
\left. \phi^{*} \phi \right|_{S_\infty^1} =1 \hspace{3cm} ,
\hspace{1cm} D_i \phi |_{S_\infty^1} =(\partial_i \phi - i A_i
\phi)|_{S_\infty^1}=0 \qquad ,
\]
such that
\[
\theta={\rm arctan}{x_2\over x_1} \qquad , \qquad \phi
|_{S_\infty^1} = e^{il \theta}\, \, , \,l\in{\mathbb Z}\qquad ,
\qquad A_i |_{S_\infty^1} =- i \phi^*\partial_i \phi|_{S_\infty^1}
\]
provides the map $\lim_{r\rightarrow +\infty}\phi(\vec{x}):
S^1_{\infty} \longrightarrow S^1_1$ between the sphere at infinity
and the vacuum orbit ${\mathbb S}_1^1$. Continuous maps between
one-dimensional spheres are classified according to the first homotopy
group and, because the temporal evolution is continuous, $
\Pi_0({\cal C})=\Pi_1(S^1_1)={\mathbb Z}$, the zero homotopy group
of ${\cal C}$ is non-trivial. Thus, $ {\cal
C}=\sqcup_{l\in{\mathbb Z}}{\cal C}_l$ is the union of
disconnected sectors characterized by an integer number $l$ and
the magnetic flux of any finite energy configuration is quantized:
$g= \int d^2 x F_{12}=2{\pi l}$.

We shall restrict ourselves to the critical point between Type I
and Type II superconductivity: $\kappa^2=1$. The energy can be
arranged in a Bogomolny splitting:
\[
E= \int \frac{d^2 x}{2} \left( |D_1 \phi \pm i D_2 \phi|^2 + [
F_{12} \pm {\textstyle\frac{1}{2}} (\phi^* \phi-1) ]^2
\right)+\frac{1}{2}|g|\qquad .
\]
We immediately realize that the solutions of the first-order
equations
\[
D_1 \phi \pm i D_2 \phi=0 \hspace{0.3cm};\hspace{0.3cm} F_{12} \pm
\frac{1}{2} (\phi^*\phi-1) =0
\]
are absolute minima of the energy, and are hence stable, in each
topological sector with a classical mass proportional to the
magnetic flux.

Because the vector field is asymptotically purely vorticial, these
solitonic solutions were christened as vortices by their discoverers
Abrikosov, Nielsen and Olesen. Also, since the first-order equations
can be derived from the self-duality equations of Euclidean 4D gauge
theory through dimensional reduction, the ANO vortices are called
self-dual at the limit $\kappa^2=1$.

\subsection{Self-dual vortices with spherical symmetry}

Another simplification is to consider the spherically symmetric
ansatz:
\begin{eqnarray*}
\phi_1(x_1,x_2) = f(r) {\rm cos}l\theta \quad &,& \quad
\phi_2(x_1,x_2) = f(r) {\rm sin}l\theta \\
A_1(x_1,x_2) =-l \frac{\alpha(r)}{r}{\rm sin}\theta \quad &,&
\quad A_2(x_1,x_2) = l \frac{\alpha(r)}{r}{\rm cos}\theta \qquad.
\end{eqnarray*}
The first-order equations reduce to
\[
{1\over r} {d \alpha \over d r}(r)= \mp \frac{1}{2 l} (f^2(r)-1)
\qquad , \qquad {d f\over d r}(r) = \pm \frac{l}{r}
f(r)[1-\alpha(r)] \qquad ,
\]
to be solved together with the boundary conditions
\begin{eqnarray*}
&&{\displaystyle \lim_{r\rightarrow\infty}} f(r) = 1
\hspace{1.5cm} , \hspace{1.5cm} {\displaystyle
\lim_{r\rightarrow\infty}}\alpha(r) = 1 \\ &&f(0) =0
 \hspace{2.5cm}
, \hspace{2cm} \alpha(0)=0
\\&&g= - \oint_{r=\infty} dx_i A_i = -l\oint_{r=\infty}{
[x_2dx_1-x_1dx_2]\over r^2}=2 \pi l \,\, ,
\end{eqnarray*}
required by energy finiteness plus regularity at the origin
(center of the vortex).

To solve this non-linear ODE system, we follow the three-step  De
Vega-Schaposnik procedure

\begin{enumerate}
\item For small values of $r$, a power series is tested in the
first-order differential equations.

\item The first-order ODE system is solved exactly for large $r$: the asymptotic behavior of the solutions is thus
found.

\item Finally, a numerical scheme can be implemented by setting a boundary
condition at a non-singular point of the ODE system, which is
obtained from the power series for small values of $r$. The behavior
of the vortex solutions for intermediate distances is then described
by means of an interpolating polynomial.
\end{enumerate}
In terms of the field profiles $f(r)$, $\alpha(r)$ the magnetic
field and the energy density are:

\[
B(r)={l\over 2r}\frac{d\alpha}{dr} \hspace{1.5cm} , \hspace{1.5cm}
\varepsilon(r)={1\over 4}(1-f^2(r))^2+{l^2\over
r^2}(1-\alpha(r))^2 f^2(r) \qquad .
\]

\noindent The field profiles, the magnetic field and the energy
density are shown in Figure 1 for $l=1,2,3,4$. $B(r)$ always has a
maximum at $r=0$ but the origin is only a maximum of the energy
density for $l=1$.

{\small\begin{center}
\includegraphics[height=2.0cm]{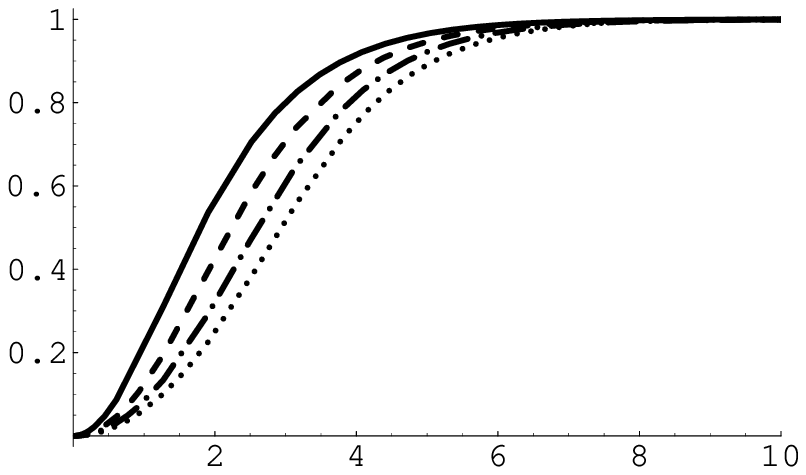
}\hspace{0.8cm}
\includegraphics[height=2.0cm]{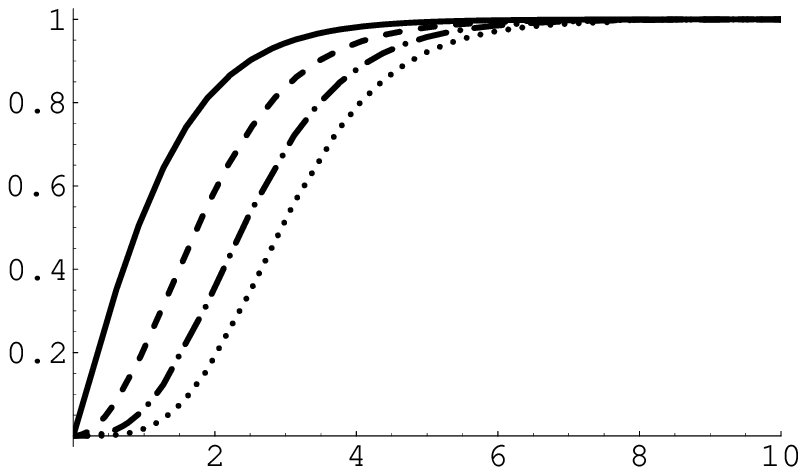}\hspace{0.8cm}
\includegraphics[height=2.0cm]{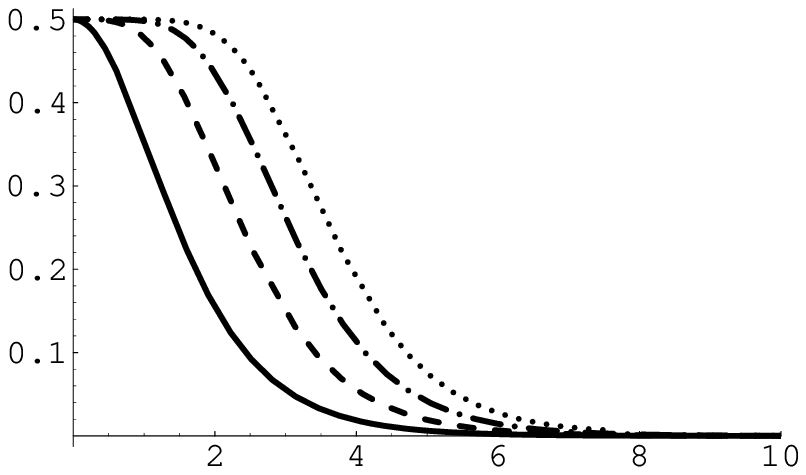}\hspace{0.8cm}
\includegraphics[height=2.0cm]{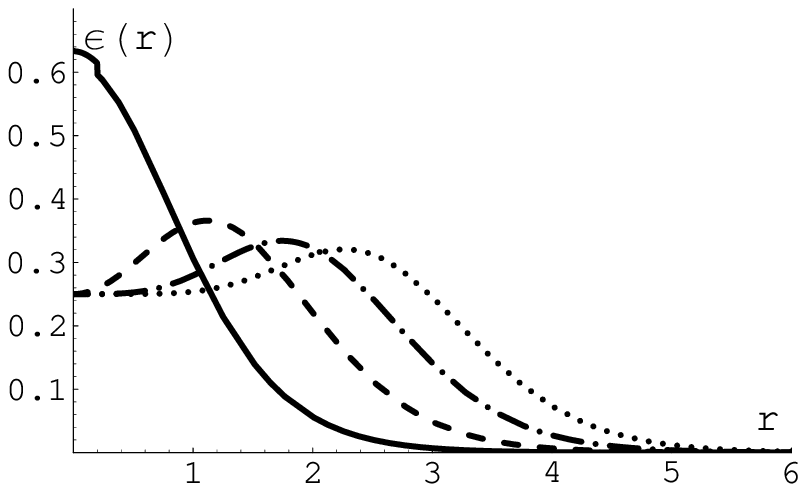} \\
{\small Figure 1. \textit{Plots of the field profiles $\alpha(r)$
(a) and $f(r)$ (b), the magnetic field $B(r)$ (c), and the energy
density $\varepsilon(r)$ for self-dual vortices with $l=1$ (solid
line), $l=2$ (broken line), $l=3$ (broken-dotted line) and $l=4$
(dotted line).}}
\end{center}}
In Figure 2, three-dimensional plots of the energy density are shown
for $l=1,2,3,4$.
\begin{center}
\includegraphics[height=4.6cm]{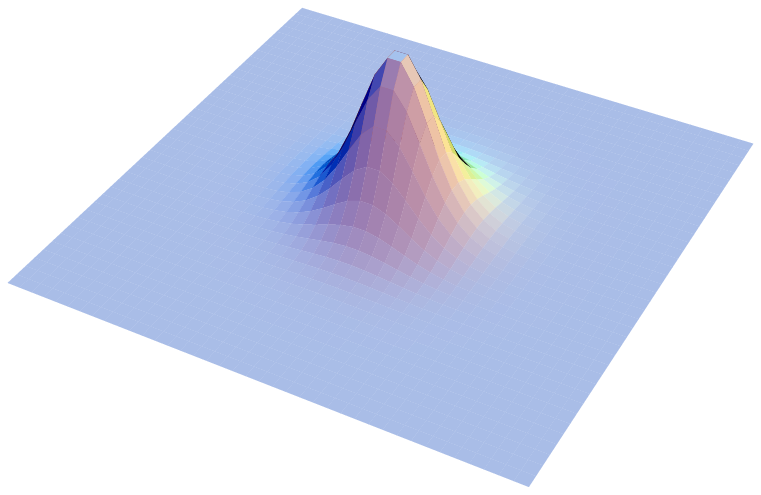}\hspace{0.8cm}
\includegraphics[height=4.6cm]{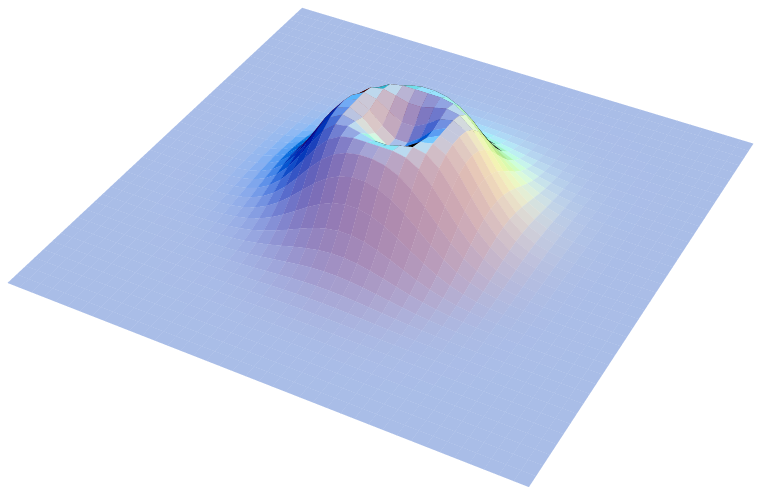}\hspace{0.8cm}
\includegraphics[height=4.6cm]{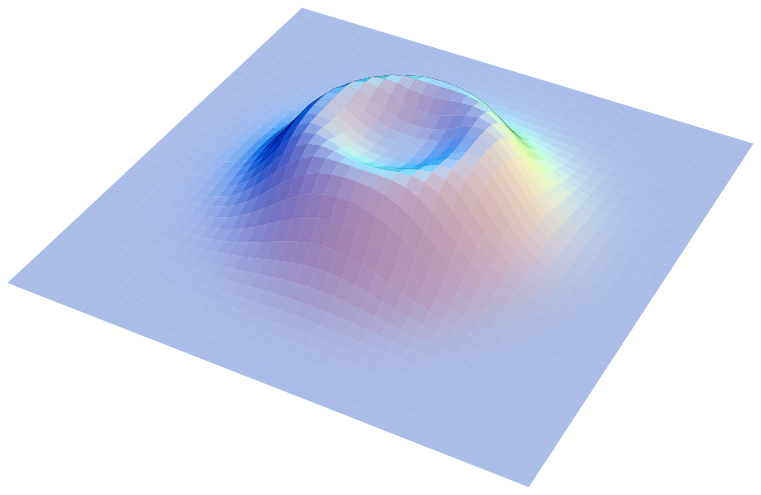}\hspace{0.8cm}
\includegraphics[height=4.6cm]{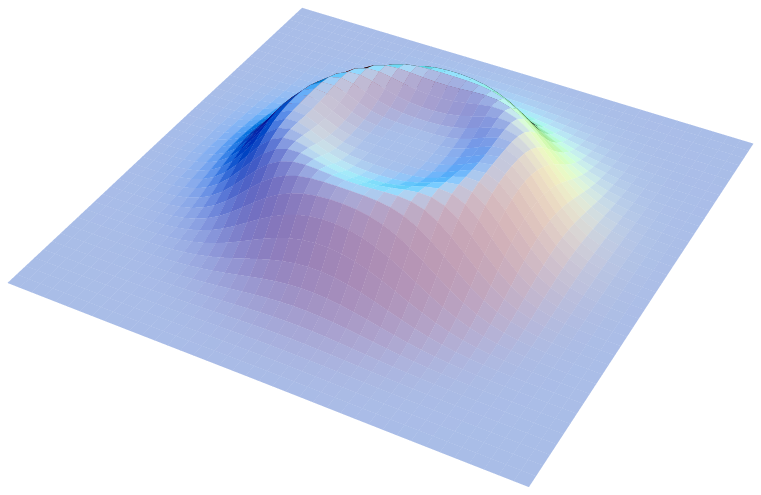} \\
{\small Figure 2. \textit{3D graphics of the energy density for
$l=1$, $l=2$, $l=3$ and $l=4$ self-dual symmetric ANO vortices.}}
\end{center}

\begin{center}
\includegraphics[height=5cm]{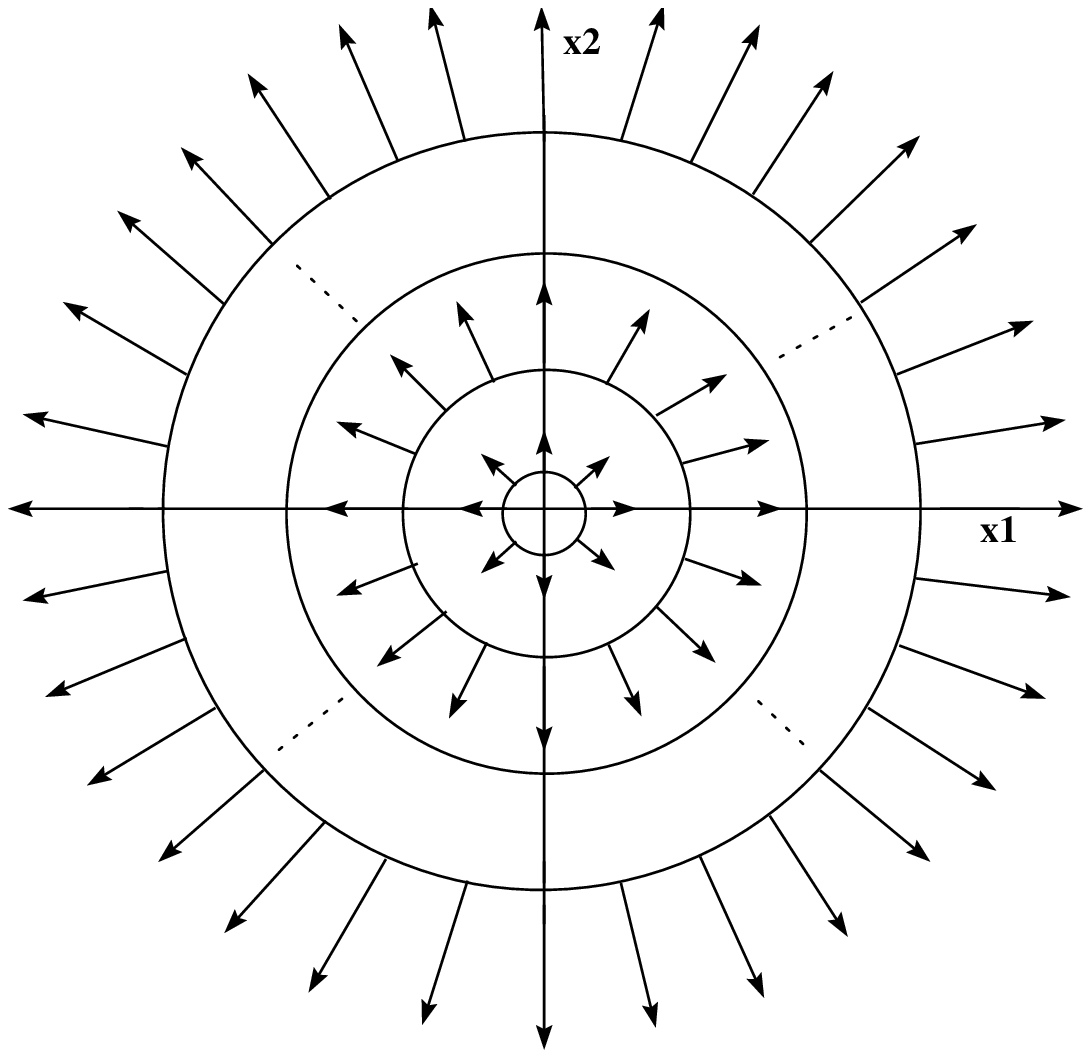}\hspace{1cm}
\includegraphics[height=5cm]{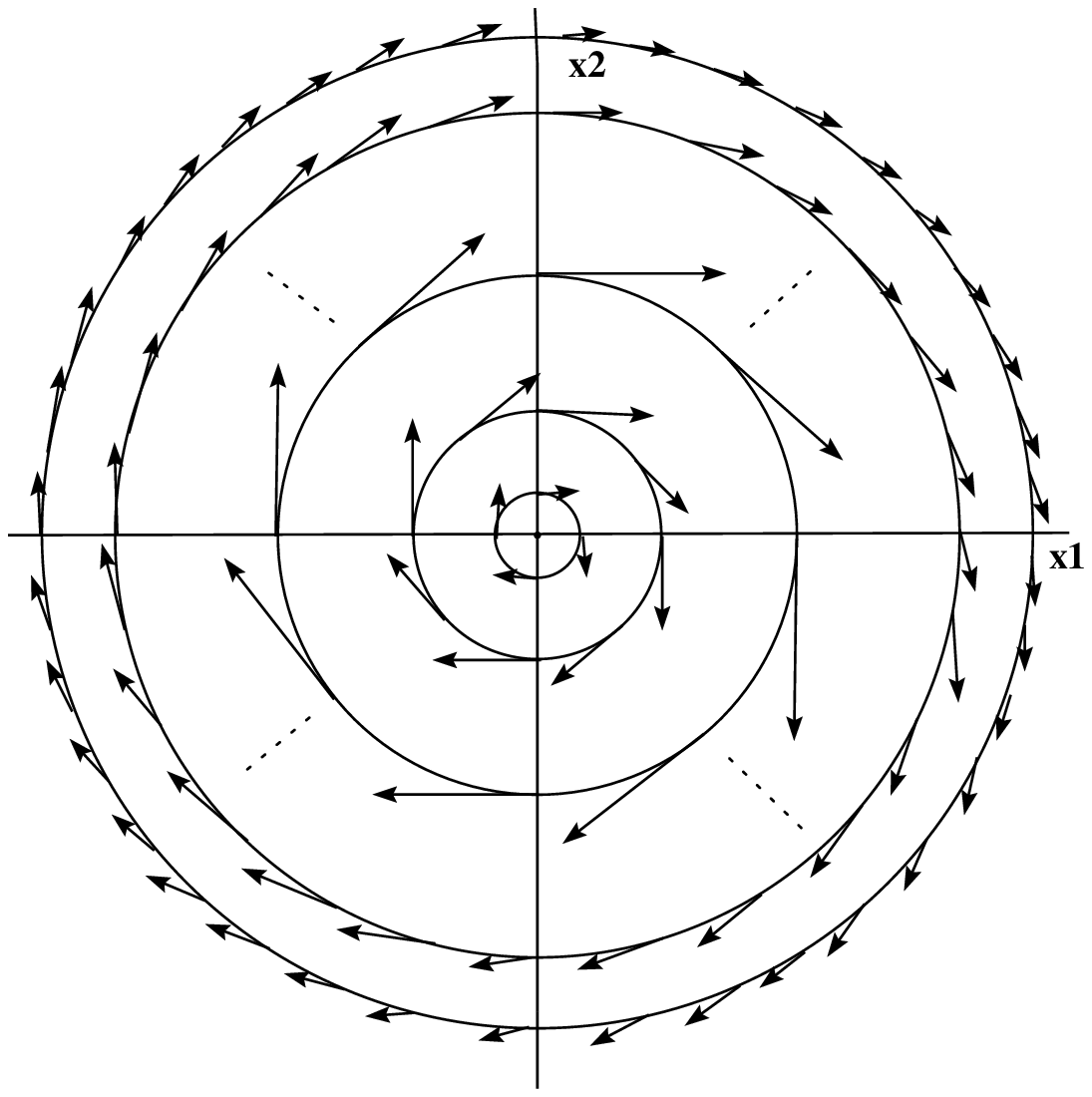}
\\
{\small Figure 3. \textit{(left) The $l=1$-vortex scalar
iso-vector field $\vec{s} (x_1, x_2) = {\rm Re} s(x_1,x_2) \vec{i}
+ {\rm Im} s(x_1,x_2) \vec{j} $ projected on the ${\bf R}^2$
spatial plane (hedgehog).} \\ \hspace{1.9cm}\textit{(right) The
$l=1$-vortex vector field ${\vec V}(x_1,x_2)= V_1(x_1,x_2) {\vec
i} + V_2(x_1,x_2) {\vec j}$. Note that
$\vec{\nabla}\cdot\vec{V}(\vec{x})=0$ and
$\vec{\nabla}\wedge\vec{V}(\vec{x})={\vec k}$. Also, ${\vec
V}(0,0)= {\vec V}(\infty {\rm cos}\theta ,\infty {\rm sin}\theta)
={\vec 0}$ (Poincar\'e-Hopf theorem) .}}
\end{center}

\subsection{Two-vortex solutions with different centers}

$l=2$ ANO self-dual solutions formed by two $l=1$ vortices with
centers separated by a distance $d$ can also be obtained
approximately. We shall implement the variational method of Jacobs
and Rebbi in two stages:

\begin{enumerate}
\item First, trial functions, depending on a single variational
parameter $w$,
\begin{eqnarray*}
\phi_\omega (z,z^*)&=& \Phi(z,z^*) \, \left[ \omega \,
f^{(1)}(|z-d/2|)\, f^{(1)}(|z+d/2|)\right.+\\&&\left.
+(1-\omega)\,
\frac{|z^2-(d/2)^2|}{|z^2|}\, f^{(2)}(|z|)\right]  \\
A^\omega (z,z^*) &=& \omega \left( \frac{i}{z^*-d/2}\,
\alpha^{(1)}(|z-d/2|)+ \frac{i}{z^*+d/2}\,
\alpha^{(1)}(|z+d/2|)\right)+\\&&+(1-\omega)\, \frac{2 i}{z^*} \,
\alpha^{(2)}(|z|)
\end{eqnarray*}
are built. Here $z=x_1+i x_2$, $
A^\omega(z,z^*)=A_1^\omega(z,z^*)+iA_2^\omega(z,z^*)$ and
\[
\Phi(z,z^*)=\sqrt{\frac{z^2-(d/2)^2}{z^{*2}-(d/2)^2}} \Rightarrow
g=4\pi
\]

$f^{(1)}$, $\alpha^{(1)}$, $f^{(2)}$ and $\alpha^{(2)}$ stand for
the functions $f$ and $\alpha$ associated with self-dual solutions
with cylindrical symmetry, respectively, with vorticity $l=1$ and
$l=2$.

Plugging this ansatz into the energy functional, expression
$E(\omega )$ is set to be minimized as a function of $\omega$.

\item
Then, a deformation is added such that: 1) the scalar field
vanishes at the two centers; 2) the gauge-invariant quantities
associated with the solution are symmetric with respect to the
reflection $z\rightarrow z^*$.

The invariant ansatz
\begin{eqnarray*}
&&\phi(z,z^*)=
\phi_\omega(z,z^*)+\\&+&\Phi(z,z^*)\left|z^2-(d/2)^2 \right|
(\cosh |z|)^{-1} \sum_{i=0}^N \sum_{j=0}^i f_{ij}
\frac{(zz^*)^i}{2} \left[ \left( \frac{z}{z^*}\right)^j +\left(
\frac{z^*}{z}\right)^j \right] \\
&&A(z,z^*)= A^\omega (z,z^*)+\\&+& \frac{1}{\cosh |z|} \left\{ z
\sum_{i=0}^N \sum_{j=0}^i a_{ij}^{I} \frac{(zz^*)^i}{2} \left[
\left( \frac{z}{z^*}\right)^j +\left( \frac{z^*}{z}\right)^j
\right]+ z^* \sum_{i=0}^N \sum_{j=0}^i a_{ij}^{II}
\frac{(zz^*)^i}{2} \left[ \left( \frac{z}{z^*}\right)^j +\left(
\frac{z^*}{z}\right)^j \right] \right\}
\end{eqnarray*}
contains $\aleph=3\frac{(N+1)(N+2)}{2}$ variational parameters:
$f_{ij}$, $a^I_{ij}$, $a^{II}_{ij}$.

\end{enumerate}
A three dimensional plot of two-vortex solutions is shown in the
next Figure for distances $d=1,2,3$.

\begin{center}
\includegraphics[height=5cm]{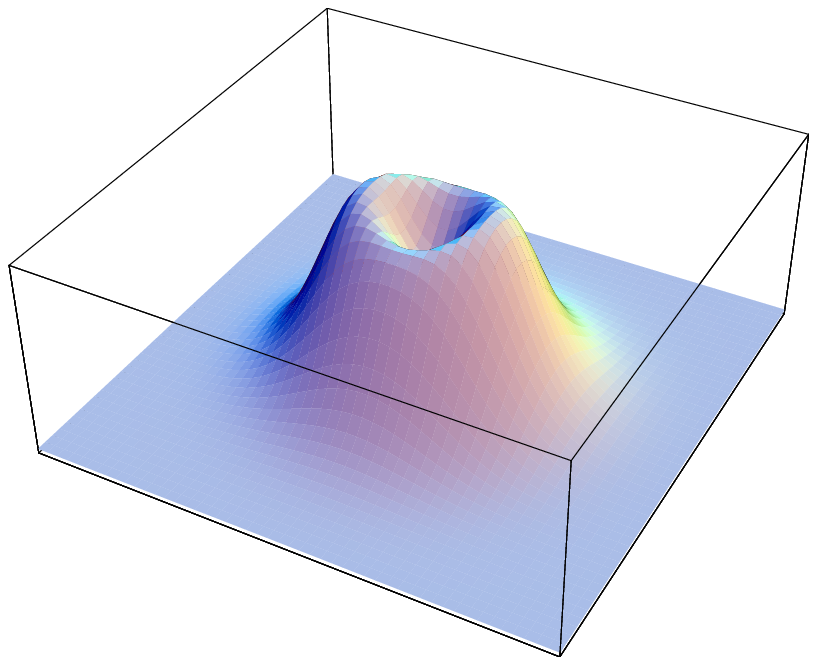}\hspace{1.5cm}
\includegraphics[height=5cm]{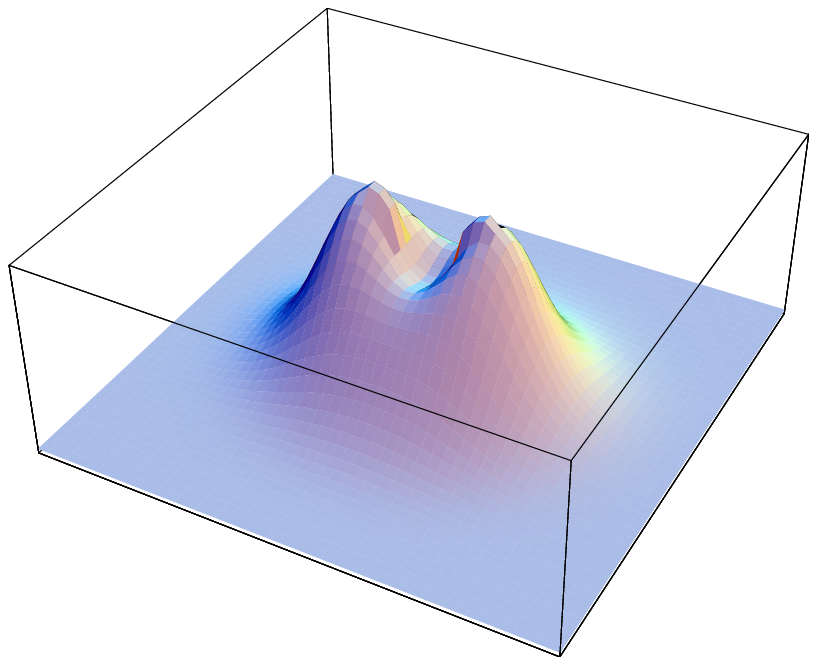}\hspace{1.5cm}
\includegraphics[height=5cm]{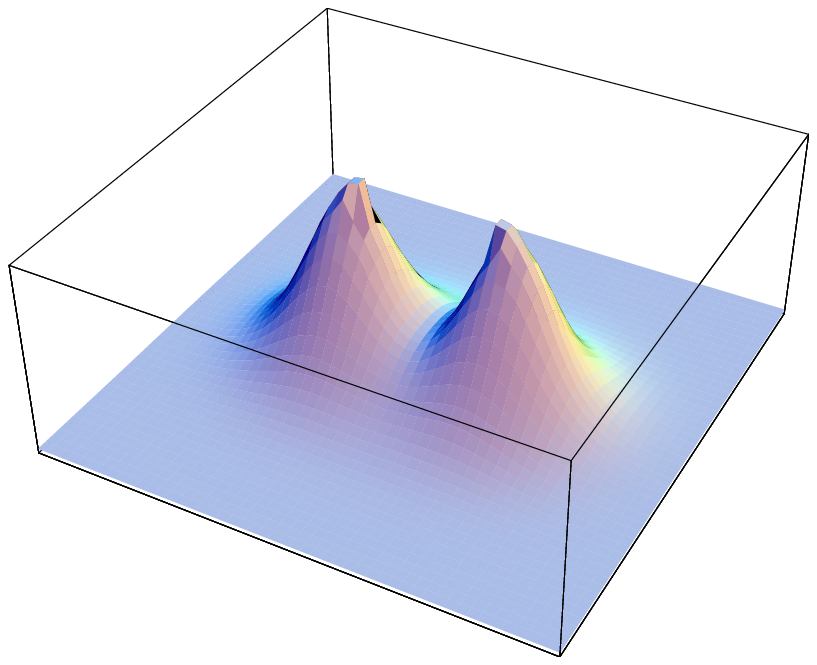}\\
{\small Figure 4. \textit{3D graphics of the energy density for
$l=2$ self-dual separate vortices with centers at distances $d=1$,
$d=2$, $d=3$. Here we have plotted the cruder approximation:
$N=1$.}}
\end{center}

\subsection{The vortex Casimir energy}

\subsubsection{Bosonic fields}
Let us consider small fluctuations around vortices
\[
\phi(x_0,\vec{x})= s(\vec{x})+\delta s(x_0,\vec{x}) \hspace{1.5cm} ,
\hspace{1.5cm} A_k(x_0,\vec{x})=V_k(\vec{x})+\delta a_k
(x_0,\vec{x}) \qquad ,
\]
where by $s(\vec{x})$ and $V_k(\vec{x})$ we respectively denote
the scalar and vector field of the vortex solutions. Working in
the Weyl/background gauge
\[
A_0(x_0,\vec{x})=0 \qquad \qquad , \qquad \qquad \partial_k\delta
a_k(x_0,\vec{x})+s_2(\vec{x})\delta
s_1(x_0,\vec{x})-s_1(\vec{x})\delta s_2(x_0,\vec{x})=0 \quad ,
\]
the classical energy up to ${\cal O}(\delta^2)$ order is:
\begin{eqnarray*}
&&H^{(2)}+H^{(2)}_{{\rm g.f.}}+H^{(2)}_{{\rm ghost}}={v^2\over
2}\int \, d^2x\left\{\frac{\partial\delta\xi^T}{\partial
x_0}(x_0,\vec{x})\frac{\partial\delta\xi^T}{\partial
x_0}(x_0,\vec{x})\right\}+\\&&+{v^2\over 2}\int \, d^2x
\left\{\delta
\xi^T(x_0,\vec{x})K\delta\xi(x_0,\vec{x})+\delta\chi^*(\vec{x})\left(-\bigtriangleup
+|s(\vec{x})|^2\right)\delta\chi(\vec{x})\right\} \qquad ,
\end{eqnarray*}
where
\[
\delta\xi(x_0,\vec{x})=\left(\begin{array}{c} \delta a_1 (x_0,\vec{x}) \\
\delta a_2 (x_0,\vec{x}) \\ \delta s_1(x_0,\vec{x}) \\ \delta
s_2(x_0,\vec{x})\end{array}\right) \qquad , \qquad
\nabla_js_a=\partial_js_a+\varepsilon_{ab}V_js_b \qquad , \qquad
\bigtriangleup=\vec{\nabla}\vec{\nabla} \quad ,
\]
and
\begin{eqnarray*}
K={\normalsize \left(\begin{array}{cccc} -\bigtriangleup +|s|^2 &
0 & -2\nabla_1s_2 & 2\nabla_1s_1
\\ 0 & -\bigtriangleup+|s|^2 & -2\nabla_2s_2 &
2\nabla_2s_1 \\ -2\nabla_1s_2 & -2\nabla_2s_2 &
-\bigtriangleup+{1\over 2}(3|s|^2+2V_kV_k-1) & -2V_k\partial_k
\\ 2\nabla_1s_1 & 2\nabla_2s_1 & 2V_k\partial_k  &
-\bigtriangleup+{1\over 2}(3|s|^2+2V_kV_k-1)
\end{array}\right)}\qquad .
\end{eqnarray*}

The linearized field equations read
\[
\frac{\partial^2\delta\xi}{\partial
x_0^2}(x_0,\vec{x})+K\delta\xi(x_0,\vec{x})=0 \hspace{2cm} ,
\hspace{2cm} \left(-\bigtriangleup
+|s(\vec{x})|^2\right)\delta\chi(\vec{x})=0 \qquad ,
\]
or, in more detailed form,
\[
\left(-\bigtriangleup +|s(\vec{x})|^2\right)\delta
a_1(x_0,\vec{x})=2\nabla_1s_2(\vec{x})\delta
s_1(x_0,\vec{x})-2\nabla_1s_1(\vec{x})\delta s_2(x_0,\vec{x})
\]
\[
\left(-\bigtriangleup +|s(\vec{x})|^2\right)\delta
a_2(x_0,\vec{x})=2\nabla_2s_2(\vec{x})\delta
s_1(x_0,\vec{x})-2\nabla_2s_1(\vec{x})\delta s_2(x_0,\vec{x})
\]
\[
\left(-\bigtriangleup +{1\over 2}(
3|s(\vec{x})|^2+2V_k(\vec{x})V_k(\vec{x})-1)\right)\delta
s_1(x_0,\vec{x})-2V_k(\vec{x})\partial_k\delta s_2(x_0,\vec{x})=
\]
\[
=2\nabla_1s_2(\vec{x})\delta
a_1(x_0,\vec{x})+2\nabla_2s_2(\vec{x})\delta a_2(x_0,\vec{x})
\]
\[
\left(-\bigtriangleup +{1\over 2}(
3|s(\vec{x})|^2+2V_k(\vec{x})V_k(\vec{x})-1)\right)\delta
s_2(x_0,\vec{x})-2V_k(\vec{x})\partial_k\delta s_1(x_0,\vec{x})=
\]
\[
=-2\nabla_1s_1(\vec{x})\delta
a_1(x_0,\vec{x})-2\nabla_2s_1(\vec{x})\delta a_2(x_0,\vec{x})\qquad .
\]

Let us assume the following spectral resolution of $K$ in the
orthogonal complement to its kernel (no bound states):
\[
\sum_{B=1}^4 \, K_{AB}\, u^{(I)}_B(\vec{x};\vec{k})=\varepsilon
(\vec{k}) u_A^{(I)}(\vec{x};\vec{k}) \hspace{0.5cm} , \, \,
I=1,2,3,4 \, \, , \hspace{0.5cm}
\varepsilon(\vec{k})=\vec{k}\vec{k}+1
\]
\[
u_A^{(I)}(\vec{x};\vec{k})=e^{i\vec{k}\vec{x}}v_A^{(I)}(\vec{x};\vec{k})
\hspace{1.5cm} , \hspace{1.5cm}
\lim_{r\rightarrow\infty}v_A^{(I)}(\vec{x};\vec{k})=v_A(\vec{k})\delta^I_A
\]
\[
\sum_{A=1}^4\int \, d^2x \,
u_A^{(I)*}(\vec{x};\vec{k})u_A^{(J)}(\vec{x};\vec{q})=e^2v^2L^2\delta_{\vec{k}\,\vec{q}}\delta^{IJ}\qquad .
\]
Expanding the small fluctuations in terms of the positive
eigenfunctions of $K$
\[
\delta\xi_A^\prime (x_0,\vec{x})={1\over
vL}\sqrt{\frac{\hbar}{ev}}\cdot
\sum_{\vec{k}}\sum_{I=1}^4\frac{1}{\sqrt{2\varepsilon(\vec{k})}}\left[a^*_I(\vec{k})e^{i\varepsilon
x_0}u^{(I)*}_A(\vec{x};\vec{k})+a_I(\vec{k})e^{-i\varepsilon
x_0}u_A^{(I)}(\vec{x};\vec{k})\right] \qquad ,
\]
one ends with the classical free Hamiltonian:
\[
H^{(2)}+H^{(2)}_{{\rm g.f.}}=\hbar {m\over
2}\sum_{\vec{k}}\sum_{I=1}^4
\,\varepsilon(\vec{k})\left[a_I^*(\vec{k})a_I(\vec{k})+a_I(\vec{k})a_I^*(\vec{k})\right]\qquad .
\]

\subsubsection{Ghost field}
Assuming also that there are no bound states in the positive
spectrum of $K^{\rm G}=-\bigtriangleup+|s(\vec{x})|^2$,
\[
K^{\rm G}\, u(\vec{x};\vec{k})=\varepsilon (\vec{k})
u(\vec{x};\vec{k}) \hspace{1.5cm} , \hspace{1.5cm}
\varepsilon(\vec{k})=\vec{k}\vec{k}+1
\]
\[
u(\vec{x};\vec{k})=e^{i\vec{k}\vec{x}}v(\vec{x};\vec{k})
\hspace{1.5cm} , \hspace{1.5cm}
\lim_{r\rightarrow\infty}v(\vec{x};\vec{k})=v(\vec{k})
\]
\[
\int \, d^2x \,
u^*(\vec{x};\vec{k})u(\vec{x};\vec{q})=e^2v^2L^2\delta_{\vec{k}\,\vec{q}}
\qquad ,
\]
small fluctuations of the ghost field can be expanded in terms of
positive eigenfunctions of $K^{\rm G}$
\[
\delta\chi_A^\prime (x_0,\vec{x})={1\over
vL}\sqrt{\frac{\hbar}{ev}}\cdot
\sum_{\vec{k}}\frac{1}{\sqrt{2\varepsilon(\vec{k})}}\left[c(\vec{k})u^*(\vec{x};\vec{k})
+d^*(\vec{k})u(\vec{x};\vec{k})\right]
\]
with Grassman variables as coefficients. The ghost classical free
energy is thus
\[
H^{(2)}_{{\rm Ghost}}=\hbar {m\over 4}\sum_{\vec{k}}
\,\varepsilon(\vec{k})\left[c^*(\vec{k})c(\vec{k})+d^*(\vec{k})d(\vec{k})-c(\vec{k})c^*(\vec{k})
-d(\vec{k})d^*(\vec{k})\right] \qquad .
\]
Two remarks are in order: (1) Because the c's are Grassman
coefficients there are no $cc$ or $c^*c^*$ terms, and
$c^*(\vec{k})d(-\vec{k})+d^*(\vec{k})c(-\vec{k})=0$. (2) Note also
that the ghost fields are static in this combined Weyl-background
gauge. Therefore, their energy is one-half with respect to the
time-dependent case.

The canonical quantization
\begin{eqnarray*}
[\hat{a}_I(\vec{k}),\hat{a}_J^\dagger(\vec{q})]=\delta_{IJ}\delta_{\vec{k}\,\vec{q}}
\hspace{1.5cm} , \hspace{1.5cm}
\{\hat{c}(\vec{k}),\hat{c}^\dagger(\vec{q})\}=\delta_{\vec{k}\,\vec{q}}
\hspace{1.5cm} , \hspace{1.5cm}
\{\hat{d}(\vec{k}),\hat{d}^\dagger(\vec{q})\}=\delta_{\vec{k}\,\vec{q}}
\end{eqnarray*}
leads to the quantum free Hamiltonian
\[
\hat{H}^{(2)}+\hat{H}^{(2)}_{{\rm g.f.}}+\hat{H}^{(2)}_{{\rm
Ghost}}=\hbar m
\cdot\sum_{\vec{k}}\varepsilon(\vec{k})\left[\sum_{I=1}^4\,\left(\hat{a}_I^\dagger(\vec{k})\hat{a}_I(\vec{k})+{1\over
2}\right)+{1\over
2}\left(\hat{c}^\dagger(\vec{k})\hat{c}(\vec{k})+\hat{d}^\dagger(\vec{k})\hat{d}(\vec{k})-1\right)\right]
\qquad ,
\]
such that the vortex Casimir energy reads
\[
\bigtriangleup E_V=\frac{\hbar m}{2}{\rm STr}^*\, K^{{1\over
2}}=\frac{\hbar m}{2}{\rm Tr}^*\, K^{{1\over 2}}-\frac{\hbar
m}{2}{\rm Tr}^*\, (K^{{\rm G}})^{{1\over 2}}\qquad   .
\]
In a similar manner we write  the vacuum energy:
\[
\bigtriangleup E_0={\hbar m\over 2}{\rm STr}\, K_0^{{1\over
2}}=\frac{\hbar m}{2}{\rm Tr}\, K_0^{{1\over 2}}-\frac{\hbar
m}{2}{\rm Tr}\, (K_0^{{\rm G}})^{{1\over 2}}
\]
\[
K_0=\left(\begin{array}{cccc} -\Delta+1 & 0 & 0 & 0
\\ 0 & -\Delta+1 & 0 & 0 \\ 0 & 0 & -\Delta+1 & 0 \\ 0 & 0 & 0 & -\Delta+1\end{array}\right) \qquad \qquad , \qquad \qquad K_0^{{\rm G}}=-\bigtriangleup + 1 \qquad   .
\]
The zero-point vacuum energy renormalization, defining the vortex
Casimir energy, is performed in the formula
\[
\bigtriangleup M_V^C=\bigtriangleup E_V-\bigtriangleup E_0={\hbar
m\over 2}\left[{\rm STr}^*\, K^{{1\over 2}}-{\rm STr}\,
K_0^{{1\over 2}}\right]\qquad   .
\]

\section{The vortex heat kernel and generalized zeta function}

\subsection{Deformation of the first-order equations}

The dimension of the kernel of the operator $K$ ruling the small
vortex fluctuations orthogonal to the gauge group is the dimension
of the moduli space of vortex solutions. Small deformations around
vortices
\[
\phi(\vec{x})=s(\vec{x})+\delta s(\vec{x}) \hspace{1.5cm} ,
\hspace{1.5cm} A_j(\vec{x})=V_j(\vec{x})+\delta a_j(\vec{x})
\]
are still solutions of the first-order equations if
\begin{eqnarray*}
F_{12}&=&{1\over 2}(1-|\phi|^2)\Leftrightarrow -\partial_2\delta
a_1+\partial_1\delta a_2+s_1\delta s_1+s_2\delta s_2=0
\\
(\partial_1\phi_1+A_1\phi_2)&-&(\partial_2\phi_2-A_2\phi_1)=0
\Leftrightarrow s_1\delta a_1-s_2\delta
a_2-(\partial_2-V_1)\delta s_1-(\partial_1+V_2)\delta s_2=0
\\
(\partial_2\phi_1+A_2\phi_2)&+&(\partial_1\phi_2-A_1\phi_1)=0
\Leftrightarrow s_2\delta a_1+s_1\delta a_2+(\partial_1+V_2)\delta
s_1-(\partial_2-V_1)\delta s_2=0 \qquad .
\end{eqnarray*}

To avoid pure gauge deformations, we set the background gauge:
\[
\partial_j\delta a_j(\vec{x})+s_2(\vec{x})\delta
s_1(\vec{x})-s_1(\vec{x})\delta s_2(\vec{x})=0\qquad   .
\]

Therefore, the tangent space to the moduli space of self-dual
vortices is the Kernel of the first-order deformation operator
${\cal D}$:

\[
{\cal D}\xi(\vec{x})= \left(\begin{array}{cccc} -\partial_2 &
\partial_1 & s_1 & s_2 \\ -\partial_1 & -\partial_2 &
-s_2 & s_1
\\ s_1 & -s_2 & -\partial_2+V_1 & -\partial_1-V_2 \\ s_2 & s_1 &
\partial_1+V_2 & -\partial_2+V_1 \end{array}\right)\left(\begin{array}{c} \delta a_1(\vec{x})\\
\delta a_2(\vec{x})\\
\delta s_1(\vec{x})\\ \delta s_2(\vec{x}) \end{array}\right)
\hspace{1cm} , \hspace{1cm}{\cal D}\xi(\vec{x})=0\qquad   .
\]
One easily checks that $K={\cal D}^\dagger{\cal D}$ has a
supersymmetric partner: $K^- ={\cal D}{\cal D}^\dagger$.

\begin{eqnarray*}
K^-={\normalsize \left(\begin{array}{cccc} -\bigtriangleup+|s|^2 &
0 & 0 & 0 \\ 0 & -\bigtriangleup+|s|^2 & 0 & 0  \\ 0 & 0 &
-\bigtriangleup+{1\over 2}(|s|^2+1)+V_kV_k & -2V_k\partial_k
\\ 0 & 0 & 2V_k\partial_k & -\bigtriangleup+{1\over
2}(|s|^2+1)+V_kV_k
\end{array}\right)} \qquad .
\end{eqnarray*}
The index of the deformation operator - ${\rm ind}\,{\cal D}={\rm
dim} {\rm Ker} {\cal D}- {\rm dim} {\rm Ker} {\cal D}^\dagger$ - is
in this case equal to the dimension of ${\rm Ker}K$ because ${\rm
dim} {\rm Ker} {\cal D}^\dagger=0$, $K^-$ being definite positive.

\subsection{The kernel of the heat equation}

The heat equation kernel of a $N\times N$ matrix differential
operator of the general form
\[
K=K_0 +Q_k(\vec{x})\partial_k+V(\vec{x})
\]
is the solution of the $K$-heat equation
\[
\left(\frac{\partial}{\partial\beta}{\mathbb I}+K
\right)K_{K}(\vec{x},\vec{y};\beta )=0 \qquad ,
\]
with initial condition: $K_{K}(\vec{x},\vec{y};0)={\mathbb I}\cdot
\delta^{(2)}(\vec{x}-\vec{y})$. From the kernel, one derives the
partition function
\[
{\rm Tr}\,e^{-\beta K}={\rm tr}\int_{{\mathbb R}^2} \, d^2\vec{x}
\, K_{K}(\vec{x},\vec{x};\beta)
\]
which, in turn, provides the dimension of the vortex moduli space
\[
{\rm ind}\, {\cal D}={\rm Tr}\, e^{-\beta K}-{\rm Tr}\, e^{-\beta
K^-} \qquad ,
\]
because ${\rm Non}$-${\rm zero}$ $\,\, {\rm Spec}K={\rm Spec}
K^-$. To find the kernel one writes
\[
K_{K}(\vec{x},\vec{y};\beta)=C_K(\vec{x},\vec{y};\beta)K_{K_0}(\vec{x},\vec{y};\beta)\qquad
,
\]
where
\[
K_{K_0}(\vec{x},\vec{y};\beta)={e^{-\beta}\over
4\pi\beta}\cdot{\mathbb I}\cdot e^{-\frac{|\vec{x}-\vec{y}|^2
}{4\beta}}
\]
is the $K_0$-heat equation kernel for small $\beta$.
$C_K(\vec{x},\vec{y};\beta)$ satisfies the transfer equation
\[
\left\{ {\partial\over\partial\beta}{\mathbb
I}+{x_k-y_k\over\beta}(\partial_k{\mathbb I}-{1\over
2}Q_k)-\bigtriangleup{\mathbb I}+Q_k\partial_k+V
\rule[-0.1in]{0.in}{0.2in} \right\}C_K(\vec{x},\vec{y};\beta)=0
\qquad ,
\]
and is the unit matrix $C_K(\vec{x},\vec{y};0)={\mathbb I}$ at
infinite temperature.

\subsection{The high-temperature expansion of the partition function}

Solving the transfer equation by means of an inverse temperature
power series expansion
\[
C_K(\vec{x},\vec{y};\beta)=\sum_{n=0}^\infty
c_n(\vec{x},\vec{y};K)\beta^n \qquad ,
\]
the PDE equation becomes tantamount to the recurrence relation
between the densities $c_n(\vec{x},\vec{y};K)$:
\[
[n{\mathbb I}+(x_k-y_k)(\partial_k{\mathbb I}-{1\over
2}Q_k)]c_n(\vec{x},\vec{y};K) =[\bigtriangleup{\mathbb I}
-Q_k\partial_k-V]c_{n-1}(\vec{x},\vec{y};K) \hspace{1cm} ,
\hspace{1cm} n\geq 1
\]
to be started from:  $c_0(\vec{x},\vec{y};K)={\mathbb I}$.

The coefficients of the asymptotic expansion for the partition
function are obtained through integration over the whole plane of
the Seeley densities:
\[
{\rm Tr}e^{-\beta K} = {e^{-\beta}\over \pi\beta}
\sum_{n=0}^\infty\beta^n c_n(K)\hspace{2cm} , \hspace{2cm}
c_n(K)=\sum_{a=1}^4 \int \, d^2x \,[c_n]_{aa}(\vec{x},\vec{x};K)\qquad   .
\]
Because
\[
c_1(\vec{x},\vec{x};K)=-V(\vec{x}) \qquad ,
\]
and since ${\rm ind}{\cal D}$ is independent of $\beta$, we find at
the $\beta=0$ -infinite temperature- limit :
\[
 {\rm ind}{\cal
D}={1\over \pi}\left\{c_1(K)-c_1(K^-)\right\}={1\over \pi}\int
d^2x \left(\frac{\partial V_2}{\partial x_1}-\frac{\partial
V_1}{\partial x_2}\right) (\vec{x})=2l\qquad   .
\]

The recurrence relation gives us the second vortex Seeley density:
\[
c_2(\vec{x},\vec{x};K)= -{1\over 6}\bigtriangleup V(\vec{x})+{1\over
12}Q_k(\vec{x})Q_k(\vec{x})V(\vec{x})- {1\over
6}\partial_kQ_k(\vec{x})V(\vec{x})+{1\over
6}Q_k(\vec{x})\partial_kV(\vec{x})+{1\over 2}V^2(\vec{x}) \qquad .
\]

The determination of higher-order densities becomes more and more
involved. To make the problem more tractable we introduce the
following notation:
\[
{}^{(\alpha_1,\alpha_2)}C_n^{ab}(\vec{x})=\lim_{\vec{y}\rightarrow
\vec{x}} \frac{\partial^{\alpha_1+\alpha_2}
[c_n]_{ab}(\vec{x},\vec{y};K)}{\partial x_1^{\alpha_1}\partial
x_2^{\alpha_2}} \hspace{1cm} , \hspace{1cm}
[{c}_n]_{ab}(\vec{x},\vec{x};K)={}^{(0,0)}C_n^{ab}(\vec{x})\qquad
.
\]
Thus, at the $\vec{y}\rightarrow\vec{x}$ limit the recurrence
relations between densities and partial derivatives of densities can
be written in the compact form:

\begin{eqnarray*} &&(k+\alpha_1+\alpha_2+1)
{}^{(\alpha_1,\alpha_2)}C_{k+1}^{ab}(\vec{x})=
{}^{(\alpha_1+2,\alpha_2)}C_{k}^{ab}(\vec{x})+
{}^{(\alpha_1,\alpha_2+2)}C_{k}^{ab}(\vec{x})-  \\
&&-\sum_{d=1}^N \sum_{r=0}^{\alpha_1}\sum_{t=0}^{\alpha_2}
{\alpha_1 \choose r} {\alpha_2 \choose t} \left[
\frac{\partial^{r+t} Q^{ad}_1}{\partial x_1^r\partial x_2^t}
{}^{(\alpha_1-r+1,\alpha_2-t)}C_{k}^{db}(\vec{x})\right.+\\&&+\left.
\frac{\partial^{r+t} Q^{ad}_2}{\partial x_1^r\partial x_2^t}
{}^{(\alpha_1-r,\alpha_2-t+1)}C_{k}^{db}(\vec{x})
\right]+  \\
&&+\frac{1}{2}\sum_{d=1}^N
\sum_{r=0}^{\alpha_1-1}\sum_{t=0}^{\alpha_2} \alpha_1{\alpha_1-1
\choose r} {\alpha_2 \choose t}  \frac{\partial^{r+t}
Q^{ad}_1}{\partial x_1^r\partial x_2^t}
{}^{(\alpha_1-1-r,\alpha_2-t)}C_{k+1}^{db}(\vec{x})+ \\
&&+\frac{1}{2}\sum_{d=1}^N
\sum_{r=0}^{\alpha_2-1}\sum_{t=0}^{\alpha_1} \alpha_2{\alpha_2-1
\choose r} {\alpha_1 \choose t}  \frac{\partial^{r+t}
Q^{ad}_2}{\partial x_1^t\partial x_2^r}
{}^{(\alpha_1-t,\alpha_2-1-r)}C_{k+1}^{db}(\vec{x})-
\\&&-\sum_{d=1}^N
\sum_{r=0}^{\alpha_2}\sum_{t=0}^{\alpha_1}{\alpha_1 \choose
t}{\alpha_2\choose r}
 \frac{\partial^{r+t}
V^{ad}}{\partial x_1^t\partial x_2^r}
{}^{(\alpha_1-t,\alpha_2-r)}C_k^{db}(\vec{x})\qquad ,
\end{eqnarray*}
to be solved starting from
\[
c_0(\vec{x},\vec{x};K)={\mathbb I} \Rightarrow\left\{
\begin{array}{c} {}^{(\alpha,\beta)}C_0^{ab}(\vec{x})=0 , \, {\rm
if}\,
\alpha \neq 0, {{\rm and}/{\rm or}}\, \beta \neq 0 \\
{}^{(0,0)}C_0^{aa}(\vec{x})=1 , \, a=1,2, \cdots ,N
\end{array}\right. \qquad .
\]

Knowledge of $c_2(\vec{x},\vec{x};K)$ requires knowledge of all the
densities and their derivatives shown below:
\[
\begin{array}{ccccccccc} & & & & {}^{(0,0)}C_0 & & & & \\ & & & {}^{(1,0)}C_0 & & {}^{(0,1)}C_0 & & & \\
 & & {}^{(2,0)}C_0 & & {}^{(1,1)}C_0 & & {}^{(0,2)}C_0 & & \\ & {}^{(3,0)}C_0 & & {}^{(2,1)}C_0 & & {}^{(1,1)}C_0 & & {}^{(1,1)}C_0 &
 \\ {}^{(4,0)}C_0 & & {}^{(3,1)}C_0 & & {}^{(2,2)}C_0 & & {}^{(1,3)}C_0 & & {}^{(0,4)}C_0\end{array}
\]
\vspace{0.5cm}
\[
\begin{array}{ccccc}  & & {}^{(0,0)}C_1 & &  \\  & {}^{(1,0)}C_1 & & {}^{(0,1)}C_1 &  \\
  {}^{(2,0)}C_1 & & {}^{(1,1)}C_1 & & {}^{(0,2)}C_1 \end{array}
\]
\vspace{0.5cm}
\[
{}^{(0,0)}C_2
\]
In general, the number of derivatives and densities required to
compute the $n^{{\rm th}}$-order density is:
\[
\aleph=16\cdot \sum_{j=1}^{n+1} \, \frac{2j(2j-1)}{2}={8\over
3}(n+1)(n+2)(4n+3)\qquad   .
\]
Evaluation of ${}^{(0,0)}C_6^{ab}(\vec{x})$ requires knowledge of
4032 local coefficients !!!. The Seeley coefficients are then
obtained by numerical integration of the Seeley densities over the
plane
\[
c_n(K)=\int \, d^2x \, \sum_{a=1}^4 \, {}^{(0,0)}C_n^{aa}(\vec{x})\qquad   .
\]
Note that the upper delta-shaped wing array is fixed by the initial
conditions set to start the recurrence relations. Thus, strictly one
could skip computing the $j=n+1$ coefficients in the sum.

\subsection{The Mellin transform of the asymptotic expansion}

A good approximation to the generalized zeta functions of both $K$
and $K^G$ is given by the Mellin transform
\[
\zeta_{K}(s)={1\over\Gamma(s)}\int_0^\infty \, d\beta \,
\beta^{s-1} \, {\rm Tr} \, e^{-\beta K} \hspace{2cm} ,
\hspace{2cm} \zeta_{K^G}(s)={1\over\Gamma(s)}\int_0^\infty \,
d\beta \, \beta^{s-1} \, {\rm Tr} \, e^{-\beta K^G}
\]
applied to the high-temperature expansion of the partition
functions
\[
{\rm Tr}e^{-\beta K} = {e^{-\beta}\over \pi\beta}
\sum_{n=0}^\infty \, \beta^n c_n(K) \hspace{1.5cm} ,
\hspace{1.5cm} {\rm Tr}e^{-\beta K^{\rm G}} = {e^{-\beta}\over
4\pi\beta} \sum_{n=0}^\infty\beta^n c_n(K^{\rm G})\qquad   .
\]
The generalized zeta functions are thus divided as sums of
meromorphic -high-temperature regime- and entire -low temperature
regime- functions of $s$:

\begin{eqnarray*}
\zeta_{K}(s)&=&{1\over\Gamma(s)}\sum_{n=0}^\infty
\sum_{a=1}^4\int_0^1 \, d\beta \,
\beta^{s+n-2}c_n(K)e^{-\beta}+{1\over\Gamma(s)}\int_1^\infty {\rm
Tr}^* e^{-\beta K} \, d\beta \, \\&=&\sum_{n=0}^\infty
c_n(K)\frac{\gamma[s+n-1,1]}{4\pi\Gamma(s)}+{1\over\Gamma(s)}B_{K}(s)
\end{eqnarray*}
\begin{eqnarray*}
\zeta_{K^G}(s)&=&{1\over\Gamma(s)}\sum_{n=0}^\infty \int_0^1 \,
d\beta \,
\beta^{s+n-2}c_n(K^G)e^{-\beta}+{1\over\Gamma(s)}\int_1^\infty \,
d\beta \, {\rm Tr}^* e^{-\beta K^{G}}\\&=&\sum_{n=0}^\infty
c_n(K^G)\frac{\gamma[s+n-1,1]}{4\pi\Gamma(s)}+{1\over\Gamma(s)}B_{K^G}(s)\qquad   .
\end{eqnarray*}
We shall neglect the entire parts $B(K)$ and $B(K^G)$ and keep a
finite number of terms, $N_0$, in future use of these generalized
zeta functions for the regularization of ultraviolet divergences.

\subsection{Zeta function regularization}

\subsubsection{Tadpole/self-energy graphs}
The contribution to the one-loop vortex mass of the counter-terms
induced by the scalar and vector fields
\[
\Delta M_{c.t.}^S=2\hbar\, m\,I(1)\int \, d^2x \,
[(1-|s(\vec{x})|^2)-1+1]\hspace{0.3cm} , \hspace{0.3cm} \Delta
M_{c.t.}^A=\hbar \, m \, I(1)\int \, d^2x \, [0_k0_k-
V_k(\vec{x})V_k(\vec{x})]
\]
is proportional to the integral $I(1)$ and, hence,
ultraviolet-divergent. To regularize this integral, we apply the
residue theorem to integration in the complex $k_0$-plane of $I(1)$
and note that on a square of area $m^2L^2$ the integral becomes an
infinite sum over discrete momenta:
\[
I(1)={1 \over 2}\int {d^2k\over (2 \pi)^2} {1\over \sqrt{\vec k
\cdot \vec k +1}}={1\over 2}{1\over
m^2L^2}\sum_{\vec{k}}\frac{1}{\sqrt{\vec{k}\vec{k}+1}}\qquad   .
\]
Thus, this integral is formally the generalized zeta function of the
Klein-Gordon operator evaluated at $s={1\over 2}$, and we shall take
\[
I(1)={1\over 2m^2L^2}\zeta_{K^{\rm G}_0}({1\over 2})={1 \over 8
\pi} {\Gamma (-{1\over 2}) \over \Gamma({1\over 2})}=-{1\over
4\pi}
\]
as the regularized value of $I(1)$. In this way, we find the
following contribution to the one-loop vortex mass shift:
\[
\Delta M_V^R=\Delta M_{c.t.}^S+\Delta M_{c.t.}^A=-\hbar\, {m\over
4\pi} \, \Sigma (s(\vec{x}),V_k(\vec{x}))
\]
\[
\Sigma (s(\vec{x}),V_k(\vec{x}))=\int \, d^2x \,
[2(1-|s(\vec{x})|^2)- V_k(\vec{x})V_k(\vec{x})]\qquad   .
\]
Contrary to the kink case, which is a one-dimensional problem, a
finite answer is obtained in the regularized integral via the
associated zeta function. The reason is that in this two-dimensional
problem the physical limit $s={1\over 2}$ is not a pole and only
finite renormalizations will be necessary. Nevertheless, to keep the
procedure as unified as possible we also define the mass
renormalization corrections as a meromorphic function in the complex
$s$-plane:
\[
\Delta M_V^R(s) = {\hbar\over 2\mu L^2}\left({\mu^2\over
m^2}\right)^s \zeta_{K^{\rm G}_0} (s) \Sigma
(s(\vec{x}),V_k(\vec{x})) \hspace{1cm} ,\hspace{1cm} \Delta
M_V^R=\lim_{s\rightarrow{1\over 2}} \Delta M_V^R(s)\qquad   .
\]

\subsubsection{Vortex Casimir energy}

The divergent vortex and vacuum energies
\[
\bigtriangleup E_V=\frac{\hbar m}{2}{\rm Tr}^*\, K^{{1\over
2}}-\frac{\hbar m}{2}{\rm Tr}^*\, (K^{{\rm G}})^{{1\over 2}}
\hspace{0.5cm} , \hspace{0.5cm} \bigtriangleup E_0=\frac{\hbar
m}{2}{\rm Tr}\, K_0^{{1\over 2}}-\frac{\hbar m}{2}{\rm Tr}\,
(K_0^{{\rm G}})^{{1\over 2}}
\]
can be regularized in a similar vein
\[
\Delta E_V (s)=\frac{\hbar\mu}{2}\left({\mu^2\over
m^2}\right)^s\left\{\zeta^*_{K}(s)-\zeta^*_{K^G}(s)\right\}\hspace{1cm},
\hspace{1cm} \Delta E_0=\frac{\hbar\mu}{2}\left({\mu^2\over
m^2}\right)^s\left\{ \zeta_{K_0}(s)-\zeta_{K^G_0}(s)\right\}\qquad   .
\]
Recall that
\[
\zeta_{K_0}(s)=\frac{m^2L^2}{\pi}\cdot \frac{\gamma[s-1,1]}{\Gamma(s)}
\hspace{1cm} , \hspace{1cm}
\zeta_{K^G_0}(s)=\frac{m^2L^2}{4\pi}\cdot
\frac{\gamma[s-1,1]}{\Gamma(s)}
\]
neglecting the entire functions $B(K)$ and $B(K^G)$. Thus,
\[
\Delta M_V^C(s)=\frac{\hbar\mu}{2}\left({\mu^2\over
m^2}\right)^s\left\{-\frac{2l}{\Gamma(s)}\int_0^1 d\beta
\beta^{s-1}+\sum_{n=1}^{N_0}\left[c_n(K)-c_n(K^G)\right]\cdot
\frac{\gamma[s+n-1,1]}{4\pi\Gamma(s)}\right\} \qquad ,
\]
where the $2l$ zero modes have been subtracted: i.e.,
\[
\Delta M_V^{(0)C}=-{\hbar m\over\sqrt{\pi}}\,\, l=-0.56419 l\hbar m
\]
is the contribution of the $2l$ zero modes to the one-loop vortex
mass shift.

The zero-point vacuum renormalization, however, amounts to throwing
away the contribution of the $c_0(K)$ and $c_0(K^G)$ coefficients.
The physical limit is
\[
\Delta M_V^C=\lim_{s\rightarrow -\frac{1}{2}}\Delta M_V^C(s)\quad ,
\quad \Delta M_V^R=\lim_{s\rightarrow \frac{1}{2}}\Delta M_V^R(s)
\qquad ,
\]
giving the vortex Casimir energy.

\subsection{The high-temperature one-loop vortex mass shift formula}

The contribution of the $c_1$ coefficients to the vortex Casimir
energy is:
\[
\Delta M_V^{(1)C} (s)  = {\hbar \over 2} \mu \left( {\mu^2 \over
m^2}\right)^s [c_1(K)-c_1(K^{\rm G})] \cdot {\gamma[s,1] \over 4
\pi \Gamma(s)} \qquad .
\]
The Seeley densities are respectively scalar,
$c_1(\vec{x},\vec{x};K^{\rm G})=1-|s(\vec{x})|^2$, and $4\times
4$-matrices:
\begin{eqnarray*}
c_1(\vec{x},\vec{x};K)={\small \left(\begin{array}{cccc}
1-|s(\vec{x})|^2 & 0 & 2\nabla_1s_2 & -2\nabla_1s_1
\\ 0 & 1-|s(\vec{x})|^2 & 2\nabla_2s_2 &
-2\nabla_2s_1 \\ 2\nabla_1s_2 & 2\nabla_2s_2 & {3\over
2}(1-|s(\vec{x})|^2)-V_kV_k & 0
\\ -2\nabla_1s_1 & -2\nabla_2s_1 & 0 &
{3\over 2}(1-|s(\vec{x})|^2)-V_kV_k)
\end{array}\right)} \qquad .
\end{eqnarray*}
Thus, the first Seeley coefficients due to normal and ghost
fluctuations are respectively
\begin{eqnarray*}
c_1(K)&=&{\rm tr}\int \, d^2x \, c_1(\vec{x},\vec{x};K)=\int \, d^2x
\, [5(1-|s(\vec{x})|^2)-2V_k(\vec{x})V_k(\vec{x})]
\\
c_1(K^{\rm G})&=&\int \, d^2x \, c_1(\vec{x},\vec{x};K^{\rm G})=\int
\, d^2x \, [1-|s(\vec{x})|^2] \qquad ,
\end{eqnarray*}
such that
\[
\Delta M_V^{(1)C} (-1/2) = - {\hbar m \over 8 \pi } \Sigma (s,V_k)
\cdot {\gamma[-1/2,1] \over \Gamma(1/2)}
\]
exactly kills the contribution of the mass renormalization
counter-terms
\[
\Delta M_V^R (1/2)  =  {\hbar m \over 8 \pi} \cdot \Sigma (s,V_k)
\cdot
 {\gamma[-1/2,1]\over \Gamma(1/2)} \qquad ,
\]
as expected. In the planar Abelian Higgs model all the particles are
massive and we have set our finite renormalization prescriptions in
such a way that the quantum corrections vanish at the limit where
the masses go to infinity.

We finally obtain the high-temperature one-loop vortex mass shift
formula:
\[
\Delta M_V= -{\hbar m \over 2} \left[ \frac{1}{8\pi\sqrt{\pi}}\cdot
\sum_{n=2}^{N_0}\, [c_n(K)-c_n(K^{\rm G})]
\cdot\gamma[n-\frac{3}{2},1]+\frac{2 l}{\sqrt{\pi}} \right] \qquad .
\]
The final form is a polynomial expressions in incomplete Gamma
functions times the heat-kernel expansion coefficients, starting
from the second-order coefficients. By cutting the expansion at a
finite number $N_0$ we admit an error - besides the rejected
entire parts - proportional to $\gamma[N_0-{1\over 2},1]\simeq
{1\over N_0-{1\over 2}}$, for $N_0$ large.

\subsection{Mathematica calculations}

\subsubsection{The mass shift of superposed vortices for
vorticities $l=1$, $l=2$, $l=3$, $l=4$}

Plugging the spherically symmetric solutions into the Seeley
densities, the coefficients can be calculated in a Mathematica
environment through numerical integration. In the next Table the
result is shown for low vorticities: $l=1,2,3,4$.
\begin{center}
\begin{tabular}{|c|cc|cc|}
\hline & \multicolumn{2}{|c|}{$l=1$} & \multicolumn{2}{|c|}{$l=2$}
\\ \hline
$n$ & $ {c}_n(K)$ & $ {c}_n(K^G)$ & $ {c}_n(K)$ & $ {c}_n(K^G)$
\\ \hline
2 & 30.36316 & 2.60773 & 61.06679 & 6.81760    \\
3 & 12.94926 & 0.31851 & 25.61572 & 1.34209    \\
4 &  4.22814 &  0.022887 & 8.21053 & 0.20481   \\
5 & 1.05116 &  0.0011928 & 2.02107 & 0.023714  \\
6 &  0.20094 & 0.00008803 & 0.40233 & 0.002212  \\
\hline
\end{tabular}
\end{center}
\begin{center}
\begin{tabular}{|c|cc|cc|}
\hline & \multicolumn{2}{|c|}{$l=3$} & \multicolumn{2}{|c|}{$l=4$}
\\ \hline
$n$ & $ {c}_n(K)$ & $ {c}_n(K^G)$ & ${c}_n(K)$ & $ {c}_n(K^G)$
\\ \hline
2 & 90.20440 & 11.51035 & 118.67540 & 16.46895 \\
3 & 36.68235 &  2.60898 &  46.01141 & 4.00762 \\
4 & 11.69979 &  0.46721 &  14.64761 & 0.77193 \\
5 & 2.86756  & 0.067279 &   3.58906 & 0.11747 \\
6 & 0.566227  & 0.0079269 & 0.667202 & 0.01620 \\
\hline
\end{tabular}
\end{center}
Note that for $N_0=6$ the differences between nearest order
coefficients are already very small, reinforcing the good
convergence properties of the high-t expansion. The general formula
thus gives the one-loop vortex mass shifts, providing the numbers
shown in the next Table as a function of $N_0$.

\begin{center}
\begin{tabular}{|c|cccc|}  \hline
$N_0$ & $\Delta M_V (N_0)$ & $\Delta M_V (N_0)$& $\Delta M_V
(N_0)$ & $\Delta M_V (N_0)$ \\
 & $l=1$ & $l=2$ & $l=3$ & $l=4$
\\ \hline
2 & -1.02951 & -2.03787  &  -3.01187 & -3.97025 \\
3 & -1.08323 & -2.14111  &  -3.15680 & -4.14891 \\
4 & -1.09270 & -2.15913  &  -3.18208 & -4.18014 \\
5 & -1.09427 & -2.16212  &  -3.18628 & -4.18534 \\
6 & -1.09449 & -2.16257  &  -3.18690 & -4.18606 \\ \hline
\end{tabular}
\end{center}

It is also remarkable to realize, as shown in the next Figure, that
the mass shift is almost linear in $l$; i.e. the mass shift for $l$
vortices is almost equal to $l$ times the mass shift for one vortex.

\begin{center}
\begin{tabular}{|c|c|}
\hline $l$ & $\Delta M_V/\hbar m$ \\ \hline 1 & -1.09449  \\
2 & -2.16257 \\ 3 & -3.18690 \\  4 & -4.18606 \\
\hline
\end{tabular} \hspace{0.7cm}
\begin{tabular}{c}\includegraphics[height=2.2cm]{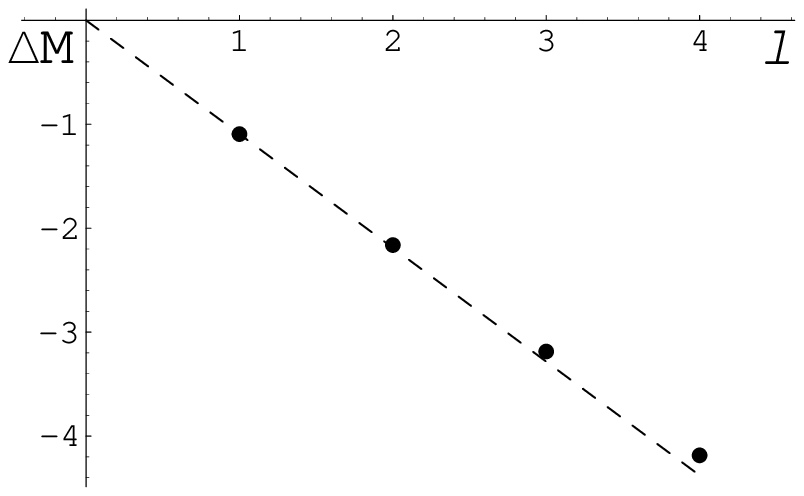}
\end{tabular}
\end{center}

\subsubsection{The mass shift for solutions with two separate vortices}

The same procedure also provides the quantum corrections for
two-vortex solutions with intermediate separations $d=1$, $d=2$, and
$d=3$ between centers, shown in the next Table
\begin{center}
\begin{tabular}{|c|cc|cc|cc|}
\hline & \multicolumn{2}{|c|}{$d=1$} & \multicolumn{2}{|c|}{$d=2$}
&  \multicolumn{2}{|c|}{$d=3$}
\\ \hline
$n$ & $ {c}_n(K)$ & $ {c}_n(K^G)$ & $ {c}_n(K)$ & $ {c}_n(K^G)$ &
$ {c}_n(K)$ & $ {c}_n(K^G)$
\\ \hline
2 & 61.0518 & 6.81277 & 58.3359 & 6.46609  & 57.3420 & 6.03872  \\
3 & 25.6137 & 1.33822 & 24.5050 & 1.23466  & 24.1187 & 1.02031  \\
\hline
\end{tabular}
\end{center}
\begin{center}
\begin{tabular}{|c|ccc|}  \hline
$N_0$ & $\Delta M_V (N_0)/\hbar m$ & $\Delta M_V (N_0)/\hbar m$&
$\Delta M_V
(N_0)/\hbar m$ \\
 & $d=1$ & $d=2$ & $d=3$
\\ \hline
2 & -2.03770 & -1.99798  &  -1.98848  \\
3 & -2.14095 & -2.09695  &  -2.08672  \\
\hline
\end{tabular}
\end{center}
In this case, the results are less precise  for two reasons: First,
the non-symmetric approximate solutions are less reliable than the
superimposed vortex solutions. Second, the first-order equations are
not enough to solve for field derivatives in terms of the vortex
field profiles. Accordingly, we have given corrections only up to
$N_0=3$.
\newpage

\section{APPENDIX I. Zeta functions and Casimir effects}

\subsection{The free scalar field in (1+1)-dimensional space-time: quantum vacuum energy}

The action for one scalar field $\phi(y^1,y^0)\in {\rm
Maps}({\mathbb R}^{1,1},{\mathbb R})$ is:
\[
S=\int \, dy^2 \, \left\{{1\over 2}\frac{\partial\phi}{\partial
y^\mu}\frac{\partial\phi}{\partial y_\mu}-{m\over
2}\phi^2(y^1,y^0)\right\}={1\over 2}\int \, dx^2 \,
\left\{\frac{\partial\phi}{\partial
x^\mu}\frac{\partial\phi}{\partial x_\mu}-\phi^2(x,t)\right\},
\]
either in dimensional $y^\mu=(y^1,y^0)={1\over m}x^\mu$ or
non-dimensional $x^\mu=(x,t)$ coordinates. The general solution on
the interval $[-{mL\over 2},{mL\over 2}]$ of non-dimensional
length $mL$ of the Klein-Gordon equation
\begin{equation}
\frac{\partial^2\phi}{\partial
t^2}(x,t)-\frac{\partial^2\phi}{\partial x^2}(x,t)+\phi(x,t)=0
\label{eq:Kg} \qquad ,
\end{equation}
with periodic boundary conditions: $\phi(-{mL\over
2},t)=\phi({mL\over 2},t)$, is the plane-wave expansion:
\[
\phi(x,t)=\sqrt{\frac{\hbar}{mL}}\sum_{n\in{\mathbb Z}}\,
\left(a_n{\rm \exp}[i\sqrt{\lambda_n}t-i{n\over R}x]+a_n^*{\rm
\exp}[-i\sqrt{\lambda_n}t+i{n\over R}x]\right) \qquad .
\]
Here, $\lambda_n={n^2\over R^2}+1$, $R={mL\over 2\pi}$, are the
eigenvalues of the differential operator $K_0=-\frac{d^2}{dx^2}+1$
in the space of functions $f:{\mathbb S}^1 \, \rightarrow {\mathbb
C}$ from a circle of radius $R$ to the complex line. The
eigenfunctions
\[
K_0 \, {\rm exp}\{i n {x\over R}\}=\lambda_n \, {\rm exp}\{i n
{x\over R}\}\quad , \quad \lambda_n={n^2\over R^2}+1 \quad , \quad
n\in{\mathbb Z}
\]
form a complete orthonormal set:
\[
f_n(x)0={1\over\sqrt{mL}}\cdot {\rm exp}[i{n\over R}x] \qquad ,
\qquad \int_{-{ml\over 2}}^{{ml\over 2}}\, dx \,
f^*_n(x)f_l(x)=\delta_{nl} \qquad .
\]
The classical energy is:
\[
H={m\over 2}\int \, dx \,\left[\frac{\partial\phi}{\partial
t}\cdot\frac{\partial\phi}{\partial
t}+\frac{\partial\phi}{\partial
x}\cdot\frac{\partial\phi}{\partial x}+\phi^2(x,t)\right]={\hbar
m\over 2}\sum_{n\in{\mathbb Z}}\,
\sqrt{\lambda_n}(a^*_na_n+a_na^*_n)
\]
because of the orthonormality relations and the fact that the
kinetic and potential energies of terms of the form $a_{-n}a_n$ or
$a^*_{-n}a^*_n$ cancel (Jeans theorem).

Canonical quantization leads us to trade Fourier coefficients by
operators satisfying the commutation rules:
\[
\, [\hat{a}_n^\dagger,\hat{a}_l^\dagger]=0 \qquad , \qquad
[\hat{a}_n^\dagger,\hat{a}_l]=\delta_{nl} \qquad , \qquad
[\hat{a}_n,\hat{a}_l]=0 \qquad .
\]
The quantum Hamiltonian operator becomes:
\[
\hat{H}=\hbar m\sum_{n\in{\mathbb Z}}\, \sqrt{\lambda_n} \,
\left(\hat{a}_n^\dagger\hat{a}_n+{1\over 2}\right) \qquad .
\]
The vacuum state is annihilated by all destruction operators,
$\hat{a}_n|0>=0, \forall n$, and the quantum vacuum energy is:
\[
E_V=<0|\hat{H}|0>=\frac{\hbar m}{2}\cdot \sum_{n\in{\mathbb Z}}\,
\sqrt{\lambda_n} \qquad \qquad .
\]
This divergent quantity can be regularized by using the zeta
function regularization method and expressed in terms of the
Epstein zeta function:
\[
E_V(s)=\frac{\hbar m}{2}\cdot\zeta_{K_0}(s)=\frac{\hbar
m}{2}\cdot\sum_{n\in\mathbb Z} \, \frac{1}{[{n^2\over
R^2}+1]^s}=\frac{\hbar m}{2}\cdot E(s,1|{1\over R^2}) \qquad .
\]

\subsection{Inserting plates: quantum Casimir energy}

The Casimir effect measures the quantum vacuum energy when two
plates are inserted at $x=0$ and $x=m a, L>>a$, with respect to the
quantum vacuum energy when the two plates are absent. Thus, we
must deal with the spectrum of $K_0$ with the Dirichlet boundary
conditions:
\[
\phi(-{mL\over 2},t)=\phi({mL\over 2},t)=0 \qquad , \qquad
\phi(0,t)=\phi(m a,t)=0 \qquad.
\]
The eigenfunctions of $K_0$ for the Dirichlet boundary conditions
of the Casimir set-up are of three types:
\begin{enumerate}
\item
\[
f_n^<(x)=\sqrt{{2\over mL}}\cdot{\rm sin}\frac{2\pi
n}{mL}x\cdot\vartheta(-x) \qquad , \qquad
\lambda_n^<=\frac{4\pi^2}{m^2L^2}\cdot n^2+1
\]
\[
\int_{-\frac{mL}{2}}^0 \, dx \, f^<_n(x)f_l^<(x)=\delta_{nl}
\qquad , \qquad n\in{\mathbb N} \qquad .
\]

\item
\[
f_n^{<>}(x)=\sqrt{{2\over ma}}\cdot{\rm sin}\frac{\pi
n}{ma}x\cdot\vartheta(ma-x)\vartheta(x) \qquad , \qquad
\lambda_n^{<>}=\frac{\pi^2}{m^2a^2}\cdot n^2+1
\]
\[
\int_{0}^a \, dx \, f^{<>}_n(x)f_l^{<>}(x)=\delta_{nl}\qquad ,
\qquad n\in{\mathbb N} \qquad .
\]
\item
\[
f_n^>(x)=\sqrt{{2\over m(L-2a)}}\cdot{\rm sin}\frac{2\pi
n}{m(L-2a)}x\cdot\vartheta(x-ma) \qquad , \qquad
\lambda_n^>=\frac{4\pi^2}{m^2(L-2a)^2}\cdot n^2+1
\]
\[
\int_0^{\frac{mL}{2}} \, dx \, f^>_n(x)f_l^>(x)=\delta_{nl}\qquad
, \qquad n\in{\mathbb N} \qquad .
\]

\end{enumerate}
Plane waves in the Casimir set up move in three disconnected
regions:
\begin{enumerate}
\item
\[
\phi^<(x,t)=\sqrt{\frac{2\hbar}{mL}}\cdot\sum_{n\in{\mathbb N}} \,
\left(a_n^<{\rm exp}[i\sqrt{\lambda_n^<}\,\,t]+(a_n^<)^*{\rm
exp}[-i\sqrt{\lambda_n^<}\,\,t] \right)\cdot f_n^<(x) \qquad .
\]

\item
\[
\phi^{<>}(x,t)=\sqrt{\frac{2\hbar}{ma}}\cdot\sum_{n\in{\mathbb N}}
\, \left(a_n^{<>}{\rm
exp}[i\sqrt{\lambda_n^{<>}}\,\,t]+(a_n^{<>})^*{\rm
exp}[-i\sqrt{\lambda_n^{<>}}\,\,t] \right)\cdot f_n^{<>}(x) \qquad
.
\]

\item
\[
\phi^>(x,t)=\sqrt{\frac{2\hbar}{m(L-2a)}}\cdot\sum_{n\in{\mathbb
N}} \, \left(a_n^>{\rm
exp}[i\sqrt{\lambda_n^>}\,\,t]+(a_n^>)^*{\rm
exp}[-i\sqrt{\lambda_n^>}\,\,t] \right)\cdot f_n^>(x) \qquad .
\]
\end{enumerate}
Accordingly, the Casimir classical energy is:
\[
H(a)=\frac{\hbar m}{2}\cdot\sum_{n\in{\mathbb N}}\,
\left[\sqrt{\lambda_n^<}(|a_n^<|^2+|(a_n^<)^*|^2)+\sqrt{\lambda_n^{<>}}(|a_n^{<>}|^2+|(a_n^{<>})^*|^2)
+\sqrt{\lambda_n^>}(|a_n^>|^2+|(a_n^>)^*|^2)\right]\qquad .
\]

Canonical quantization proceeds by requiring the commutation
rules:
\[
[(\hat{a}_n^<)^\dagger,\hat{a}_l^<]=\delta_{nl} \qquad , \qquad
[(\hat{a}_n^{<>})^\dagger,\hat{a}_l^{<>}]=\delta_{nl} \qquad ,
\qquad [(\hat{a}_n^>)^\dagger,\hat{a}_l^>]=\delta_{nl}
\]
and any other commutator between the creation and annihilation
operators equal to zero. The quantum Hamiltonian for the Casimir
set up is therefore:
\[
\hat{H}(a)=\hbar m\cdot\sum_{n\in{\mathbb N}}\,
\left[\sqrt{\lambda_n^<}\left((\hat{a}_n^<)^\dagger\hat{a}_n^<+{1\over
2}\right)+\sqrt{\lambda_n^{<>}}\left((\hat{a}_n^{<>})^\dagger\hat{a}_n^{<>}+{1\over
2}\right) +\sqrt{\lambda_n^>}\left((\hat{a}_n^>)^\dagger
\hat{a}_n^>+{1\over 2}\right)\right]\qquad .
\]
The vacuum state is annihilated by all the annihilation operators:
\[
\hat{a}_n^<|0(a)>=\hat{a}_n^{<>}|0(a)>=\hat{a}_n^>|0(a)>=0\qquad ,
\quad \forall n \qquad ,
\]
and the quantum Casimir energy
$E_C(a)=E_V(a)-E_V=<0(a)|\hat{H}(a)|0(a)>-<0|\hat{H})|0>$ is:
\[
E_C(a)=\frac{\hbar m}{2}\cdot \left(\sum_{n\in{\mathbb
N}}\,\,(\sqrt{\lambda_n^<}+\sqrt{\lambda_n^{<>}}+\sqrt{\lambda_n^>})-\sum_{n\in{\mathbb
Z}}\,\, \sqrt{\lambda_n}\right) \qquad .
\]

\subsection{Zeta function regularization}

We regularize the divergent quantity $E_V(a)=<0(a)|\hat{H}(a)|0(a)>$
by means of generalized zeta functions {\footnote {By the notation
$\zeta_{K/D}(s)$ we wish to stress the fact that the spectrum of the
differential operator $K$ is considered with Dirichlet boundary
conditions.}}:
\[
E_V(a,s)=\frac{\hbar m}{2}\,
\left(\zeta_{K_0^</D}(s)+\zeta_{K_0^{<>}/D}(s)+\zeta_{K_0^>/D}(s)\right)
\qquad , \qquad \zeta_{K_0^</D}(s)=\sum_{n=1}^\infty \,\,
\frac{1}{[{4\pi^2\over m^2L^2}\cdot n^2+1]^s}
\]
\[
\zeta_{K_0^{<>}/D}(s)=\sum_{n=1}^\infty \,\, \frac{1}{[{\pi^2\over
m^2a^2}\cdot n^2+1]^s} \qquad , \qquad
\zeta_{K_0^>/D}(s)=\sum_{n=1}^\infty \,\, \frac{1}{[{4\pi^2\over
m^2(L-2a)^2}\cdot n^2+1]^s} \qquad .
\]
Defining $R^2={m^2L^2\over 4\pi^2}$, $z^2=4m^2a^2$, and ${\cal
R}^2={m^2(L-2a)^2\over 4\pi^2}$, the three generalized zeta function
are given, via Mellin transform, by Epstein zeta functions:
\begin{eqnarray*}
\zeta_{K_0}(s)&=&{1\over\Gamma(s)}\cdot \int_0^\infty \, d\beta \,
\beta^{s-1}\cdot \sum_{n=-\infty}^{\infty}\, e^{-\beta({n^2\over
R^2}+1)}=E[s,1|{1\over R^2}] \qquad ,
\\
\zeta_{K_0^</D}(s)&=&{1\over 2}\left({1\over\Gamma(s)}\cdot
\int_0^\infty \, d\beta \, \beta^{s-1}\cdot
\sum_{n=-\infty}^{\infty}\, e^{-\beta({n^2\over
R^2}+1)}-1\right)={1\over 2}\left(E[s,1|{1\over
R^2}]-1\right)\qquad ,
\\
\zeta_{K_0^{<>}/D}(s)&=&{1\over 2}\left({1\over\Gamma(s)}\cdot
\int_0^\infty \, d\beta \, \beta^{s-1}\cdot
\sum_{n=-\infty}^{\infty}\, e^{-\beta({4\pi^2\over z^2}\cdot
n^2+1)}-1\right)={1\over 2}\left(E[s,1|{4\pi^2\over
z^2}]-1\right)\qquad ,
\\
\zeta_{K_0^>/D}(s)&=&{1\over 2}\left({1\over\Gamma(s)}\cdot
\int_0^\infty \, d\beta \, \beta^{s-1}\cdot
\sum_{n=-\infty}^{\infty}\, e^{-\beta({n^2\over {\cal
R}^2}+1)}-1\right)={1\over 2}\left(E[s,1|{1\over{\cal
R}^2}]-1\right) \qquad .
\end{eqnarray*}
Application of the Poisson summation formula provides the
expressions:
\begin{eqnarray*}
\zeta_{K_0}(s)&=&{\sqrt{\pi}R\over\Gamma(s)}\cdot \int_0^\infty \,
d\beta \, \beta^{s-{3\over 2}}\cdot
e^{-\beta}\cdot\sum_{l=-\infty}^{\infty}\, e^{-{\pi^2
R^2l^2\over\beta}}\qquad ,
\\
\zeta_{K_0^</D}(s)&=&{1\over
2}\left({\sqrt{\pi}R\over\Gamma(s)}\cdot \int_0^\infty \, d\beta
\, \beta^{s-{3\over 2}}\cdot e^{-\beta}
\sum_{l=-\infty}^{\infty}\, e^{-{\pi^2
R^2l^2\over\beta}}-1\right)\qquad ,
\\
\zeta_{K_0^{<>}/D}(s)&=&{1\over 2}\left({z\over 2\Gamma(s)}\cdot
\int_0^\infty \, d\beta \, \beta^{s-{3\over 2}}\cdot
e^{-\beta}\cdot \sum_{l=-\infty}^{\infty}\, e^{-{z^2l^2\over
4\beta}}-1\right) \qquad ,
\\
\zeta_{K_0^>/D}(s)&=&{1\over 2}\left({\sqrt{\pi}{\cal
R}^2\over\Gamma(s)}\cdot \int_0^\infty \, d\beta \,
\beta^{s-{3\over 2}}\cdot e^{-\beta}\cdot
\sum_{l=-\infty}^{\infty}\, e^{-{\pi^2{\cal
R}^2l^2\over\beta}}-1\right) \qquad .
\end{eqnarray*}
Therefore,
\[
\zeta_{K_0^</D}(s)+\zeta_{K_0^>/D}(s)-\zeta_{K_0}(s)=\frac{\sqrt{\pi}}{2\Gamma(s)}\int_0^\infty
\, d\beta \, \beta^{s-{3\over 2}}e^{-\beta}\left[{\cal
R}\sum_{l=-\infty}^\infty \, e^{-\frac{\pi^2 {\cal
R}^2l^2}{\beta}}- R\sum_{l=-\infty}^\infty \, e^{-\frac{\pi^2
R^2l^2}{\beta}}\right]-1
\]
and the limit when the length of the line goes to infinity is:
\begin{eqnarray*}
\lim_{L\rightarrow\infty}\left(\zeta_{K_0^</D}(s)+\zeta_{K_0^>/D}(s)-\zeta_{K_0}(s)\right)&=&
\frac{\sqrt{\pi}}{2\Gamma(s)}\cdot \lim_{L\rightarrow\infty}[{\cal
R}-R]\cdot\int_0^\infty \, d\beta \, \beta^{s-{3\over
2}}e^{-\beta}-1 \\&=&
-\frac{ma}{2\sqrt{\pi}}\cdot\frac{\Gamma(s-{1\over
2})}{\Gamma(s)}-1 \qquad .
\end{eqnarray*}
Addition of
\[
\zeta_{K_0^{<>}/D}(s)=\frac{z}{4\sqrt{\pi}}\cdot\frac{\Gamma(s-{1\over
2})}{\Gamma(s)}+\frac{2}{\sqrt{\pi}\Gamma(s)}\cdot\sum_{l=1}^\infty
\left(\frac{zl}{2}\right)^{s+{1\over 2}}\cdot{1\over l}\cdot
K_{{1\over 2}-s}(z l) \qquad ,
\]
where
\[
K_\nu(z)={1\over 2}\cdot\left({z\over 2}\right)^{-\nu}\cdot\int \,
dt \, t^{\nu-1}e^{-t-\frac{z^2}{4t}}
\]
is the integral representation of the Kelvin functions, provides
the result:
\[
E_C(a,s)=E_V(a,s)-E_V(s)=-\frac{3\hbar m}{4}+\frac{\hbar
m}{\sqrt{\pi}\Gamma(s)}\cdot\sum_{l=1}^\infty
\left(\frac{zl}{2}\right)^{s+{1\over 2}}\cdot{1\over l}\cdot
K_{{1\over 2}-s}(z l) \qquad .
\]

The physical limit $s=-{1\over 2}$ of the renormalized and
regularized quantity $E_C(a,s)$ is finite:
\[
E_C(a)=E_C(a,-{1\over 2})=-\frac{3\hbar m}{4}-\frac{\hbar
m}{2\pi}\cdot\sum_{l=1}^\infty \, {1\over l}\cdot K_1(2ma l)
\qquad .
\]
Therefore, the one-dimensional Casimir force is:
\[
F_C^{{\rm 1D}}(a)=\frac{E_C}{da}(a)=-\frac{\hbar
m}{2\pi}\cdot\sum_{l=1}^\infty \, {1\over l}\cdot \frac{d
K_1}{da}(2ma l) \qquad .
\]
The asymptotic behavior of the Kelvin/Bessel function for a large,
\[
K_1(2mal){{\,\atop\simeq}\atop{a\rightarrow\infty}}
\sqrt{\frac{\pi}{4mal}}\cdot e^{-2aml} \qquad ,
\]
tells us that:
\begin{eqnarray*}
E_C(a)&{{\, \atop\simeq}\atop{a\rightarrow\infty }}&-\frac{3\hbar
m}{4}-\hbar\sqrt{\frac{m}{16\pi a}}\cdot\sum_{l=1}^\infty \,
\frac{e^{-2aml}}{l^{{3\over 2}}}=-\frac{3\hbar
m}{4}-\hbar\sqrt{\frac{m}{16\pi a}}\cdot{\rm Li}_{{3\over
2}}[e^{-2am}] \qquad ,
\\
F_C^{{\rm 1D}}(a)&{{\, \atop\simeq}\atop{a\rightarrow\infty }}&
+\hbar\sqrt{\frac{m}{4\pi a}}\cdot\left(m\cdot{\rm Li}_{1\over
2}[e^{-2am}]+{1\over 4a}\cdot{\rm Li}_{3\over 2}[e^{-2am}]\right)
\qquad ,
\end{eqnarray*}
where ${\rm Li}_\nu[z]=\sum_{l=1}^\infty \, \frac{z^l}{l^\nu}, \,
|z|<1$, is the Polylogarithm function.

\subsection{Three-dimensional Casimir forces}

The real Casimir effect is measured in a three-dimensional space.
As in the one-dimensional toy model, the quantum Casimir energy is
the difference between the energies of the vacuum state of a free
field when two plates divide the space in three regions or there
are no plates. With no plates and periodic boundary conditions
chosen to solve the Klein-Gordon equations for one scalar field,
the space is a three-torus ${\mathbb T}^3={\mathbb
S}^1\times{\mathbb S}^1\times{\mathbb  S}^1$ and the quantum
vacuum energy is:
\[
E_V^{3D}=\frac{\hbar m}{2}\cdot \sum_{\vec{n}\in{\mathbb Z}^3} \,
\sqrt{{n_1^2\over R_1^2}+{n_2^2\over R_2^2}+{n_3^2\over R_3^2}+1}
\qquad , \qquad n_1, n_2, n_3 \in{\mathbb Z} \quad , \quad
\vec{n}=n_1\vec{e}_1+n_2\vec{e}_2+n_3\vec{e}_3 \quad ,
\]
\[
\vec{e}_k\cdot\vec{e}_j=\delta_{kj} \, , \quad k.j=1,2,3 \qquad ,
\qquad R_1={mL_1\over 2\pi}\, \, , \,\, R_2={mL_1\over 2\pi} \,\,
, \,\, R_3={mL_3\over 2\pi} \qquad .
\]

If two impenetrable plates are located at the two-dimensional
sub-spaces $(x_1,x_2,0)$ and $(x_1,x_2,ma)$, the cylindrical space
${\mathbb M}^3={\mathbb S}^1\times{\mathbb S}^1\times{\mathbb I}$ is
divided into three zones by the plates. We choose periodic boundary
conditions in the $x_1,x_2$ coordinates, and Dirichlet boundary
condition as in the one-dimensional Casimir effect in the $x_3$
coordinate to solve the free field equations. Therefore, the quantum
vacuum energy in the Casimir device is:
\begin{eqnarray*}
&&E_V^{{\rm 3D}}(a)=\\&&\frac{\hbar m}{2}\cdot
\sum_{\vec{n}\in{\mathbb Z}^2\times{\mathbb N}} \,
\left\{\sqrt{{n_1^2\over R_1^2}+{n_2^2\over R_2^2}+{n_3^2\over
R_3^2}+1} +\sqrt{{n_1^2\over R_1^2}+{n_2^2\over
R_2^2}+{4\pi^2\over z^2}n_3^2+1}+\sqrt{{n_1^2\over
R_1^2}+{n_2^2\over R_2^2}+{n_3^2\over{\cal R}_3^2}+1}\right\}\,\,
,
\end{eqnarray*}
$n_1, n_2\in{\mathbb Z}\, , \, n_3\in{\mathbb N}$, ${\cal
R}_3=\frac{m(L_3-2a)}{2\pi}$, whereas the Casimir energy reads:
$E_C^{{\rm 3D}}(a)=E_V^{{\rm 3D}}(a)-E_V^{{\rm 3D}}$.

The divergences can be regularized in terms of Epstein zeta
functions,
\[
E_V^{{\rm 3D}}(s)=\frac{\hbar
m}{2}E[s,1|\frac{\vec{e}_1}{R_1^2}+\frac{\vec{e}_2}{R_2^2}+\frac{\vec{e}_3}{R_3^2}]
\]
\begin{eqnarray*}
E_V^{{\rm 3D}}(a,s)&=&\frac{\hbar m}{4}\left[
E[s,1|\frac{\vec{e}_1}{R_1^2}+\frac{\vec{e}_2}{R_2^2}+\frac{\vec{e}_3}{R_3^2}]+
E[s,1|\frac{\vec{e}_1}{R_1^2}+\frac{\vec{e}_2}{R_2^2}+4\pi^2\frac{\vec{e}_3}{z^2}]\right.\\
&+&\left.
E[s,1|\frac{\vec{e}_1}{R_1^2}+\frac{\vec{e}_2}{R_2^2}+\frac{\vec{e}_3}{{\cal
R}_3^2}]
-3E[s,1|\frac{\vec{e}_1}{R_1^2}+\frac{\vec{e}_2}{R_2^2}]\right]
\qquad ,
\end{eqnarray*}
better written, via the Mellin transform, as:
\[
E[s,1|\frac{\vec{e}_1}{R_1^2}+\frac{\vec{e}_2}{R_2^2}]={1\over\Gamma(s)}\cdot\sum_{\vec{n}\in{\mathbb
Z}^2}\,\int_0^\infty \, d\beta \, \beta^{s-1} \, {\rm
exp}[-\beta(\frac{n_1^2}{R_1^2}+\frac{n_2^2}{R_2^2}+1)]
\]
\[
E[s,1|\frac{\vec{e}_1}{R_1^2}+\frac{\vec{e}_2}{R_2^2}+\frac{\vec{e}_3}{R_3^2}]={1\over\Gamma(s)}\cdot\sum_{\vec{n}\in{\mathbb
Z}^3}\,\int_0^\infty \, d\beta \, \beta^{s-1} \, {\rm
exp}[-\beta(\frac{n_1^2}{R_1^2}+\frac{n_2^2}{R_2^2}+\frac{n_3^2}{R_3^2}+1)]
\]
\[
E[s,1|\frac{\vec{e}_1}{R_1^2}+\frac{\vec{e}_2}{R_2^2}+4\pi^2\frac{\vec{e}_3}{z^2}]={1\over\Gamma(s)}\cdot\sum_{\vec{n}\in{\mathbb
Z}^3}\,\int_0^\infty \, d\beta \, \beta^{s-1} \, {\rm
exp}[-\beta(\frac{n_1^2}{R_1^2}+\frac{n_2^2}{R_2^2}+4\pi^2\frac{n_3^2}{z^2}+1)]
\]
\[
E[s,1|\frac{\vec{e}_1}{R_1^2}+\frac{\vec{e}_2}{R_2^2}+\frac{\vec{e}_3}{R_3^2}]={1\over\Gamma(s)}\cdot\sum_{\vec{n}\in{\mathbb
Z}^3}\,\int_0^\infty \, d\beta \, \beta^{s-1} \, {\rm
exp}[-\beta(\frac{n_1^2}{R_1^2}+\frac{n_2^2}{R_2^2}+\frac{n_3^2}{{\cal
R}_3^2}+1)] \qquad .
\]
Before taking the limit of large $L_3$, it is convenient use the
Poisson summation formula to obtain:
\[
E[s,1|\frac{\vec{e}_1}{R_1^2}+\frac{\vec{e}_2}{R_2^2}]={\pi
R_1R_2\over\Gamma(s)}\cdot\left(\Gamma(s-1)+\sum_{\vec{l}\in{\mathbb
Z}^2-\{\vec{0}\}}\,\int_0^\infty \, d\beta \, \beta^{s-2} \,
e^{-\beta}\,{\rm
exp}[-\frac{\pi^2(l_1^2R_1^2+l_2R_2^2)}{\beta}]\right) \qquad ,
\]
$l_1,l_2\in{\mathbb Z}$, but $l_1=l_2=0$ is excluded. Also, since
$l_3\in{\mathbb Z}$ and $l_1=l_2=l_3=0$ does not count, we have:
\begin{eqnarray*}
&&E[s,1|\frac{\vec{e}_1}{R_1^2}+\frac{\vec{e}_2}{R_2^2}+\frac{\vec{e}_3}{R_3^2}]=\\&=&{\pi^{3\over
2}R_1R_2R_3\over\Gamma(s)}\cdot\left(\Gamma(s-{3\over
2})+\sum_{\vec{l}\in{\mathbb Z}^3-\{\vec{0}\}}\,\int_0^\infty \,
d\beta \, \beta^{s-{5\over 2}} \, e^{-\beta}\,{\rm
exp}[-\frac{\pi^2(l_1^2R_1^2+l_2R_2^2+l_3^2R_3^2)}{\beta}]\right)
\end{eqnarray*}
\begin{eqnarray*}
&&E[s,1|\frac{\vec{e}_1}{R_1^2}+\frac{\vec{e}_2}{R_2^2}+4\pi^2\frac{\vec{e}_3}{z^2}]=\\&=&{\pi^{3\over
2}R_1R_2z\over 2\pi\Gamma(s)}\cdot\left(\Gamma(s-{3\over
2})+\sum_{\vec{l}\in{\mathbb Z}^3-\{\vec{0}\}}\,\int_0^\infty \,
d\beta \, \beta^{s-{5\over 2}} \, e^{-\beta}\,{\rm
exp}[-\frac{\pi^2(l_1^2R_1^2+l_2R_2^2)}{\beta}]{\rm
exp}[-\frac{l_3^2z^2}{4\beta}]\right)
\end{eqnarray*}
\begin{eqnarray*}
&&E[s,1|\frac{\vec{e}_1}{R_1^2}+\frac{\vec{e}_2}{R_2^2}+\frac{\vec{e}_3}{{\cal
R}_3^2}]=\\&=&{\pi^{3\over 2}R_1R_2{\cal
R}_3\over\Gamma(s)}\cdot\left(\Gamma(s-{3\over
2})+\sum_{\vec{l}\in{\mathbb Z}^3-\{\vec{0}\}}\,\int_0^\infty \,
d\beta \, \beta^{s-{5\over 2}} \, e^{-\beta}\,{\rm
exp}[-\frac{\pi^2(l_1^2R_1^2+l_2R_2^2+l_3^2{\cal
R}_3^2)}{\beta}]\right) \qquad .
\end{eqnarray*}
Thus, the Epstein zeta functions are given in terms of
Kelvin/Bessel functions:
\[
E[s,1|\frac{\vec{e}_1}{R_1^2}+\frac{\vec{e}_2}{R_2^2}]={\pi
R_1R_2\over\Gamma(s)}\cdot\left(\Gamma(s-1)+\sum_{\vec{l}\in{\mathbb
Z}^2-\{\vec{0}\}} \, [\pi^2(R_1^2l_1^2
+R_2^2l_2^2)]^{\frac{s-1}{2}}\cdot K_{1-s}(2\pi\sqrt{R_1^2l_1^2
+R_2^2l_2^2}\,)\right)
\]
\begin{eqnarray*}
&&E[s,1|\frac{\vec{e}_1}{R_1^2}+\frac{\vec{e}_2}{R_2^2}+\frac{\vec{e}_3}{R_3^2}]=\\&=&{\pi^{3\over
2}R_1R_2R_3\over\Gamma(s)}\cdot\left(\Gamma(s-{3\over
2})+\sum_{\vec{l}\in{\mathbb Z}^3-\{\vec{0}\}}\, [\pi^2(R_1^2l_1^2
+R_2^2l_2^2+R_3l_3^2)]^{{1\over 2}(s-{3\over 2})}\cdot K_{{3\over
2}-s}(2\pi\sqrt{R_1^2l_1^2 +R_2^2l_2^2+R_3l_3^2}\,)\right)
\end{eqnarray*}
\begin{eqnarray}
&&E[s,1|\frac{\vec{e}_1}{R_1^2}+\frac{\vec{e}_2}{R_2^2}+4\pi^2\frac{\vec{e}_3}{z^2}]=\label{eq:zet}\\&=&{\pi^{3\over
2}R_1R_2z\over 2\pi\Gamma(s)}\cdot\left(\Gamma(s-{3\over
2})+\sum_{\vec{l}\in{\mathbb Z}^3-\{\vec{0}\}}\,[\pi^2(R_1^2l_1^2
+R_2^2l_2^2+{z^2\over 4\pi^2}l_3^2)]^{{1\over 2}(s-{3\over
2})}\cdot K_{{3\over 2}-s}(2\pi\sqrt{R_1^2l_1^2
+R_2^2l_2^2+{z^2\over 4\pi^2}l_3^2}\,)\right)\nonumber
\end{eqnarray}
\begin{eqnarray*}
&&E[s,1|\frac{\vec{e}_1}{R_1^2}+\frac{\vec{e}_2}{R_2^2}+\frac{\vec{e}_3}{{\cal
R}_3^2}]=\\&=&{\pi^{3\over 2}R_1R_2{\cal
R}_3\over\Gamma(s)}\cdot\left(\Gamma(s-{3\over
2})+\sum_{\vec{l}\in{\mathbb Z}^3-\{\vec{0}\}}\,[\pi^2(R_1^2l_1^2
+R_2^2l_2^2+{\cal R}_3l_3^2)]^{{1\over 2}(s-{3\over 2})}\cdot
K_{{3\over 2}-s}(2\pi\sqrt{R_1^2l_1^2 +R_2^2l_2^2+{\cal
R}_3l_3^2}\,) \right)\qquad .
\end{eqnarray*}
A magic cancelation occurs in the $L_3\rightarrow\infty$ limit:
\begin{eqnarray*}
&&\lim_{L_3\rightarrow\infty}{1\over
2}\left\{E[s,1|\frac{\vec{e}_1}{R_1^2}+\frac{\vec{e}_2}{R_2^2}+\frac{\vec{e}_3}{{\cal
R}_3^2}]-E[s,1|\frac{\vec{e}_1}{R_1^2}+\frac{\vec{e}_2}{R_2^2}+\frac{\vec{e}_3}{R_3^2}]\right\}=\\&=&\frac{\pi^{{3\over
2}}R_1R_2}{2}\cdot \frac{\Gamma(s-{3\over
2})}{\Gamma(s)}\cdot\left({\cal
R}_3-R_3\right)=-\frac{\pi^{{3\over 2}}R_1R_2}{2}\cdot
\frac{\Gamma(s-{3\over 2})}{\Gamma(s)}\cdot \frac{ma}{\pi}
\end{eqnarray*}
exactly cancels with half the first summand in (\ref{eq:zet}).
There are no divergences left in the physical limit $s=-{1\over
2}$ and we finally obtain the renormalized, regularized and
physical, 3D Casimir energies:
\begin{eqnarray*}
&&E_C^{{\rm 3D}}(a,s)= -\frac{3\hbar m}{4}\cdot\pi
R_1R_2\cdot\frac{\Gamma(s-1)}{\Gamma(s)}\\&-&\frac{3\hbar
m}{4}\cdot{\pi R_1R_2\over\Gamma(s)}\cdot\sum_{\vec{l}\in{\mathbb
Z}^2-\{\vec{0}\}} \, [\pi^2(R_1^2l_1^2
+R_2^2l_2^2)]^{\frac{s-1}{2}}\cdot K_{1-s}(2\pi\sqrt{R_1^2l_1^2
+R_2^2l_2^2}\,)\\&+&\frac{\hbar m}{4}\cdot{\pi^{3\over
2}R_1R_2z\over 2\pi\Gamma(s)}\cdot\sum_{\vec{l}\in{\mathbb
Z}^3-\{\vec{0}\}}\,[\pi^2(R_1^2l_1^2 +R_2^2l_2^2+{z^2\over
4\pi^2}l_3^2)]^{{1\over 2}(s-{3\over 2})}\cdot K_{{3\over
2}-s}(2\pi\sqrt{R_1^2l_1^2 +R_2^2l_2^2+{z^2\over 4\pi^2}l_3^2}\,)
\end{eqnarray*}
\begin{eqnarray*}
&&E_C^{{\rm 3D}}(a)= \frac{\hbar m}{2}\cdot\pi
R_1R_2\\&+&\frac{3\hbar m}{8}\cdot\sqrt{\pi}
R_1R_2\cdot\sum_{\vec{l}\in{\mathbb Z}^2-\{\vec{0}\}} \,
[\pi^2(R_1^2l_1^2 +R_2^2l_2^2)]^{-\frac{3}{4}}\cdot K_{{3\over
2}}(2\pi\sqrt{R_1^2l_1^2 +R_2^2l_2^2}\,)\\&-&\frac{\hbar
m}{16}\cdot R_1R_2z \cdot\sum_{\vec{l}\in{\mathbb
Z}^3-\{\vec{0}\}}\,[\pi^2(R_1^2l_1^2 +R_2^2l_2^2+{z^2\over
4\pi^2}l_3^2)]^{-1}\cdot K_2(2\pi\sqrt{R_1^2l_1^2
+R_2^2l_2^2+{z^2\over 4\pi^2}l_3^2}\,) \qquad .
\end{eqnarray*}
Isolation of the part depending on $a$ provides the exact result
in a very long, $L_3\rightarrow\infty$, cylinder:
\begin{equation}
\bar{E}_C(a)=-\frac{\hbar m}{16\pi^2}\cdot
L_1L_2\cdot\sum_{\vec{l}\in{\mathbb
Z}^3-\{\vec{0}\}}\,\frac{1}{L_1^2l_1^2+L_2l_2^2+4a^2l_3^2}\cdot
K_2(m\sqrt{L_1^2l_1^2+L_2l_2^2+4a^2l_3^2}\, \,)
\label{eq:case}\quad .
\end{equation}
This expression, involving the Kelvin or modified Bessel functions
of order two, is not very useful.

\subsection{The infinite-plate area limit}

Taking also the $L_1\rightarrow\infty$ and $L_2\rightarrow\infty$
limits it is possible to obtain a numerical answer. The key idea is
to perform sums only over $n_3$ and postpone sums in $n_1$ and $n_2$
until the end. It is then possible to trade the sum of series by the
computation of integrals at the limit where the plate areas go to
infinity. Thus, we have:
\[
E[s,1|\frac{\vec{e}_1}{R_1^2}+\frac{\vec{e}_2}{R_2^2}]=\sum_{n_1=-\infty}^\infty\sum_{n_2=-\infty}^\infty
\, P^{-2s}(n_1,n_2) \qquad , \qquad
P^2(n_1,n_2)=\frac{n_1^2}{R_1^2}+\frac{n_2^2}{R_2^2}+1
\]
\begin{eqnarray*}
&&E[s,1|\frac{\vec{e}_1}{R_1^2}+\frac{\vec{e}_2}{R_2^2}+\frac{\vec{e}_3}{
R_3^2}]={\pi^{1\over 2}
R_3\over\Gamma(s)}\cdot\sum_{n_1=-\infty}^\infty\sum_{n_2=-\infty}^\infty\,\int_0^\infty
\, d\beta \, \beta^{s-{3\over 2}}\, e^{-\beta
P^2}\sum_{l_3=-\infty}^\infty \,
e^{-\frac{\pi^2R_3^2l_3^2}{\beta}}\\&=&{\pi^{1\over 2}
R_3\over\Gamma(s)}\cdot\left\{\sum_{n_1=-\infty}^\infty\sum_{n_2=-\infty}^\infty\,\left(\frac{\Gamma(s-{1\over
2})}{P^{2s-1}}+\frac{4}{P^{s-{1\over 2}}}\cdot\sum_{l_3=1}^\infty \,
(\pi R_3l_3)^{s-{1\over 2}}\cdot K_{{1\over 2}-s}(2\pi
R_3l_3P)\right)\right\}
\end{eqnarray*}
\[
E[s,1|\frac{\vec{e}_1}{R_1^2}+\frac{\vec{e}_2}{R_2^2}+4\pi^2\frac{\vec{e}_3}{z^2}]=
{z\over 2\pi^{1\over
2}\Gamma(s)}\cdot\left\{\sum_{n_1=-\infty}^\infty\sum_{n_2=-\infty}^\infty\,\left(\frac{\Gamma(s-{1\over
2})}{P^{2s-1}}+\frac{4}{P^{s-{1\over 2}}}\cdot\sum_{l_3=1}^\infty \,
({zl_3\over 2})^{s-{1\over 2}}\cdot K_{{1\over
2}-s}(zl_3P)\right)\right\}
\]
\[
E[s,1|\frac{\vec{e}_1}{R_1^2}+\frac{\vec{e}_2}{R_2^2}+\frac{\vec{e}_3}{{\cal
R}_3^2}]={\pi^{1\over 2} {\cal
R}_3\over\Gamma(s)}\cdot\left\{\sum_{n_1=-\infty}^\infty\sum_{n_2=-\infty}^\infty\,\left(\frac{\Gamma(s-{1\over
2})}{P^{2s-1}}+\frac{4}{P^{s-{1\over 2}}}\cdot\sum_{l_3=1}^\infty \,
(\pi {\cal R}_3l_3)^{s-{1\over 2}}\cdot K_{{1\over 2}-s}(2\pi {\cal
R}_3l_3P)\right)\right\}\, .
\]
At the $L_3\rightarrow\infty$ limit, the cancelation between
infinities explained above takes place and we are left with the
result:
\[
E_C^{{\rm 3D}}(a,s)=\frac{\hbar
m}{2}\cdot\left\{\sum_{n_1=-\infty}^\infty\sum_{n_2=-\infty}^\infty\,\frac{2z}{\sqrt{\pi}\Gamma(s)P^{s-{1\over
2}}}\cdot\sum_{l_3=1}^\infty \, ({zl_3\over 2})^{s-{1\over 2}}\cdot
K_{{1\over 2}-s}(zl_3P)-\frac{3}{2P^{2s}}\right\} \qquad ,
\]
which at the physical limit $s=-{1\over 2}$ is:
\[
E_C^{{\rm 3D}}(a)=-\frac{\hbar
m}{2}\cdot\sum_{n_1=-\infty}^\infty\sum_{n_2=-\infty}^\infty\,P\left[\frac{2}{\pi}\sum_{l_3=1}^\infty
\, {1\over l_3}\cdot K_1(2aml_3P)+{3\over 2}\right]\qquad .
\]
At the limit of very large areas of the plates the discrete momenta
become continuous and the series become integrals according to very
well known prescriptions:
\[
\frac{n_1^2}{R_1^2}\,\,{{\, \atop\simeq}\atop{mL_1\rightarrow\infty
}}p_1^2 \qquad , \qquad \frac{n_2^2}{R_2^2}\,\,{{\,
\atop\simeq}\atop{mL_2\rightarrow\infty }}p_2^2 \qquad ,
\]
\[
\sum_{n_1=-\infty}^\infty\sum_{n_2=-\infty}^\infty{{\,
\atop\simeq}\atop{m^2L_1L_2\rightarrow\infty
}}\frac{m^2L_1L_2}{4\pi^2}\cdot\int_{-\infty}^\infty \, dp_1
\,\int_{-\infty}^\infty \, dp_2 \qquad .
\]
Therefore, the part of the Casimir energy depending on $a$ is:
\[
\bar{E}_C^{{\rm 3D}}(a)=-\frac{\hbar
m}{2}\cdot\frac{m^2L_1L_2}{4\pi^2}\cdot{2\over\pi}\cdot\int_{-\infty}^\infty
\, dp_1 \,\int_{-\infty}^\infty \, dp_2 \, P
\cdot\sum_{l_3=1}^\infty \, {1\over l_3}\cdot K_1(2aml_3P)\qquad .
\]
Because $\int_{-\infty}^\infty \, dp_1 \,\int_{-\infty}^\infty \,
dp_2 =2\pi\int_0^\infty \, pdp=2\pi\int_0^\infty \, PdP$,
$p^2=p_1^2+p_2^2$, the integral of the modified Bessel function can
be performed to find,
\[
\bar{E}_C^{{\rm
3D}}(a)=-\frac{\hbar}{8\pi^2}\cdot\frac{L_1L_2}{a^3}\cdot\sum_{l_3=1}^\infty
\, {1\over
l_3^4}=-\frac{\hbar}{8\pi^2}\cdot\frac{L_1L_2}{a^3}\cdot\zeta(4)=-\frac{\hbar\pi^2}{720}\cdot\frac{L_1L_2}{a^3}\qquad
,
\]
and the Casimir three-dimensional force is:
\[
\bar{F}_C^{{\rm 3D}}=\frac{d\bar{E}_C^{{\rm
3D}}}{da}(a)=\frac{\hbar\pi^2}{240}\cdot\frac{L_1L_2}{a^4} \qquad .
\]

\section{APPENDIX II. Kinks: d=1, N=1}

$K_0=-\frac{d^2}{dx^2}+4$ acts on functions $f:{\mathbb S}^1 \,
\rightarrow {\mathbb C}$ from a circle of radius
$R=\frac{mL}{2\pi\sqrt{2}}$, (PBC), to the complex plane. The
spectral resolution of $K_0$ is:
\[
K_0 \, {\rm exp}\{i n {x\over R}\}=\lambda_n \, {\rm exp}\{i n
{x\over R}\}\quad , \quad \lambda_n={n^2\over R^2}+4 \quad , \quad
n\in{\mathbb Z} \qquad ,
\]
and the vacuum energy on the circle is:
\[
\bigtriangleup E_0=\frac{\hbar m}{2}\,\cdot {1\over R} \,\cdot
\sum_{n=-\infty}^\infty \, \left[n^2+4R^2\right]^{{1\over 2}}\qquad   .
\]
Regularization of this divergent quantity by means of the
generalized zeta function leads to:
\[
\bigtriangleup E_0(s)={\hbar\over 2}\cdot \left({2\mu^2\over m^2}
\right)^s \cdot\mu \, \cdot \zeta_{K_0}(s) \qquad , \qquad
s\in{\mathbb C}
\]
\[
\zeta_{K_0}(s)={\rm Tr}\left[-{d^2\over
dx^2}+4\right]^{-s}=R^{2s}\cdot \sum_{n=-\infty}^\infty \,
\frac{1}{(n^2+4R^2)^s}={1\over 4^s}+2R^{2s}\cdot \sum_{n=1}^\infty
\, \frac{1}{(n^2+4R^2)^s}
\]

\subsection{The kink generalized zeta function versus Riemann zeta
functions}

To express the generalized zeta function in terms of ordinary
Riemann zeta functions one can use the binomial series:
\[
\sum_{n=1}^\infty \, \frac{1}{(n^2+R^2)^s}=\sum_{l=0}^\infty \,
\left(\begin{array}{c}-s \\ l \end{array}\right)\cdot
4^lR^{2l}\cdot \sum_{n=1}^\infty \, n^{-2s-2l} \qquad , \qquad
l\in{\mathbb Z}^+
\]
\[
\sum_{n=1}^\infty \, \frac{1}{(n^2+4R^2)^s}=\sum_{l=0}^\infty \,
\frac{\Gamma(1-s)}{\Gamma(1+l)\Gamma(1-s-l)}\cdot 4^lR^{2l}\cdot
\zeta(2s+2l)
\]
\[
\zeta_{K_0}(s)={1\over 4^s}+2\sum_{l=0}^\infty \,
\frac{\Gamma(1-s)}{\Gamma(1+l)\Gamma(1-s-l)}\cdot
4^lR^{2s+2l}\cdot \zeta(2s+2l)\qquad .
\]
\subsection{The kink generalized zeta function versus Epstein zeta
functions}

Alternatively, the generalized zeta function is an Epstein zeta
function,
\[
E(s,4|A)=\sum_{n=-\infty}^\infty \, \frac{1}{(An^2+4)^s}\Rightarrow
\zeta_{K_0}(s)=E(s,4|{1\over R^2})\qquad ,
\]
which via the Mellin transform
\[
E(s,4|A)={1\over\Gamma(s)}\cdot \sum_{n=-\infty}^\infty \,
\int_0^\infty \, d\beta \, \beta^{s-1}e^{-\beta\, (An^2+4)}
\]
and use of the Poisson summation formula
\[
\sum_{n=-\infty}^\infty \, e^{-\beta A n^2}=\sqrt{\frac{\pi}{\beta
A}}\cdot \sum_{l=-\infty}^\infty \, {\rm exp}\{-\frac{\pi^2
l^2}{A\beta}\}
\]
reads:
\[
E(s,4|A)={1\over\Gamma(s)}\cdot\sqrt{{\pi\over
A}}\cdot\left\{\int_0^\infty \, d\beta \, \beta^{s-{3\over 2}}\,
e^{-4\beta}+2\sum_{l=1}^\infty \, \int_0^\infty \, d\beta \,
\beta^{s-{3\over 2}}\, {\rm exp}\{-4\beta-\frac{\pi^2
l^2}{A\beta}\}\right\} \qquad .
\]
When $R\rightarrow\infty$, $A={1\over R^2}\rightarrow 0$, the result
of Section \S 3 is obtained:
\[
E(s,4|0)=\lim_{A\rightarrow 0}{1\over\Gamma(s)}\cdot\sqrt{{\pi\over
A}}\cdot\int_0^\infty \, d\beta \, \beta^{s-{3\over 2}}\,
e^{-4\beta}=\lim_{L\rightarrow\infty} \,
\frac{mL}{\sqrt{8\pi}}\cdot\frac{1}{4^{s-{1\over
2}}}\cdot\frac{\Gamma(s-{1\over 2})}{\Gamma(s)}\qquad .
\]
In sum,
\begin{enumerate}

\item The result is the same when we work with periodic boundary
conditions on a finite interval and allow the length to go to
infinity as when physicist's techniques are used to cope with
continuous spectra.

\item In both approaches there are contributions proportional to
the volume: infrared divergences which must somehow be renormalized.
\end{enumerate}

\subsection{The high-temperature expansion of the kink heat equation
kernel}

Let us now consider the differential operator
\[
K=-{d^2\over dx^2}+4-V(x) \qquad , \qquad V(x)=V(x+2\pi R)
\]
acting on functions from a circle of radius $R$ to ${\mathbb C}$.
If $V(x)=0$, we find the $K_0$-heat kernel:
\begin{eqnarray*}
K_{K_0}(x,y;\beta)&=&\sum_{n=-\infty}^\infty \, {\rm
exp}\{-\beta({n^2\over R^2}+4)\}\cdot {\rm exp}\{i{n\over
R}(x-y)\}\\&=& e^{-4\beta}{\rm exp}\{-\frac{(x-y)^2}{4\beta}\} \,
\cdot \, \sum_{n=-\infty}^\infty \, {\rm exp}\{-{\beta\over
R^2}[n+i{R\over 2\beta}(y-x)]^2\} \qquad .
\end{eqnarray*}
Using again the Poisson summation formula,
\[
\sum_{n=-\infty}^\infty \, e^{-t(n+v)^2}=\sqrt{{\pi\over t}} \cdot
\sum_{l=-\infty}^\infty \, {\rm exp}\{-{\pi^2 l^2\over t^2}-2\pi
ilv\} \qquad ; \qquad t={\beta\over R^2} \quad , \quad v=i{R\over
2\beta} (y-x) \qquad ,
\]
the $K_0$-heat kernel can be written as:
\[
K_{K_0}(x,y;\beta)=e^{-4\beta}\, \cdot \, \sqrt{{\pi
R^2\over\beta}}\, \cdot \, {\rm exp}\{-\frac{(y-x)^2}{4\beta}\}\,
 \cdot \, \sum_{l=-\infty}^\infty \, {\rm exp}\{-\frac{\pi Rl[\pi
Rl-(y-x)]}{\beta}\} \qquad .
\]
For $\beta<1$, (high-temperature), we obtain the asymptotic
formula,
\[
K_{K_0}(x,y;\beta)\simeq e^{-4\beta}\, \cdot \, \sqrt{{\pi
R^2\over\beta}}\, \cdot \, {\rm exp}\{-\frac{(y-x)^2}{4\beta}\}\,
 \cdot \, \left(1+{\cal O}({\rm
 exp}\{-\frac{C}{\beta}\})\right)\qquad ,
\]
and from this, the asymptotic high-temperature expansion for the
$K$-heat kernel is derived:
\[
K_K(x,y;\beta)\simeq e^{-4\beta}\, \cdot \, \sqrt{{\pi
R^2\over\beta}}\, \cdot \, {\rm exp}\{-\frac{(y-x)^2}{4\beta}\}\,
 \cdot \,\sum_{n=0}^\infty \, c_n(x,y) \, \beta^n \qquad .
\]

\subsection{Kink Seeley densities}

We now describe the iterative procedure that gives the
coefficients $c_n(x,x;K)$ used in the text. For an interesting
interpretation of these coefficients as invariants of the
Korteweg-de Vries equation, see the book on Quantum Mechanics by
Perelomov and Zeldovich.

The recurrence relation
\begin{equation}
(n+1) \, c_{n+1}(x,y)+(x-y) \frac{\partial c_{n+1}(x,y)}{\partial
x}+V(x) c_n(x,y)=\frac{\partial^2 c_n(x,y)}{\partial x^2}
\label{eq:recu8}
\end{equation}
comes from plugging the high-temperature expansion into the transfer
equation. In order to take the limit $y\rightarrow x$ properly, we
introduce the notation
\[
{^{(k)}C}_n(x)=\lim_{y \rightarrow x} \frac{\partial^k
C_n(x,y)}{\partial x^k} \qquad ,
\]
and, after differentiating (\ref{eq:recu8}) $k$ times, we find
\[
{^{(k)} C}_n(x) =\frac{1}{n+k} \left[ \rule{0cm}{0.6cm} \right.
{^{(k+2)} C}_{n-1}(x) - \sum_{j=0}^k {k \choose j}
\frac{\partial^j V(x)}{\partial x^j}\, \, {^{(k-j)} C}_{n-1}(x)
\left. \rule{0cm}{0.6cm} \right].
\]
From this equation and ${^{(k)} C}_0(x)=\lim_{y\rightarrow x}
\frac{\partial^k c_0}{\partial x^k}= \delta^{k0}$, all the
${^{(k)} C}_n(x)$ can be generated recursively . We finally obtain
a well-defined recurrence relation
\[
c_{n+1}(x,x)=\frac{1}{n+1} \left[ {^{(2)} C}_n(x)-V(x) \, c_n(x,x)
\right]
\]
suitable for our purposes.

We give the explicit expressions of the first eight $c_n(x,x)$
kink coefficients. The abbreviated notation is $u_k=\frac{d^k
V}{dx^k}(x)$, $u_k^n=\left(\frac{d^k V}{dx^k}(x)\right)^n$ :
\begin{eqnarray*}
c_1(x,x)&=&u_0 \\\\ c_2(x,x)&=&\frac{1}{2}u_0^2 +
\frac{1}{6}u_2\\\\ c_3(x,x)&=&\frac{1}{6} u_0^3 + \frac{1}{6}u_2
u_0 + \frac{1}{12} u_1^2 +
  \frac{1}{60} u_4\\\\
c_4(x,x)&=&\frac{1}{24}u_0^4 + \frac{1}{12}u_2 u_0^2 +
  \frac{1}{12} u_1^2 u_0 + \frac{1}{60}u_4 u_0 +
  \frac{1}{40}u_2^2+ \frac{1}{30} u_1u_3 +
  \frac{1}{840} u_6\\\\
c_5(x,x)&=&\frac{1}{120} u_0^5 + \frac{1}{36} u_2 u_0^3 +
  \frac{1}{24}u_1^2 u_0^2 + \frac{1}{120} u_4 u_0^2 +
  \frac{1}{40}u_2^2 u_0 +
  \frac{1}{30} u_1u_3 u_0+
  \frac{1}{840} u_6u_0+
  \frac{11}{360}u_1^2 u_2\\&+&
  \frac{23}{5040}u_3^2 +
  \frac{19}{2520}u_2 u_4+
  \frac{1}{280} u_1u_5 +\frac{1}{15120} u_8\\\\
c_6(x,x)&=&\frac{1}{720}u_0^6 + \frac{1}{144} u_2u_0^4 +
  \frac{1}{72} u_1^2 u_0^3 + \frac{1}{360} u_4u_0^3+
  \frac{1}{80} u_2^2 u_0^2 +
  \frac{1}{60} u_1u_3 u_0^2+
  \frac{11}{360} u_1^2 u_2 u_0+
  \frac{1}{280} u_1u_5 u_0\\&+&
  \frac{1}{288} u_1^4 + \frac{1}{15120}u_8 u_0 +
  \frac{61}{15120}u_2^3+
  \frac{43} {2520}u_1u_2 u_3 +
  \frac{23}{5040} u_0 u_3^2+
  \frac{5}{1008}u_1^2 u_4+
  \frac{19}{2520}u_0 u_2 u_4\\ &+&
  \frac{23}{30240} u_4^2+
  \frac{19}{15120} u_3 u_5+
  \frac{1}{1680}u_0^2u_6+
  \frac{11}{15120} u_2 u_6+
  \frac{1}{3780}u_1u_7 + \frac{1}{332640}u_{10}\\\\
c_7(x,x)&=&\frac{1}{5040}u_0^7 + \frac{1}{720} u_2u_0^5 +
  \frac{1}{288} u_1^2 u_0^4 + \frac{1}{240} u_2^2 u_0^3+
  \frac{1}{180} u_1 u_3 u_0^3+
  \frac{11}{720} u_1^2 u_2 u_0^2 +
  \frac{1}{560} u_1 u_5 u_0^2\\&+&
  \frac{1}{288} u_1^4 u_0+ \frac{61}{15120} u_2^3 u_0+
  \frac{43}{2520} u_1 u_2 u_3u_0 + \frac{5}{1008} u_1^2 u_4u_0 +
  \frac{1}{332640}u_{10} u_0 +
  \frac{23}{10080} u_3^2 u_0^2\\&+&
  \frac{19}{5040} u_2 u_4 u_0^2+
  \frac{1}{5040}u_6 u_0^3+
  \frac{83}{10080}u_1^2 u_2^2 +
  \frac{1}{252} u_1^3u_3+
  \frac{31}{10080} u_2 u_3^2+
  \frac{1}{280} u_1 u_3 u_4+
  \frac{1}{1440}u_0^4 u_4\\ &+&
  \frac{5}{2016} u_2^2 u_4+
  \frac{23}{30240} u_0 u_4^2+
  \frac{1}{420} u_1 u_2 u_5+
  \frac{19}{15120} u_0 u_3u_5 +
  \frac{71}{665280} u_5^2+
  \frac{1}{2016}u_1^2 u_6\\&+&
  \frac{11}{15120} u_0 u_2 u_6+
  \frac{61}{332640}u_4u_6+
  \frac{1}{3780}u_0 u_1 u_7+
  \frac{19}{166320} u_3u_7 +
  \frac{1}{30240}u_0^2 u_8+
  \frac{17}{332640}u_2u_8\\ &+&
  \frac{1}{66528}u_1 u_9+
  \frac{1}{8648640}u_{12}\\\\
c_8(x,x)&=&\frac{1}{40320}u_0^8 + \frac{1}{960} u_2^2 u_0^4 +
  \frac{1}{720} u_1 u_3 u_0^4 +
  \frac{1}{576} u_1^4 u_0^2 +
  \frac{1}{252} u_1^3 u_3 u_0+
  \frac{1}{280} u_1 u_3 u_4 u_0+
  \frac{1}{420} u_1 u_2 u_5 u_0\\ &+&
  \frac{31}{10080} u_2u_3^2 u_0 +
  \frac{5}{2016} u_2^2 u_4 u_0+
  \frac{1}{2016}u_1^2 u_6 u_0+
  \frac{1}{8648640}u_{12} u_0 +
  \frac{23}{60480} u_4^2 u_0^2 +
  \frac{19}{30240} u_3 u_5 u_0^2\\ &+&
  \frac{11}{30240}u_2u_6 u_0^2+
  \frac{1}{7560}u_1 u_7u_0^2+
  \frac{11}{2160} u_1^2 u_2 u_0^3 +
  \frac{1}{90720}u_8 u_0^3+ \frac{1}{7200}u_4 u_0^5+
 \frac{1}{1440}u_0^5 u_1^2\\ &+& \frac{1}{4320}u_0^6 u_2+
  \frac{17}{8640} u_1^4 u_2+
  \frac{83}{10080} u_0 u_1^2 u_2^2+
  \frac{61}{30240} u_0^2 u_2^3+
  \frac{1261}{1814400} u_2^4+
  \frac{43}{5040} u_0^2 u_1u_2u_3\\&+& \frac{227}{37800} u_1 u_2^2
    u_3+ \frac{23}{30240} u_0^3 u_3^2 +
  \frac{659}{302400} u_1^2 u_3^2+
  \frac{5}{2016} u_0^2 u_1^2 u_4+
  \frac{19}{15120} u_0^3 u_2u_4+
  \frac{527}{151200} u_1^2 u_2u_4\\&+& \frac{7939}{9979200} u_3^2u_4+
  \frac{6353}{9979200}u_2u_4^2+
  \frac{1}{1680}u_0^3 u_1 u_5+
  \frac{17}{30240} u_1^3 u_5+
  \frac{13}{12320}u_2u_3u_5 +
  \frac{3067}{4989300} u_1 u_4u_5\\&+& \frac{71}{665280} u_0 u_5^2+
  \frac{1}{20160}u_0^4 u_6+
  \frac{3001}{9979200} u_2^2u_6 +
  \frac{13}{29700} u_1 u_3u_6+
  \frac{61}{332640} u_0 u_4u_6\\&+&
  \frac{3433}{259459200} u_6^2+
  \frac{109}{498960} u_1 u_2u_7 +
  \frac{19}{166320} u_0 u_3u_7 +
  \frac{1501}{64864800} u_5u_7+
  \frac{71}{1995840} u_1^2 u_8\\&+&
  \frac{17}{332640} u_0 u_2 u_8 +
  \frac{2003}{129729600} u_4u_8 +
  \frac{1}{66528}u_0 u_1 u_9+
  \frac{5}{648648} u_3u_9+
  \frac{1}{665280}u_0^2u_{10}\\&+&
  \frac{73}{25945920} u_2u_{10}+
  \frac{1}{1441440}u_1 u_{11}+
  \frac{1}{259459200}u_{14}
\end{eqnarray*}

\section{APPENDIX III. Two-component kinks: d=1, N=2}
The operator
\[
K_0=\left(\begin{array}{cc} -\frac{d^2}{dx^2}+4 & 0 \\ 0 &
-\frac{d^2}{dx^2}+\sigma^2 \end{array}\right)
\]
acts on functions $f:{\mathbb S}^1 \, \rightarrow {\mathbb
C}\oplus{\mathbb C}$ from a circle of radius $R=\frac{mL}{2\pi}$,
(PBC), to complex isospinors. The spectral resolution of $K_0$ is:
\[
K_0 \, \left(\begin{array}{c} {\rm exp}\{i {n^{(1)}\over R}\cdot x \} \\
0 \end{array}\right)=\lambda_{n^{(1)}} \,\left( \begin{array}{c}
{\rm exp}\{i {n^{(1)}\over R}\cdot x \} \\ 0
\end{array}\right)\quad , \quad
\lambda_{n^{(1)}}={(n^{(1)})^2\over R^2}+4 \quad , \quad
n^{(1)}\in {\mathbb Z} \qquad ,
\]
\[
K_0 \, \left(\begin{array}{c} 0 \\ {\rm exp}\{i {n^{(2)}\over
R}\cdot x \} \end{array}\right)=\lambda_{n^{(2)}} \,\left(
\begin{array}{c} 0 \\ {\rm exp}\{i {n^{(2)}\over R}\cdot x \}
\end{array}\right)\quad , \quad
\lambda_{n^{(2)}}={(n^{(2)})^2\over R^2}+\sigma^2 \quad , \quad
n^{(2)}\in {\mathbb Z} \qquad ,
\]
and the vacuum energy on the circle is:
\[
\bigtriangleup E_0=\frac{\hbar m}{2}\,\cdot {1\over R} \,\cdot\left[
\sum_{n^{(1)}=-\infty}^\infty \,
\left[(n^{(1)})^2+4R^2\right]^{{1\over
2}}+\sum_{n^{(2)}=-\infty}^\infty \,
\left[(n^{(2)})^2+\sigma^2R^2\right]^{{1\over 2}}\right] \qquad .
\]
Regularization of this divergent quantity by means of the
generalized zeta function affords:
\[
\bigtriangleup E_0(s)={\hbar\over 2}\cdot \left({\mu^2\over m^2}
\right)^s \cdot\mu \, \cdot \zeta_{K_0}(s) \qquad , \qquad
s\in{\mathbb C}
\]
\begin{eqnarray*}
\zeta_{K_0}(s)&=&{\rm Tr}\left[\left(\begin{array}{cc} -{d^2\over
dx^2}+4 & 0 \\ 0 & -{d^2\over
dx^2}+\sigma^2\end{array}\right)\right]^{-s}\\&=&R^{2s}\cdot\left(
\sum_{n^{(1)}-\infty}^\infty \,
\frac{1}{((n^{(1)})^2+4R^2)^s}+\sum_{n^{(2)}=-\infty}^\infty \,
\frac{1}{((n^{(2)})^2+\sigma^2R^2)^s}\right)\\&=&{1\over
4^s}+2R^{2s}\cdot \sum_{n^{(1)}=1}^\infty \,
\frac{1}{((n^{(1)})^2+4R^2)^s}+{1\over\sigma^{2s}}+2R^{2s}\cdot
\sum_{n^{(2)}=1}^\infty \, \frac{1}{((n^{(2)})^2+\sigma^2R^2)^s}
\qquad .
\end{eqnarray*}

\subsection{The TK2 kink generalized zeta function and Epstein zeta functions}

The generalized zeta function is the sum of Epstein zeta
functions,
\[
\zeta_{K_0}(s)=E(s,4|{1\over R^2})+E(s,\sigma^2|{1\over
R^2})\qquad ,
\]
which via the Mellin transform,
\[
E(s,4|A)={1\over\Gamma(s)}\cdot \sum_{n=-\infty}^\infty \,
\int_0^\infty \, d\beta \, \beta^{s-1}e^{-\beta\, (An^2+4)}\qquad
,
\]
and use of the Poisson summation formula
\[
\sum_{n=-\infty}^\infty \, e^{-\beta A n^2}=\sqrt{\frac{\pi}{\beta
A}}\cdot \sum_{l=-\infty}^\infty \, {\rm exp}\{-\frac{\pi^2
l^2}{A\beta}\}\qquad ,
\]
reads:
\[
E(s,4|A)={1\over\Gamma(s)}\cdot\sqrt{{\pi\over
A}}\cdot\left\{\int_0^\infty \, d\beta \, \beta^{s-{3\over 2}}\,
e^{-4\beta}+2\sum_{l=1}^\infty \, \int_0^\infty \, d\beta \,
\beta^{s-{3\over 2}}\, {\rm exp}\{-4\beta-\frac{\pi^2
l^2}{A\beta}\}\right\}\qquad   .
\]
When $R\rightarrow\infty$, $A={1\over R^2}\rightarrow 0$, the
result of Section \S 4 is obtained:
\begin{eqnarray*}
E(s,4|0)+E(s,\sigma^2|0)&=&\lim_{A\rightarrow
0}{1\over\Gamma(s)}\cdot\sqrt{{\pi\over A}}\cdot\int_0^\infty \,
d\beta \, \beta^{s-{3\over 2}}\,
[e^{-4\beta}+e^{-\sigma^2\beta}]\\&=&\lim_{L\rightarrow\infty} \,
\frac{mL}{\sqrt{4\pi}}\cdot [\frac{1}{4^{s-{1\over
2}}}+\frac{1}{\sigma^{2s-1}}]\cdot\frac{\Gamma(s-{1\over
2})}{\Gamma(s)}\qquad .
\end{eqnarray*}
\subsection{The high-temperature expansion of the TK2 kink heat equation
kernel}

We now consider the differential operator
\[
K=K_0+V(x)=\left(\begin{array}{cc} -{d^2\over dx^2}+4-V^{11}(x) &
0 \\ 0 & -{d^2\over dx^2}+\sigma^2-V^{22}(x) \end{array}\right)
\qquad , \,\, V(x)=V(x+2\pi R)
\]
acting on functions from a circle of radius $R$ to ${\mathbb
C}\oplus{\mathbb C}$. If $V(x)=0$, we find the $K_0$-heat kernel:
\begin{eqnarray*}
&&K_{K_0}(x,y;\beta)=\\&=&{\small\left(\begin{array}{cc}
e^{-4\beta}\cdot \, \sum_{n^{(1)}=-\infty}^\infty \, {\rm
exp}\{-{\beta\over R^2}[n^{(1)}+i{R\over 2\beta}(y-x)]^2\} & 0
\\ 0 & e^{-\sigma^2\beta}\, \cdot
\, \sum_{n^{(2)}=-\infty}^\infty \, {\rm exp}\{-{\beta\over
R^2}[n^{(2)}+i{R\over 2\beta}(y-x)]^2\}\end{array}\right)}\\&=&{\rm
exp}\{-\frac{(x-y)^2}{4\beta}\}\times
\\&&{\small\left(\begin{array}{cc} \sum_{n^{(1)}=-\infty}^\infty \,
{\rm exp}\{-\beta({(n^{(1)})^2\over R^2}+4)\}\cdot {\rm
exp}\{i{n^{(1)}\over R}(x-y)\} & 0 \\ 0 &
\sum_{n^{(2)}=-\infty}^\infty \, {\rm exp}\{-\beta({(n^{(2)})^2\over
R^2}+\sigma^2)\}\cdot {\rm exp}\{i{n^{(2)}\over
R}(x-y)\}\end{array}\right)}\,\, \, .
\end{eqnarray*}
Using again the Poisson summation formula,
\[
\sum_{n=-\infty}^\infty \, e^{-t(n+v)^2}=\sqrt{{\pi\over t}} \cdot
\sum_{l=-\infty}^\infty \, {\rm exp}\{-{\pi^2 l^2\over t^2}-2\pi
ilv\} \qquad ; \qquad t={\beta\over R^2} \quad , \quad v=i{R\over
2\beta} (y-x) \qquad ,
\]
the $K_0$-heat kernel can be written as:
\begin{eqnarray*}
K_{K_0}(x,y;\beta)&=& \sqrt{{\pi R^2\over\beta}}\, \cdot \, {\rm
exp}\{-\frac{(y-x)^2}{4\beta}\}\,
 \times \\ && {\small\left(\begin{array}{cc} e^{-4\beta}\,\sum_{l^{(1)}=-\infty}^\infty \,
{\rm exp}\{-\frac{\pi Rl^{(1)}[\pi Rl^{(1)}-(y-x)]}{\beta}\} & 0 \\
0 & e^{-\sigma^2\beta}\, \cdot \, \sum_{l^{(2)}=-\infty}^\infty \,
{\rm exp}\{-\frac{\pi Rl^{(2)}[\pi
Rl^{(2)}-(y-x)]}{\beta}\}\end{array}\right)}\quad .
\end{eqnarray*}
For $\beta<1$, (high-temperature), we obtain the asymptotic formula:
\[
K_{K_0}(x,y;\beta)\simeq  \sqrt{{\pi R^2\over\beta}}\, \cdot \, {\rm
exp}\{-\frac{(y-x)^2}{4\beta}\}\,
 \cdot \,\left(\begin{array}{cc} e^{-4\beta} & 0 \\ 0 & e^{-\sigma^2 \beta}\end{array}\right)\cdot \left(1+{\cal O}({\rm
 exp}\{-\frac{C}{\beta}\})\right)\qquad ,
\]
and from this, the asymptotic high-temperature expansion for the
$K$-heat kernel is derived:
\[
K_K(x,y;\beta)\simeq \sqrt{{\pi R^2\over\beta}}\, \cdot \, {\rm
exp}\{-\frac{(y-x)^2}{4\beta}\}\, \cdot \left(\begin{array}{cc}
e^{-4\beta} & 0 \\ 0 & e^{-\sigma^2 \beta}\end{array}\right)
 \cdot \,\sum_{n=0}^\infty \, c_n(x,y;K) \, \beta^n \qquad .
\]

\subsection{Conserved charges of the $N\times N$ matrix KdV equation }

We now describe the iterative procedure that allows us to compute
the coefficients $c_n^{AB}(x,x;K(c))$ used in the text. We shall
present the diagonal coefficients as the invariant charges of a
$N\times N$ matrix generalization of the Korteweg-de Vries equation.
In this subsection we shall derive the formulas for arbitrary
dimension $N$ of the target space. We shall also deal with a
slightly more general situation in which all the particles have
different masses, i.e. the vacuum fluctuation operator is of the
form:
\[
K_0=\left(\begin{array}{cccc}
-\frac{d^2}{dx^2}+v^2_{(1)} & 0 & \cdots & 0 \\ 0 & -\frac{d^2}{dx^2}+v^2_{(2)} & \cdots & 0 \\
\cdots & \cdots & \cdots &  \cdots \\ 0 & 0 & \cdots &
-\frac{d^2}{dx^2}+v^2_{(N)}
\end{array}\right) \qquad .
\]
The kink moduli space for $N$ scalar fields is of dimension $N$; by
$c$ we shall denote collectively the $N-1$ integration constants
that determine the kink orbits.

We start from the recurrence relations
\begin{eqnarray*}
(n+1)c_{n+1}^{AB}(x,y;K(c))+(x-y)\frac{\partial
c_{n+1}^{AB}}{\partial x}(x,y;K(c))&=&\frac{\partial^2
c_n^{AB}}{\partial
x^2}(x,y;K(c))+(v^2_{(A)}-v^2_{(B)})c_n^{AB}(x,y;K(c))-\\&-&\sum_{C=1}^NV^{AC}(x)c_n^{CB}(x,y;K(c))\label{eq:recur}
\end{eqnarray*}
coming from plugging in the high-temperature expansion
\[
C^{AB}(x,y)={1\over\sqrt{4\pi\beta}}\cdot\sum_{n=0}^\infty
c_n^{AB}(x,y;K(c))\beta^n
\]
in the transfer equation:
\[
\left\{\frac{\partial}{\partial\beta}+\frac{x-y}{\beta}\cdot\frac{\partial}{\partial
x}-\frac{\partial^2}{\partial
x^2}+v^2_{(A)}-v^2_{(B)}\right\}\cdot C^{AB}(x,y)+\sum_{C=1}^N
V^{AC}(x)C^{CB}(x,y)=0^{AB}\qquad .
\]

In order to take the limit $y\rightarrow x$ properly, we introduce
the notation
\[
^{(k)}C_n^{AB}(x)=\lim_{y\rightarrow x}\frac{\partial^k
c_n^{AB}}{\partial x^k}(x,y;K)\qquad , \qquad
 ^{(k)}C_0^{AB}(x)=\lim_{y\rightarrow
x}\frac{\partial^kc_0^{AB}}{\partial
x^k}(x,y;K)=\delta^{k0}\delta^{AB}
\]
and, after differentiating (\ref{eq:recur}) $k$ times, we find:
\[
^{(k)}C_{n+1}^{AB}(x)={1\over
n+k+1}\left[^{(k+2)}C_{n}^{AB}(x)+(v^2_{(A)}-v^2_{(B)})^{(k)}C_n^{AB}(x)-
\sum_{j=0}^k\sum_{C=1}^N\left(\begin{array}{c} k
\\ j
\end{array}\right)\frac{d^jV^{AC}(x)}{d x^j}\cdot
^{(k-j)}C_{n}^{CB}(x)\right] \, .
\]
The recurrence relations become
\[
^{(0)}C_{n+1}^{AB}(x)=\frac{1}{n+1}\left[^{(2)}C_n^{AB}(x)+(v^2_{(A)}-v^2_{(B)})^{(0)}C_n^{AB}(x)-\sum_{C=1}^N
V^{AC}(x)^{(0)}C_n^{CB}(x)\right]
\]
when $y\rightarrow x$. From this equation and knowledge of
$^{(k)}C_0^{AB}(x)$, all the Seeley coefficients can be computed
recursively. For instance, the lowest-order Seeley densities are:
\[
c_0^{AB}(x,x;K(c))=\delta^{AB} \qquad , \qquad
c_1^{AB}(x,x;K(c))=-V^{AB}(x)
\]
\[
c_2^{AB}(x,x;K(c))=-{1\over 6}\frac{d^2V^{AB}}{dx^2}(x)+{1\over
2}\sum_{C=1}^NV^{AC}(x)V^{CB}(x)+{1\over
2}(v^2_{(A)}-v^2_{(B)})V^{AB}(x)
\]
\begin{eqnarray*}
&&c_3^{AB}(x,x;K(c))=-{1\over 60}\frac{d^4V^{AB}}{dx^4}(x)+{1\over
12}\sum_{C=1}^N\left(V^{AC}(x)\frac{d^2V^{CB}}{dx^2}(x)+\frac{d^2V^{AC}}{dx^2}(x)V^{CB}(x)\right)+\\&+&{1\over
12}\sum_{C=1}^N\frac{dV^{AC}}{dx}(x)\frac{dV^{CB}}{dx}(x)-{1\over
6}\sum_{C=1}^N\sum_{D=1}^NV^{AC}(x)V^{CD}(x)V^{DB}(x)+{1\over
6}\sum_{C=1}^N(v^2_{(B)}-v^2_{(C)})V^{AC}(x)V^{CB}(x)+\\&+&
{1\over
12}(v^2_{(A)}-v^2_{(B)})\left(\frac{d^2V^{AB}}{dx^2}(x)+2\sum_{C=1}^N
V^{AC}(x)V^{CB}(x)\right)-{1\over
6}(v^2_{(A)}-v^2_{(B)})^2V^{AB}(x) \, .
\end{eqnarray*}
The diagonal terms $c_n^{AB}(x,x;K)$ are the densities giving the
infinite conserved charges of a $N\times N$ matrix Korteweg-de
Vries equation, namely:
\begin{equation}
\frac{\partial V}{\partial t}(x,t)-3\left(V(x,t)\frac{\partial
V}{\partial x}(x,t)+\frac{\partial V}{\partial
x}(x,t)V(x,t)\right)+\frac{\partial^3 V}{\partial x^3}(x,t)=0
\qquad , \label{eq:KdVN}
\end{equation}
where the matrix potential now evolves in \lq\lq time " $t$,
$V=V(x,t)$. The reason is that this equation can be written in the
 form:
\[
L_t+[L,M]=0 \qquad , \qquad L=-\frac{\partial^2}{\partial
x^2}+V(x,t) \qquad , \qquad M=4\frac{\partial^3}{\partial
x^3}+3V(x,t)\frac{\partial}{\partial x}+3\frac{\partial V}{\partial
x}(x,t)+B(t) \qquad ,
\]
with $B(t)$ an arbitrary matrix of functions of time. Therefore,
standard arguments guarantee that the time evolution ruled by
(\ref{eq:KdVN}) produces a uniparametric isospectral transformation
of the Schrodinger operator $L$. Because the integrals $c_n^{AA}(K)$
are determined by the spectrum of $L$, their invariance follows.

\subsection{TK2 kink Seeley densities }

From the transfer equation for $C(x,y;\beta)$ in a scalar field
theory with two phonon branches of gaps $4$ and $\sigma^2$

{\small
\begin{eqnarray}
\left( \frac{\partial}{\partial \beta}+\frac{x-y}{\beta} \frac{\partial}{\partial x}-\frac{\partial^2}{\partial x^2}+V^{11}(x)-4 \right)C_K^{11}(x,y;\beta)+V^{12}(x) C_K^{21}(x,y;\beta)=0&& \label{eq:cu1} \\
\left( \frac{\partial}{\partial \beta}+\frac{x-y}{\beta} \frac{\partial}{\partial x}-\frac{\partial^2}{\partial x^2}+V^{11}(x)-\sigma^2 \right)C_K^{12}(x,y;\beta)+V^{12}(x) C_K^{22}(x,y;\beta)=0&& \label{eq:cu2} \\
\left( \frac{\partial}{\partial \beta}+\frac{x-y}{\beta} \frac{\partial}{\partial x}-\frac{\partial^2}{\partial x^2}+V^{22}(x)-\sigma^2 \right)C^{21}(x,y;\beta)+V^{12}(x) C_K^{11}(x,y;\beta)=0&& \label{eq:cu3} \\
\left( \frac{\partial}{\partial \beta}+\frac{x-y}{\beta}
\frac{\partial}{\partial x}-\frac{\partial^2}{\partial
x^2}+V^{22}(x)-\sigma^2\right)C_K^{22}(x,y;\beta)+V^{12}(x)
C_K^{12}(x,y;\beta)=0&& \qquad ,
\end{eqnarray}}
we derive the recurrence relations for the Seeley densities:
{\footnotesize
\begin{eqnarray*}
&(n+1) c_{n+1}^{11}(x,y;K)+(x-y) \frac{\partial c_{n+1}^{11}(x,y;K)}{\partial x}-\frac{\partial^2 c_n^{11}(x,y;K)}{\partial x^2}+(V^{11}-4) c_n^{11}(x,y;K)+V^{12} c_n^{21}(x,y;K)=0& \\[0.1cm]
&(n+1) c_{n+1}^{12}(x,y;K)+(x-y) \frac{\partial c_{n+1}^{12}(x,y;K)}{\partial x}-\frac{\partial^2 c_n^{12}(x,y;K)}{\partial x^2}+(V^{11}-\sigma^2) c_n^{12}(x,y;K)+V^{12} c_n^{22}(x,y;K)=0 &\\[0.1cm]
&(n+1) c_{n+1}^{21}(x,y;K)+(x-y) \frac{\partial c_{n+1}^{21}(x,y;K)}{\partial x}-\frac{\partial^2 c_n^{21}(x,y;K)}{\partial x^2}+(V^{22}-4) c_n^{21}(x,y;K)+V^{12} c_n^{11}(x,y;K)=0& \\[0.1cm]
&(n+1) c_{n+1}^{22}(x,y;K)+(x-y) \frac{\partial
c_{n+1}^{22}(x,y;K)}{\partial x}-\frac{\partial^2
c_n^{22}(x,y;K)}{\partial x^2}+(V^{22}-\sigma^2)
c_n^{22}(x,y;K)+V^{12} c_n^{12}(x,y;K)=0& \qquad ,
\end{eqnarray*}}
Denoting
\[
{^{(k)} C}_n^{ij}(x) =\lim_{y\rightarrow x} \frac{\partial^k
c_n^{ij}(x,y;K)}{\partial x^k} \qquad ,
\]
differentiating the recurrence relations above $k$ times, and taking
the $y\rightarrow x$ limit, we obtain new recurrence relations
between the derivatives of the Seeley densities when $y=x$:

{\footnotesize\begin{eqnarray*} {^{(k)} C}_{n+1}^{11}(x) =
\frac{1}{m_{n,k}} \left\{ {^{(k+2)} C}_{n}^{11}(x)-\sum_{j=0}^k {k
\choose j}
\left[ \frac{\partial^j (V^{11}-4)}{\partial x^j} {^{(k-j)} C}_{n}^{11}(x)+\frac{\partial^j V^{12}}{\partial x^j} {^{(k-j)} C}_{n}^{21}(x) \right] \right\} && \\
{^{(k)} C}_{n+1}^{12}(x) = \frac{1}{m_{n,k}} \left\{ {^{(k+2)} C}_{n}^{12}(x)- \sum_{j=0}^k {k \choose j}\left[ \frac{\partial^j (V^{11}-\sigma^2)}{\partial x^j} {^{(k-j)} C}_{n}^{12}(x)+\frac{\partial^j V^{12}}{\partial x^j} {^{(k-j)} C}_{n}^{22}(x) \right] \right\} &&\\
{^{(k)} C}_{n+1}^{21}(x) = \frac{1}{m_{n,k}} \left\{ {^{(k+2)} C}_{n}^{21}(x)- \sum_{j=0}^k {k \choose j} \left[ \frac{\partial^j (V^{22}-4)}{\partial x^j} {^{(k-j)} C}_{n}^{21}(x)+\frac{\partial^j V^{12}}{\partial x^j} {^{(k-j)} C}_{n}^{11}(x) \right] \right\} &&\\
{^{(k)} C}_{n+1}^{22}(x) = \frac{1}{m_{n,k}} \left\{ {^{(k+2)}
C}_{n}^{22}(x)- \sum_{j=0}^k {k \choose j} \left[ \frac{\partial^j
(V^{22}-\sigma^2)}{\partial x^j} {^{(k-j)}
C}_{n}^{22}(x)+\frac{\partial^j V^{12}}{\partial x^j} {^{(k-j)}
C}_{n}^{12}(x) \right] \right\}&& \qquad ,
\end{eqnarray*}}
where $m_{n,k}=n+k+1$. Moreover, the infinite temperature
condition requires that: ${^{(k)} C}_{0}^{ij}(x)=\delta^{k0}
\delta^{ij}$. Thus, the recurrence relations become:
\begin{eqnarray*}
c_{n+1}^{11}(x,x)&=&\frac{1}{n+1} \left[ {^{(2)} C}_{n}^{11}(x)-(V^{11}(x)-4) \, c_n^{11}(x,x)-V^{12}(x) \, c_n^{21}(x,x)  \right] \\
c_{n+1}^{12}(x,x)&=&\frac{1}{n+1} \left[ {^{(2)} C}_{n}^{12}(x)-(V^{11}(x)-\sigma^2) \, c_n^{12}(x,x)-V^{12}(x) \, c_n^{22}(x,x)  \right] \\
c_{n+1}^{21}(x,x)&=&\frac{1}{n+1} \left[ {^{(2)} C}_{n}^{21}(x)-(V^{22}(x)-4) \, c_n^{21}(x,x)-V^{12}(x) \, c_n^{11}(x,x)  \right] \\
c_{n+1}^{22}(x,x)&=&\frac{1}{n+1} \left[ {^{(2)}
C}_{n}^{22}(x)-(V^{22}(x)-\sigma^2) \, c_n^{22}(x,x)-V^{12}(x) \,
c_n^{12}(x,x)  \right] \qquad ,
\end{eqnarray*}
which must be solved iteratively. Up to third order, the Seeley
densities are:

{\small\[ c_0^{AB}(x,x)=\delta^{AB}
\]
\begin{eqnarray*}
c_1^{11}(x,x)&=&-({V}^{11}(x)-4) \\
c_1^{12}(x,x)&=&-{V}^{12}(x)\\
c_1^{21}(x,x)&=&-{V}^{12}(x) \\
c_1^{22}(x,x)&=&-({V}^{22}(x)-\sigma^2)
\end{eqnarray*}
\begin{eqnarray*}
c_2^{11}(x,x)&=&-\frac{1}{6} \frac{\partial^2 V^{11}}{\partial
x^2}+\frac{1}{2}
(V^{11}(x)-4)^2 +\frac{1}{2}V^{12}(x) V^{12}(x) \\
c_2^{12}(x,x)&=&-\frac{1}{6} \frac{\partial^2 V^{12}}{\partial
x^2}+\frac{1}{2} V^{12}(x)\left[ V^{11}(x)+ V^{22}(x)-2 \sigma^2\right]\\
c_2^{21}(x,x)&=&-\frac{1}{6} \frac{\partial^2 V^{12}}{\partial
x^2}+\frac{1}{2} V^{12}(x)\left[
(V^{11}(x)+ V^{22}(x)-8 \right]\\
c_2^{22}(x,x)&=&-\frac{1}{6} \frac{\partial^2 V^{22}}{\partial
x^2}+\frac{1}{2} (V^{22}(x)-\sigma^2)^2 +\frac{1}{2}V^{12}(x)
V^{12}(x)
\end{eqnarray*}
\begin{eqnarray*}
c_3^{11}(x,x) &=& -\frac{1}{60} \frac{\partial^4 V^{11}}{\partial
x^4} +\frac{1}{6} (V^{11}(x)-4) \frac{\partial^2 V^{11}}{\partial
x^2}+\frac{1}{6} V^{12}(x) \frac{\partial V^{12}}{\partial x^2} +
\frac{1}{12} \frac{\partial V^{11}}{\partial x} \frac{\partial
V^{11}}{\partial x}+\\
&&+  \frac{1}{12} \frac{\partial V^{12}}{\partial x}
\frac{\partial V^{12}}{\partial x}-\frac{1}{6} (V^{12}(x))^2
(V^{22}(x)-4)-\frac{1}{6} (V^{11}(x)-4)^3 \\
&& -\frac{1}{3} (V^{12}(x))^2 (V^{11}(x)-4)\\
c_3^{22}(x,x) &=& -\frac{1}{60} \frac{\partial^4 V^{22}}{\partial
x^4} +\frac{1}{6} (V^{22}(x)-\sigma^2) \frac{\partial^2
V^{22}}{\partial x^2}+\frac{1}{6} V^{12}(x) \frac{\partial
V^{12}}{\partial x^2} + \frac{1}{12} \frac{\partial
V^{22}}{\partial x} \frac{\partial V^{22}}{\partial x}+\\ &&+
\frac{1}{12} \frac{\partial V^{12}}{\partial x} \frac{\partial
V^{12}}{\partial x}-\frac{1}{6} (V^{12}(x))^2
(V^{11}(x)-\sigma^2)-\frac{1}{6}
(V^{22}(x)-\sigma^2)^3 \\
&& -\frac{1}{3} (V^{12}(x))^2 (V^{22}(x)-\sigma^2)
\end{eqnarray*}}

\section{APPENDIX IV. Self-dual Vortices: d=2, N=4}

At the self-dual limit $\kappa^2=1$ there are two PD operators
ruling the small fluctuations around the vacuum of the Higgs,
Goldstone, vector, and ghost fields. In the $R$-gauge these
operators are:
\[
K_0=\left(\begin{array}{cccc} -{\partial^2\over\partial
x_1^2}-{\partial\over\partial x_2^2}+1 & 0 & 0 & 0 \\ 0 &
-{\partial^2\over\partial x_1^2}-{\partial\over\partial x_2^2}+1 &
0 & 0 \\ 0 & 0 & -{\partial^2\over\partial
x_1^2}-{\partial\over\partial x_2^2}+1 & 0 \\ 0 & 0 & 0 &
-{\partial^2\over\partial x_1^2}-{\partial\over\partial x_2^2}+1
\end{array}\right) \qquad \qquad ,
\]
\[
K_0^G=-{\partial^2\over\partial x_1^2}-{\partial\over\partial
x_2^2}+1 \hspace{9.4cm}\qquad \qquad .
\]
$K_0$ acts on isospinors $f:S^1\times S^1 \longrightarrow {\mathbb
C}^4$ from a torus of area $A=4\pi R^2=m^2L^2$ to ${\mathbb C}^4$,
whereas $K_0^G$ acts on scalar functions $f: S^1\times S^1
\longrightarrow {\mathbb C}$ from the same torus to ${\mathbb C}$.
The eigenfunctions of both $K_0$ and $K_0^G$ are plane waves of
discrete momenta:
\[
K_0^{AA}{\rm exp}\left\{i{n_1^{(A)}\over R}x_1+i{n_2^{(A)}\over
R}x_2\right\}u^A=\lambda_{\vec{n}^{(A)}}{\rm
exp}\left\{i{n_1^{(A)}\over R}x_1+i{n_2^{(A)}\over
R}x_2\right\}u^A \quad ,
\]
\[
K_0^G{\rm exp}\left\{i{n_1\over R}x_1+i{n_2\over
R}x_2\right\}=\lambda_{\vec{n}}{\rm exp}\left\{i{n_1\over
R}x_1+i{n_2\over R}x_2\right\} \qquad .
\]
The eigenvalues are
\[
\lambda_{\vec{n}^{(A)}}={(n_1^{(A)})^2\over
R^2}+{(n_2^{(A)})^2\over R^2}+1 \qquad \qquad , \qquad \qquad
\lambda_{\vec{n}}={n_1^2\over R^2}+{n_2^2\over R^2}+1
\]
and the vacuum energy on the finite torus (no infrared
divergences) is:
\[
\Delta E_0={\hbar m\over 2}\cdot{\rm Tr} K_0^{1\over 2}-{\hbar
m\over 2}\cdot{\rm Tr} (K_0^G)^{1\over 2}=3{\hbar m\over
2}\sum_{n_1=-\infty}^\infty\sum_{n_2=-\infty}^\infty[{n_1^2\over
R^2}+{n_2^2\over R^2}+1]^{1\over 2} \qquad ,
\]
coming from small fluctuations on the vacuum of the Higgs field and
the two physical polarizations of the vector field. The fluctuations
of the temporal polarization of the vector field and the Goldstone
field are canceled by the ghost fluctuations, thus restoring the
unitarity, that was lost in the combined Weyl/$R$-gauge.

Zeta function regularization of this ultraviolet divergent quantity
leads us to replace $\Delta E_0$ by the meromorphic function of the
complex variable $s$:
\[
\Delta E_0(s)={\hbar\over 2}\left({\mu^2\over m^2}\right)^s
\cdot\mu \cdot\zeta_{K_0}(s)-\hbar \left({\mu^2\over m^2}\right)^s
\cdot \mu \cdot \zeta_{K_0^G}(s)=3{\hbar \over 2}\left({\mu^2\over
m^2}\right)^s \cdot\mu \cdot\sum_{\vec{n}\in{\mathbb
Z}^2}\,{1\over [{n_1^2\over R^2}+{n_2^2\over R^2}+1]^s}
\]
and take as the regularized finite value of the vacuum energy the
value obtained by analytic continuation of the zeta functions.

\subsection{The vortex generalized zeta function versus Epstein zeta functions}

Therefore, the vacuum energy is regularized by means of Epstein
zeta functions:
\[
\zeta_{K_0}(s)=\sum_{A=1}^4\sum_{\vec{n}^{(A)}\in{\mathbb Z}^2} \,
\frac{1}{[\frac{n_1^{(A)}n_1^{(A)}}{R^2}+\frac{n_2^{(A)}n_2^{(A)}}{R^2}+1]^s}=\sum_{A=1}^4\,
E(s,1|(\frac{\vec{e}_1}{R^2}+\frac{\vec{e}_2}{R^2})u^A)
\]
\[
\zeta_{K_0^G}(s)=\sum_{\vec{n}\in{\mathbb Z}^2} \,
\frac{1}{[\frac{n_1^2}{R^2}+\frac{n_2^2}{R^2}+1]^s}=E(s,1|\frac{\vec{e}_1}{R^2}+\frac{\vec{e}_2}{R^2})
\qquad , \qquad \Delta E_0(s)=3{\hbar \over 2}\left({\mu^2\over
m^2}\right)^s \cdot\mu \cdot
E(s,1|\frac{\vec{e}_1}{R^2}+\frac{\vec{e}_2}{R^2}) \quad .
\]
Via the Mellin transform
\[
E(s,1|\frac{\vec{e}_1}{R^2}+\frac{\vec{e}_2}{R^2})={1\over\Gamma(s)}\,\sum_{\vec{n}\in{\mathbb
Z}^2}\, \int \, d\beta \, \beta^{s-1}\,
e^{-\beta\sum_{\vec{n}\in{\mathbb Z}^2}({\vec{n}\cdot\vec{n}\over
R^2}+1)} \qquad , \qquad \vec{n}=n_1\vec{e}_1+n_2\vec{e}_2
\]
and use of the Poisson summation formula
\[
\sum_{\vec{n}\in{\mathbb Z}^2}\, e^{-\beta {\vec{n}\cdot\vec{n}\over
R^2}}={R^2\pi\over \beta}\cdot \sum_{\vec{l}\in{\mathbb Z}^2}\,
e^{-{R^2\pi^2 \vec{l}\cdot\vec{l}\over \beta}} \qquad , \qquad
\vec{l}=l_1\vec{e}_1+l_2\vec{e}_2 \quad , \quad l_1, l_2\in{\mathbb
Z} \qquad ,
\]
the Epstein zeta function reads:
\[
E(s,1|\frac{\vec{e}_1}{R^2}+\frac{\vec{e}_2}{R^2})=R^2\pi{1\over\Gamma(s)}\cdot
\int_0^\infty \, d\beta \, \beta^{s-2}\,
e^{-\beta}+2\sum_{\vec{l}\in{\mathbb Z}^+\otimes{\mathbb Z}^+}\,
\int \, d\beta \, \beta^{s-2}\, e^{-\beta}\,e^{-{R^2\pi^2
\vec{l}\cdot\vec{l}\over \beta}} \qquad .
\]
At the limit of infinite area only the first term survives and we
obtain:
\[
E(s,1|\vec{0})=\lim_{R\rightarrow\infty}R^2\pi{\Gamma(s-1)\over\Gamma(s)}=\lim_{L\rightarrow\infty}{m^2L^2\over
4\pi}\cdot{\Gamma(s-1)\over\Gamma(s)} \qquad .
\]
Thus, in the Euclidean plane the regularized vacuum energy is:
\[
\Delta E_0=\lim_{s\rightarrow -{1\over 2}}3{\hbar \over
2}\left({\mu^2\over m^2}\right)^s \cdot\mu \cdot
\lim_{L\rightarrow\infty}{m^2L^2\over
4\pi}\cdot{\Gamma(s-1)\over\Gamma(s)} \qquad .
\]
Note that ${\Gamma(-{3\over 2})\over\Gamma(-{1\over 2})}=-{2\over
3}$ is a regular value of $\Delta E_0(s)$.

\subsection{The high-temperature expansion of the vortex heat equation kernel}

Let us consider now the PD differential operators
$K=K_0+Q_k(\vec{x}){\partial\over\partial x^k}+V(\vec{x})$ and
$K^G=K_0^G+V^G(\vec{x})$:
\[
Q_k(x_1,x_2)=Q_k(x_1+2\pi R,x_2+2\pi R)  \qquad , \qquad
Q_k(\vec{x}){\partial\over\partial x^k}=\left(\begin{array}{cccc}
0 & 0 & 0 & 0 \\ 0 & 0 & 0 & 0 \\ 0 & 0 & 0 &
-2V_k(\vec{x}){\partial\over\partial x^k} \\ 0 & 0 &
-2V_k(\vec{x}){\partial\over\partial x^k} & 0 \end{array}\right)
\]
\[
V(\vec{x})=\left(\begin{array}{cccc} |s(\vec{x})|^2-1 & 0 &
-2\nabla_1s_2(\vec{x}) & 2\nabla_1s_1(\vec{x})\\ 0 & |s(\vec{x})|^2-1 & -2\nabla_2s_2(\vec{x}) & 2\nabla_2s_1(\vec{x})\\
-2\nabla_1s_2(\vec{x}) & -2\nabla_2s_2(\vec{x}) & {3\over
2}(|s(\vec{x})|^2-1)+V_k(\vec{x})V_k(\vec{x}) & 0 \\
2\nabla_1s_1(\vec{x}) & 2\nabla_2s_1(\vec{x}) & 0 & {3\over
2}(|s(\vec{x})|^2-1)+V_k(\vec{x})V_k(\vec{x})
\end{array}\right)
\]
\[
V(x_1,x_2)=V(x_1+2\pi R,x_2+2\pi R) \quad ; \quad
V^G(\vec{x})=|s(\vec{x})|^2-1 \quad , \quad
V^G(x_1,x_2)=V^G(x_1+2\pi R,x_2+2\pi R)\qquad .
\]
$K$ acts on isospinors $f:S^1\times S^1 \longrightarrow {\mathbb
C}^4$ from a torus of area $A=4\pi R^2=m^2L^2$ to ${\mathbb C}^4$
whereas $K^G$ acts on scalar functions $f: S^1\times S^1
\longrightarrow {\mathbb C}$ from the same torus to ${\mathbb C}$.
The $K_0$ and $K_0^G$ heat kernels are:
\begin{eqnarray*}
K_{K_0}^{AA}(\vec{x},\vec{y};\beta)&=&
\sum_{\vec{n}^{(A)}\in{\mathbb Z}^2}{\rm
exp}\{-\beta(\frac{\vec{n}^{(A)}\cdot\vec{n}^{(A)}}{R^2}+1)\}\cdot{\rm
exp}\{i\frac{\vec{n}^{(A)}}{R}\cdot
(\vec{x}-\vec{y})\}\\&=&e^{-\beta}{\rm
exp}\{-\frac{|\vec{x}-\vec{y}|^2}{4\beta}\}\cdot\sum_{\vec{n}^{(A)}\in{\mathbb
Z}^2}{\rm exp}\{-{\beta\over R^2}\cdot (\vec{n}^{(A)}+i{R\over
2\beta}(\vec{x}-\vec{y}))\cdot (\vec{n}^{(A)}+i{R\over
2\beta}(\vec{x}-\vec{y}))\}
\end{eqnarray*}
\begin{eqnarray*}
K_{K_0^G}(x,y;\beta)&=&\sum_{\vec{n}\in{\mathbb Z}^2}\, {\rm
exp}\{-\beta({|\vec{n}|^2\over R^2}+1)\}\cdot {\rm
exp}\{i{\vec{n}\over R}(\vec{x}-\vec{y})\}\\&=& e^{-\beta}{\rm
exp}\{-\frac{|\vec{x}-\vec{y}|^2}{4\beta}\} \, \cdot \,
\sum_{\vec{n}\in{\mathbb Z}^2}{\rm exp}\{-{\beta\over R^2}\cdot
(\vec{n}+i{R\over 2\beta}(\vec{x}-\vec{y}))\cdot (\vec{n}+i{R\over
2\beta}(\vec{x}-\vec{y}))\}  \qquad .
\end{eqnarray*}

Using again the Poisson summation formula,
\[
\sum_{\vec{n}\in{\mathbb Z}^2}\, e^{-t|\vec{n}+\vec{v}|^2}={\pi\over
t}\sum_{\vec{l}\in{\mathbb Z}^2}{\rm
exp}\left\{-\frac{\pi^2\vec{l}\cdot\vec{l}}{t^2}-2\pi
i\vec{l}\cdot\vec{v}\right\} \quad ; \quad t={\beta\over R^2} \quad
, \quad \vec{v}=i{R\over 2\beta}(\vec{y}-\vec{x})\qquad ,
\]
the $K_0$ and $K_0^G$ heat kernels can be written as:
\[
K_{K_0}^{AA}(\vec{x},\vec{y};\beta)={\pi
R^2\over\beta}e^{-\beta}{\rm
exp}\left\{-\frac{|\vec{x}-\vec{y}|^2}{4\beta}\right\}\cdot
\sum_{\vec{l}^{(A)}\in{\mathbb Z}^2}\, {\rm exp}\left\{-\frac{\pi
R\vec{l}^{(A)}\cdot(\pi
R\vec{l}^{(A)}-(\vec{y}-\vec{x}))}{\beta}\right\}\qquad .
\]
\[
K_{K_0^G}(\vec{x},\vec{y};\beta)={\pi R^2\over\beta}e^{-\beta}{\rm
exp}\left\{-\frac{|\vec{x}-\vec{y}|^2}{4\beta}\right\}\cdot
\sum_{\vec{l}\in{\mathbb Z}^2}\, {\rm exp}\left\{-\frac{\pi
R\vec{l}\cdot(\pi
R\vec{l}-(\vec{y}-\vec{x}))}{\beta}\right\}\qquad .
\]

For $\beta<1$, (high-temperature), we obtain the asymptotic
formulas
\[
K_{K_0}(\vec{x},\vec{y};\beta)={\pi R^2\over\beta}e^{-\beta}{\rm
exp}\left\{-\frac{|\vec{x}-\vec{y}|^2}{4\beta}\right\}\cdot
\left({\mathbb I}+{\cal O}({\rm exp}(-{C\over\beta}){\mathbb
I})\right)\qquad ,
\]
\[
K_{K_0^G}(\vec{x},\vec{y};\beta)={\pi R^2\over\beta}e^{-\beta}{\rm
exp}\left\{-\frac{|\vec{x}-\vec{y}|^2}{4\beta}\right\}\cdot
\left(1+{\cal O}({\rm exp}(-{C\over\beta}))\right)\qquad ,
\]
and from these, the asymptotic high-temperature expansions for the K
and $K^G$ heat kernels are derived:
\[
K_{K}^{AB}(\vec{x},\vec{y};\beta)={\pi
R^2\over\beta}e^{-\beta}{\rm
exp}\left\{-\frac{|\vec{x}-\vec{y}|^2}{4\beta}\right\}\cdot
\sum_{n=0}^\infty \, c_n^{AB}(\vec{x},\vec{y};K)\beta^n \qquad ,
\]
\[
K_{K^G}(\vec{x},\vec{y};\beta)={\pi R^2\over\beta}e^{-\beta}{\rm
exp}\left\{-\frac{|\vec{x}-\vec{y}|^2}{4\beta}\right\}\cdot
\sum_{n=0}^\infty \, c_n^G(\vec{x},\vec{y};K)\beta^n \qquad .
\]

\subsection{Spherically symmetric vortex Seeley densities}

Defining $\theta_l=(l-1)\theta$ and $\Lambda(r)=[1-\alpha(r)]$, the
Hessians $K$ and $K^G$ for spherically symmetric vortex solutions
read:
\[
K= \left(\begin{array}{cccc} K^{11} & K^{12} & K^{13} & K^{14} \\
K^{21} & K^{22} & K^{23} & K^{24} \\ K^{31} & K^{32} & K^{33} &
K^{34} \\ K^{41} & K^{42} & K^{43} & K^{44}
\end{array}\right) \quad , \quad K^G=-\frac{\partial^2}{\partial r^2}-{1\over
r^2}\frac{\partial^2}{\partial\theta^2}+f^2(r)
\]
\[
K^{11}=K^{22}=-\frac{\partial^2}{\partial r^2}-{1\over
r^2}\frac{\partial^2}{\partial\theta^2}+f^2(r) \qquad , \qquad
K^{12}=K^{21}=0
\]
\[
K^{24}=K^{42}=2\frac{lf(r)}{r}\Lambda(r){\rm
sin}\theta_l=-K^{13}=-K^{31}\quad , \quad
K^{14}=K^{41}=2\frac{lf(r)}{r}\Lambda(r){\rm
cos}\theta_l=-K^{23}=-K^{32}
\]
\[
K^{33}=K^{44}=-\frac{\partial^2}{\partial r^2}-{1\over
r^2}\frac{\partial^2}{\partial\theta^2}+{3\over
2}f^2(r)+\frac{l^2\alpha(r)}{r^2}-{1\over 2} \qquad , \qquad
K^{34}=K^{43}=-2\frac{lf(r)}{r^2}{\partial\over\partial\theta}
\qquad ,
\]
where the first-order equations have been used to write field
derivatives in terms of field profiles
\begin{eqnarray*} \frac{\partial s_1}{\partial x_1}&=&\frac{l
f(r)}{r}\left[\cos\theta \cos l\theta(1-\alpha(r))+\sin\theta \sin
l\theta \right]\quad , \quad \frac{\partial s_2}{\partial
x_2}=\frac{l f(r)}{r}\left[\sin\theta \sin l\theta(1-\alpha(r))+
\cos\theta
\cos l\theta \right]\\
\frac{\partial s_1}{\partial x_2}&=&\frac{l f(r)}{r}\left[\sin\theta
\cos l\theta(1-\alpha(r))- \cos\theta \sin l\theta \right]\quad ,
\quad \frac{\partial V_1}{\partial x_1}=\sin\theta \cos
\theta\left[\frac{2l f(r)\alpha(r)}{r}+{1\over 2}(f^2(r)-1)\right]\\
\frac{\partial s_2}{\partial x_1}&=&\frac{l f(r)}{r}\left[\cos\theta
\sin l\theta(1-\alpha(r))- \sin\theta \cos l\theta \right]\quad ,
\quad \frac{\partial V_1}{\partial x_2}=-l \cos
2\theta\frac{\alpha(r)}{r^2}+{1\over 2}\sin^2\theta(f^2(r)-1)\\
\frac{\partial V_2}{\partial x_1}&=&-l \cos
2\theta\frac{\alpha(r)}{r^2}-{1\over 2}
\cos^2\theta(f^2(r)-1)\hspace{1.35cm} , \quad  \frac{\partial
V_2}{\partial x_2}=-\sin\theta \cos \theta\left[\frac{2l
f(r)\alpha(r)}{r}+{1\over 2}(f^2(r)-1)\right]
\end{eqnarray*}

The Seeley densities for spherically symmetric vortex and ghosts
are:

{\normalsize\begin{eqnarray*}
{\rm tr} [c_1](\vec{x},\vec{x};K)&=&5[1-f^2(r)]-\frac{2}{r^2} l^2 \alpha^2(r)
\\
{\rm tr}[c_2](\vec{x},\vec{x};K)&=& \frac{1}{12r^4}\left\{ 37 r^4+
4 l^4 \alpha^4 (r)+8(7l^2r^2-8r^4)f^2(r)+27r^4 f^4(r)- \right.
\\
 &-&\left. 8lr^2 \alpha(r)[-1+(1+13l)f^2(r)]+8l^2\alpha^2(r)(-2-3r^2+9r^2f^2(r))\right\}
\\
{\rm tr}[c_3](\vec{x},\vec{x};K)&=& \frac{1}{120 r^6} \left\{
-4l^6 \alpha^6(r)-4 l^3 r^2 \alpha^3(r) [14 +(-132+167
l)f^2(r)]+4l^4 \alpha^4(r)(20+9r^2+32r^2f^2(r))\right.\\
&-&2lr^2
\alpha(r) [-4(16+9r^2)+
 + (64+96 l-472 l^2+344 l^3+88l^2+243 l r^2)
f^2(r)\\&+&(-52+109l)r^2 f^4(r)]+ l^2 \alpha^2(r) [-256-144
r^2-117r^4\\&+&2r^2(88-548 l+516l^2+183r^2)f^2(r)+99r^4
f^4(r)]+r^2[r^2(-16+151r^2)\\&+&(-320l^3+160l^4+32r^2+48lr^2-321 r^4 +
8l^2(20+39r^2))f^2(r)\\&+&\left. r^2(-16-48
l+44l^2+199r^2)f^4(r)-29r^4f^6(r)] \right\}
\\
c_1(\vec{x},\vec{x};K^G)&=& 1-f^2(r)\\
c_2(\vec{x},\vec{x};K^G)&=&-\frac{1}{6r^2}\left\{ [4 l^2+ 5 r^2-8
l^2
\alpha(r) +4 l^2 \alpha^2(r)]f^2(r)
 -3 r^2-2r^2f^4(r) \right\} \\
c_3(\vec{x},\vec{x};K^G)&=& \frac{1}{60 r^4} \left\{ 10 r^4-[-32
l^3+16 l^4+8lr^2+23 r^4+  16 l^2 (1+r^2)\right. \\&-&
8l(-12l^2+8l^3+r^2+4l(1+r^2))\alpha(r)
+ 16
l^2(1-6l+6l^2+r^2)\alpha^2(r)\\&+& 32(1-2l)l^3\alpha^3(r)+
16l^4\alpha^4(r)]f^2(r)+ r^2[8l+16l^2+17r^2+16l^2\alpha^2(r)-
\\ &&-8l(1+4l)\alpha(r)\left. ] f^4(r)-4r^4f^6(r)
\right\} \qquad .
\end{eqnarray*}}
The Seeley coefficients are obtained from numerical integration over
the whole plane of the above densities when the field profiles
$f(r)$ and $\alpha(r)$ of the vortex solutions are plugged in.

\subsection{Renormalization of one-loop divergent graphs in the
planar Abelian Higgs model}

In this last Appendix we shall present detailed calculations of
Feynman amplitudes for one-loop divergent graphs in the Abelian
Higgs model. The results obtained here have been used in sub-section
\S. 7.3.

\subsubsection{The Higgs boson tadpole}
\begin{itemize}

\item Application of the Feynman rules gives the following
Feynman amplitude to the contribution of a Higgs loop to the Higgs
tadpole:

\parbox{3.5cm}{\epsfig{file=qsdv15.ps,width=2.5cm}}
$\displaystyle \equiv \hspace{1cm}-{3 \over 2}i\kappa^2\cdot\int
\frac{d^3 k }{(2\pi)^3} \frac{i}{k^2-\kappa^2 + i
\epsilon}=-{3\over 2}i\kappa^2\cdot I(\kappa^2)$ \qquad .

A combinatorial factor for this graph of ${1\over 2}$ has been
taken into account. The Mc Laurin expansion
\[
{1\over x+y}=\sum_{n=0}^\infty \, (-1)^n \frac{x^n}{y^{n+1}}
\]
of
\[
\frac{1}{k^2-\kappa^2+i\varepsilon}=\frac{1}{k^2-1+i\varepsilon}\sum_{n=0}^\infty
\, (-1)^n \frac{(1-\kappa^2)^n}{[k^2-1+i\varepsilon]^n} \qquad ,
\qquad x=1-\kappa^2 \quad , \quad y=k^2-1+i\varepsilon
\]
allows us to write:
\[
I(\kappa^2)=I(1)+\sum_{n=1}^\infty \, (\kappa^2-1)^n \cdot \int
\frac{d^3 k }{(2\pi)^3} \cdot \frac{i}{[k^2-\kappa^2 + i
\epsilon]^{n+1}} \qquad \qquad .
\]
Because all the integrals in the sum - except $I(1)$ - are
convergent, we can safely conclude:
\[
I(\kappa^2)=I(1)+ {\rm finite}\hspace{0.2cm} {\rm part} \qquad ,
\]
which is a useful relation if a minimal subtraction scheme is to be
used.

\item Similar calculations show that the Feynman amplitude of the
Higgs tadpole due to a Goldstone loop is:

\parbox{3.5cm}{\epsfig{file=qsdv16.ps,width=2.5cm}}
$\displaystyle \equiv \hspace{1cm} -{1 \over 2}i\kappa^2 \cdot
\int \frac{d^3 k }{(2\pi)^3}\frac{i}{k^2-1 + i\epsilon}=-{1\over
2}i\kappa^2\cdot I(1)$ \qquad .

There is also a combinatorial factor of ${1\over 2}$ and the two
differences are that the trivalent Higgs-Goldstone-Goldstone vertex
does not have a factor of three and the Goldstone propagator has
poles at $k_0=\pm \sqrt{\vec{k}\vec{k}+1}$, contrary to the Higgs
propagator with poles at $k_0=\pm \sqrt{\vec{k}\vec{k}+\kappa^2}$.

\item The Feynman amplitude of the Higgs tadpole due to a ghost
loop is:

\parbox{3.5cm}{\epsfig{file=qsdv17.ps,width=2.5cm}}
$\displaystyle \equiv \hspace{1cm}-(-1)i\cdot \int \frac{d^3
k}{(2\pi)^3} \frac{i}{k^2-1+i \epsilon}=i\cdot I(1)$ \qquad .

There is no combinatorial factor and a minus sign is included, as
corresponds to all fermionic loops.

\item The Feynman amplitude of the Higgs tadpole due to a vector
boson loop is:

\parbox{3.5cm}{\epsfig{file=qsdv18.ps,width=2.5cm}}
$\displaystyle\equiv \hspace{1cm} {2  \over 2} ig^{\mu
\nu}\cdot\int \frac{d^3 k }{(2\pi)^3}\frac{- i g_{\nu \mu}}{k^2-1+
i \epsilon}=-3i\cdot I(1)$ \qquad .

There is a combinatorial factor of ${1\over 2}$ and
$g^{\mu\nu}g_{\nu\mu}=3$ in $(2+1)$-dimensional Minkowski
space-time.

\end{itemize}
Adding the four summands, the result used in subsection \S. 7.3 is
obtained: $-2i(\kappa^2+1)\cdot I(1)+{\rm finite}\hspace{0.2cm} {\rm
part}$.

\subsubsection{The Higgs boson self-energy}

\begin{itemize}

\item Identical calculations to those performed in the previous
subsection give the Feynman amplitude for the self-energy of the
Higgs boson due to a Higgs loop:

\parbox{3.5cm}{\epsfig{file=qsdv19.ps,width=2.5cm}}
$\displaystyle \equiv \hspace{1cm} -{3 \over 2}i \kappa^2\cdot\int
\frac{d^3 k }{(2\pi)^3} \frac{i}{k^2-\kappa^2 + i \epsilon}=-{3
\over 2}i \kappa^2\cdot I(\kappa^2)$ \qquad .

There is also a ${1\over 2}$ combinatorial factor and now a
four-valent vertex of four Higgs particles replaces the former
three-valent vertex.

\item The Feynman amplitude for the Higgs self-energy produced by
Goldstone loops is almost the same:

\parbox{3.5cm}{\epsfig{file=qsdv20.ps,width=2.5cm}}
$\displaystyle\equiv\hspace{1cm}-{1 \over 2}i \kappa^2\cdot \int
\frac{d^3 k }{(2\pi)^3}\frac{i}{k^2-1 + i\epsilon}=-{1 \over 2}i
\kappa^2\cdot I(1)$ \qquad .

The only differences come from a different four-valent vertex of two
Higgs and two Goldstone particles and a different propagator:
Goldstone instead Higgs.

\item  A vector boson loop in the Higgs self-energy is easily
dealt with. The Feynman amplitude reads:

\parbox{3.5cm}{\epsfig{file=qsdv21.ps,width=2.5cm}}
$\displaystyle\equiv\hspace{1cm} {2  \over 2}i g^{\mu \nu}\cdot
\int \frac{d^3 k }{(2\pi)^3}\frac{- i g_{\nu \mu}}{k^2-1+ i
\epsilon}=-3i\cdot I(1)$ \qquad .

\item The computation of the Feynman amplitude of the Higgs self-energy due to the only divergent
two-vertex graph with two external Higgs legs is more difficult:

\parbox{3.5cm}{\epsfig{file=qsdv22.ps,width=2.5cm}}
$\displaystyle \equiv\hspace{1cm}   \int \frac{d^3
k}{(2\pi)^3}\cdot \frac{ (p^\mu+k^\mu)\cdot -ig_{\mu\nu}\cdot
i\cdot (-k^\nu-p^\nu) }{[(p-k)^2-1+i \epsilon] [k^2 -1 + i
\epsilon]}$

\[
\hspace{2.cm}\displaystyle\equiv \hspace{1cm}i  \int \frac{d^3
k}{(2\pi)^3} \frac{i (p+k)^2 }{[(p-k)^2-1+i \epsilon] [k^2 -1 + i
\epsilon]}=i\cdot I(p ;1) \qquad .
\]
Note, however, that:
\begin{eqnarray*}
I(p ;1)=I(1)&+&I_2(p)+I_3(p ;1)\\=I(1)&+&\int \frac{d^3
k}{(2\pi)^3}\frac{4i(pk)}{[(p-k)^2-1+i \epsilon] [k^2 -1 + i
\epsilon]}\\&+&\int \frac{d^3
k}{(2\pi)^3}\frac{i(1-i\varepsilon)}{[(p-k)^2-1+i \epsilon] [k^2
-1 + i \epsilon]} \,\, \qquad .
\end{eqnarray*}
$I_3(p;1)$ is convergent but it is not obvious that $I_2(p)$ is also
convergent {\footnote{It could be guessed that $I_2(p)$ is indeed
convergent because the integrand is odd in the loop momenta $k$.}}.
We use the Feynman parametrization
\[
{1\over ab}=\int_0^1 \frac{dx}{[ax+b(1-x)]^2}
\]
to write $I_2(p)$ in the form:
\[
I_2(p)=\int \frac{d^3 k}{(2\pi)^3}\int_0^1 \, dx
\frac{4i(pk)}{[k^2-2x(pk)+xp^2-1+i\varepsilon]^2} \qquad .
\]
Changing variables to $q=k-xp$,
\[
I_2(p)=\int_0^1 \, dx \,\int \frac{d^3 q}{(2\pi)^3}\cdot
\left[\frac{4i(pq)}{[q^2+\mu^2+i\varepsilon]^2}+\frac{4ixp^2}{[q^2+\mu^2+i\varepsilon]^2}\right]
\quad , \quad \mu^2=x(1-x)p^2-1 \qquad ,
\]
one sees that $I_2(p)$ is indeed convergent. Therefore,
\[
I(p;1)=I(1)+{\rm finite}\hspace{0.2cm}{\rm part} \qquad .
\]

\end{itemize}
Adding the four summands, the result used in subsection \S. 7.3 is
obtained: $-2(\kappa^2+1)I(1)+{\rm finite}\hspace{0.2cm}{\rm part}$.

\subsubsection{The Goldstone boson self-energy}

\begin{itemize}

\item The Feynman amplitude of the Goldstone boson self-energy
caused by one loop of Higgs particles is:

\parbox{3.5cm}{\epsfig{file=qsdv23.ps,width=2.5cm}}
$\displaystyle\equiv \hspace{1cm} -{1 \over 2}i\kappa^2 \cdot\int
\frac{d^3 k }{(2\pi)^3} \frac{i}{k^2-\kappa^2 + i
\varepsilon}=-{1\over 2}i\kappa^2\cdot I(\kappa^2)$

$\hspace{3.70cm}\equiv \hspace{1cm} -{1\over 2}i\kappa^2\cdot
I(1)+{\rm finite}\hspace{0.2cm}{\rm part}\qquad \qquad .$

\item The Feynman amplitude of Goldstone boson self-energy from one loop
of Goldstone particles is:

\parbox{3.5cm}{\epsfig{file=qsdv24.ps,width=2.5cm}}
$\displaystyle \equiv \hspace{1cm} -{3 \over 2}i \kappa^2\cdot
\int \frac{d^3 k }{(2\pi)^3}\frac{i}{k^2-1 +
i\varepsilon}=-{3\over 2}i\kappa^2\cdot I(1)\qquad .$

\item The Feynman amplitude of Goldstone boson self-energy from one loop of vector
boson particles is:

\parbox{3.5cm}{\epsfig{file=qsdv25.ps,width=2.5cm}}
$\displaystyle \equiv\hspace{1cm} {2  \over 2}i g^{\mu \nu}\cdot
\int \frac{d^3 k }{(2\pi)^3}\frac{- i g_{\nu \mu}}{k^2-1+ i
\varepsilon}=-3i\cdot I(1)\qquad .$

\item The Feynman amplitude of the two-vertex Goldstone boson self-energy graph
is:

\parbox{3.5cm}{\epsfig{file=qsdv26.ps,width=2.5cm}}
$\displaystyle \equiv\hspace{1cm}   \int \frac{d^3
k}{(2\pi)^3}\cdot \frac{ (p^\mu+k^\mu)\cdot -ig_{\mu\nu}\cdot
i\cdot (-k^\nu-p^\nu) }{[(p-k)^2-1+i \epsilon] [k^2 -\kappa^2 + i
\varepsilon]}$

\[
\hspace{3cm}\displaystyle\equiv \hspace{1cm} i  \int \frac{d^3
k}{(2\pi)^3} \frac{i (p+k)^2 }{[k^2-\kappa^2+i \varepsilon]
[(p-k)^2 -1 +i\varepsilon]}= i\cdot I(p;\kappa^2,1) \qquad .
\]

As in the fourth item of the previous subsubsection \S. 12.4.2, the
divergent integral is the sum of three integrals:
\begin{eqnarray*}
I(p ;1)=I(\kappa^2)&+&I_2(p;\kappa^2)+I_3(p
;\kappa^2,1)\\=I(\kappa^2)&+&\int \frac{d^3
k}{(2\pi)^3}\frac{4i(pk)}{[(p-k)^2-1+i \epsilon] [k^2 -\kappa^2 +
i \varepsilon]}\\&+&\int \frac{d^3
k}{(2\pi)^3}\frac{i(1-i\varepsilon)}{[(p-k)^2-1+i \varepsilon]
[k^2 -\kappa^2 + i \varepsilon]} \,\,\qquad  .
\end{eqnarray*}
Again $I_3(p ;\kappa^2,1)$ is convergent and routine Feynman
parametrization shows that $I_2(p ;\kappa^2)$ is also finite.
Therefore,
\[
I(p;\kappa^2,1)=I(1)+{\rm finite}\hspace{0.2cm}{\rm part} \qquad .
\]
\end{itemize}
Adding the four summands the result used in subsection \S. 7.3 is
obtained: $-2(\kappa^2+1)I(1)+{\rm finite}\hspace{0.2cm}{\rm
part}$.

\subsubsection{The vector boson self-energy}

\begin{itemize}

\item In the computation of the Feynman amplitude for the vector
boson self-energy graph with one Higgs loop there is a
combinatorial factor of ${1\over 2}$ and we include a factor of
three to take into account the three possible polarizations of the
ingoing and outgoing vector bosons:

\parbox{3.5cm}{\epsfig{file=qsdv27.ps,width=2.5cm}}
$\displaystyle \equiv\hspace{1cm} 3\cdot{2  \over 2}i  g^{\mu
\nu}\cdot \int \frac{d^3 k }{(2\pi)^3}\frac{i}{k^2-\kappa^2 +
i\varepsilon}=3ig^{\mu\nu}\cdot I(\kappa^2)$

$\hspace{3.70cm}\equiv\hspace{1cm} 3ig^{\mu\nu}\cdot I(1)+{\rm
finite}\hspace{0.2cm}{\rm part}\qquad \qquad .$

\item The Feynman amplitude for the vector boson self-energy graph
with one Goldstone loop is identical except that the Goldstone
propagator replaces the Higgs propagator:

\parbox{3.5cm}{\epsfig{file=qsdv28.ps,width=2.5cm}}
$\displaystyle  \equiv\hspace{1cm} 3\cdot{2 \over 2}i g^{\mu
\nu}\cdot \int \frac{d^3 k }{(2\pi)^3}\frac{i}{k^2-1 +
i\varepsilon}=3ig^{\mu\nu}\cdot I(1)$ \qquad .

\item The Feynman amplitude for the vector boson self-energy graph
with two vertices is more complicated:

\parbox{3.5cm}{\epsfig{file=qsdv29.ps,width=2.5cm}}
$\displaystyle  \equiv\hspace{1cm} -3i \cdot \int \frac{d^3
k}{(2\pi)^3} \cdot\frac{i (p^{\mu} - 2 k^{\mu} ) (p^{\nu} - 2
k^{\nu} ) }{[k^2-1+i \epsilon] [(p-k)^2 - \kappa^2 + i
\epsilon]}=-i\cdot I^{\mu\nu}(p)$ \qquad .

Again, there is a factor of three due to the three possible
polarizations of the vector bosons on the external legs but the
important fact is that the tensorial amplitude can be written as the
sum of three integrals:
\begin{eqnarray*}
 I^{\mu\nu}(p)&=& I_1^{\mu\nu}(p)+I_2^{\mu\nu}(p)+I_3^{\mu\nu}(p)\\&=&\int \frac{d^3 k }{(2\pi)^3}\cdot\frac{ 12 i k^{\mu}
k^{\nu} }{[k^2-1+i \epsilon] [(p-k)^2 - \kappa^2 + i \varepsilon]}
- \int \frac{d^3 k}{(2\pi)^3}\cdot \frac{ 6 i (p^{\mu}
k^{\nu}+k^{\mu} p^{\nu}) \,}{[(p-k)^2-\kappa^2+i
\varepsilon] [k^2 -1 + i \varepsilon]} \\
 &+& \int \frac{d^3 k}{(2\pi)^3}\cdot \frac{3 i p^{\mu} p^{\nu}
}{[(p-k)^2-\kappa^2+i \varepsilon] [k^2 -1 + i \varepsilon]}
\qquad .
\end{eqnarray*}

$I_1^{\mu\nu}(p)$ is in turn the sum of two integrals:
\begin{eqnarray*}
I_1^{\mu\nu}(p) &=& \int \frac{d^3 k }{(2\pi)^3}\cdot\frac{ 12 i
k^{\mu} k^{\nu}
}{[k^2-1+i \varepsilon] [(p-k)^2 - 1 + i \varepsilon]} \\
&+& \int \frac{d^3 k }{(2\pi)^3}\cdot\frac{ 12 i k^{\mu} k^{\nu}
(\kappa^2-1) }{[k^2-1+i \varepsilon] [(p-k)^2 -1 + i \varepsilon]
[(p-k)^2 - \kappa^2 + i \varepsilon]}\\ &=& \int \frac{d^3 k
}{(2\pi)^3}\cdot\frac{ 4 i k^2 g^{\mu\nu}
}{[k^2-1+i \varepsilon] [(p-k)^2 - 1 + i \varepsilon]} \\
&+& \int \frac{d^3 k }{(2\pi)^3}\cdot\frac{ 4 i k^2 g^{\mu\nu}
(\kappa^2-1) }{[k^2-1+i \varepsilon] [(p-k)^2 -1 + i \varepsilon]
[(p-k)^2 - \kappa^2 + i \varepsilon]} \qquad ,
\end{eqnarray*}
where the identity $k^{\mu} k^{\nu} = {1\over 3} k^2 g^{\mu \nu}$,
valid under the integral symbol, has been used. The second integral
is convergent but the first integral can be written as:
\begin{eqnarray*}
\int \frac{d^3 k }{(2\pi)^3}\cdot\frac{ 4 i k^2 g^{\mu\nu}
}{[k^2-1+i \varepsilon] [(p-k)^2 - 1 + i \varepsilon]}&=&\int
\frac{d^3 k }{(2\pi)^3}\cdot\frac{ 4 i  g^{\mu\nu} }{[k^2-1+i
\varepsilon] }\\&-&\int \frac{d^3 k }{(2\pi)^3}\cdot\frac{ 4 i
g^{\mu\nu}(p^2-2pk-1+i\varepsilon)}{[k^2-1+i \varepsilon][(p-k)^2
-1 + i \varepsilon]} \qquad .
\end{eqnarray*}
Therefore,
\[
I_1^{\mu\nu}(p)=4g^{\mu\nu}\cdot I(1)+{\rm
finite}\hspace{0.2cm}{\rm part}
\]
because the second integral above can be shown to be finite using
the Feyman parametrization as in previous subsections.

To study the $I_2^{\mu\nu}(p)$ integral we use Feynman
parametrization:

\begin{eqnarray*}
I_2^{\mu\nu}(p)&=&\int \frac{d^3 k}{(2\pi)^3}\cdot \frac{ -6 i
(p^{\mu} k^{\nu}+k^{\mu} p^{\nu}) \,}{[(p-k)^2-\kappa^2+i
\varepsilon] [k^2 -1 + i \varepsilon]} \\&=&\int_0^1 \, dx \, \int
\frac{d^3 k}{(2\pi)^3}\cdot \frac{ -6 i (p^{\mu} k^{\nu}+k^{\mu}
p^{\nu}) \,}{[(k-xp)^2+\mu^2+i \varepsilon]^2}  \qquad , \qquad
\mu^2=(xp^2-1)(1-x)-\kappa^2x \qquad .
\end{eqnarray*}
Therefore,
\[
I_2^{\mu\nu}(p)=\int_0^1 \, dx \, \int \frac{d^3 q}{(2\pi)^3}\cdot
\frac{ -6 i (p^{\mu} q^{\nu}+q^{\mu} p^{\nu}+2xp^\mu p^\nu)
\,}{[q^2+\mu^2+i \varepsilon]^2 }=\int_0^1 \, dx \, \int \frac{d^3
q}{(2\pi)^3}\cdot \frac{ -12 i xp^\mu p^\nu  \,}{[q^2+\mu^2+i
\varepsilon]^2 }
\]
is finite. Here, we have taken into account that integrals with an
odd integrand in $q^\mu$ are zero. $I_3^{\mu\nu}(p)$ is obviously
convergent.

\end{itemize}

Adding the three summands, the result used in subsection \S. 7.3 is
obtained: $2ig^{\mu\nu}\cdot I(1)+{\rm finite}\hspace{0.2cm}{\rm
part}$.
\section*{ACKNOWLEDGEMENTS}
All of us are grateful to the Physics Department of the Universidade
Federal de Paraiba (Joao Pessoa, Brazil) for giving us the
opportunity to gather and organize the material elaborated over the
past seven years in our research into quantum corrections to
topological soliton masses. JMG is specially indebted to Dionisio
Bazeia and Laercio Losano for inviting him to lecture in such a
stimulating scientific atmosphere, a miracle itself within the
tempting offer of beautiful tropical beaches, caipirinhas, and
Brazilian music. Finally, we also thank A. Rebhan, P. van
Nieuwenhuizen, and R. Wimmer for electronic mail exchange, helping
us to improve our understanding of very subtle issues in this rather
difficult subject.


\end{document}